\documentclass[fleqn,onecolumn,usenatbib]{mnras}

\usepackage{newtxtext,newtxmath}

\usepackage[T1]{fontenc}
\usepackage{comment}

\DeclareRobustCommand{\VAN}[3]{#2}
\let\VANthebibliography\thebibliography
\def\thebibliography{\DeclareRobustCommand{\VAN}[3]{##3}\VANthebibliography}

\usepackage{graphicx}	
\usepackage{amsmath}	
\usepackage{amssymb}	
\usepackage{multicol} 
\usepackage{bm}		
\usepackage{pdflscape}	
\usepackage{natbib}
\usepackage{adjustbox}
\usepackage{booktabs}

\usepackage{deluxetable}
\usepackage{longtable}
\usepackage[caption=false]{subfig}
\usepackage{hyperref}

\usepackage[left]{lineno}
\graphicspath{{./}{figures/}}
\usepackage{caption}
\usepackage[justification=centering]{caption}
\usepackage{float}
\usepackage{enumitem}

\title{The Gamma-ray Bursts fundamental plane correlation as a cosmological tool.}

\author[Dainotti et al.]{
M. G. Dainotti$^{1,2,3}$ \thanks{E-mail: maria.dainotti@nao.ac.jp },
A. Ł. Lenart $^{4}$,
A. Chraya,$^{5}$,
G. Sarracino$^{6,7}$,
S. Nagataki$^{8,9}$,
N. Fraija$^{10}$,
\newauthor
S. Capozziello$^{6,7,11}$,
M. Bogdan$^{12}$\\
$^{1}$National Astronomical Observatory of Japan, 2 Chome-21-1 Osawa, Mitaka, Tokyo 181-8588, Japan\\
$^{2}$ The Graduate University for Advanced Studies (SOKENDAI),
2-21-1 Osawa, Mitaka, Tokyo 181-8588, Japan \\
$^{3}$Space Science Institute, Boulder, Colorado\\
$^{4}$Astronomical Observatory, Jagiellonian University, ul. Orla 171, 31-501 Krak{\'o}w, Poland\\
$^{5}$Department of Physical Sciences, Indian Institute of Science Education and Research (IISER) Mohali, Sector 81, SAS Nagar, Punjab 140306, India\\
$^{6}$SDipartimento di Fisica, ``E. Pancini''
Universit\`{a} ``Federico II'' di Napoli, 
Compl. Univ. Monte S. Angelo Ed. G, Via Cinthia, I-80126
Napoli (Italy)\\
$^{7}$INFN Sez. di Napoli, Compl. Univ. 
Monte S. Angelo Ed. G, Via Cinthia, I-80126 Napoli (Italy)\\
$^{8}$Interdisciplinary Theoretical \& Mathematical Science Program, RIKEN (iTHEMS), 2-1 Hirosawa, Wako, Saitama, Japan 351-0198\\
$^{9}$ RIKEN Cluster for Pioneering Research, Astrophysical Big Bang Laboratory (ABBL), 2-1 Hirosawa, Wako, Saitama, Japan 351-0198\\
$^{10}$ Instituto de Astronom\'ia, Universidad Nacional Aut\'{o}noma de M\'{e}xico, Apdo. Postal 70-264, Cd. Universitaria, Ciudad de M\'{e}xico 04510\\
$^{11}$ Scuola Superiore Meridionale, Università di Napoli Federico II
Largo San Marcellino 10, 80138 Napoli (Italy)\\
$^{12}$ University of Wroclaw, plac Grunwaldzki 2/4, Wroclaw, Lower Silesia Province, 50-384, Poland.
}

\date{Accepted XXX. Received YYY; in original form ZZZ}

\pubyear{2021}
\begin{document}
\maketitle

\begin{abstract}

Cosmological models and their corresponding parameters are widely debated because of the current discrepancy between the results of the Hubble constant, $H_{0}$, obtained by SNe Ia, and the Planck data from the Cosmic Microwave Background Radiation. Thus, considering high redshift probes like Gamma-Ray Bursts (GRBs) is a necessary step. However, using GRB correlations between their physical features to infer cosmological parameters is difficult because GRB luminosities span several orders of magnitude. In our work, we use a 3-dimensional relation between the peak prompt luminosity, the rest-frame time at the end of the X-ray plateau, and its corresponding luminosity in X-rays: the so-called 3D Dainotti fundamental plane relation. We correct this relation by considering the selection and evolutionary effects with a reliable statistical method, obtaining a lower central value for the intrinsic scatter, $\sigma_{int}=0.18 \pm 0.07$ (47.1 \%) compared to previous results, when we adopt a particular set of GRBs with well-defined morphological features, called the platinum sample.
We have used the GRB fundamental plane relation alone with both Gaussian and uniform priors on cosmological parameters and in combination with SNe Ia and BAO measurements to infer cosmological parameters like $H_{0}$, the matter density in the universe ($\Omega_{M}$), and the dark energy parameter $w$ for a $w$CDM model. Our results are consistent with the parameters given by the $\Lambda$CDM model but with the advantage of using cosmological probes detected up to $z=5$, much larger than the one observed for the furthest SNe Ia.

\end{abstract}


\section{Introduction} \label{Introduction}

Gamma-Ray Bursts (GRBs) are incredibly powerful phenomena: they are the brightest objects after the Big Bang, as well as one of the farthest astrophysical objects ever detected \citep{Paczynski,Piran,Kumar}. These features allow us to use them as cosmological tools, similar to what has been achieved for Supernovae Type Ia (SNe Ia, \cite{Riess}). Because of their high luminosity, GRBs can be observed up to very large distances, corresponding to high redshifts. Indeed, GRBs have been observed up to $z=8.2$ and $z=9.4$ \citep{Tanvir2009,Cucchiara}, while SNe Ia has only been observed up to $z=2.26$ \citep{Rodney}. 
Using GRBs as cosmological tools requires a full understanding of their physical mechanisms. Both their energy emission mechanisms and progenitors are still being studied by the scientific community. For their birth, there is a general consensus on two different main scenarios: the explosion of a very massive star at the end of its lifetime \citep{Narayan1992, Woosley, MacFadyen, Nagataki2007,Nagataki2009, Nagataki2011}, followed by a core-collapse SNe \citep{Stanek,MacFadyen2001} or the coalescence of two compact objects, like black holes (BHs) or neutron stars (NSs) \citep{Lattimer, Eichler1989,Li1998, Rowlinson, Rea2015, Stratta}. The most probable frameworks for the central engine that powers the GRB consider the following astrophysical objects: BHs, NSs, or fast spinning newly born highly magnetized NSs magnetars, (\citet{Usov, Liang2018, Ai2018,Komissarov2007, Barkov2008}). 

To identify the different possible natures of their origin, it is necessary to classify GRBs according to their observable features. A general paradigm divides GRB light curves (LCs) into a rapid prompt energy emission followed by a longer emission phase called the afterglow. The afterglow is usually detected in X-ray, optical, and also radio wavelengths \citep{Sari,O'Brien, Sakamoto2007, Perley, Li2015, Morsony, Warren2017, Warren2018}. We usually detect the prompt emission of GRBs in high-energy bands, like from X-rays up to $\ge$ 100 MeV $\gamma$-rays, but sometimes they have been observed in the optical band as well \citep{Fraija18, Panaitescu2011, Fraija20b}.

A first categorization divides GRBs into Short and Long, depending on the duration of their prompt emission: $T_{90}\leq 2$ s or $T_{90} \ge 2$ s \footnote{$T_{90}$ is the time during which a GRB ejects from $5\%$ to $95\%$ of its total measured photons during the prompt phase}, respectively \citep{Mazets, Kouveliotou}. There is a very strong association between the prompt duration and the progenitor of GRBs: indeed, the majority of Long GRBs originate from the core collapse of a very massive star, while Short GRBs are born by the merging of two compact objects \citep{Abbott,Troja, Zhang2006, Ito2015, Ito2021}. A new classification of GRBs according to their progenitor mechanism has been proposed \citep{Zhang2006}: Type I GRBs are the ones born by the merging of two compact objects, while Type II are the ones born by the core collapse of very massive stars. Their progenitors can be inferred from morphological and physical characteristics.

The plateau phase of GRBs is a flat part of the GRB LC following the prompt phase, and it was discovered by the {\it Neil Gehrels Swift Observatory} ({\it Swift}) \citep{O'Brien, Sakamoto2007, W07}. The duration of this plateau usually ranges from $10^2$ to $10^5$ s, after which a power-law (PL) decay phase is observed.
Several scenarios describe the plateau, such as the external shock model, according to which the shock front between the ejecta of the emission and the interstellar medium is powered by a long-lasting energy emission from the central engine \citep{Zhang2006}, or due to the spin-down of a new-born magnetar \citep{Stratta,Fraija20}. 

In the past decades many efforts have been performed by the scientific community in order to find possible correlations between physical features of GRBs. Regarding correlations involving only the prompt features we cite, among the others, the relation between the peak in the ${\nu F_\nu}$ spectrum, ${E_{peak}}$ the isotropic energy in the prompt emission, ${E_{iso}}$ \citep{Amati2002}; and the one between $E_{peak}$ and the isotropic prompt luminosity ${L_{iso}}$ \citep{Yonetoku, Ito2019}. We also mention the correlations between the collimated-corrected energy ${E_{jet} = E_{iso}\times (1-cos \theta)}$ where ${\theta}$ is the jet opening angle and ${E_{peak}}$ found by \cite{Ghirlanda}; the one found by \cite{Liang2005} between ${E_{p}}$, ${E_{iso}}$, and the break time of the optical afterglow LCs, ${t_b}$. The last two correlations, even if they involve prompt features, introduce the jet-break time, which can also be inferred from the X-ray afterglow, which in some cases can include a plateau.

Several other correlations directly involving the plateau \citep{Dainotti2008, Dainotti2013b, Dainotti2015, Dainotti2016, Liang, Bernardini2012a, Xu2012, Margutti, Zaninoni,Shun-Kun2018, Tang2019, Zhao2019, Srinivasaragavan, Wen2020} and their applications as cosmological probes \citep{Cardone, Postnikov, Dainotti2013a, Izzo2015} have been presented. For a more extensive discussion on the prompt, prompt-afterglow relations, their selection biases and the application as cosmological tools see \citet{DaiDel2017, DaiAm2018, Dainotti2018}. One of these correlations 
is the so-called Dainotti relation, which links the time at the end of the plateau emission measured in the rest frame, $T^{*}_X$, with the corresponding X-Ray luminosity of the LC, $L_X$ \citep{Dainotti2008}, see Equation \ref{Lpeak}. This correlation is theoretically supported by the magnetar model \citep{Dall'Osso, Bernardini2012b,Rowlinson}. Its extension in three dimensions has been discovered by adding the prompt peak luminosity, $L_{\rm peak}$ \citep{Dainotti2016, Dainotti2017c} and is known as the fundamental plane correlation or the 3D Dainotti relation. \footnote{We note that we are referring to the fundamental plane correlation related to GRBs, and not to other definitions of fundamental planes used in astronomy, such as the fundamental plane of elliptical galaxies \citep{Djorgovski}}

To use this relation as a cosmological tool we selected a GRB sample with well-defined morphological properties and almost flat plateaus, called the platinum sample, which was introduced in \cite{Dainotti2020a}, and whose properties are detailed in Sec. \ref{sample selection}. We clarify that, following a well-established approach in the realm of the SNe Ia cosmology in which only the golden SNe Ia LCs are taken (see \cite{Scolnic}), we choose a well-defined sample. This is built in the observer frame and not in the rest-frame - namely, the LCs are in the fluxes versus time parameter space. This means that there is no involvement of cosmological parameters in this selection of the LCs. Therefore there is no circularity problem involved in the application of this sample for cosmological use.
We correct this correlation for evolutionary effects due to the redshift and selection biases, as done in \cite{Dainotti2020a}, to infer cosmological parameters, such as the Hubble constant ${H_{0}}$, the current mass density of the universe, ${\Omega_{ M}}$, and the dark energy parameter, ${w}$, for a ${w}$CDM model, together with other cosmological probes like the SNe Ia and the Baryon Acoustic Oscillations (BAO). Indeed, the evolution of the cosmological parameters is a vital topic and it has been discussed especially in relation to ${H_{0}}$. It has been highlighted even for the SNe Ia by \cite{Dainotti2021a} and \cite{dainotti2022d} that there is an evolutionary trend on $H_0$ as a function of the redshift, which can be possibly explained either with selection biases or with a new physics (i.e., invoking the so-called ${f(R)}$-gravity theory). For a general review on the Hubble constant tension see \cite{2022abdalla}.

In Sec. \ref{sample selection} the criteria used for the sample selection of the GRB data are detailed. In Sec. \ref{3D correlation} we show the GRB fundamental plane both with and without correcting for evolutionary effects and selection biases. In Sec. \ref{section4} we study the evolutionary parameters as a function of cosmology. In Sec. \ref{section5} we apply the fundamental plane as a cosmological tool. Our results are shown in Sec. \ref{Results}. In Sec. \ref{comparison} we present a comparison between the results obtained using GRBs alone versus the SNe Ia and SNe Ia + BAO sets. In Sec. \ref{standalone} we discuss the future use of GRBs as standalone probes. Finally, our conclusions are discussed in Sec. \ref{conclusions}.

\begin{figure} 
\centering
\subfloat[]{\label{fig1_a}
\includegraphics[width=0.48\hsize,height=0.3\textwidth,angle=0,clip]{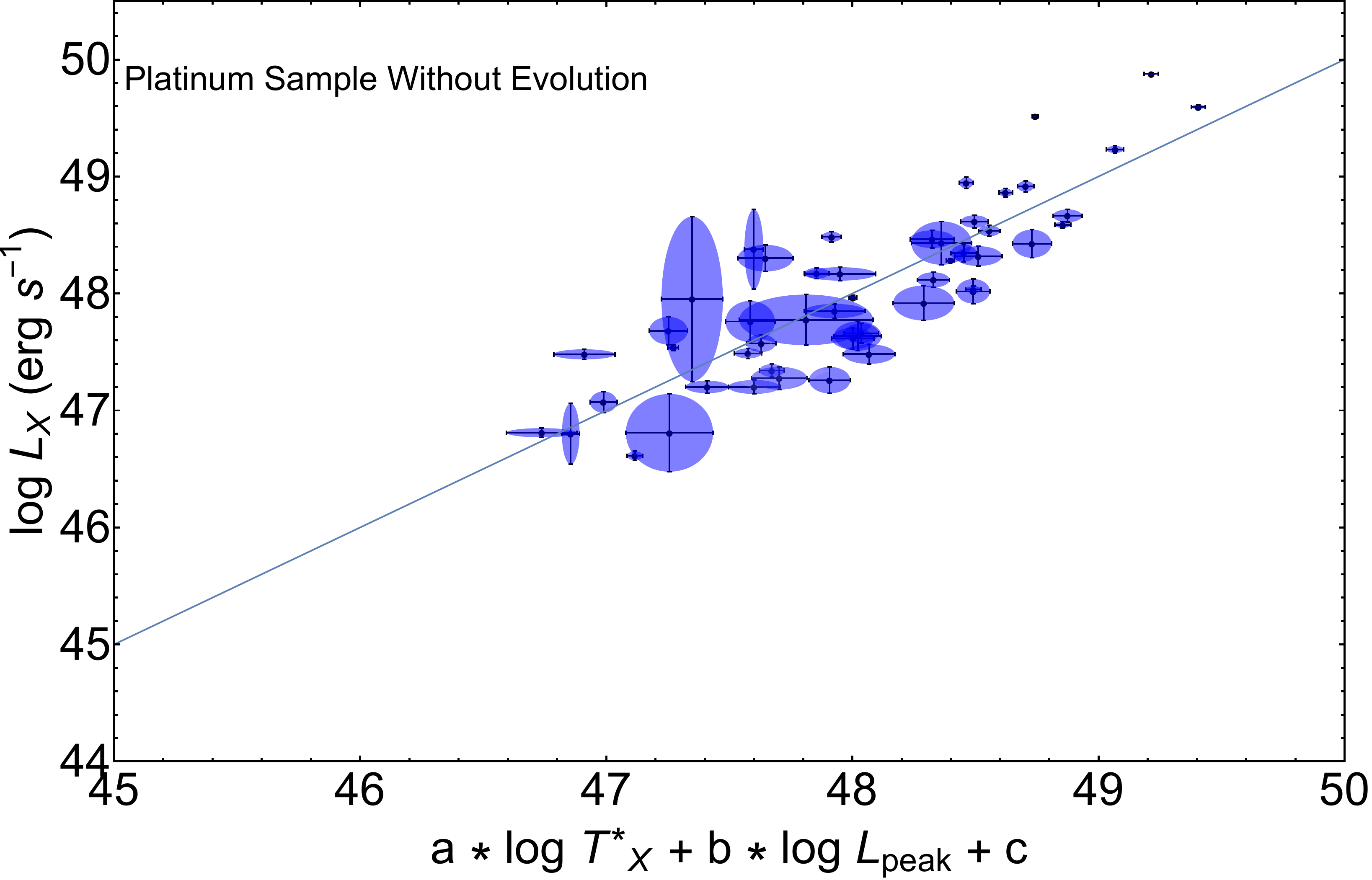}}
\subfloat[]{\label{fig1_b}
\includegraphics[width=0.48\hsize,height=0.3\textwidth,angle=0,clip]{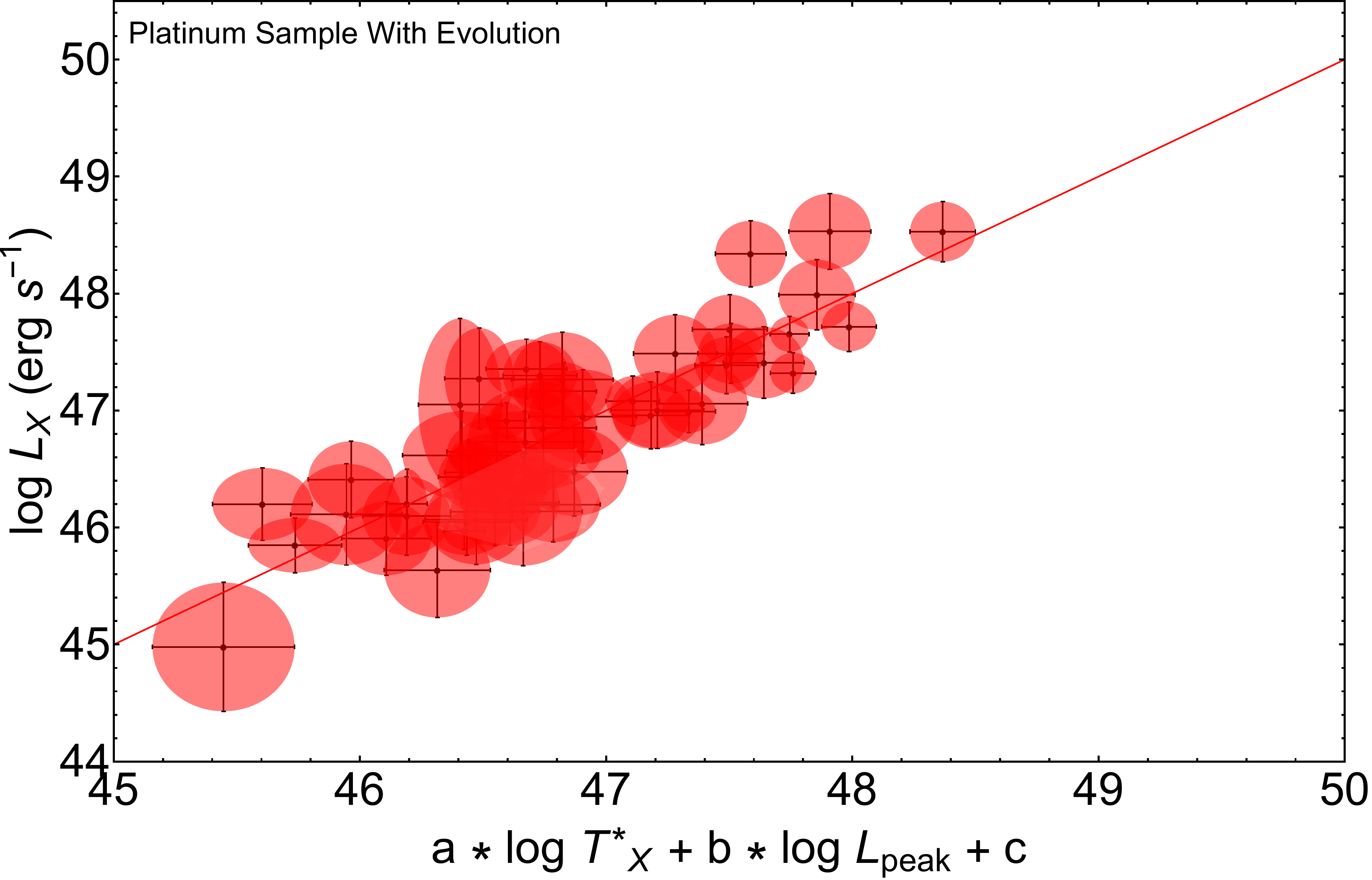}}\\\hspace{0cm}

\subfloat[]{\label{fig1_c}
\includegraphics[width=0.49\hsize,height=0.3\textwidth,angle=0,clip]{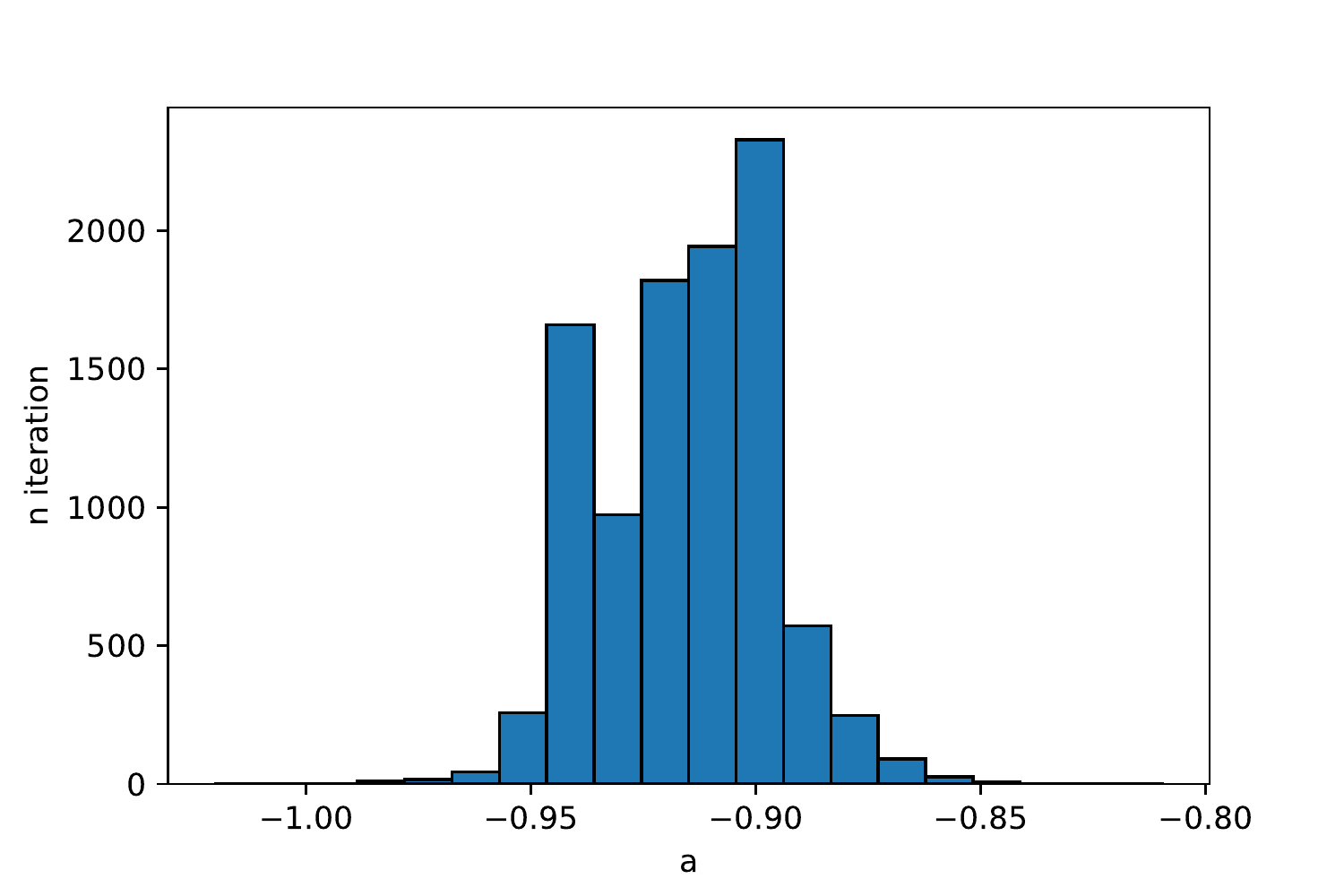}}
\subfloat[]{\label{fig1_d}
\includegraphics[width=0.49\hsize,height=0.3\textwidth,angle=0,clip]{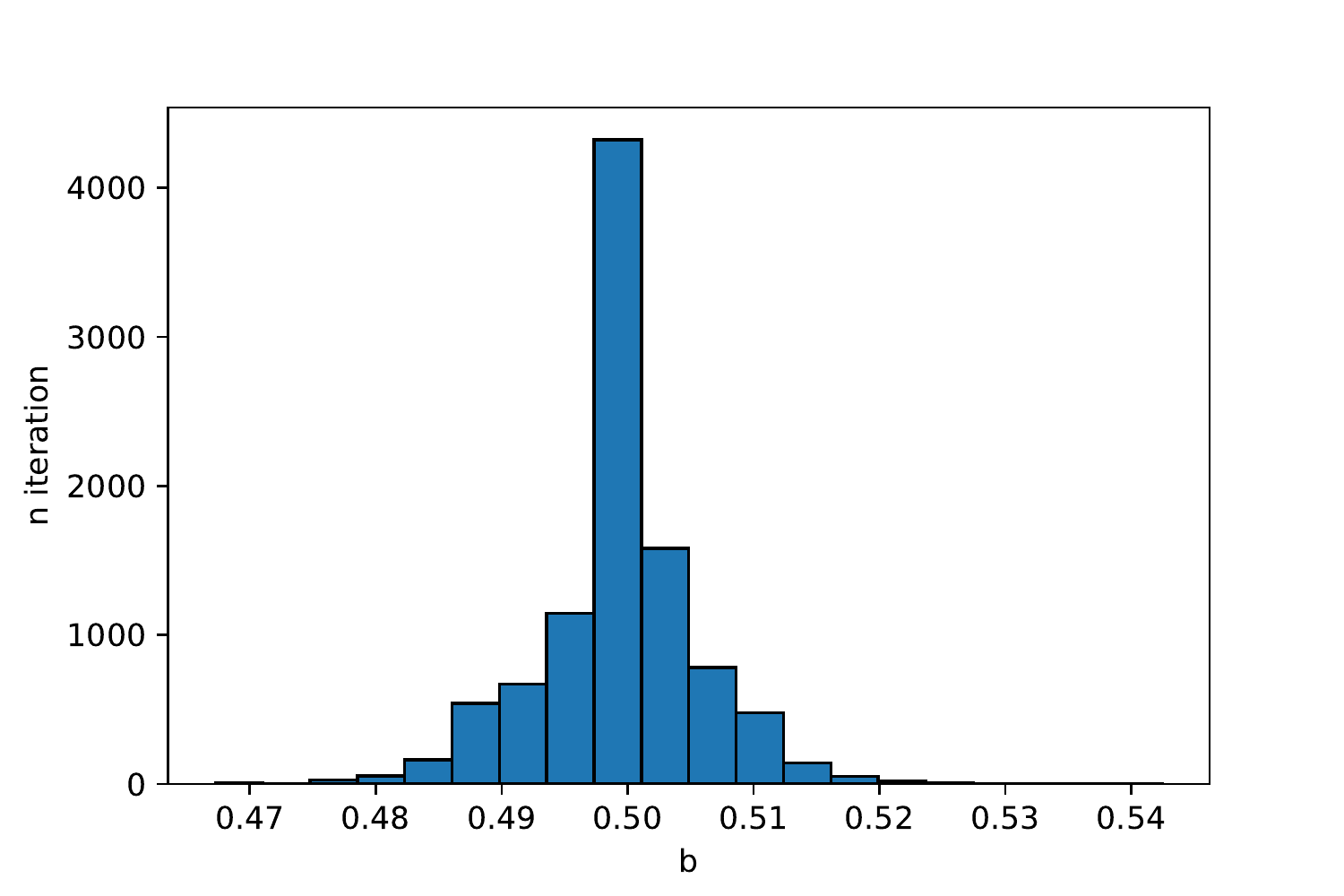}}\\\hspace{0cm}

\subfloat[]{\label{fig1_e}
\includegraphics[width=0.49\hsize,height=0.3\textwidth,angle=0,clip]{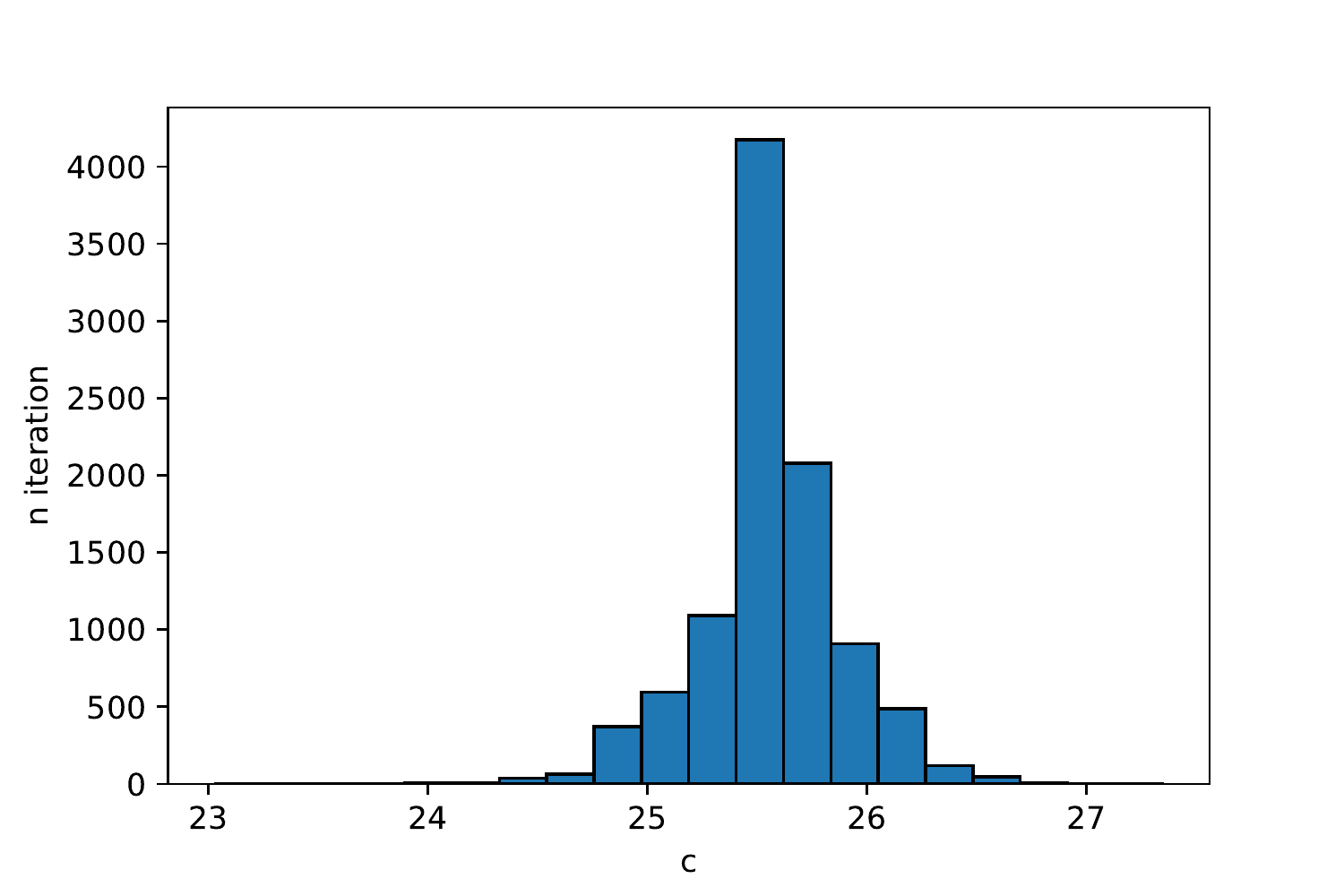}}
\subfloat[]{\label{fig1_f}
\includegraphics[width=0.49\hsize,height=0.3\textwidth,angle=0,clip]{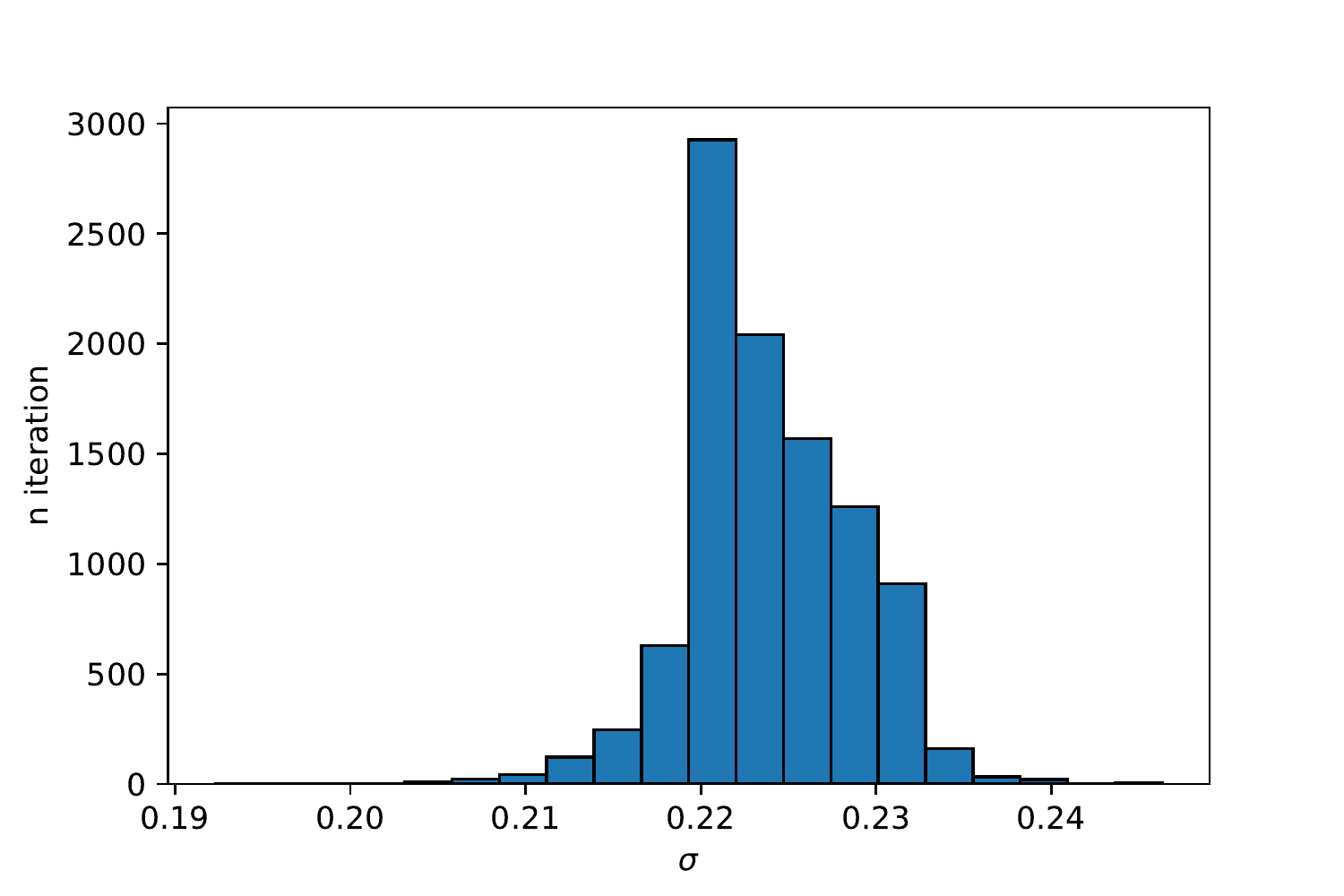}}

\caption{Top panels: the 2D projections of the fundamental plane related to the platinum sample without correcting for redshift evolution (\ref{fig1_a}), and with the corrections for selection and evolutionary effects (\ref{fig1_b}). Panels \ref{fig1_c}, \ref{fig1_d}, \ref{fig1_e} and \ref{fig1_f} : Histograms of the parameters $a, b, c$, and $ \sigma_{int}$ evaluated from the simulation of evolutionary coefficients taken from the 1 $\sigma$ range.}
\label{fig1}
\end{figure}

\section{Sample Selection}\label{sample selection}
We take into account the GRBs which can be described by the \citet{W07} phenomenological model using the BAT + XRT LCs gathered from the Swift web page repository \citep{Evans2009,Evans2010} \footnote{http://www.swift.ac.uk/burst\texttt{\_}analyser. We follow the criteria for the GRB sample selection considered in \citet{Srinivasaragavan} and \citet{Dainotti2020a}, and we use the platinum sample detailed in \citet{Dainotti2020a}. 
}. 
 We fit this sample with the Willingale functional form for ${f(t)}$, which reads as follows:
\begin{linenomath*}
\begin{equation}
f(t) = \begin{cases}
F_i \exp{\left( \alpha_i \left( 1 - \frac{t}{T_i} \right) \right)} \exp{\left(-\frac{t_i}{t} \right)} & {\rm for} \ \ t < T_i \\
F_i \left(\frac{t}{T_i} \right)^{-\alpha_i}\exp{\left( -\frac{t_i}{t} \right)} & {\rm for} \ \ t \ge T_i \, ,\newline
\end{cases}
\label{eq: fc}
\end{equation}
\end{linenomath*}
modelled for both the prompt (index `i=\textit{p}') ${\gamma}$\,-\,ray and initial hard X-ray decay and for the afterglow (`i=\textit{X}'), so that the complete LC ${f_{tot}(t) = f_p(t) + f_X(t)}$ contains two sets of four free parameters ${(T_{i},F_{i},\alpha_i,t_i)}$, where ${\alpha_{i}}$ is the temporal power law (PL) decay index and $T_i$ is the end time of the prompt and the plateau emission, respectively, while the time ${t_{i}}$ is the initial rise timescale. The transition from the exponential to PL occurs at the point ${(T_{i},F_{i}e^{-t_i/T_i})}$, where the two functions have the same value and this point marks the beginning of the plateau.
Using these criteria, we fit 222 LCs. The peak prompt luminosity at 1 second, $L_{peak}$, and the X-ray luminosity measured in the final part of the plateau phase, $L_X$, have been calculated as follows:
\begin{linenomath*}
\begin{equation}
L= 4 \pi D_L^2(z) \, F (E_{min},E_{max},T^{*}_{X}) \cdot K.
\label{Lpeak}
\end{equation}
\end{linenomath*}
To calculate $L_{peak}$ one substitutes the flux $F$ with $F_{peak}$ which is the $\gamma$-ray flux in $1$ s interval ($erg$ $cm^{-2} s^{-1}$) measured at the peak of the prompt emission, while to calculate $L_{\rm X}$, one uses the flux $F_{X}$, measured in X-rays at the end of the plateau; $D_L(z)$ is the luminosity distance computed for a particular redshift in the flat $\Lambda$CDM cosmological model, according to which we have an energy equation of state $w=-1$, $\Omega_M=0.3$, and $H_0=70$ $Km$ $s^{-1}$ $Mpc^{-1}$; $T^{*}_{X}$ is the time measured in the rest frame at the end of the plateau, and $K$ is the $K$-correction for the cosmic expansion \citep{Bloom}. For GRBs whose spectrum is fitted by a simple PL this correction is given by $K=(1+z)^{(\beta - 1)}$, where $\beta$ is the spectral index of the plateau in the X-ray band \citep{Evans2009, Evans2010}.

The Platinum Sample \citep{Dainotti2020a} is a subset of the Gold Sample, the latter being defined in \cite{Dainotti2016} and inspired by similar samples presented in the literature \citep{Xu2012,Tang2019}. To define the Gold Sample, we consider the following requirements for the plateau: 1) its beginning, defined by the quantity $T_t$, must have at least five data points; 2) its inclination must be $< 41^{\circ}$, this latter criterion is adopted in \cite{Dainotti2016} on the Gold Sample, where a Gaussian distribution fit the plateau angles, and the outliers are beyond the threshold of $ 41^{\circ} $.

To build the Platinum Sample, we also add the following requirements for the plateau:
3) its end time $T_{X}$ must not fall within observational gaps of the LCs to allow us the determination of this quantity directly from the data and not from the LC fitting; 4) it should last at least 500 $s$; and 5) should not present flares at its start or during the entire duration of the plateau itself (refining the idea of \cite{Xu2012}). 

Using these criteria, the platinum sample is composed of 50 GRBs out of the 222 plateaus analyzed. The furthest GRB in this sample is at $z=5$.

\begin{figure*} 
\includegraphics[width=0.49\hsize,height=0.335\textwidth,angle=0,clip]{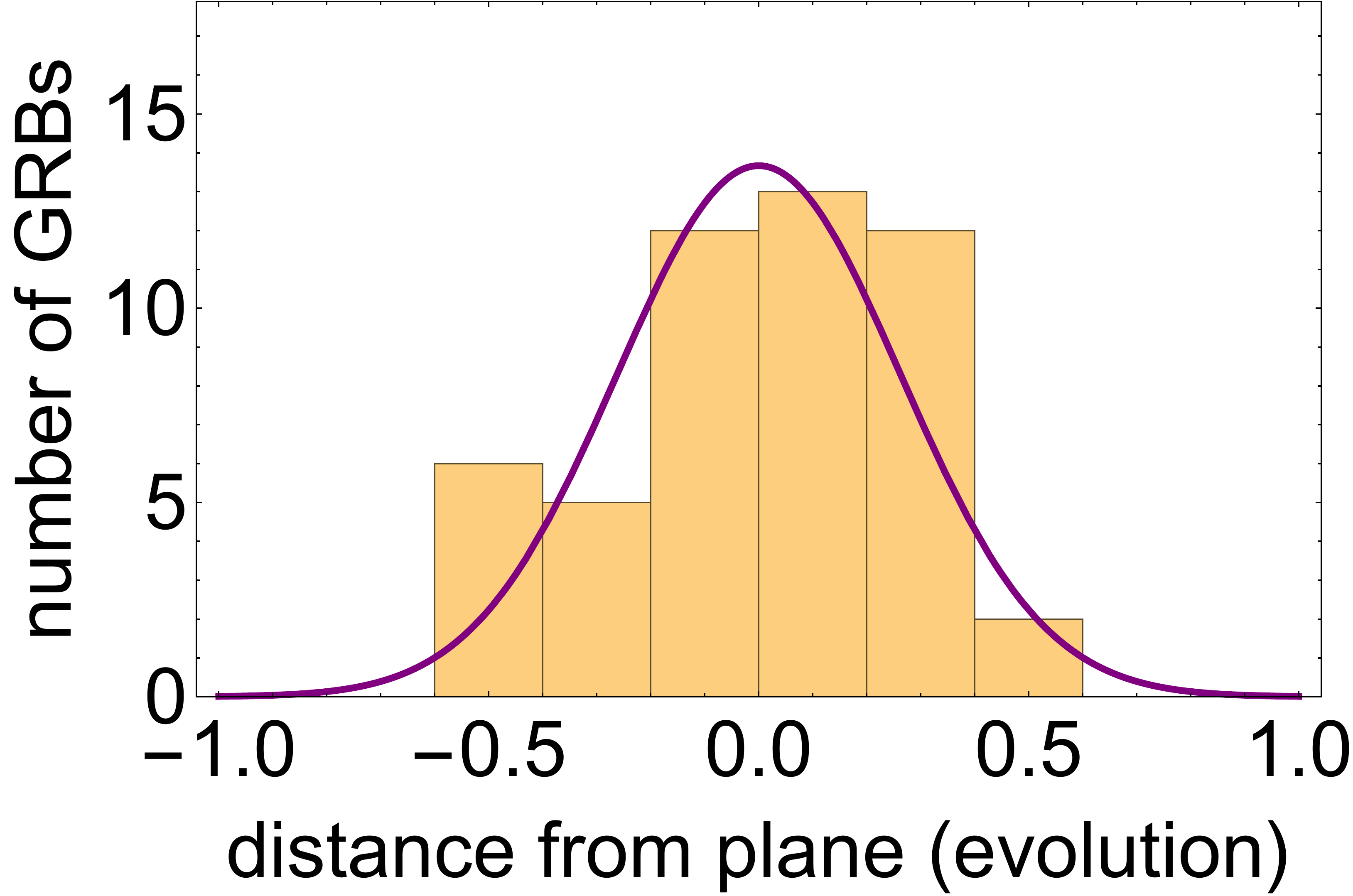}
\includegraphics[width=0.49\hsize,height=0.335\textwidth,angle=0,clip]{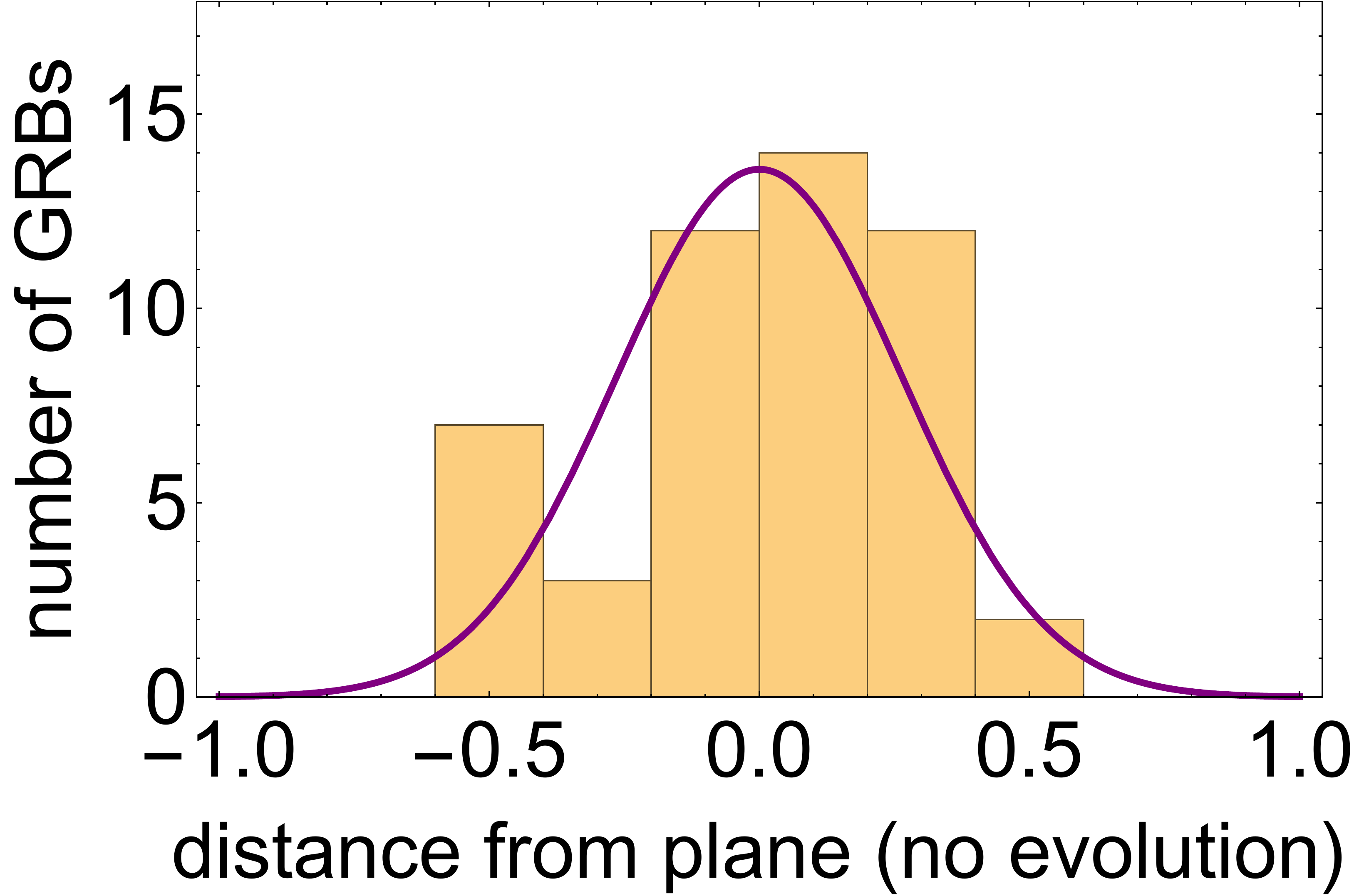}
\caption{The distributions of distances of the Platinum sample from the 3D fundamental plane with and without correction for evolution with their fitted Gaussian distributions.}
\label{histdistanceplane}
\end{figure*}
 
Regarding the SNe Ia data, we use the Pantheon sample 
 \citep{Scolnic}, a set composed of $1048$ SNe Ia collected by different surveys spanning from $z=0.01$ up to $z=2.26$.

{It is important to note that the criteria defining our sample are objectively determined before the construction of the correlations sought; sample cuts are introduced strictly following either data quality or physical class constraints.}

\section{The 3D Relation for The Platinum Gamma-ray Bursts}\label{3D correlation} 
In order to robustly apply any GRB correlation as a cosmological tool we need to have a reliable model supporting the theoretical scenario, like what has been done for SNe Ia. We also have to note that even if there is a very clear idea on the birth of the SNe Ia after the complete disruption of the accreted white dwarf in a binary system, there is still a debate on the particular mechanism that originates the SN explosion \citep{Livio}.
A possible example of a model that can satisfactorily explain the plateau, which we have pinpointed in Sec. \ref{Introduction}, is the magnetar model. Indeed, the intrinsic scatter, $\sigma_{int}$, of the correlation, and the errors on the parameters can be derived directly from the values of the periods of the spin and the magnitude of the magnetic fields representative of the magnetar. Thus, the slope and the intercept of the $L_X-T^{*}_{X}$ relation are naturally derived from the equation of the magnetar \citep{Rowlinson,Stratta}, which links $L_X$ to $T^{*}_{X}$ using physical quantities of the astrophysical object, such as the moment of inertia and the spin period through the following equations:
\begin{linenomath*}
\begin{equation}
L_{0,49} \propto B_{p,15}^2 P_{0,3}^{-4} R_{6}^6,
\label{magnetarlum}
\end{equation}
\end{linenomath*}
\begin{linenomath*}
\begin{equation}
T_{em,3} = 2.05 I_{45} B_{p,15}^{2} P_{0,-3}^{^2} R_6^{-6}.
\label{magnetartime}
\end{equation}
\end{linenomath*}
\noindent In Equations \ref{magnetarlum} and \ref{magnetartime}, ${L_{0,49}}$ is the plateau luminosity in ${10^{49} erg s^{-1}}$, ${I_{45}}$ is the moment of inertia in units of ${10^{45}}$ g ${cm^{2}}$, ${B_{p,15}}$ is the magnetic field strength at the poles in units of $10^{15}$ G, $R_6$ is the radius of the neutron star in ${10^6}$ cm and ${P_{0,-3}}$ is the spin period in milliseconds. If we substitute in Equation \ref{magnetarlum} the radii from Equation \ref{magnetartime} we obtain the following:
\begin{linenomath*}
\begin{equation}
\log L_0 \propto (\log (10^{52} I_{45}^{ -1} P_{0,-3}^{2}) - \log(T_{em})),
\end{equation}
\end{linenomath*}
where it is possible to see immediately that the first term is a constant for a given fixed period of the magnetar and momentum of inertia, and the luminosity is inversely correlated with the rest frame time at the end of the plateau emission.
However, there are additional explanations for the plateau emission and the existence of the $L_{peak}-L_a$ relation, which can be ascribed to the external forward shock model by changing the microphysical parameters \citep{Hascoet}. 
In addition, in \cite{Stratta} the properties of the Dainotti 3D relation are explained through the anti-correlation between $L_{peak}$ and the spin period within the model of the pulsar spin-down described in \citet{Contopoulos}.

Having stressed that this correlation and the 3D extension can be supported reliably within the magnetar scenario, we can safely proceed with the description of the procedure for using this correlation as a cosmological tool.
We leave the parameters $a$, $b$, $c$ of the fundamental plane free to vary and 
we fit the correlation using the \cite{Dago05} Bayesian method. In the paper the uncertainties on our computed values are always be quoted in 1 $\sigma$.
The luminosities and times carry error bars which are comparable, namely the $\frac{\Delta_{x}}{x}$ and $\frac{\Delta_{y}}{y}$ are of the same order of magnitude, where $\Delta_{x}$ is the error on the x-axis (error on time in our case) and $\Delta_{y}$ is the error on y axis (luminosity at the end of the plateau phase in our case). Thus, it is necessary to adopt methods which take into account both error bars like the \cite{Dago05}.
\noindent The fundamental plane relation has the following form:
\begin{linenomath*}
\begin{equation}
\log L_X = c + a\cdot \log T^{*}_{X} + b \cdot( \log L_{peak}),
\label{isotropic}
\end{equation} 
\end{linenomath*}
\noindent where $a$ and $b$ are the best-fit parameters given by the \cite{Dago05} procedure linked to $\log T^{*}_{X}$ and $\log L_{peak}$, respectively, while $c$ is the normalization. The best fit results are: $ a =-0.86 \pm 0.13$,
$b =0.56 \pm 0.12$, $c = 21.8 \pm 6.3$, and $\sigma_{int}=0.34 \pm 0.04$. 
We stress that if we had an ad hoc choice of the sample we would not have had any outliers from the distribution of the geometric distances of any point from the fundamental plane, while it is clear from the distribution of the distance from the platinum plane that we have several GRBs at distance -0.5 which are outliers from the distribution of the GRBs from the plane itself. 


The role of the corrections due to selection biases and evolutionary effects has been studied for the $L_X-T^{*}_{X}$ \citep{Dainotti2013b} and for the $L_X-L_{peak}$ relations \citep{dainotti2015b, Dainotti2017b}. 
Indeed, each physical feature, $L_X$, $T^{*}_{X}$ and $L_{peak}$, is affected by selection biases due to instrumental thresholds and redshift evolution of the variables involved in the correlations.
To correct these effects for each variable, we employ the \citet{Efron} method, which tests the statistical dependence among $L_X$, $T^{*}_{X}$ and $L_{peak}$, see \citet{Dainotti2013b, dainotti2015b, Dainotti2017b,Petrosian}.

\begin{figure*} 
\includegraphics[width=0.99\hsize,height=0.55\textwidth,angle=0,clip]{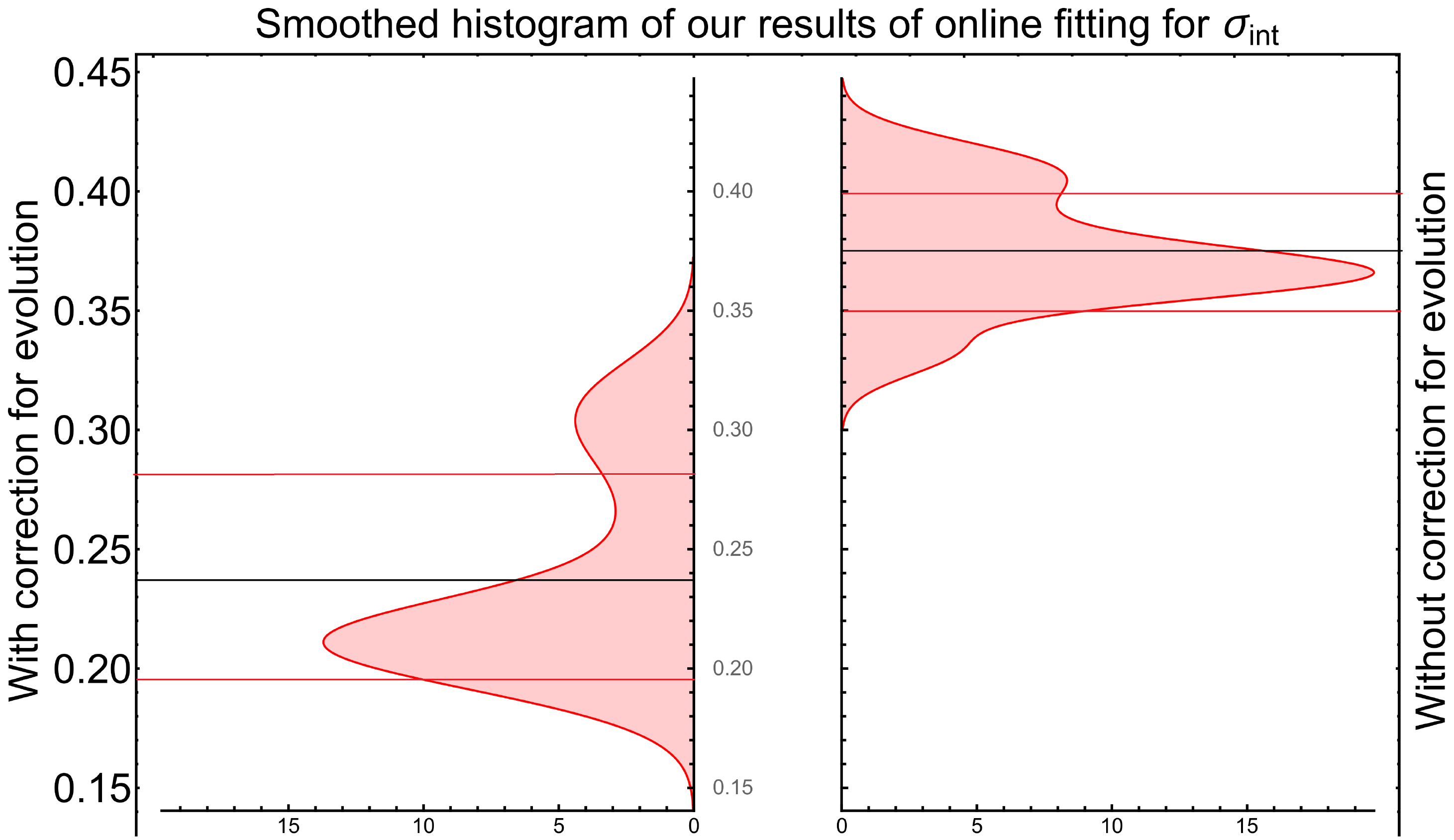}
\caption{Paired smoothed histograms of the $\sigma_{int} $ obtained for cases with and without evolution with different methods using the HyperFit online routine. The thin black horizontal lines indicate the central value of the $\sigma_{int}$ parameter from the D'Agostini fitting, while the red ones correspond to the 1 $\sigma$ error bars.} 
\label{hist:paisim}
\end{figure*}

The fundamental plane correlation, once the selection effects are considered, becomes:
\begin{linenomath*}
\begin{equation}
\log L_X-k_{L_{a}} \log(z+1) = a_{ev} \cdot(\log T^{*}_{X}- k_{T_{X}} \log (z+1))+
b_{ev} \cdot(\log L_{peak}- k_{L_{peak}} \log (z+1))+c_{ev},
\label{planeev}
\end{equation}
\end{linenomath*}
where $a_{\rm ev}$, $b_{\rm ev}$, and $c_{\rm ev}$ denote the parameters with redshift evolution. With evolution we define the dependence of the parameters on the redshift. The $k_{L_{peak}}$, $k_{T_{X}}$, and $k_{L_{a}}$ are the evolutionary coefficients computed by us for the whole sample of 222 GRBs: $k_{L_{peak}}=2.24^{+0.30}_{-0.30}$, $k_{T_{X}}=-1.25^{+0.28}_{-0.27}$, $k_{L_{a}}=2.42^{+0.41}_{-0.74}$. Comparing these results with the ones obtained in the literature, we note that the evolution on $L_{peak}$ is compatible within 1 $\sigma$, the evolution on $T^{*}_a$ is compatible within 1.6 $\sigma$, while the $L_a$ evolution is compatible within 1.8 $\sigma$ with the ones taken from \cite{Dainotti2017b} and used in \cite{Dainotti2020a}. 

To verify the reliability of our results, we simulated random values of the power-law coefficients of the evolution, $k_{L_{peak}}, k_{T_{X}}, k_{L_{a}}$ drawn from uniform distributions within the 1 $\sigma$ error range.
For our best-fit computations of the Dainotti 3D relation we used the \cite{Dago05} and \cite{Reichart} methods (the latter considers a slightly different likelihood, but still includes $\sigma_{int}$, like the D'Agostini one) with the same minimization algorithm. To further ensure the reliability of our results, we repeated this procedure 1000 times finding that the results from these two computations are compatible within 1 $\sigma$. The results of minimization of the D'Agostini likelihood are shown in the central and bottom panels of Fig. \ref{fig1}. 
Considering these effects, the new best-fit parameters for the fundamental plane of the platinum sample are $a_{\rm ev}=-0.85 \pm 0.12$, $b_{\rm ev}=0.49 \pm 0.13$, $c_{\rm ev} = 25.4 \pm 6.9$, and $\sigma_{int}=0.18 \pm 0.09$. We note that \cite{Rowlinson} predicts that the $a$ coefficient of the $\log L_X-\alpha \log(z+1) \propto \log T^{*}_{X}- \beta \log (z+1)$ relation should be $a_{theoretical}=-1$, which is compatible with our results within $1.3\; \sigma$ error range. The central value of the intrinsic scatter is $47.1 \%$ smaller than the one computed for the original fundamental plane. 

We compare the two intrinsic scatters obtained with and without considering the corrections due to the EP method using the following formula, adapted from \cite{Dainotti2020a}: 
\begin{linenomath*}
\begin{equation}
 x=\frac{\sigma_{int,NoEV}-\sigma_{int,EV}}{\sqrt{\sigma_{\sigma_{int,NoEV}}^2+\sigma_{\sigma_{int,EV}}^2}}.
 \label{discrepancies}
\end{equation}
\end{linenomath*}
We obtain $x=1.98$, meaning that the evolutionary effects do indeed reduce the intrinsic scatter on the platinum fundamental plane correlation in a significant way. We also show in Fig. \ref{histdistanceplane} the distances of each data point belonging to the Platinum sample with respect to the best fit of the fundamental plane, both with the evolutionary effects (left panel) and without them (right panel).

The 2D projections of both fundamental planes are shown in the top panels of Fig. \ref{fig1}. These figures allow us to show how the error bars for each GRB shown with the ellipses are placed around the plane when we correct for selection biases and redshift evolution (right upper panel) and when we do not correct for them (left upper panel). We note that when the correction for evolution is applied, the data points are closer to the plane including the errorbars, and fewer outliers are present compared to the situation in which the evolution is not taken into account. To compare these two relations we computed for both cases the Akaike information criterion (AIC) and the model weight: $B_{i}=e^{\frac{AIC_{min}-AIC_{i}}{2}}$ for each relation, where $AIC_{min}=MIN(AIC_{evolution},AIC_{noev})$, and $AIC_i$ is the AIC value corresponding to the relation for which the $B$ parameter is computed. 
For each model we computed the ``relative likelihood": $P_i=\dfrac{B_i}{\sum \limits _j B_j}$, obtaining $P_{evolution}=0.99$ and $P_{noev}=0.01$. Thus, the model with evolution is favoured compared to the one without evolution. 

To further confirm the reliability of our results and their independence for the particular Bayesian method adopted we performed other best fit procedures for the fundamental plane, both with and without evolution. We also used the \cite{Reichart} method, which we recall is another Bayesian approach which takes also into account of both error bars, obtaining best fit results compatible within 1 $\sigma$ to the D'Agostini ones, and an online fitting procedure called “HyperFit" (https://hyperfit.icrar.org/) based on \cite{Robotham}, built to obtain the best-fit of linear models that consider heteroscedastic errors for multidimensional data using Bayesian inference. The latter tool offers the possibility to employ different algorithms and methods, which we used to compute a smoothed paired histogram of the $\sigma_{int}$ obtained for each case, with and without evolution, presented in Fig. \ref{hist:paisim}. 
The values obtained with the \cite{Dago05} method are consistent with these histograms. Specifically, we add a black line indicating the mean value of ${\sigma_{int}}$ and red lines indicating the error on ${\sigma_{int}}$ obtained by the \cite{Dago05} method.
We also applied other best fit methods: the Principal Component Analysis, PCA, the PC Regression (PCR, \cite{Liu}), and the Partial Least Squares (PLS), where the latter two are regression methods based on PCA. For PCA we found: $a=-1.19$, $b=0.44$, $c=28.87$ and $a_{\rm ev}=-1.17$, $b_{\rm ev}=0.49$, $c_{\rm ev}=26.75$ for the no evolution and the evolution cases, respectively.
When comparing the PCA results with the \cite{Dago05} ones for the non-evolution case the parameters $a$, $b$, and $c$ are within $2.5$, $1$, and $1.1$ $\sigma$, respectively; for the evolution case $b_{ev}$ and $c_{ev}$ are consistent in $1\; \sigma$, while $a_{ev}$ is consistent in $2.7\; \sigma$. The PCA fitting does not account for the error bars, thus does not consider the intrinsic scatter that is instead computed by the Bayesian methods of \cite{Dago05} and \cite{Reichart}. This drives the difference in the results.
For PCR and PLS we used the bootstrapping technique to infer the errors on the best fit parameters. These methods are consistent with the D'Agostini ones in 1 $\sigma$, thus giving more reliability to our conclusions. 


\section{The Study of the Evolutionary Parameters as a Function of Cosmology}
\label{section4}

The reliability of this procedure has already been proven via Monte Carlo simulations \citep{Dainotti2013b}.
To correct for the evolution we use $g(z)=1/(1+z)^{k}$, where the $k$ parameter mimics the evolution due to the redshift.
As addressed in \cite{dainotti2015b}, the functional form for the evolution can be a power law or a more complex function, and the results for these functions are compatible within 2 $\sigma$ for the luminosities and 1 $\sigma$ for the time evolutions. Here, we detail our results of the EP method for the whole sample of 222 GRBs for the studied parameters. 
The EP method takes into account these effects by using an adaptation of the Kendall $\tau$ test, according to which $\tau$ has the following definition:

\begin{figure}
 \centering
 \includegraphics[width=0.45\hsize,height=0.3\textwidth,angle=0,clip]{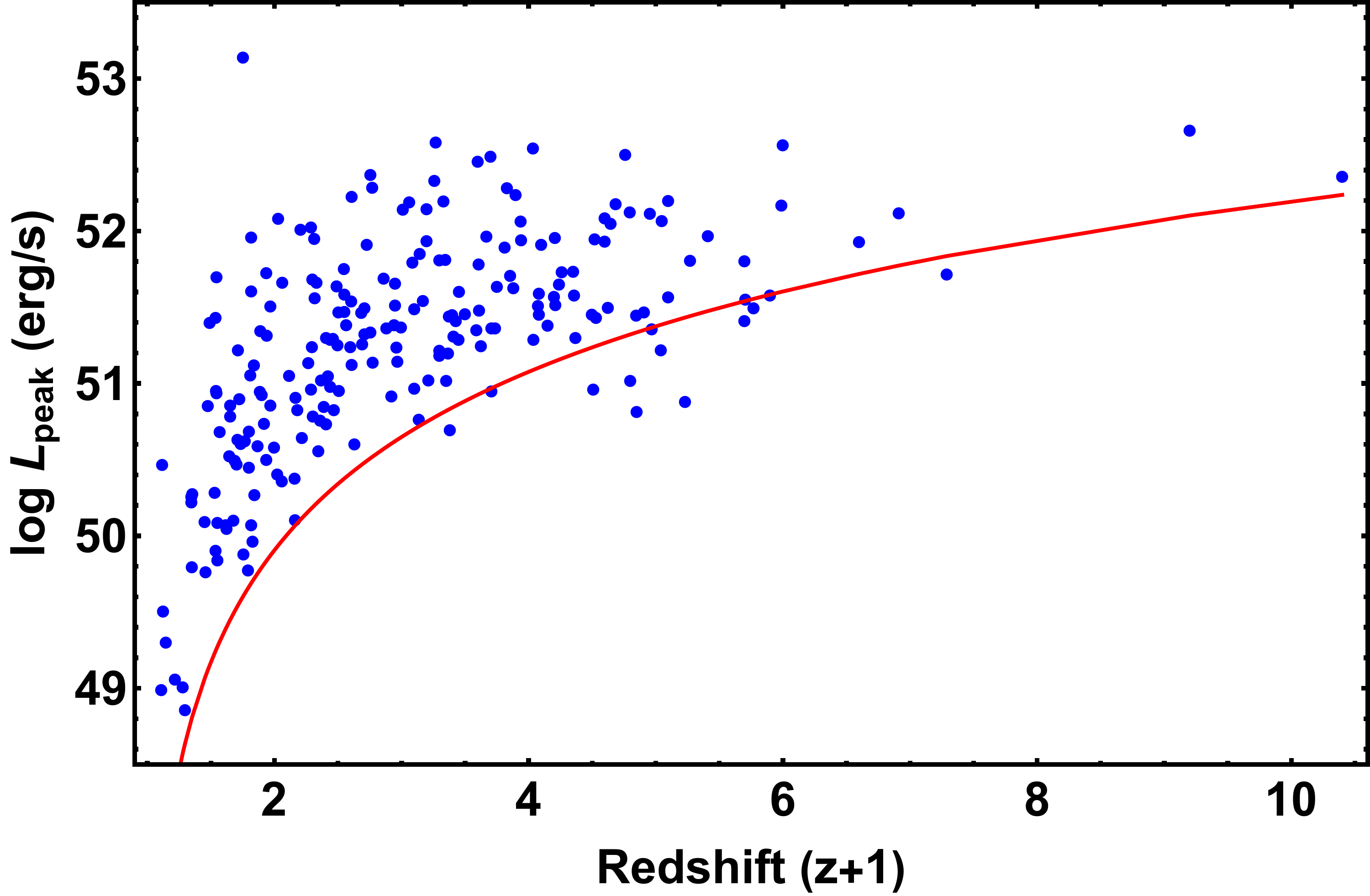}
 \includegraphics[width=0.45\hsize,height=0.3\textwidth,angle=0,clip]{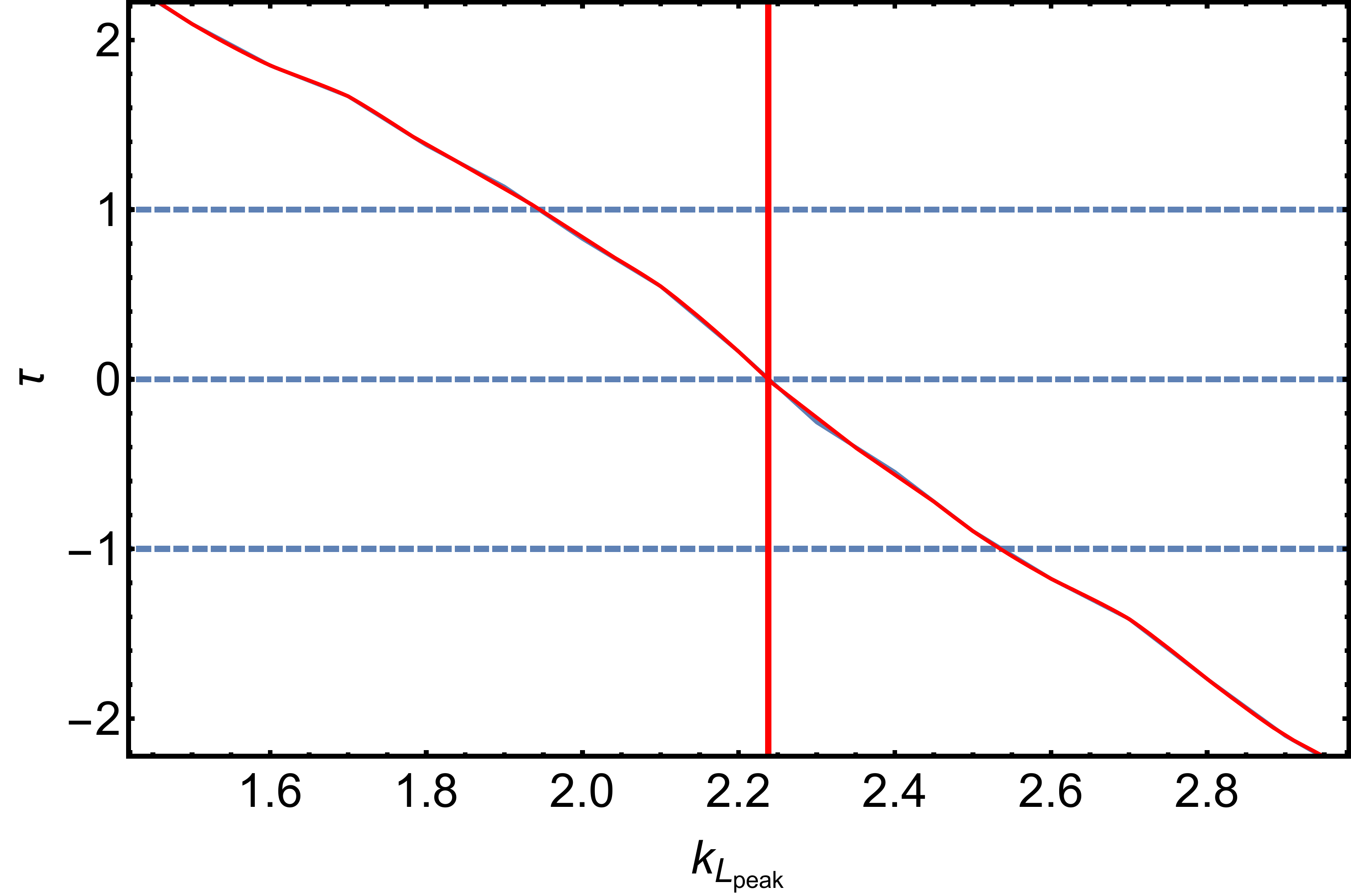}
 \includegraphics[width=0.45\hsize,height=0.3\textwidth,angle=0,clip]{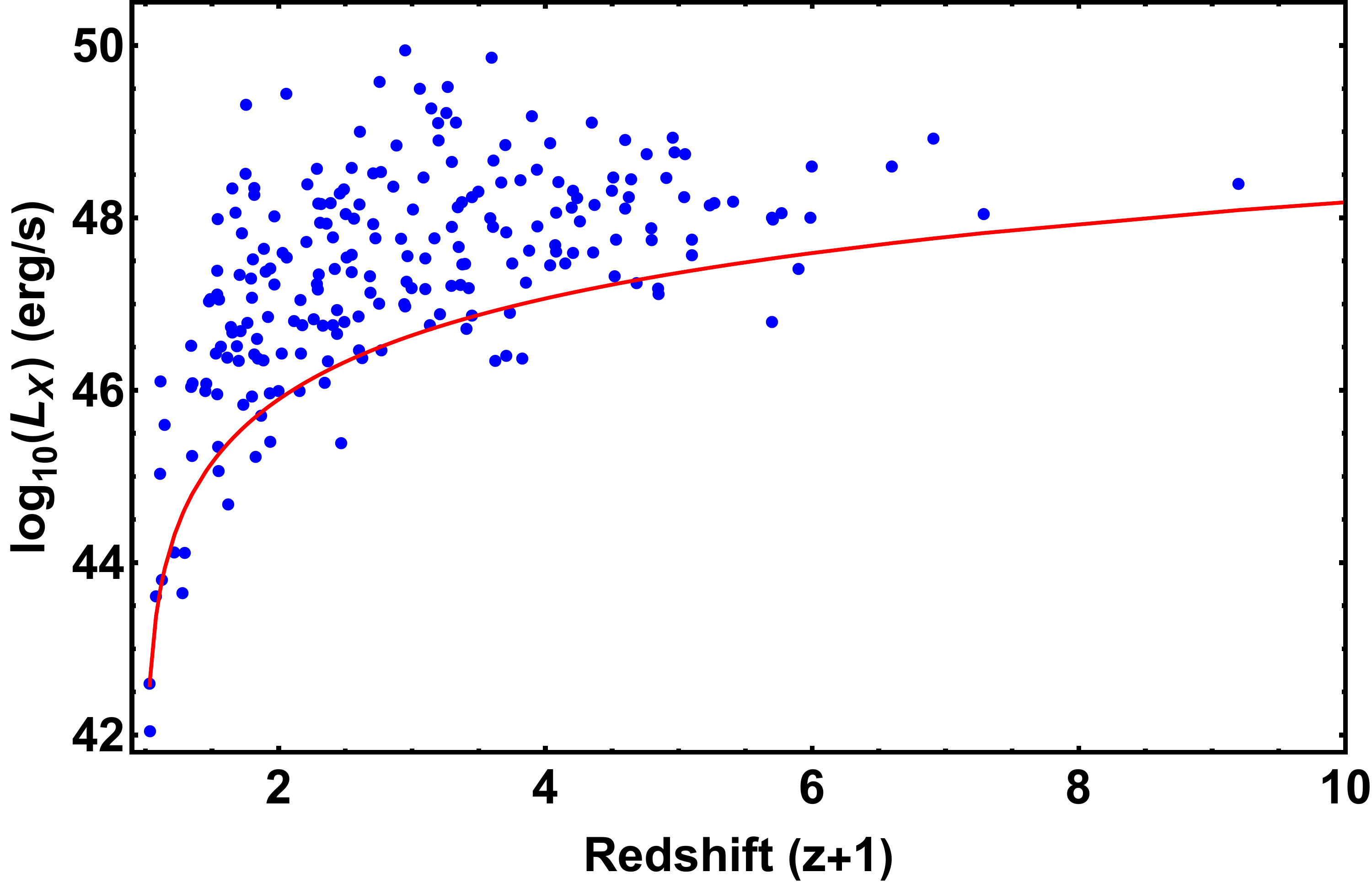}
 \includegraphics[width=0.45\hsize,height=0.3\textwidth,angle=0,clip]{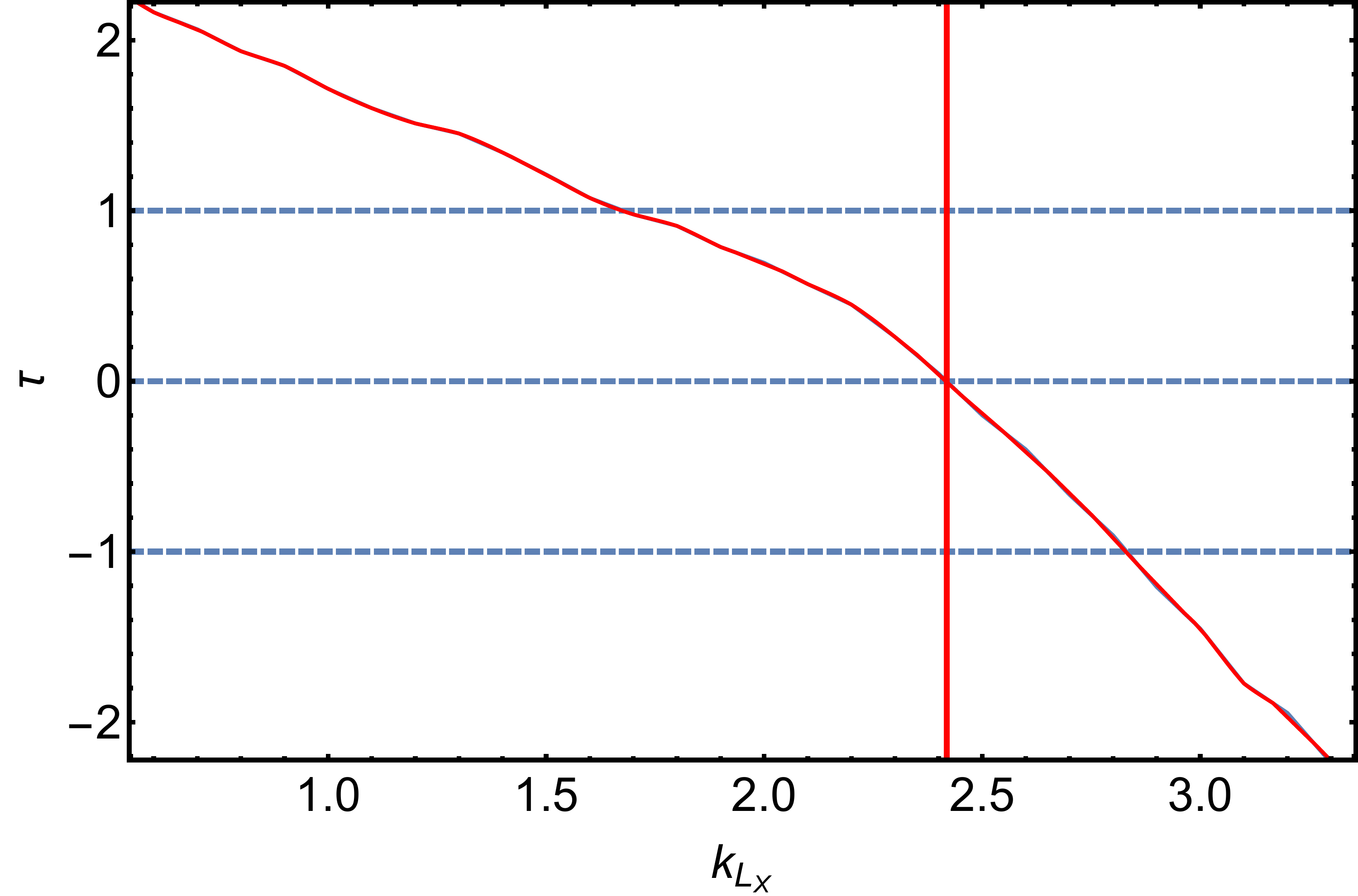}
 \includegraphics[width=0.45\hsize,height=0.3\textwidth,angle=0,clip]{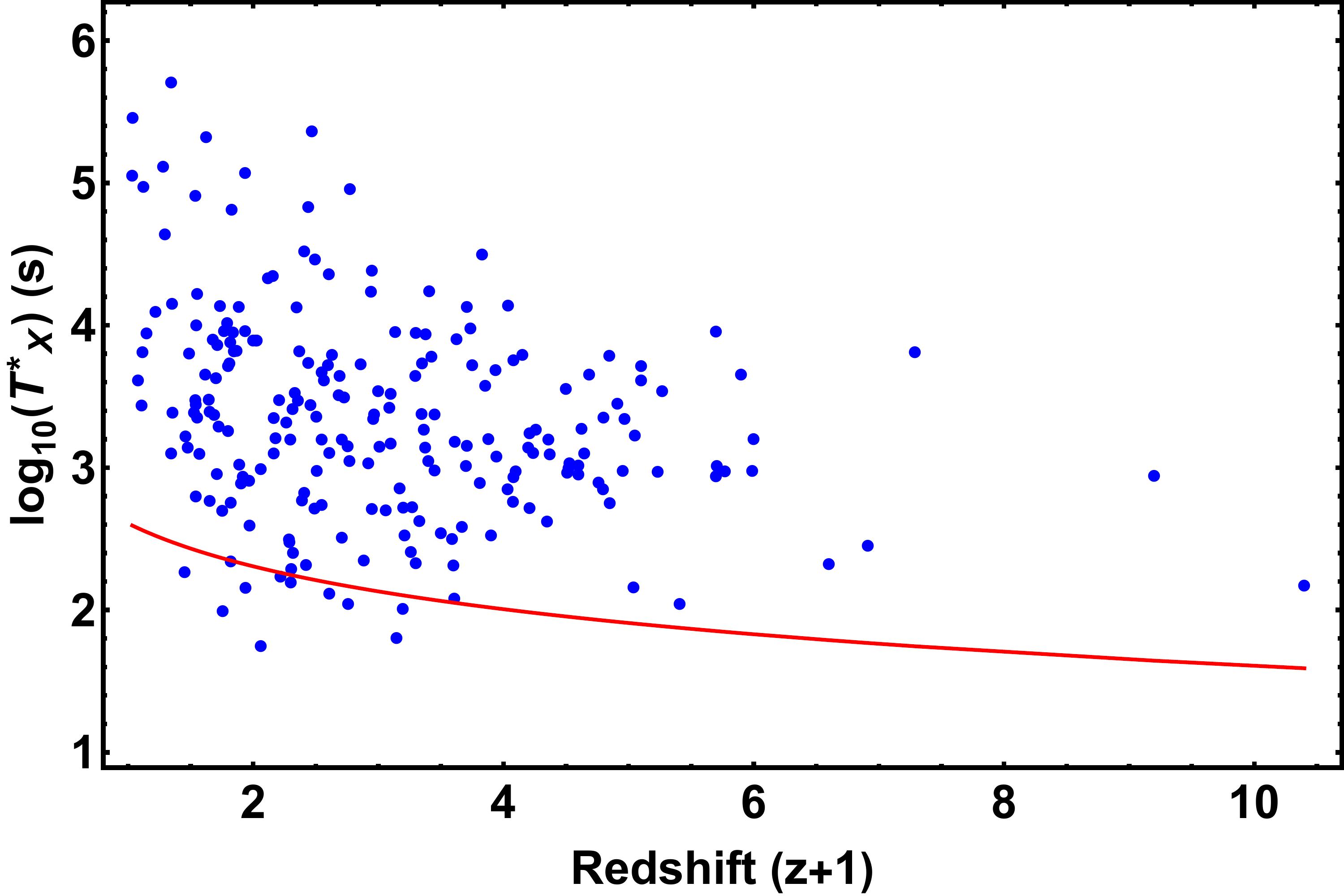}
 \includegraphics[width=0.45\hsize,height=0.3\textwidth,angle=0,clip]{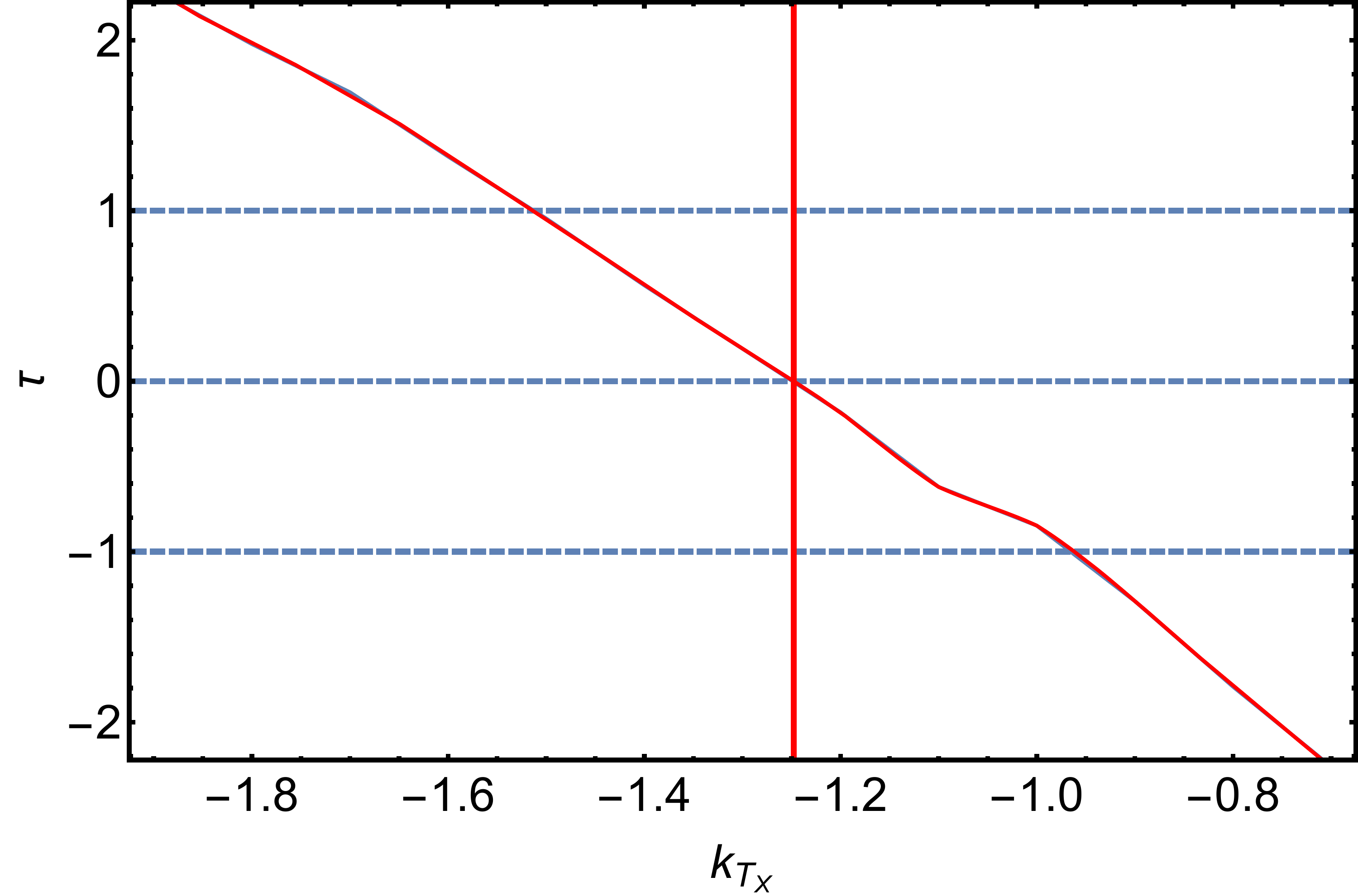}
 \caption{The application of the EP method to our entire sample for the parameters involved in the fundamental plane correlation. The limiting lines chosen for the EP method are visible in red. The left panels show the distribution of studied parameters versus $redshift+1$, while the right panels show the relation between $\tau$ and the evolutionary coefficients in red. The vertical red solid lines indicate the value for which $\tau=0$ and thus the evolution is removed. The dashed blue lines represent the 1 $\sigma$ for the evolution, which is determined for $\tau \leq 1$.}
 \label{fig:evoL}
\end{figure}

\begin{linenomath*}
\begin{equation}
\tau =\frac{\sum_{i}{(\mathcal{R}_i-\mathcal{E}_i)}}{\sqrt{\sum_i{\mathcal{V}_i}}},
\label{tau}
\end{equation}
\end{linenomath*}

\noindent where $\mathcal{E}_i=(1/2)(i+1)$ is the expectation value, $R_i$ is the rank, and $\mathcal{V}_i=(1/12)(i^{2}+1)$ is the variance. To eliminate the impact of the redshift on our data we demand $\tau=0$. $R_i$ is computed for each data point considering the position of the data in the so-called associated sets, which are samples that include all the objects that can be detected considering a particular observational limit \citep{Dainotti2013b, dainotti2015b, Dainotti2017c}. The computations to derive the evolutionary coefficients follow the same procedure for $L_{peak}$, $L_{X}$, and $T^*_X$. The limiting values for these quantities are shown in the left panels of Fig. \ref{fig:evoL}, while the evolutionary coefficients are shown in the right panels. Here, for simplicity we detail only the computation for $L_X$, given that for the other parameters is similar. For this luminosity we compute the flux limit at the end of the plateau phase, $f_{lim}$, and then we compute the correspondent luminosity $L_{min}(z_i)$, that would allow us to detect that object at a given $z_i$. The associated set for $z_i$ contains all GRBs with $L_{min} \leq L(z_j)$ and $z_j \leq z_i$, where with $j$ and $i$ we denote the objects of the associated set and of the GRB sample, respectively. According to the EP method, the samples used to derive the evolutionary effects should not be less than $90 \%$ of the original ones, so conservative choices regarding the limiting values are needed. The chosen coefficients for the evolutionary functions ($g(z)$, $f(z)$, and $h(z)$ for the $L_{peak}$, $L_X$ and $T^*_X$, respectively) are the ones for which $\tau=0$ as shown with the red vertical lines in Fig. \ref{fig:evoL}, while the dashed blue lines correspond to the 1 $\sigma$ for the evolutionary coefficients, which is determined for $\tau\leq 1$. We have used this method on our sample of 222 GRBs to compute new evolutionary parameters, thus updating the values with respect to the ones used in \cite{Dainotti2020a}.
We also would like to point out that \cite{Dainotti2021b} has shown that this method is reliable regardless of the choice of the limiting values for several sample sizes for Short GRBs (samples of 56, 32 and 34 GRBs). Thus, the discussion of \cite{2021MNRAS.504.4192B} on the EP method and its applicability are not a concern given the approach and the reliability of the results in \cite{Dainotti2021b}. 

\begin{figure}
 \centering
 \includegraphics[width=0.49\hsize,height=0.34\textwidth,angle=0,clip]{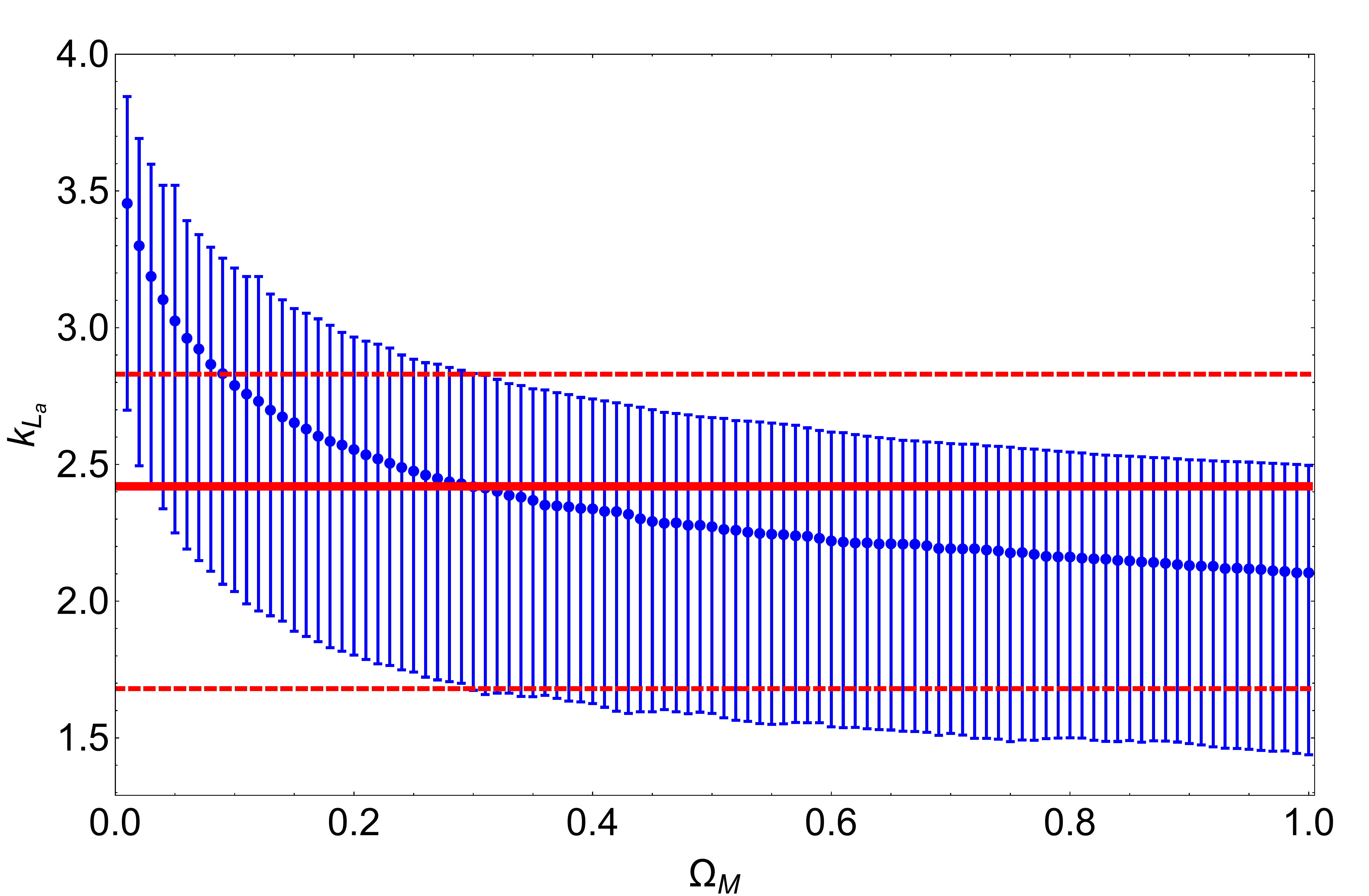}
 \includegraphics[width=0.49\hsize,height=0.34\textwidth,angle=0,clip]{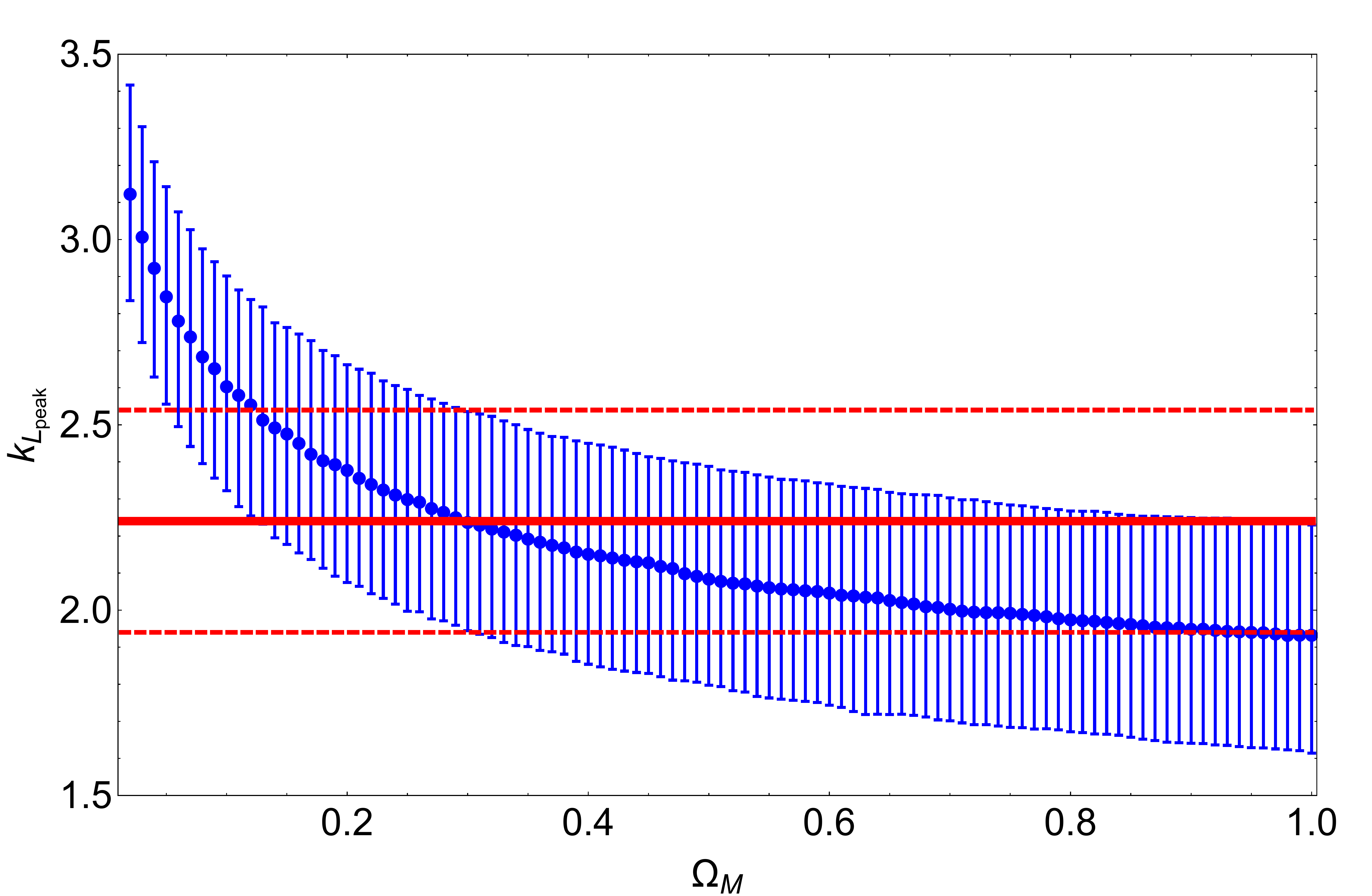}
 \includegraphics[width=0.49\hsize,height=0.34\textwidth,angle=0,clip]{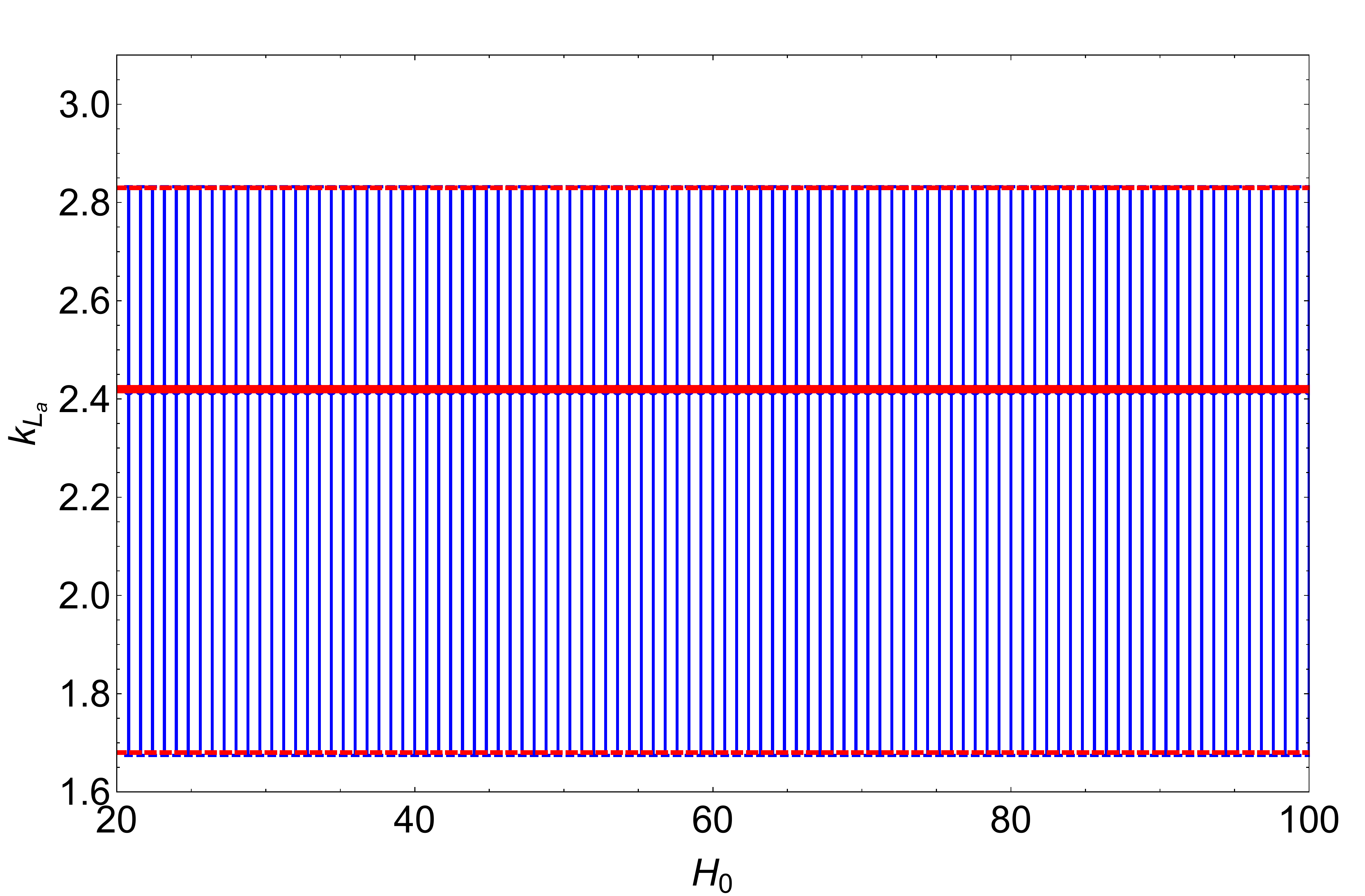}
 \includegraphics[width=0.49\hsize,height=0.34\textwidth,angle=0,clip]{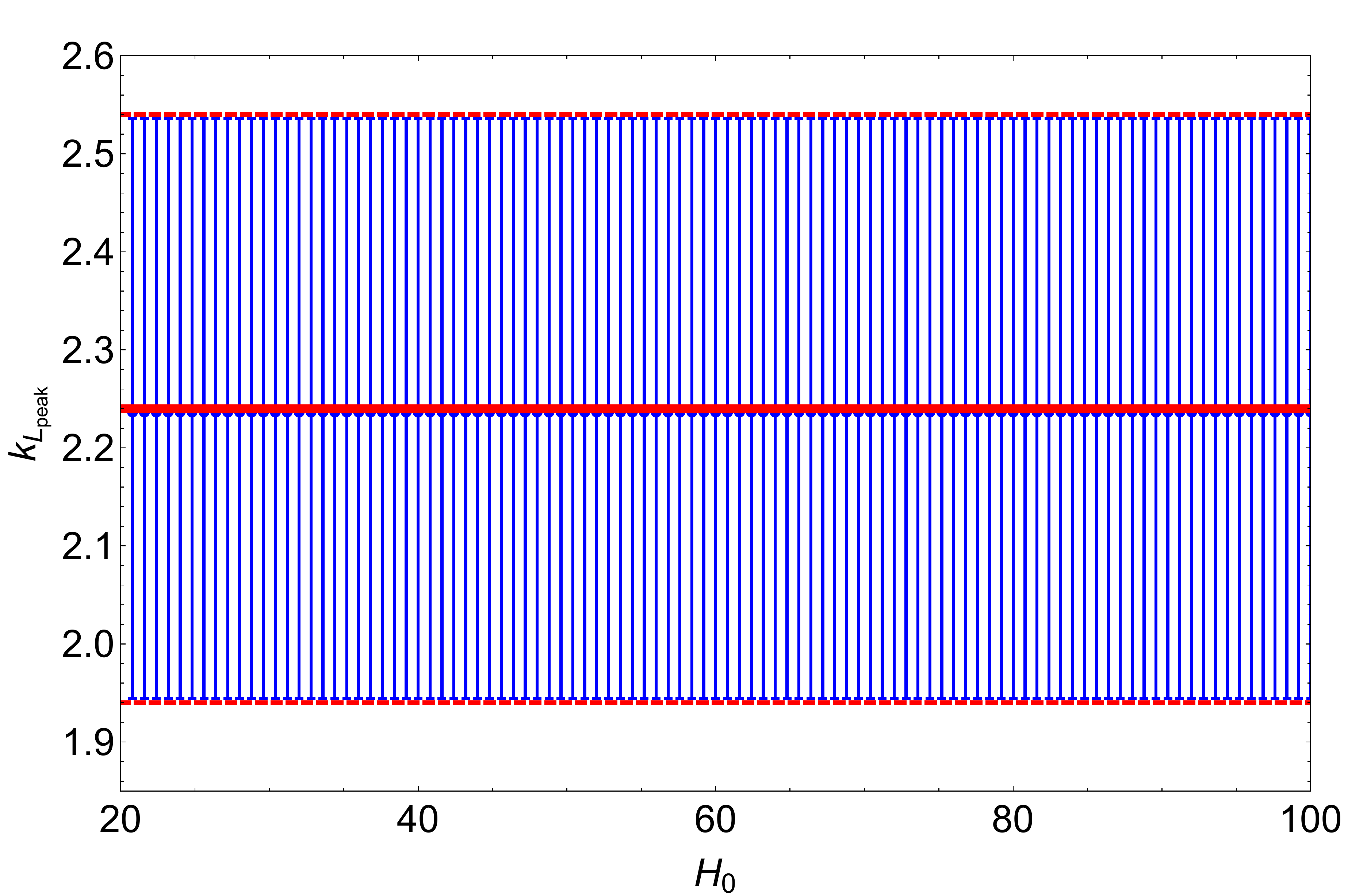}
 \includegraphics[width=0.49\hsize,height=0.36\textwidth,angle=0,clip]{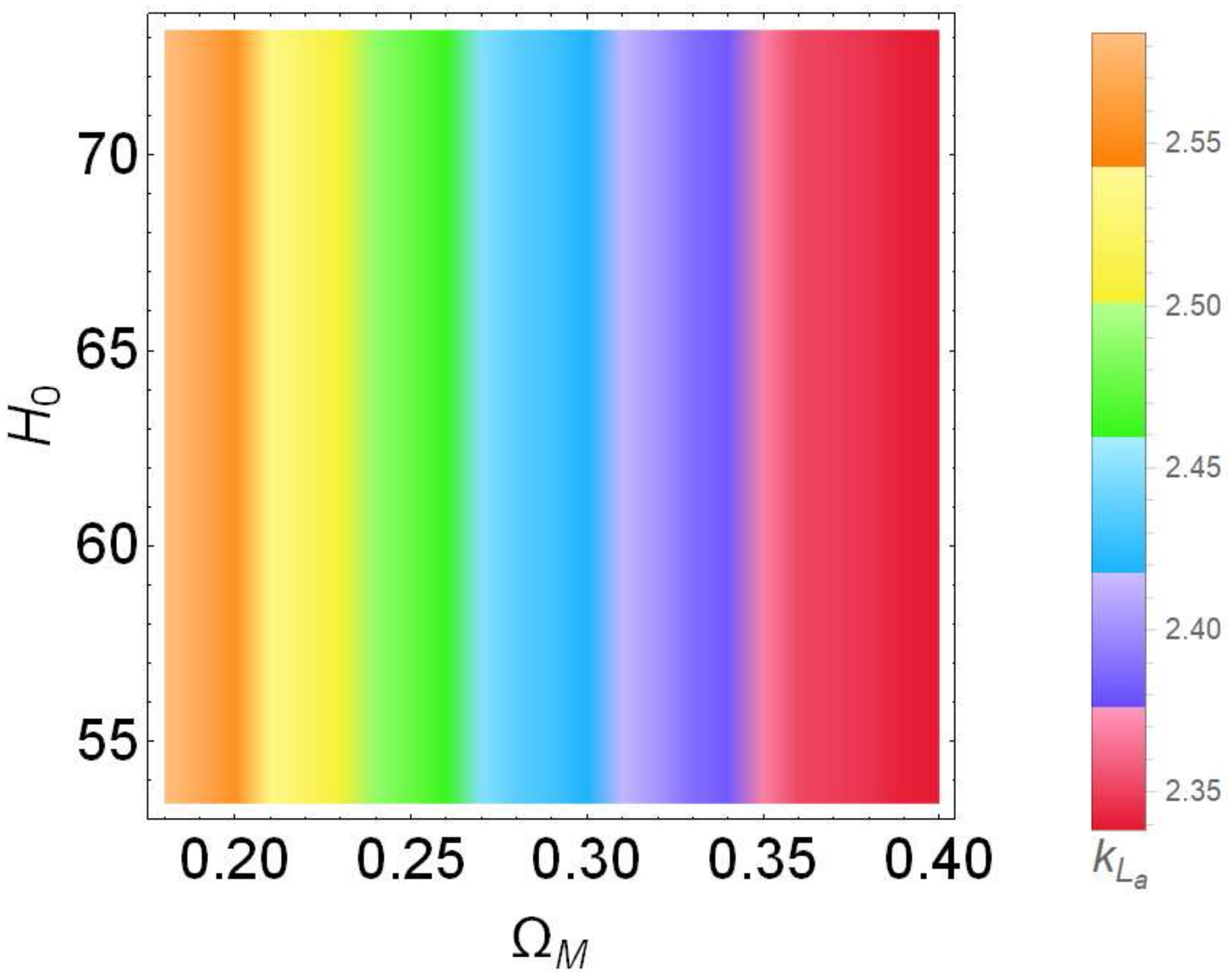}
 \includegraphics[width=0.49\hsize,height=0.36\textwidth,angle=0,clip]{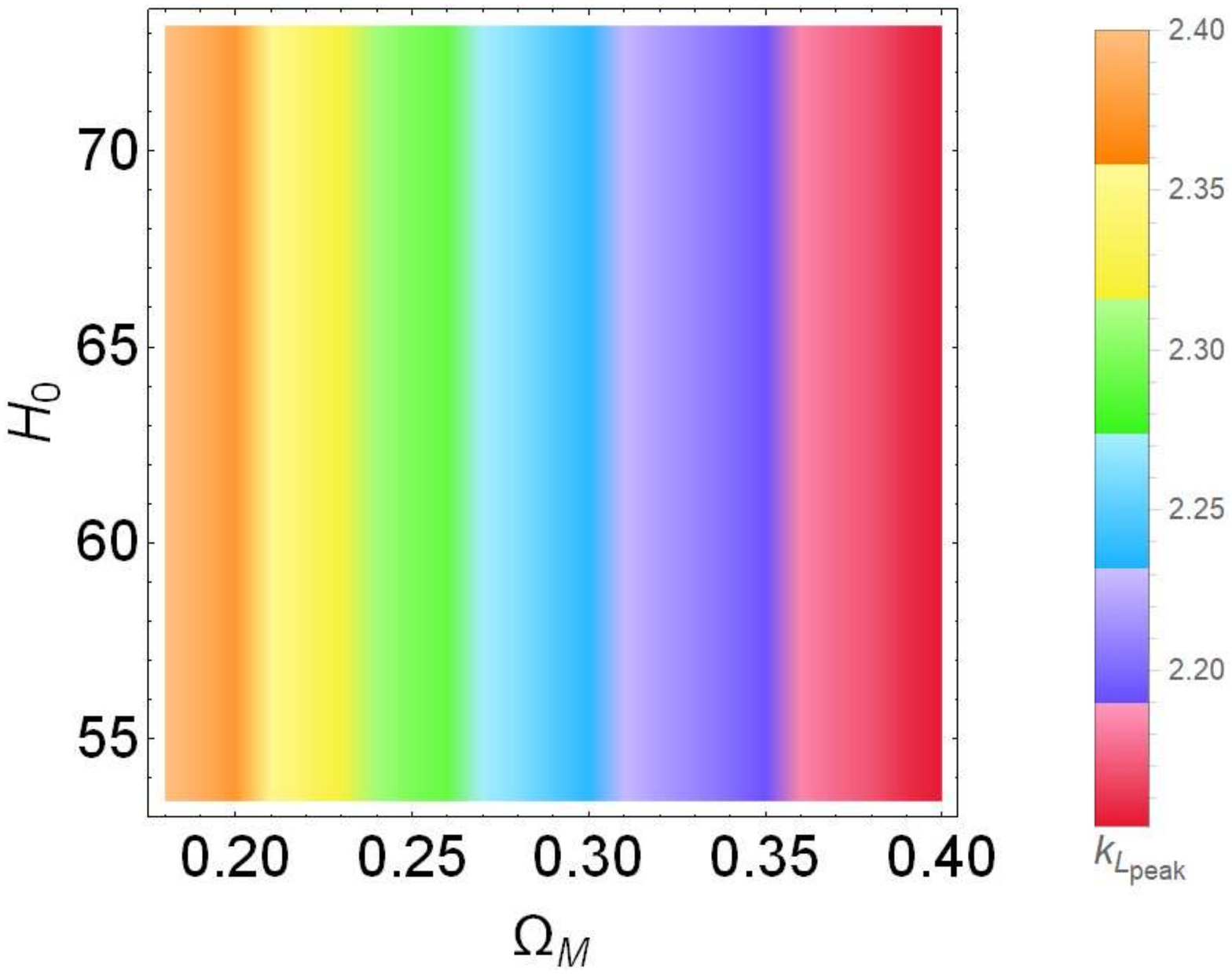}
 \caption{{The ${k_{L_{a}}}$ (left) and ${k_{L_{peak}}}$ (right) as a function of ${\Omega_{M}}$ and ${H_{0}}$. In the first four pictures the 1-${\sigma}$ error bars are shown with the thin red line together with the thick central line that represents the value of the slope of the function for which the evolution is removed. In the two bottom pictures the contour plots of ${\Omega_{M}}$ and ${H_{0}}$ as a function of ${k_{L_{a}}}$ (left) and ${k_{L_{peak}}}$ (right). The different colours indicate the different values of the ${k}$ parameters.}}
 \label{fig:evoL-k}
\end{figure}

\begin{figure}
 \centering
 \includegraphics[width=0.49\hsize,height=0.34\textwidth,angle=0,clip]{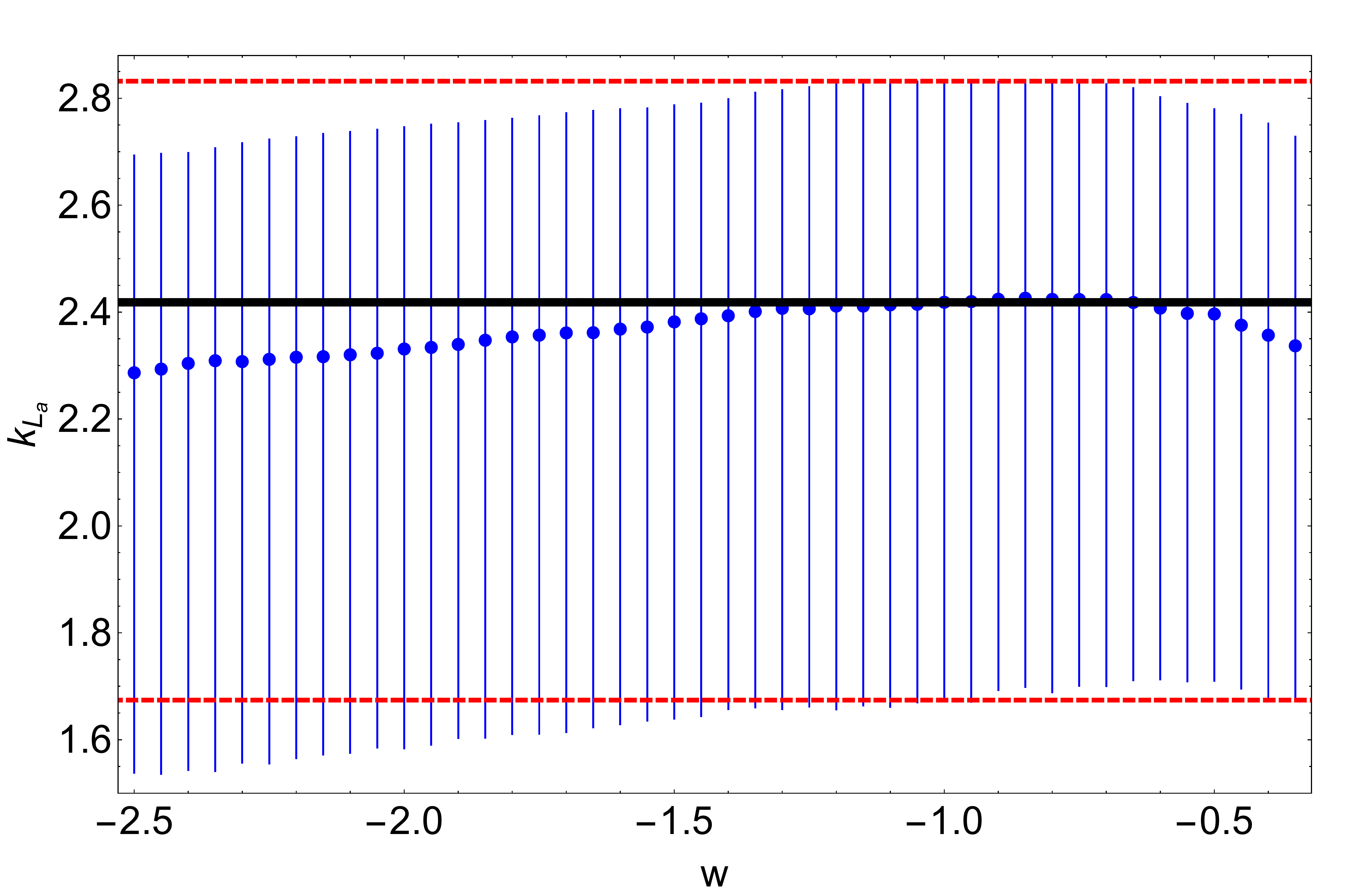}
 \includegraphics[width=0.49\hsize,height=0.34\textwidth,angle=0,clip]{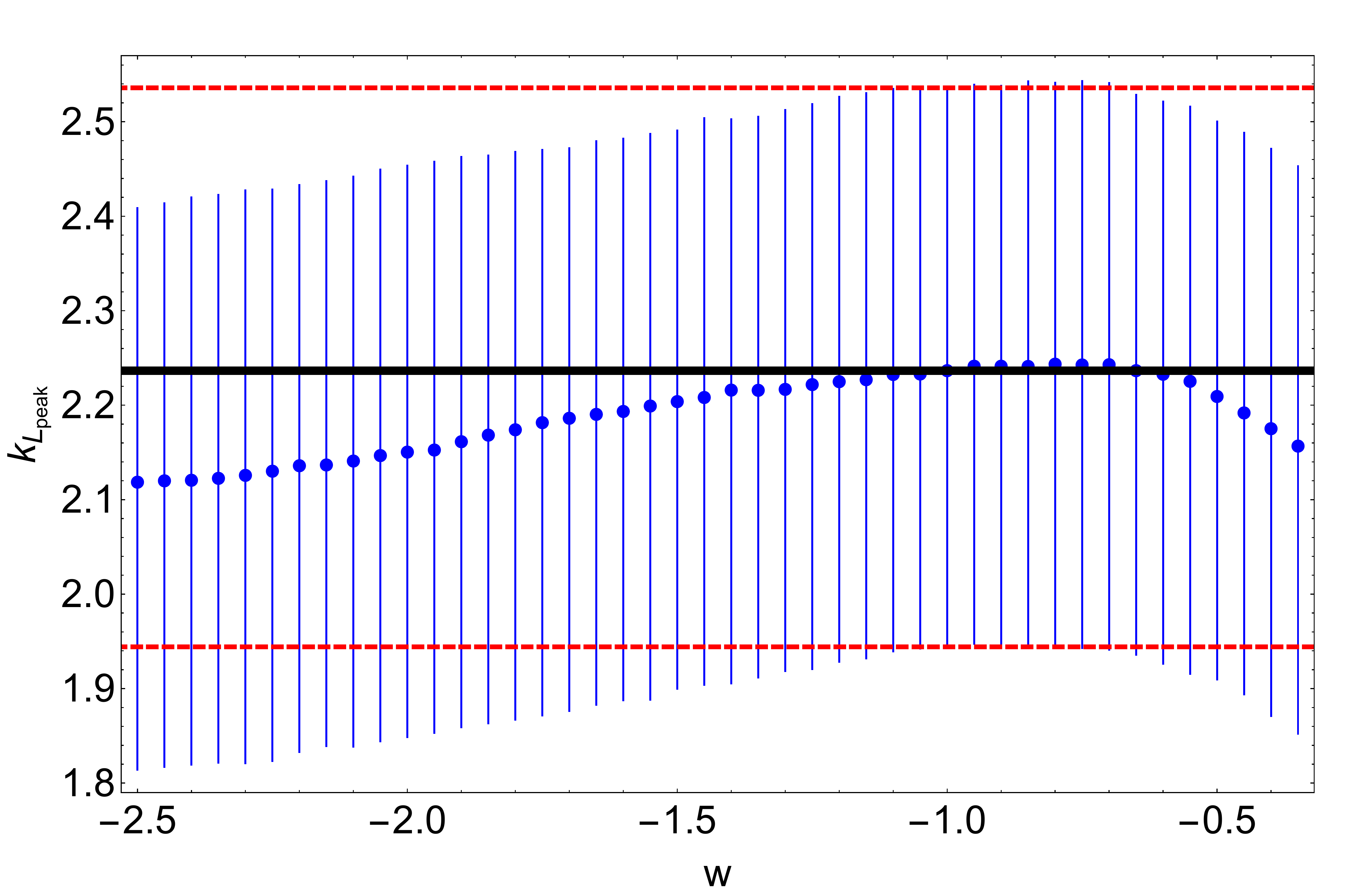}
 \includegraphics[width=0.49\hsize,height=0.34\textwidth,angle=0,clip]{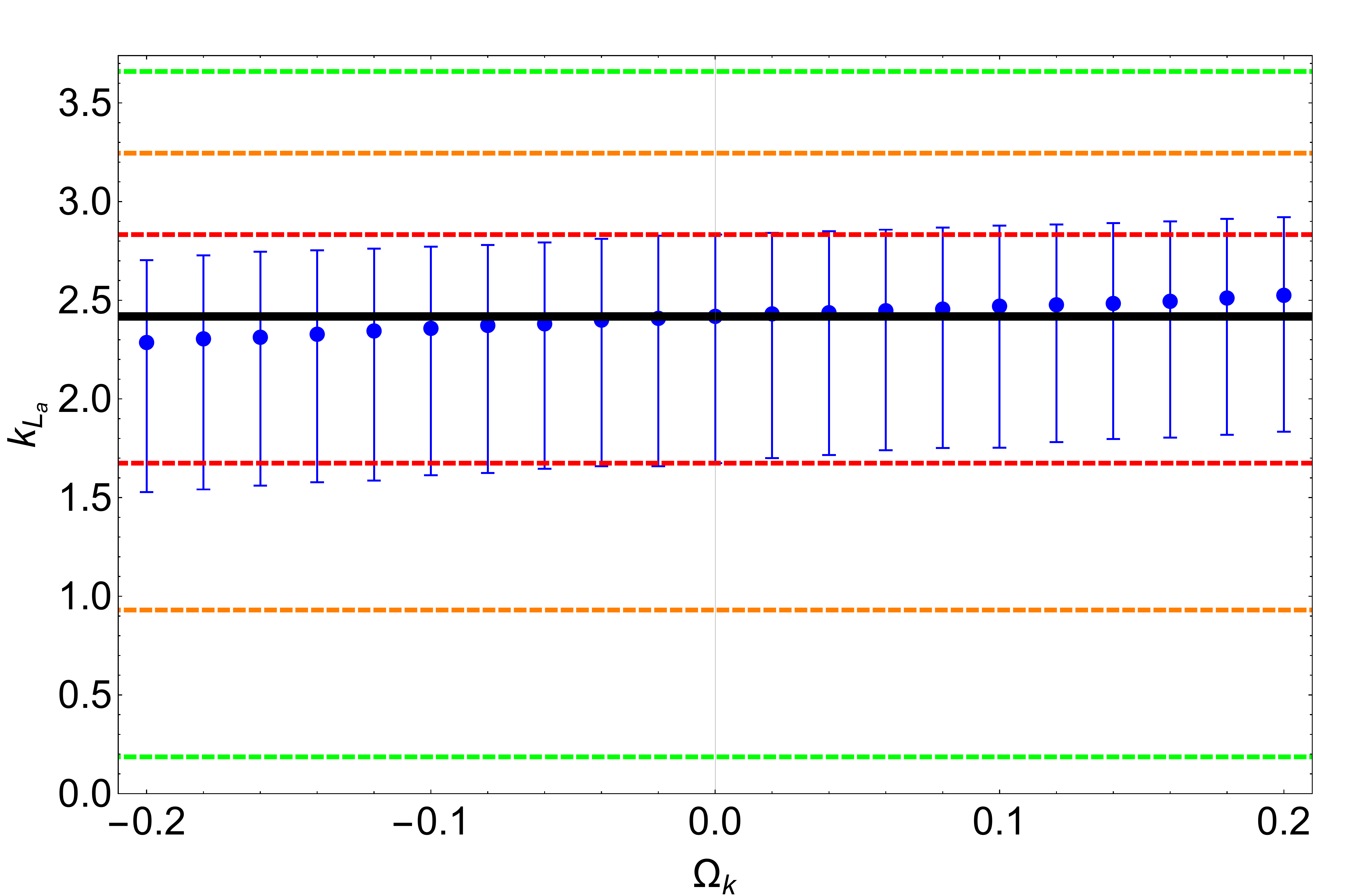}
 \includegraphics[width=0.49\hsize,height=0.34\textwidth,angle=0,clip]{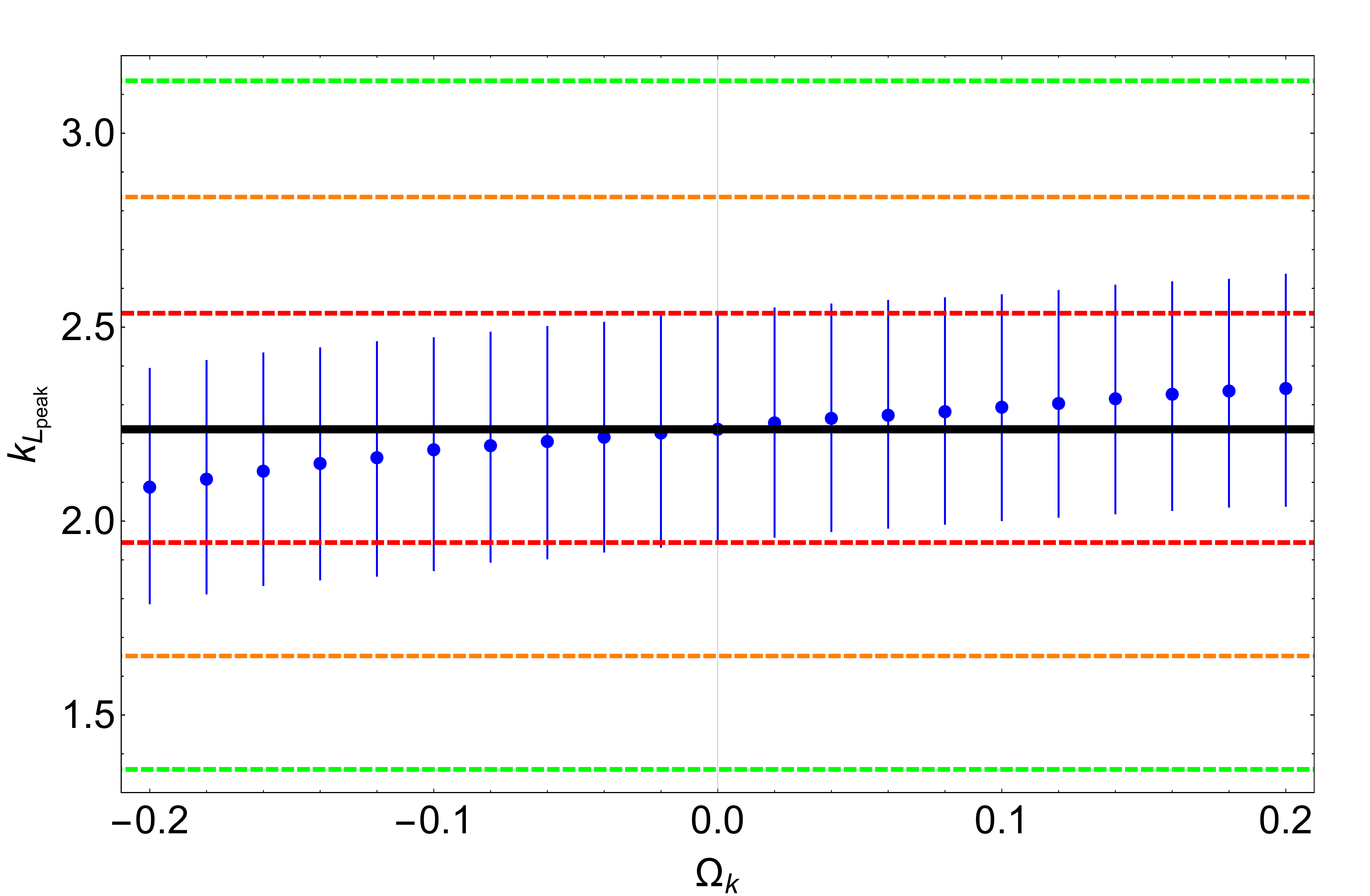}
 \caption{{\bf The ${k_{L_{a}}}$ (left) and ${k_{L_{peak}}}$ (right) as a functions of $w$ and ${\Omega_{k}}$. In the pictures, the 1-${\sigma}$, 2-${\sigma}$ and 3-${\sigma}$ error bars are shown with thin red, orange and green lines, respectively. The thick central line represents the value of the slope of the function for which the evolution is removed assuming $w = -1$ and ${\Omega_{k}} = 0$. }}
 \label{fig:evoL-kwOk}
\end{figure}

The previous analysis in this paper fixes the value of an evolutionary parameter at a given ${\Omega _{M}}$, ${H_{0}}$, ${\Omega _{k}}$ and $w$. One may wonder how this influences the cosmological results. To verify the impact of the cosmological parameters when ${k_{L_{a}}}$ and ${k_{L_{peak}}}$ depend on ${\Omega _{M}}$, ${H_{0}}$, ${\Omega _{k}}$ and $w$, we repeat the EP method with luminosity distances computed over a grid of different ${\Omega _{M}}$, ${H_{0}}$, ${\Omega _{k}}$ and $w$ values. Similar analysis of the dependence of evolutionary parameters on cosmology was performed on quasars in \cite{dainottigiada2022}. This allows us to determine how the evolutionary functions vary when the cosmological parameters change. The results of this procedure are presented in Fig. \ref{fig:evoL-k} and Fig. \ref{fig:evoL-kwOk}. We note that there are no changes of the evolutionary parameter, $k$, as $H_{0}$ varies. Regarding ${\Omega_{M}}$ and ${\Omega_{k}}$, there is a mild evolution of the evolutionary parameter, although at very low values of ${\Omega_{M}}$ (between 0 and 0.2) the evolution is noticeable, but still within 1 $\sigma$ when we account for the errorbars. {\bf When we consider the behaviour of the evolutionary slope, $k$, with ${\Omega_{k}}$, it is compatible in 1-$\sigma$ both for $k_{L_{a}}$ and $k_{L_{peak}}$ over the whole considered range of ${\Omega_{k}}$ values. When we consider even a very wide range of the $w$ parameter, all obtained values of the $k$ are compatible with each other in less than 1-$\sigma$, thus again in this case the relation of $k$ with $w$ is negligible.} To account for those results, we created a numerical function ($k=k(\Omega_{M})$) with a linear interpolation method and varied the values and errors on the $k$ parameters with $\Omega_{M}$. This is the first time in the literature that such a complete treatment has been performed, which completely overcomes the circularity problem. {\bf In this approach we neglect the $k$ as a function of $\Omega_{k}$ and $w$ because over a reasonable range of parameters the change of $k$ is insignificant. \cite{dainottigiada2022} showed that the change of $k$ with $\Omega_{M}$ is much more significant for QSOs, this is a result derived using a much larger sample. Thus, in future for a large sample we may encounter a more significant relation of $k$ with $\Omega_{k}$ and $w$. We postulate that all present cosmological efforts should investigate the impact of selection biases and redshift evolution, contrary to assuming a lack of such effects.}

\section{The GRB cosmology with the Fundamental Plane Relation}\label{section5}

In order to check the reliability of the fundamental plane for cosmological purposes, we here perform a series of tests. First of all we plot for a fixed fiducial cosmology the distance moduli ($\mu_{GRB}$) derived by the fundamental plane relation, with the distance moduli obtained by the SNe Ia, $\mu_{SNe}$ in Fig. 6. The left and right panels respectively show GRBs without accounting for selection biases and the right panel the ones accounting for selection biases.

\begin{figure}
\includegraphics[width=0.53\hsize,height=0.38\textwidth,angle=0,clip]{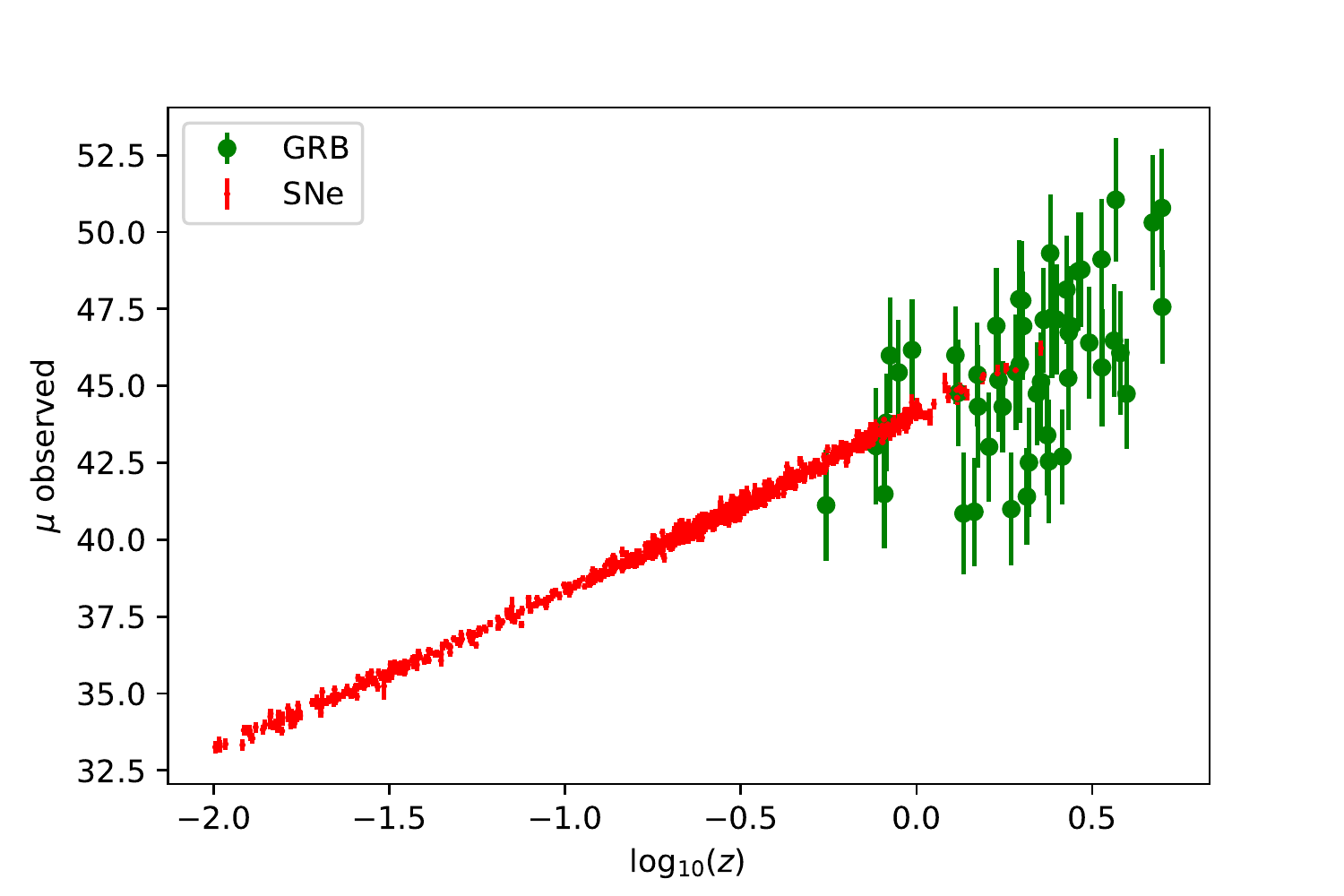}
\includegraphics[width=0.53\hsize,height=0.38\textwidth,angle=0,clip]{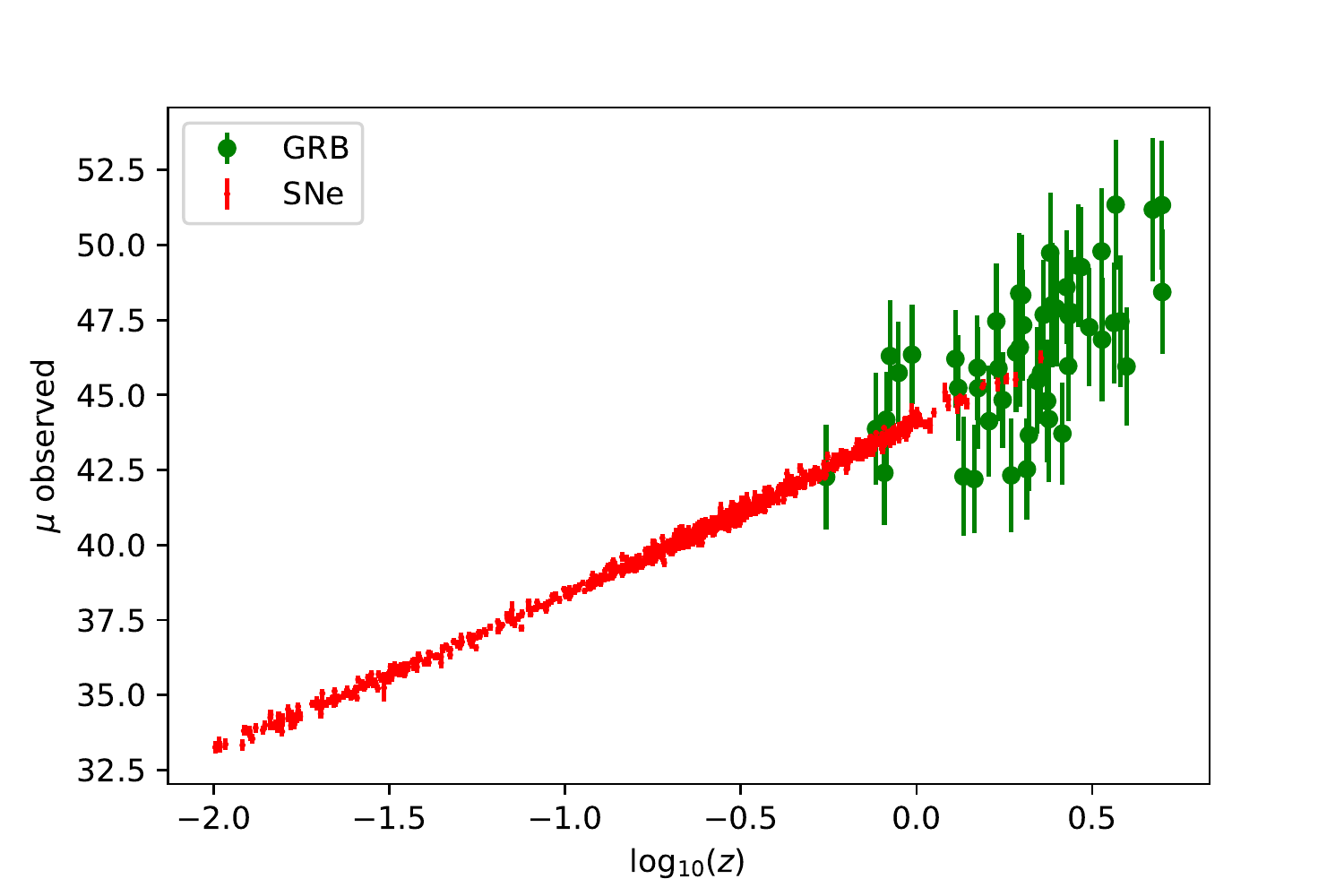}
\label{SNeMuGRB}
\caption{The distance moduli versus logarithm-ed redshift of SNe Ia ($\mu_{SNe}$) and GRBs ($\mu_{GRB}$) belonging to the platinum sample assuming the fundamental plane relation and {\bf assuming $\Lambda CDM$}. On the left for the case without correction for evolution, while on the right with correction.}
\end{figure}

To solve the so-called circularity problem, we compute the cosmological parameters together with the coefficients of the fundamental plane relation, starting from the peak fluxes and the fluxes at the end of the plateau emission which are the observer frame quantities of the corresponding peak prompt luminosity and the luminosity at the end of the plateau emission, respectively. Thus, this procedure does not involve any fixed a priori cosmological models, and the results of this computation leads to the best-fit cosmological parameters together with the coefficients of the fundamental plane correlation using the right hand side of Equation 6 in which the luminosity is defined in Equation 2. More specifically, in our computation we run MCMC simulations using either uniform or Gaussian priors on ${\Omega_M, H_0}$ and ${{w}}$, and compute the corresponding distance luminosity ${D_L(z, \Omega_M, H_0, w)}$ and the corresponding ${L_{peak}}$, ${{L_X}}$ for each value of this grid. Then, for each value of this grid, we compute the best fit parameters of the plane. Thus, 
this procedure completely avoids the circularity problem. 
This method does not need any calibration of the fundamental plane relation on other local probes; the correlation's parameters are free to vary following \cite{Dainotti2013a}. 
For the flat $\Lambda$CDM cosmological model, in which $w=-1$, and where we neglect the radiation contribution, the luminosity distance used in 
Equation \ref{Lpeak} 
 is : 
 \begin{linenomath*}
\begin{equation} \label{flatdistanceluminosity}
 D_L(z)=(1+z)\frac{c}{H_{0}}\int_{0}^{z} \frac{dz'}{\sqrt{\Omega_{M}(1+z')^3+(1-\Omega_{M}})},
 \end{equation}
\end{linenomath*}
\noindent where $c$ is the speed of light. 
For simplicity we can also write that integrand of Equation \ref{flatdistanceluminosity} as the following:
\begin{linenomath*}
\begin{equation}
E(z)=\frac{1}{\sqrt{\Omega_{M}(1+z')^3+(1-\Omega_{M}})}.
\label{E(z)}
\end{equation}
\end{linenomath*}

We combine the GRB Platinum sample, the SNe Ia Pantheon Sample, and the BAO data presented in \cite{SharovBAO}. We note that even if according to \cite{Riess} $\Omega_M$ and $H_0$ are kinematically independent, we still have chosen to take into account the separate case of varying both of them together as a check of how the precision reached by us on these quantities depends on the parameter space's dimension.
We use the fundamental plane correlation both with and without the correction computed by the EP method to see if this correction may carry a reduction on $\sigma_{int}$, and consequently on the cosmological parameters. 
 We derive $\mu_{obs,GRBs}$ in such a way that it is completely independent from the $\mu_{SNe}$, by manipulating the fundamental plane relation corrected for evolution:
\begin{linenomath*}
\begin{equation}
\begin{split}
\mu_{obs,GRBs}= 5 (b_{1} \log F_{p,cor}+a_{1} \log F_{X,cor} + c_{1}+d_{1} \log T^{*}_X)+25
\end{split},
\label{equmu}
\end{equation}
\end{linenomath*}

\noindent where $\log F_{p,cor}$ and $\log F_{X,cor}$ are the prompt and afterglow emission fluxes, respectively, corrected by the $K$-correction and the evolutionary functions. We show some of the algebraic computations performed to obtain Equation \ref{isotropic} starting from Equation \ref{Lpeak}. For the case where we do not take into account the evolutionary effects, considering the relation between fluxes and luminosities given by Equation 2 we obtain:
 
\begin{linenomath*}
 \begin{equation}
 \log_{10}(4\pi d_L^2)+\log_{10}K_{X}-b \cdot (\log_{10}(4\pi d_L^2)+\log_{10}K_{peak})=b \cdot \log_{10} F_{peak}+a \cdot \log_{10} T^{*}_X+C-\log_{10} F_{X},
 \end{equation}
 \end{linenomath*}
\noindent where $K_{peak}$ and $K_{X}$ are the $K$-corrections computed for the prompt and the afterglow, respectively, and $a$, $b$, and $C$ are the coefficients of the fundamental plane correlation. Isolating the luminosity distance from the previous equation we obtain:
\begin{linenomath*}
\begin{equation}
 \log_{10}(d_L)=-\frac{\log_{10} F_{X}+\log_{10}K_{X}}{2 (1-b)}+\frac{b \cdot (\log_{10} F_{peak}+\log_{10}K_{peak})}{2 (1-b)} -\frac{(1-b)\log_{10}(4\pi)+C}{2 (1-b)}+ \frac{a \log_{10} T^{*}_X}{2 (1-b)}.
 \label{distancemoduli}
\end{equation}
\end{linenomath*}
With new definitions of the coefficients and the fluxes we finally reproduce Equation \ref{equmu} considering also the relation between luminosity distance and distance modulus.


We compare both $\mu_{obs, GRBs, SNe}$ with the theoretical $\mu_{th}$, defined as: 

\begin{linenomath*}
\begin{equation}
\mu_{th}=5\cdot \log\ D_L(z,\Omega_{M}, H_0, w) +25.
\label{modulus}
\end{equation}
\end{linenomath*}
We now present the constraints given by BAO measurements used in our computations. 

The data comes from \cite{SharovBAO} who refer to the equation for the $d_{z}(z')$ function defined as: 

\begin{equation}
 d_{z}(z') = \frac{r_{s}(z_{d})}{D_{V}(z')}, \label{dzformula}
\end{equation}

\begin{equation}
 \text{where} \hspace{2ex} D_{V}(z') =\frac{c}{H_{0}}\left[ \frac{z'}{E(z')}\times \left(\int_{0}^{z'}\frac{dz}{E(z)} \right)^{2}\right]^{\frac{1}{3}}, \hspace{1.5ex} \& \hspace{1.5ex} r_s(z_d) =\frac{55.514 \cdot e^{[72.3(\Omega_{\nu}h^{2}+0.0006)^2]}}{(\Omega_{M}h^{2})^{0.25351}(\Omega_{b}h^{2})^{0.12807}}Mpc. \label{eq_rsfiducialtrue}
\end{equation}

\noindent The value $z_d$ corresponds to the decoupling of photons in the comoving sound horizon scale $r_s(z_d)$ using the fitting formula from \cite{SharovBAO}.

For this approach we combine the likelihoods and write the logarithm-ed equation as: 

\begin{align}
\begin{split}
 \mathcal{L}(GRBs+SNe Ia+BAO) =&
 \sum_{i}\biggl[\log \biggl( \frac{1}{\sqrt{2\pi}\sigma_{\mu,i}} \biggr)- \frac{1}{2}\biggl( \frac{\mu_{th,GRB,i}-\mu_{obs,GRB,i}}{\sigma_{\mu,i}}\biggr)^{2} \biggr] \\ &-\frac{1}{2}[(\mu_{th,SNe}-\mu_{obs,SNe})^T\times C_{inv} \times (\mu_{th,SNe}-\mu_{obs,SNe}) + (\Delta d_{z})^{T} \times C_{inv}^{BAO} \times \Delta d_{z}],
 \label{likelihood} 
\end{split}
\end{align}

\noindent where the first term relates to GRBs' distance moduli \citep{Dainotti2013a, Amati2019}, the second to the SNe Ia's, where $C_{inv}$ is the inverse of the covariance matrix of the SNe Ia data taken from \cite{Scolnic}, and the third to the BAO, where $C_{inv}^{BAO}$ is the inverse of the covariance matrix of the BAO data taken from \cite{SharovBAO} and the $\Delta d_{z}$ is defined as: $\Delta d_{z,i} = d_{z}^{obs}(z_{i}) - d_{z}^{th}(z_{i})$; $d_{z}^{obs}(z_{i})$ is taken from the data, while $d_{z}^{th}(z_{i})$ is computed with the Equation \ref{dzformula}.

We have also computed the cosmological parameters using only SNe Ia data as well as SNe Ia+BAO, to verify if adding GRBs would confirm the results and to what extent we could enhance the precision on the cosmological parameters. 

\section{Results} \label{Results}
The results presented here are divided in three major steps. First, we show the capability of GRBs as alone standardizable candles with the fundamental plane using Equations \ref{isotropic} and \ref{planeev}, as well as the Equation for $\mu_{GRB}$, \ref{equmu}, without calibration using Gaussian priors based on the values of SNe Ia in \cite{Scolnic} (see \ref{GRBs alone without calibration}). Then, we show the calibration on SNe Ia using Gaussian priors (see \ref{GRB alone with calibration}). Then, we derive the cosmological parameters with uniform priors, both with and without calibrating GRBs on the SNe Ia (see \ref{uniformprior}) and we compare with the results with Gaussian priors. The analysis has the scope of showing the precision of GRBs in constraining cosmological parameters in a flat $\Lambda$CDM model. Second, we use GRBs together with SNe Ia and BAO to verify the usefulness of GRBs in combination with other probes, see \ref{SN+BAO+GRB}). These results will entail both the observational data of GRBs with no correction for evolution as well as accounting for these corrections. The third step is instead the analysis of the open cosmological model with the GRB fundamental plane relation both corrected and uncorrected for selection biases and redshift evolution (see \ref{flatnessUniverse}).

\subsection{GRBs alone with no calibration with Gaussian priors}\label{GRBs alone without calibration}
 We here clarify that in all computations we do not minimize the relation of the evolutionary parameters as a function of $\Omega_M$, but we use the evolutionary function $k(\Omega_M)$. This indeed is an independent computation, see sec. \ref{section4}, which shows how the evolutionary functions depend on $\Omega_M$. There is no minimization involved in this computation.

\begin{table}
\addtolength{\tabcolsep}{-2pt}
\centering
\title{By Distance Modulus and Fundamental Plane}
\maketitle
\begin{tabular}{p{35mm}|l|c|c|c|c|c|c}
 \toprule[1.2pt]
 \toprule[1.2pt]
\textbf{No calibration on SNe Ia, Equation \ref{equmu}} & {\textbf{parameters varied}} & \textbf{Model} & $\boldsymbol{\Omega_{M}}$ & $\boldsymbol{H_{0}}$ & $w$ & $z-score_{SN}$ & $z-score_{SN+BAO}$ \\ 
\midrule
without evolution & $\Omega_{M}$ & $\Lambda$CDM & $0.316 \pm 0.063$ & \bf{70} & \bf{-1} & 0.268 & 0.190 \\\hline
without evolution & $H_0$ & $\Lambda$CDM & \bf{0.30} & $73.225\pm 3.307$ & \bf{-1} & 0.984 & 0.987 \\\hline
without evolution & $\Omega_{M}$ and $H_0$ & $\Lambda$CDM & $0.320 \pm 0.068 $ & $73.149\pm 3.026$ & \bf{-1} & 0.307, 1.025 & 0.137, 1.09 \\\hline
without evolution & $w$ & $w$CDM & \bf{0.30} & \bf{70} & $-0.673 \pm 0.717$ & 0.460 & 0.484\\\hline
\midrule
with fixed evolution & $\Omega_{M}$ & $\Lambda$CDM & $0.308 \pm 0.063$ & \bf{70} & \bf{-1} & 0.142 & 0.063 \\\hline
with fixed evolution & $H_0$ & $\Lambda$CDM & \bf{0.30} & $72.869\pm 2.921$ & \bf{-1} & 0.988 & 0.992 \\\hline
with fixed evolution & $\Omega_{M}$ and $H_0$ & $\Lambda$CDM & $0.304 \pm 0.064 $ & $73.128\pm 3.008$ & \bf{-1} & 0.089, 1.024 & 0.109, 1.10 \\\hline
with fixed evolution & $w$ & $w$CDM & \bf{0.30} & \bf{70} & $-0.977 \pm 0.620$ & 0.037 & 0.064\\\hline
\midrule
with $k=k(\Omega_{M})$ & $\Omega_{M}$ & $\Lambda$CDM & $0.295 \pm 0.062$ & \bf{70} & \bf{-1} & 0.064 & 0.144 \\\hline
with $k=k(\Omega_{M})$ & $\Omega_{M}$ and $H_0$ & $\Lambda$CDM & $0.297 \pm 0.065 $ & $73.036\pm 3.139$ & \bf{-1} & 0.015, 0.955 & 0.214, 1.02\\\hline
 \toprule[1.2pt]
 \toprule[1.2pt]
\textbf{No calibration on SNe Ia, Equations \ref{isotropic} and \ref{planeev}} & {\textbf{parameters varied}} & \textbf{Model} & $\boldsymbol{\Omega_{M}}$ & $\boldsymbol{H_{0}}$ & $w$ & $z-score_{SN}$ & $z-score_{SN+BAO}$ \\ 
\midrule
without evolution & $\Omega_{M}$ & $\Lambda$CDM & $0.302 \pm 0.061$ & \bf{70} & \bf{-1} & 0.049 & 0.033 \\\hline
without evolution & $H_0$ & $\Lambda$CDM & \bf{0.30} & $73.152\pm 3.113$ & \bf{-1} & 1.021 & 1.024 \\\hline
without evolution & $\Omega_{M}$ and $H_0$ & $\Lambda$CDM & $0.302 \pm 0.064 $ & $73.074\pm 3.145$ & \bf{-1} & 0.163, 0.99 & 0.031, 1.09\\\hline
without evolution & $w$ & $w$CDM & \bf{0.30} & \bf{70} & $-1.133 \pm 1.048$ & 0.127 & 0.111\\\hline
\midrule
with fixed evolution & $\Omega_{M}$ & $\Lambda$CDM & $0.299 \pm 0.065$ & \bf{70} & \bf{-1} & 0.000 & 0.077 \\\hline
with fixed evolution & $H_0$ & $\Lambda$CDM & \bf{0.30} & $73.073\pm 3.126$ & \bf{-1} & 0.992 & 0.995 \\\hline
with fixed evolution & $\Omega_{M}$ and $H_0$ & $\Lambda$CDM & $0.294 \pm 0.065 $ & $72.785\pm 3.049$ & \bf{-1} & 0.058, 0.900 & 0.260, 0.971 \\\hline
with fixed evolution & $w$ & $w$CDM & \bf{0.30} & \bf{70} & $-0.978 \pm 0.662$ & 0.033 & 0.059 \\\hline
\midrule
with $k=k(\Omega_{M})$ & $\Omega_{M}$ & $\Lambda$CDM & $0.305 \pm 0.063$ & \bf{70} & \bf{-1} & 0.095 & 0.016 \\\hline
with $k=k(\Omega_{M})$ & $\Omega_{M}$ and $H_0$ & $\Lambda$CDM & $0.305 \pm 0.064 $ & $73.126\pm 3.101$ & \bf{-1} & 0.103, 0.996 & 0.093, 1.065 \\\hline
\end{tabular}
\caption{Results of the fitting of the cosmological parameters without calibration on SNe Ia, using GRBs alone and using Gaussian priors with $\mu_{GRB}$ (first part) and with the Fundamental plane Equation, \ref{isotropic} (2nd part) without evolution correction, with fixed evolution, and with evolution correction as a function of $\Omega_{M}$. In the first column we define the studied case and the type of evolution correction. In the second column we define which cosmological parameters are left free to vary. In the subsequent columns we present the values of parameters with error bars obtained in the computation for the corresponding cases. The fixed values are presented in bold. In the last column we present a comparison of each of the case results with the ones obtained with SNe Ia alone, present in Table \ref{Table5}. For this purpose we compute the z-score given by: $z = \frac{|X_{SN}-X_{GRB}|}{\sqrt{\sigma_{X,SN}^2+\sigma_{X,GRB}^2}}$, where $X$ is a computed value of the considered cosmological parameter for SNe Ia and GRBs, respectively, while $\sigma_{X}$ is its error.}
\label{Table1}
\end{table}

We have tested two approaches to derive the cosmological parameters with GRBs. For each approach we vary a) both $\Omega_M$ and $H_0$, b) only 
$\Omega_M$, c) only $H_0$ and d) only $w$.
For our computations related to GRBs we consider the Gaussian priors of 3 $\sigma$ based on the results and the uncertainties computed in \cite{Scolnic}. Although this procedure does not allow to test if there are deviations beyond the 3 $\sigma$ limit, it still allows us to understand the role and the impact of GRBs as standalone cosmological probes, and what are the uncertainties we can achieve considering the current state of the art.
We use the following likelihoods:

\begin{figure} 
\centering
\subfloat[Varying only $\Omega_M$ without correction for evolution]{\label{fig9_a}
\includegraphics[width=0.49\hsize,height=0.49\textwidth,angle=0,clip]{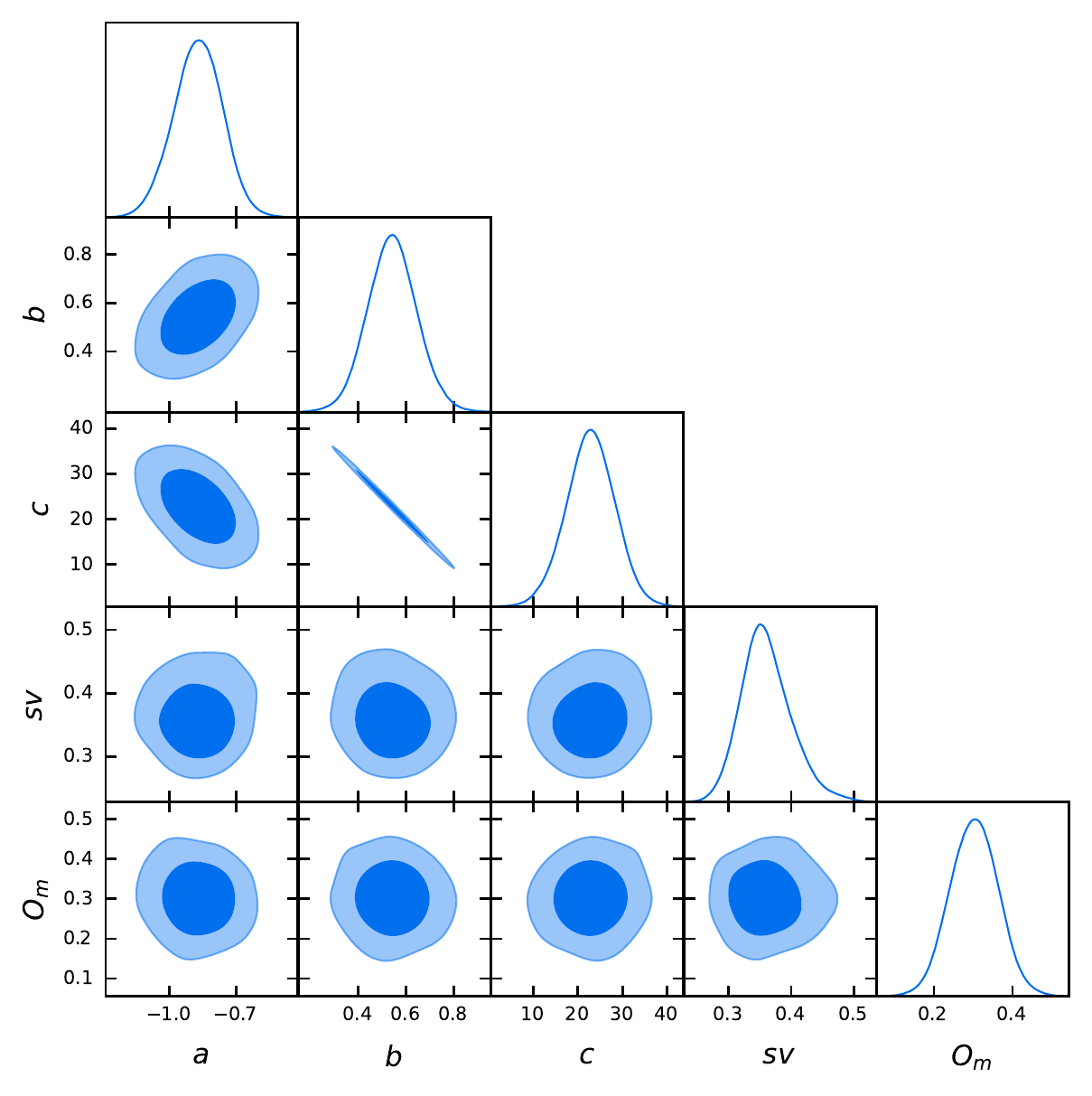}}
\subfloat[Varying only $H_0$ without correction for evolution]{\label{fig9_b}
\includegraphics[width=0.49\hsize,height=0.49\textwidth,angle=0,clip]{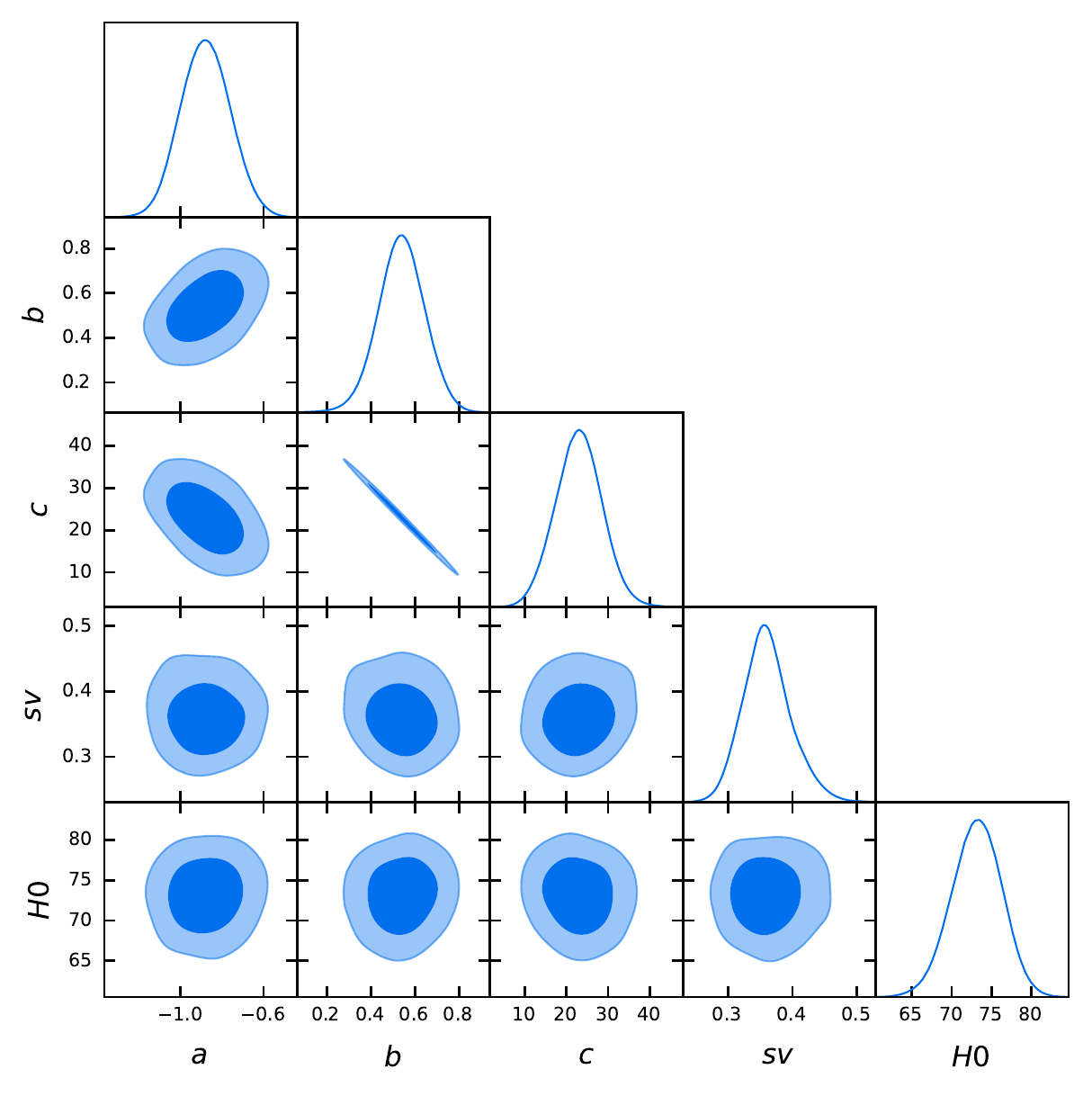}}\\
\subfloat[Varying both $\Omega_M$ and $H_0$ without correction for evolution]{\label{fig9_c}
\includegraphics[width=0.49\hsize,height=0.49\textwidth,angle=0,clip]{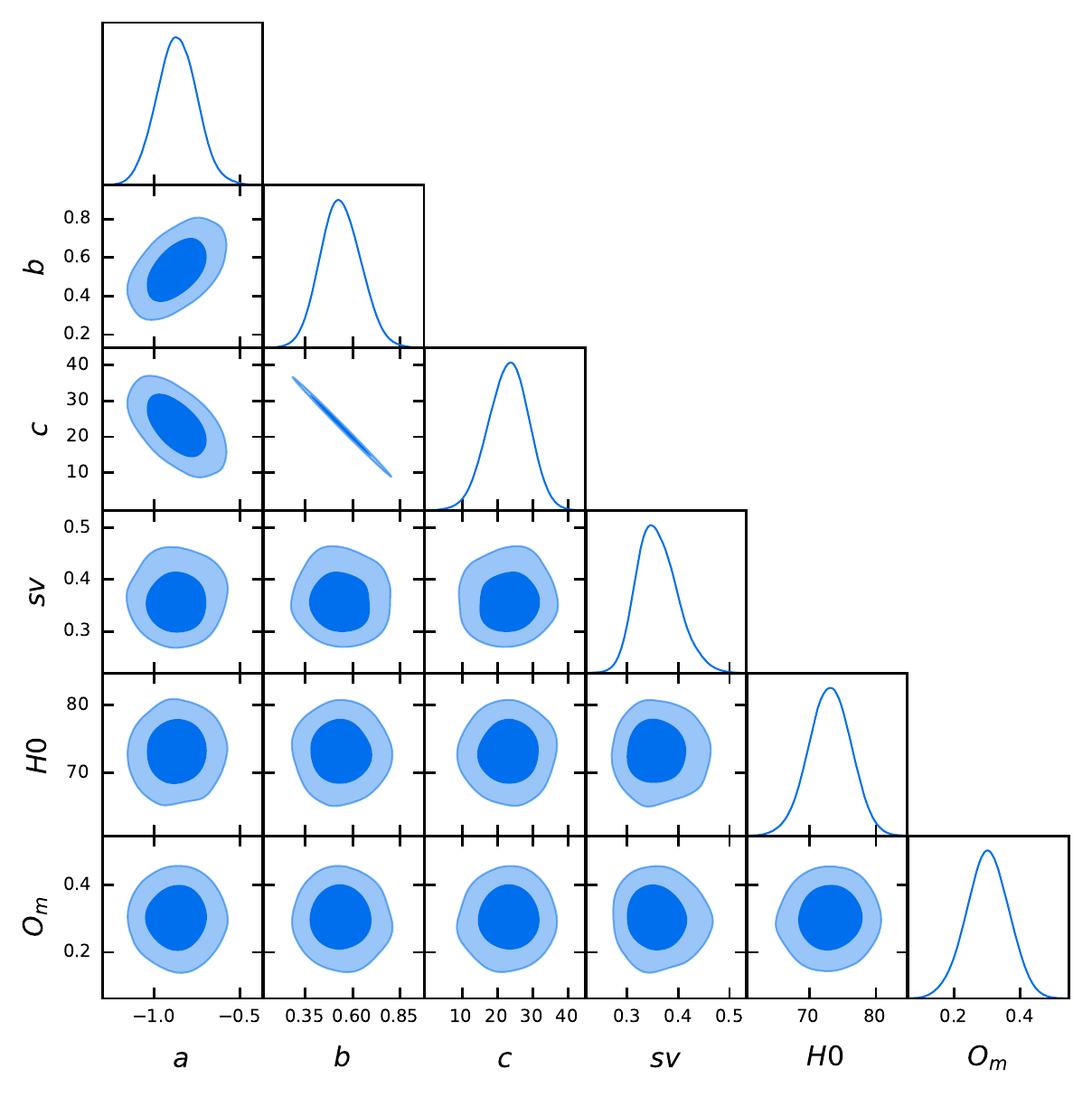}}
\subfloat[Varying only $w$ without correction for evolution]{\label{fig9_d}
\includegraphics[width=0.49\hsize,height=0.49\textwidth,angle=0,clip]{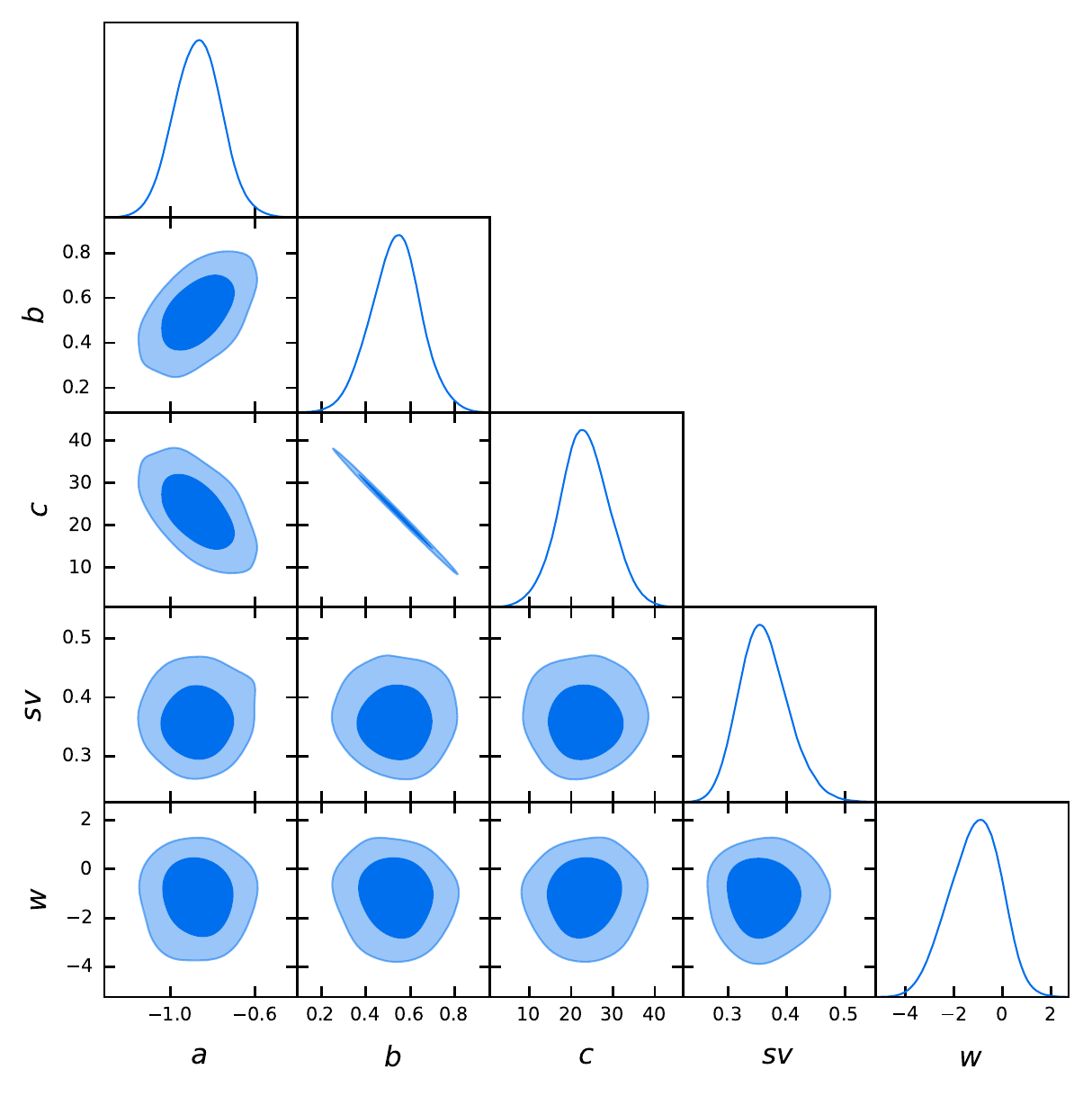}}
\caption{Cosmological results for the GRBs alone (with no calibration) without evolution using the equation of the fundamental plane, Equation \ref{isotropic} and using the 3 $\sigma$ Gaussian priors on the cosmological results reported in \citet{Scolnic}. We derive in the sub-panels $a$, $b$, $c$ and $d$ the values of $\Omega_{M}$, $H_{0}$, $\Omega_{M}$ and $H_{0}$ contemporaneously, and $w$, respectively.}
\label{fig7}
\end{figure}

\begin{figure} 
\centering
\subfloat[Varying only $\Omega_{M}$ with fixed evolution]{\label{fig10_a}
\includegraphics[width=0.50\hsize,height=0.50\textwidth,angle=0,clip]{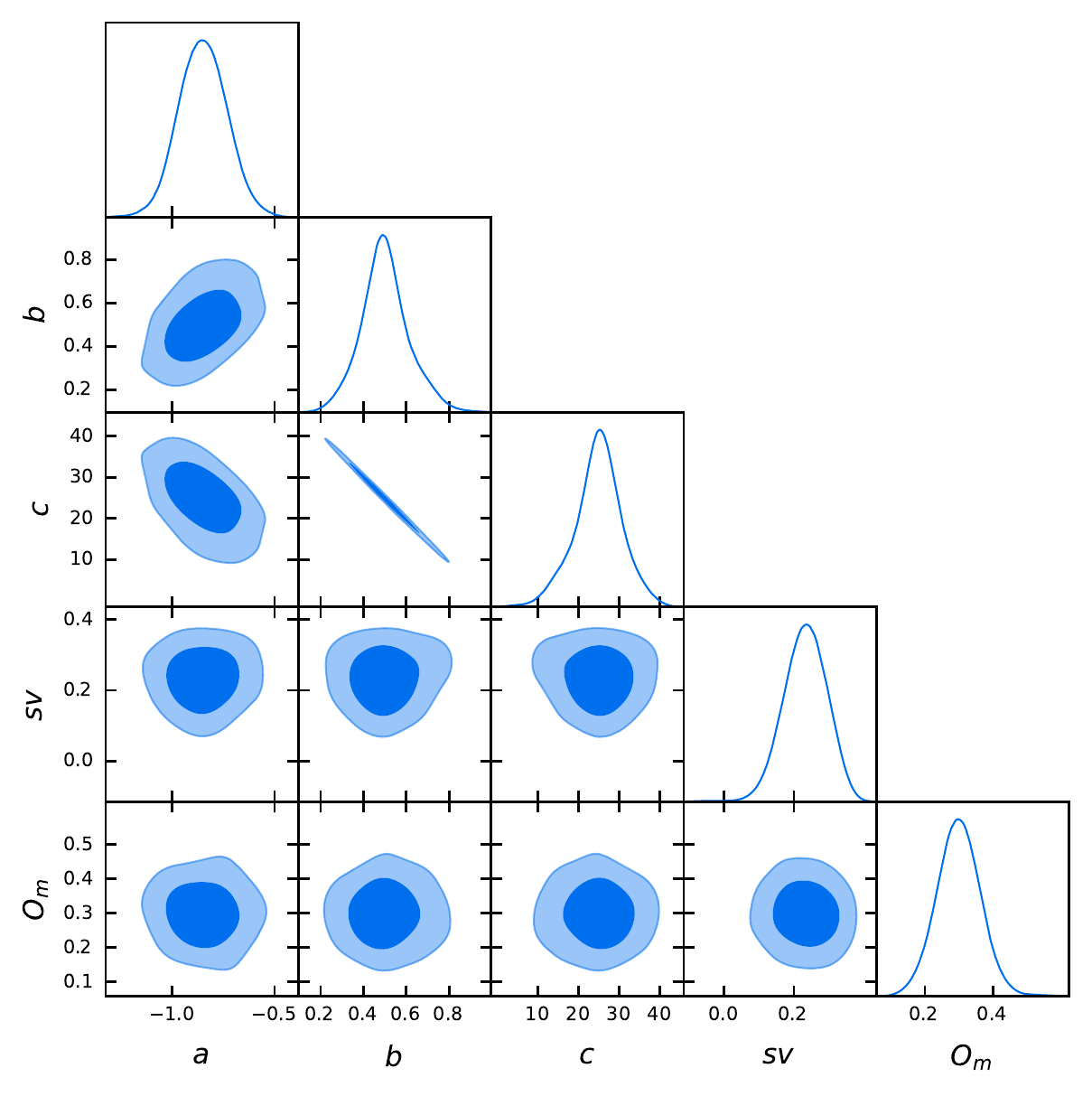}}
\subfloat[Varying only $H_0$ with fixed evolution]{\label{fig10_b}
\includegraphics[width=0.50\hsize,height=0.50\textwidth,angle=0,clip]{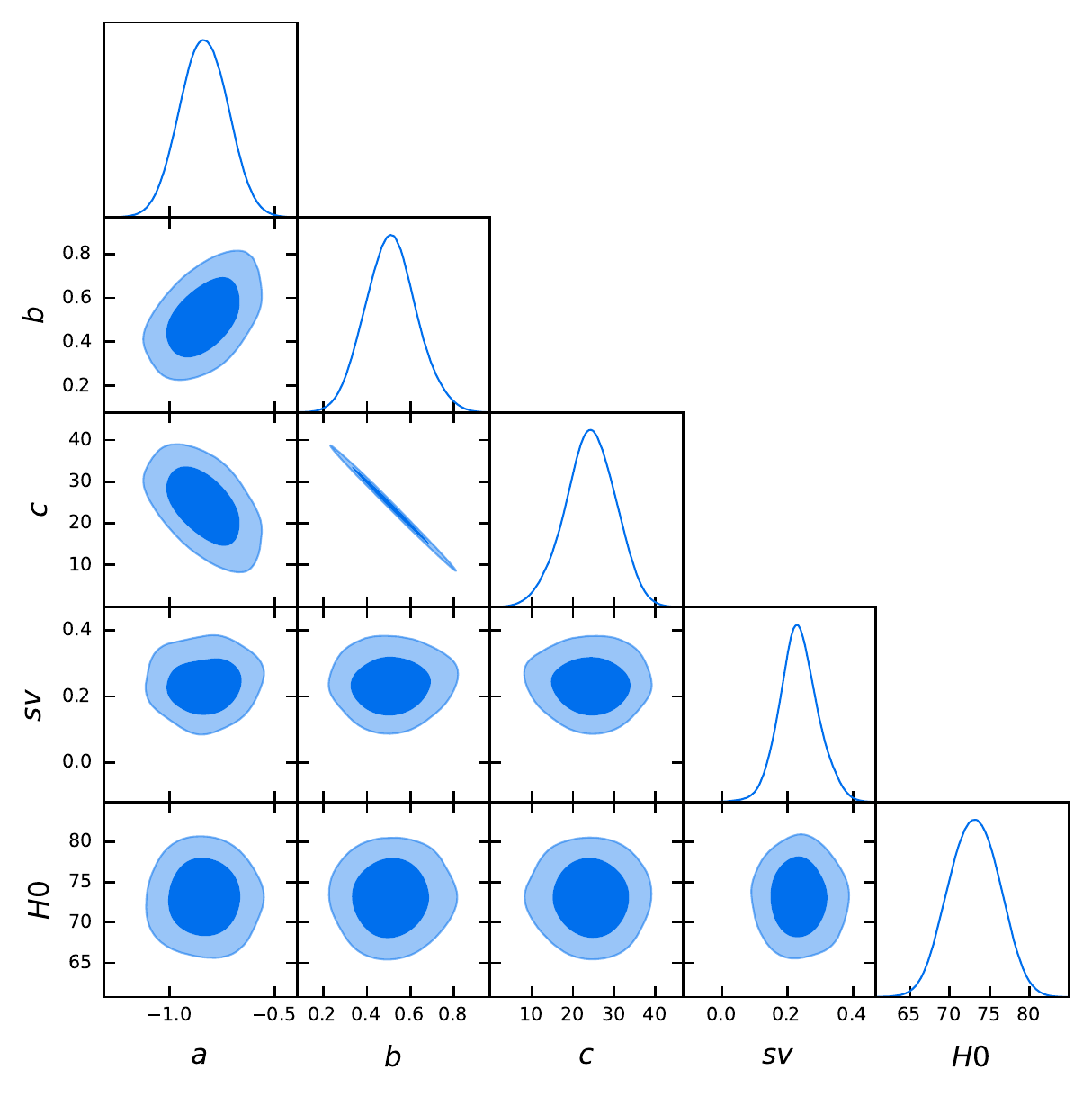}}\\
\subfloat[Varying both $\Omega_M$ and $H_0$ with fixed evolution]{\label{fig10_c}
\includegraphics[width=0.50\hsize,height=0.50\textwidth,angle=0,clip]{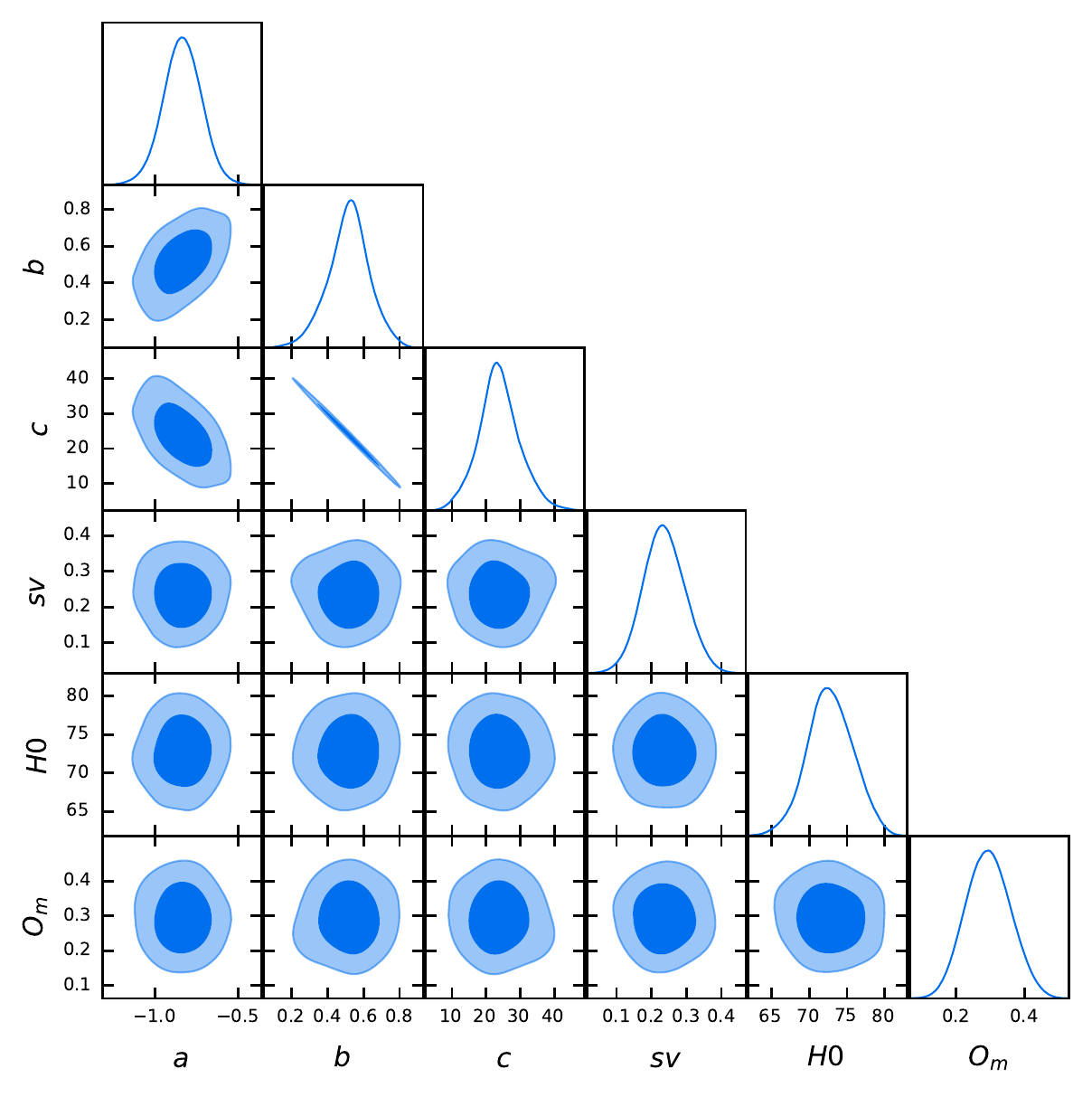}}
\subfloat[Varying only $w$ with fixed evolution]{\label{fig10_d}
\includegraphics[width=0.50\hsize,height=0.50\textwidth,angle=0,clip]{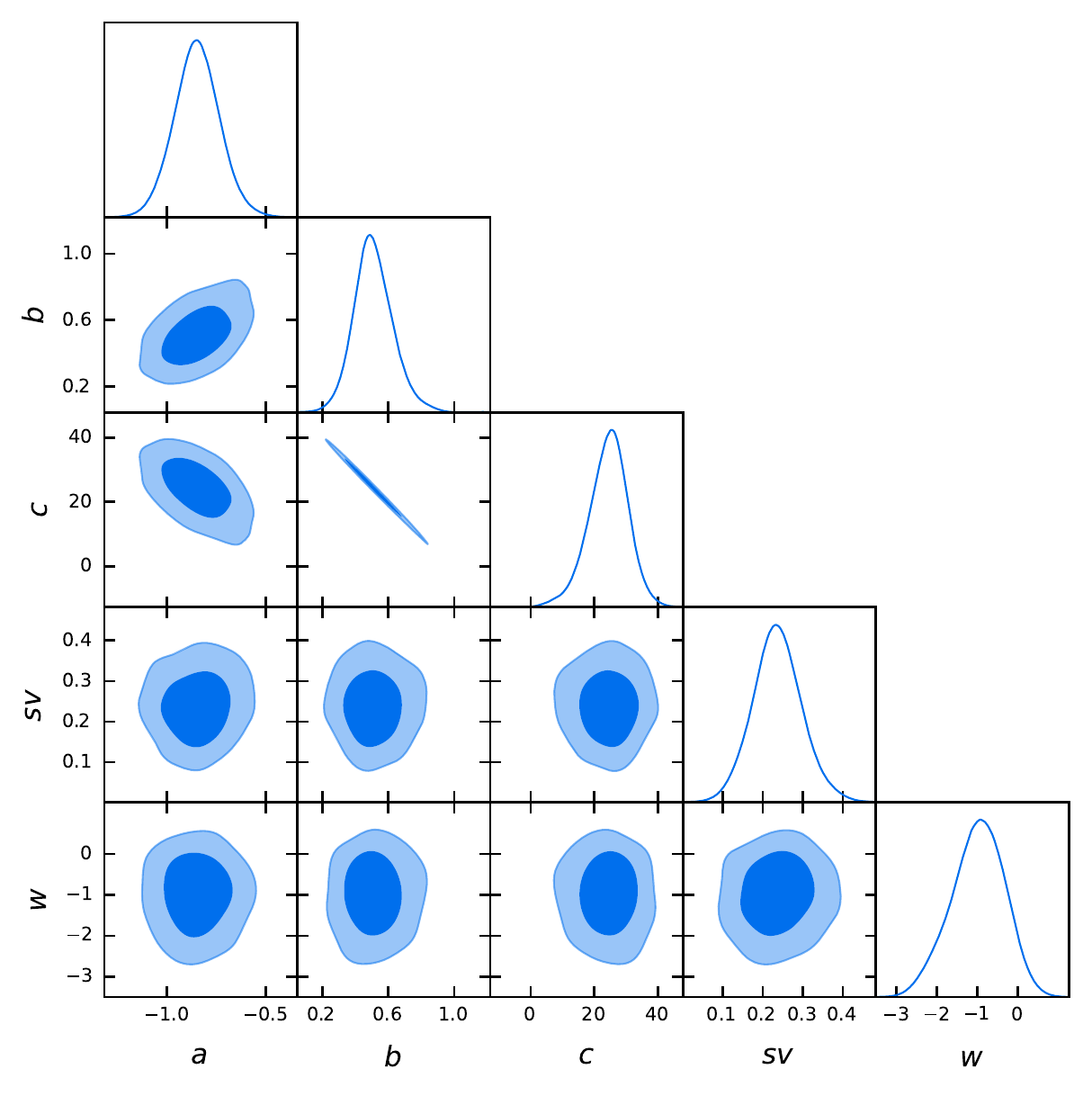}}
\caption{Cosmological results for the GRBs alone with no calibration with Fundamental Plane using fixed evolution on 3 $\sigma$ Gaussian priors on the cosmological parameters investigated following \citet{Scolnic}. Panels a), b) c) and d) show the contours from the case (ii) for the case of $\Omega_M$, $H_0$, $\Omega_M$ and $H_0$ together, and $w$.}
\label{fig8}
\end{figure}
\begin{table}
\centering
\scalebox{0.84}{\begin{tabular}{p{35mm}|l|c|c|c|c|c|c|c|c}
 \toprule[1.2pt]
 \toprule[1.2pt]
\textbf{Calibration with SNe Ia, Equation \ref{equmu}} & {\textbf{parameters varied}} & \textbf{Model} & $\boldsymbol{\Omega_{M}}$ & $\boldsymbol{H_{0}}$ & $\boldsymbol{w}$ & $\boldsymbol{\Delta^{GRB_{ C}}_{GRB_{NO-C}}}\%$ & $z-score_{SN}$ & $z-score_{SN+BAO}$ \\ 
\midrule
without evolution &$\Omega_{M}$ & $\Lambda$CDM & $0.292 \pm 0.068$ & \bf{70} & \bf{-1} & 7.94 & 0.102 & 0.176 \\\hline
without evolution & $H_0$ & $\Lambda$CDM & \bf{0.30} & $73.286\pm 3.007$ & \bf{-1} & -9.07 & 1.102 & 1.105\\\hline
without evolution & $\Omega_{M}$ and $H_0$ & $\Lambda$CDM & $0.295 \pm 0.064 $ & $73.358\pm 3.006$ & \bf{-1} & -5.88, -0.66 & 0.044, 1.103 & 0.249,1.176 \\\hline
without evolution & $w$ & $w$CDM & \bf{0.30} & \bf{70} & $-1.094 \pm 0.673$ & -6.13 & 0.140 & 0.114 \\\hline
\midrule
with fixed evolution & $\Omega_{M}$ & $\Lambda$CDM & $0.316 \pm 0.068$ & \bf{70} & \bf{-1} & 7.94 & 0.249 & 0.176 \\\hline
with fixed evolution & $H_0$ & $\Lambda$CDM & \bf{0.30} & $72.762 \pm 3.227$ & \bf{-1} & 10.48 & 0.864 & 0.868 \\\hline
with fixed evolution & $\Omega_{M}$ and $H_0$ & $\Lambda$CDM & $0.306 \pm 0.060 $ & $73.264\pm 3.082$ & \bf{-1} & -6.25, 2.46 & 0.125, 1.046 & 0.083, 1.116 \\\hline
with fixed evolution & $w$ & $w$CDM & \bf{0.30} & \bf{70} & $-0.743 \pm 0.694$ & 11.94 & 0.370 & 0.395 \\\hline
\midrule
with $k=k(\Omega_{M})$ & $\Omega_{M}$ & $\Lambda$CDM & $0.298 \pm 0.063$ & \bf{70} & \bf{-1} &1.61 & 0.016 & 0.095 \\\hline
with $k=k(\Omega_{M})$ & $\Omega_{M}$ and $H_0$ & $\Lambda$CDM & $0.295 \pm 0.064 $ & $73.159\pm 3.134$ & \bf{-1} & -1.54, -0.16 & 0.044, 0.996 & 0.249, 1.064 \\\hline
\toprule[1.2pt]
 \toprule[1.2pt]
\textbf{Calibration with SNe Ia, Equations \ref{isotropic} and \ref{planeev}} & {\textbf{parameters varied}} & \textbf{Model} & $\boldsymbol{\Omega_{M}}$ & $\boldsymbol{H_{0}}$ & $\boldsymbol{w}$ & $\boldsymbol{\Delta^{GRB_{ C}}_{GRB_{NO-C}}}\%$ & $z-score_{SN}$ & $z-score_{SN+BAO}$ \\ 
\midrule
without evolution & $\Omega_{M}$ & $\Lambda$CDM & $0.306 \pm 0.069$ & \bf{70} & \bf{-1} & 13.11 & 0.101 & 0.029 \\\hline
without evolution & $H_0$ & $\Lambda$CDM & \bf{0.30} & $73.519\pm 3.119$ & \bf{-1} & 0.19 & 1.137 & 1.140 \\\hline
without evolution & $\Omega_{M}$ and $H_0$ & $\Lambda$CDM & $0.301 \pm 0.065 $ & $73.089\pm 3.251$ & \bf{-1} & 1.56, 3.37 & 0.044, 0.939 & 0.083, 1.004\\\hline
without evolution & $w$ & $w$CDM & \bf{0.30} & \bf{70} & $-0.906 \pm 0.697$ & -33.49 & 0.159 & 0.159\\\hline
\midrule
with fixed evolution & $\Omega_{M}$ & $\Lambda$CDM & $0.295 \pm 0.060$ & \bf{70} & \bf{-1} & -7.69 & 0.066 & 0.149 \\\hline
with fixed evolution & $H_0$ & $\Lambda$CDM & \bf{0.30} & $73.272 \pm 3.143$ & \bf{-1} & 0.54 & 1.050 & 1.140 \\\hline
with fixed evolution & $\Omega_{M}$ and $H_0$ & $\Lambda$CDM & $0.296 \pm 0.066 $ & $73.201\pm 3.062$ & \bf{-1} & 1.54, 0.43 & 0.029, 1.033 & 0.226, 1.103 \\\hline
with fixed evolution & $w$ & $w$CDM & \bf{0.30} & \bf{70} & $-0.959 \pm 0.631$ & -4.68 & 0.065 & 0.0918 \\\hline
\midrule
with $k=k(\Omega_{M})$ & $\Omega_{M}$ & $\Lambda$CDM & $0.300 \pm 0.073$ & \bf{70} & \bf{-1} & 15.87 & 0.014 & 0.055\\\hline
with $k=k(\Omega_{M})$ & $\Omega_{M}$ and $H_0$ & $\Lambda$CDM & $0.296 \pm 0.064 $ & $73.024\pm 3.073$ & \bf{-1} & 0, -0.90 & 0.030, 0.972 & 0.233,1.042 \\\hline
\end{tabular}}
\caption{Cosmological parameters obtained using GRB alone calibrated on SNe Ia (indicated with the subscript C) using Gaussian priors with distance modulus equation (first part) and with the Fundamental plane equation (2nd part) without evolution, with evolution with fixed parameters, and with evolution correction as a function of $\Omega_{M}$. In the first column we define the studied type of the correction for the evolution. In the second column we define which cosmological parameters are left free to vary, while in the third column we define the investigated cosmological model. In the next three columns we present the values of parameters with error bars obtained in computation for each given case, namely with the fixed cosmological parameters present in bold.
In the 7th column, we show the percentage increase/decrease between the uncertainties of GRBs alone without calibration (indicated with No-C) and with Gaussian priors from Table \ref{Table1} (the ${\Delta_{GRB} \%}$) vs the results from this Table with calibration (indicated with C) and with Gaussian priors. The formula for the percentage change is: $\frac{\Delta_{\text{comparing}}-\Delta_{\text{reference}}}{\Delta_{\text{reference}}}$, where $\Delta_{\text{reference}}$ is the error obtained with GRBs without calibration. With the negative sign we indicate a percentage decrease in the error, while with the positive one, we indicate a percentage increase in the error. Finally, in the last two columns we show the z-scores computed with respect to the SNe Ia alone and SNe Ia+ BAO results, respectively.}
\label{Table2}
\end{table}
\begin{figure} 
\centering
\subfloat[Varying only $\Omega_M$ with evolutionary function]{\label{fig11_a}
\includegraphics[width=0.50\hsize,height=0.50\textwidth,angle=0,clip]{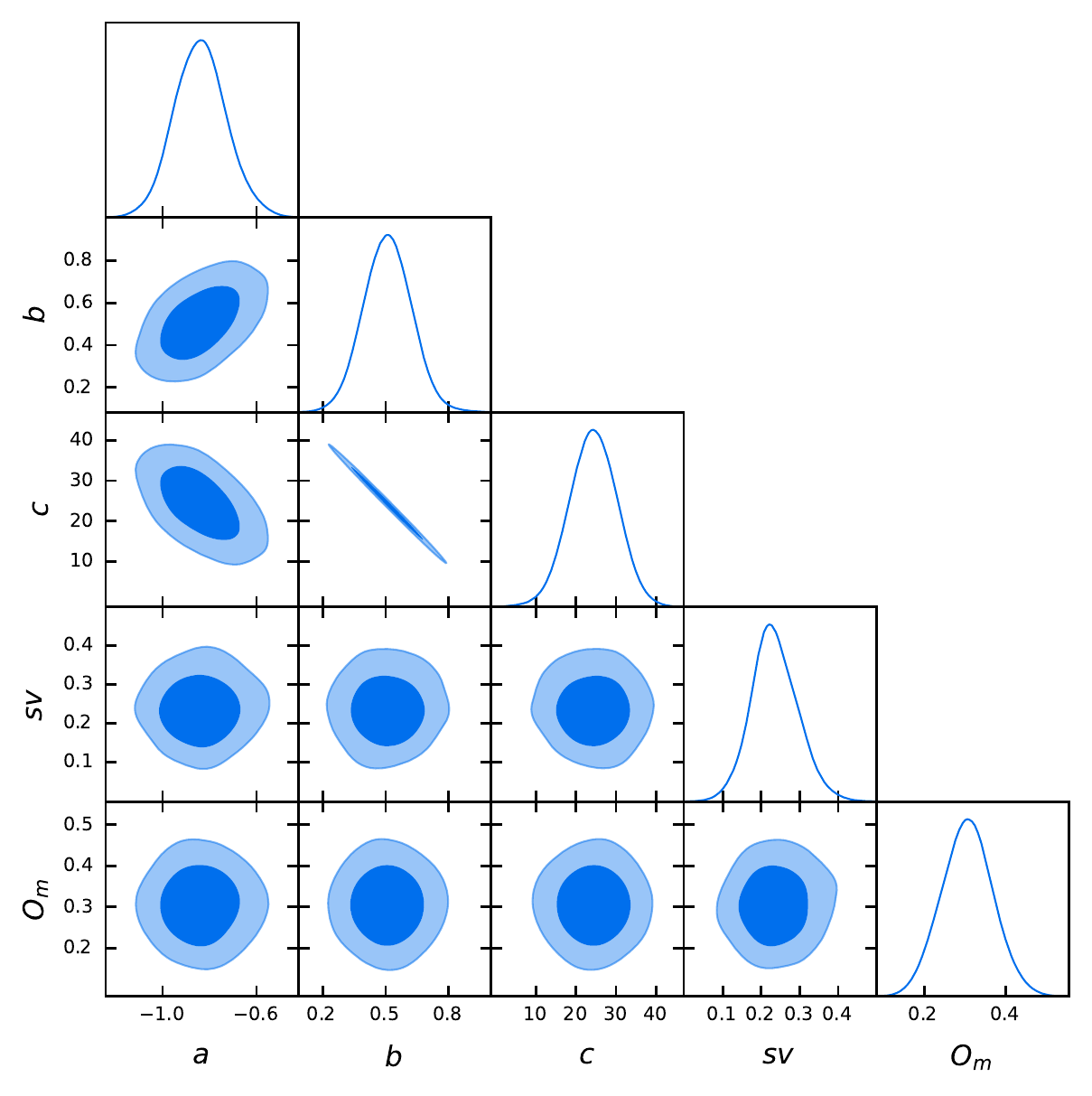}}
\subfloat[Varying both $\Omega_M$ and $H_0$ with evolutionary function]{\label{fig11_c}
\includegraphics[width=0.50\hsize,height=0.50\textwidth,angle=0,clip]{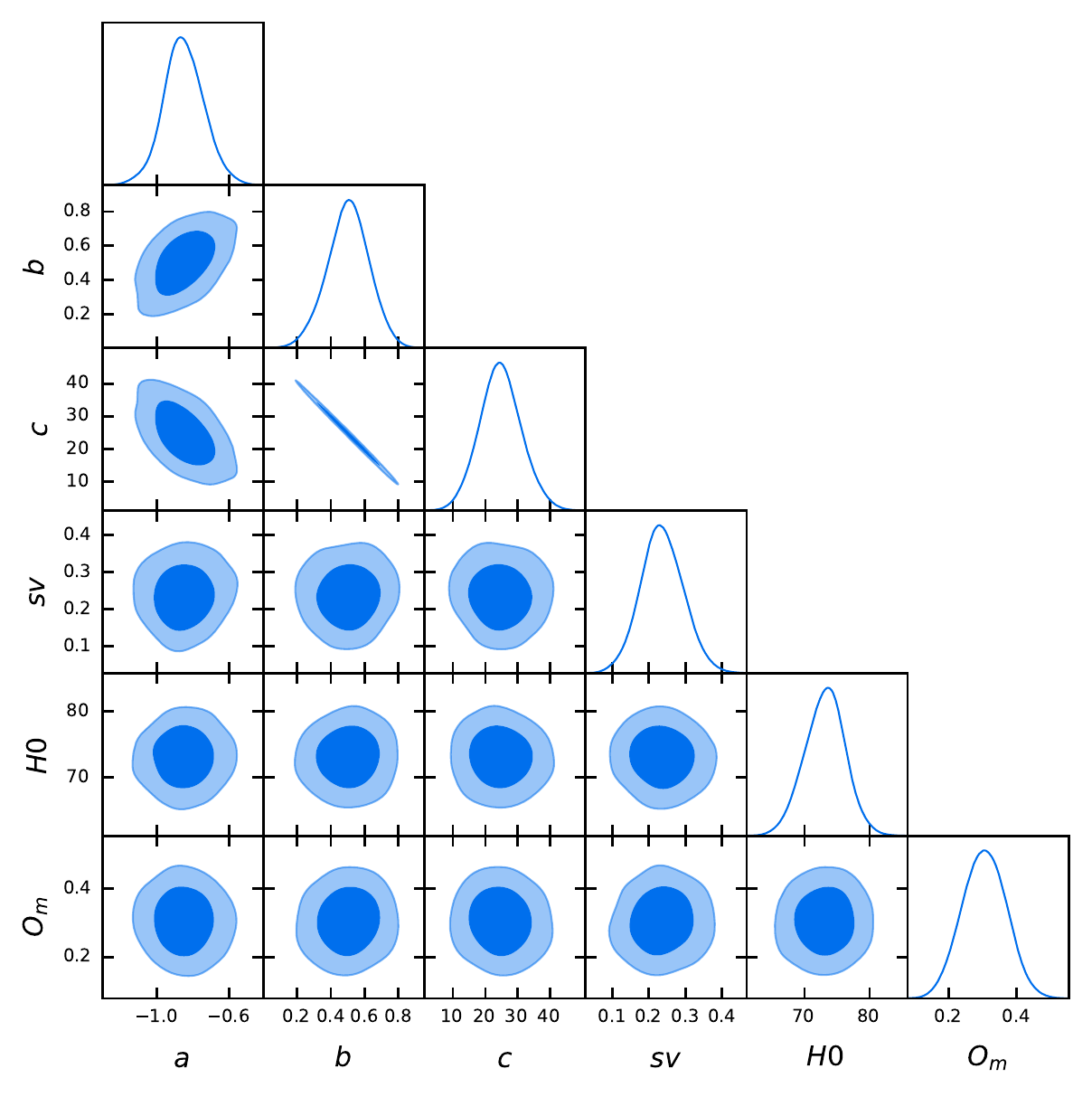}}
\caption{Cosmological results for the GRBs alone (with no calibration) with Fundamental Plane using evolutionary functions and the assumptions of 3 $\sigma$ Gaussian priors on the cosmological parameters investigated following \citet{Scolnic}. Panels a) and b) show the contours from case (iii) for the case of $\Omega_M$ and the case of $\Omega_M$ and $H_0$ together, respectively. }
\label{fig9}
\end{figure}

\begin{figure} 
\centering
\subfloat[Varying only $\Omega_M$ without correction for evolution]{\label{fig12_a}
\includegraphics[width=0.50\hsize,height=0.50\textwidth,angle=0,clip]{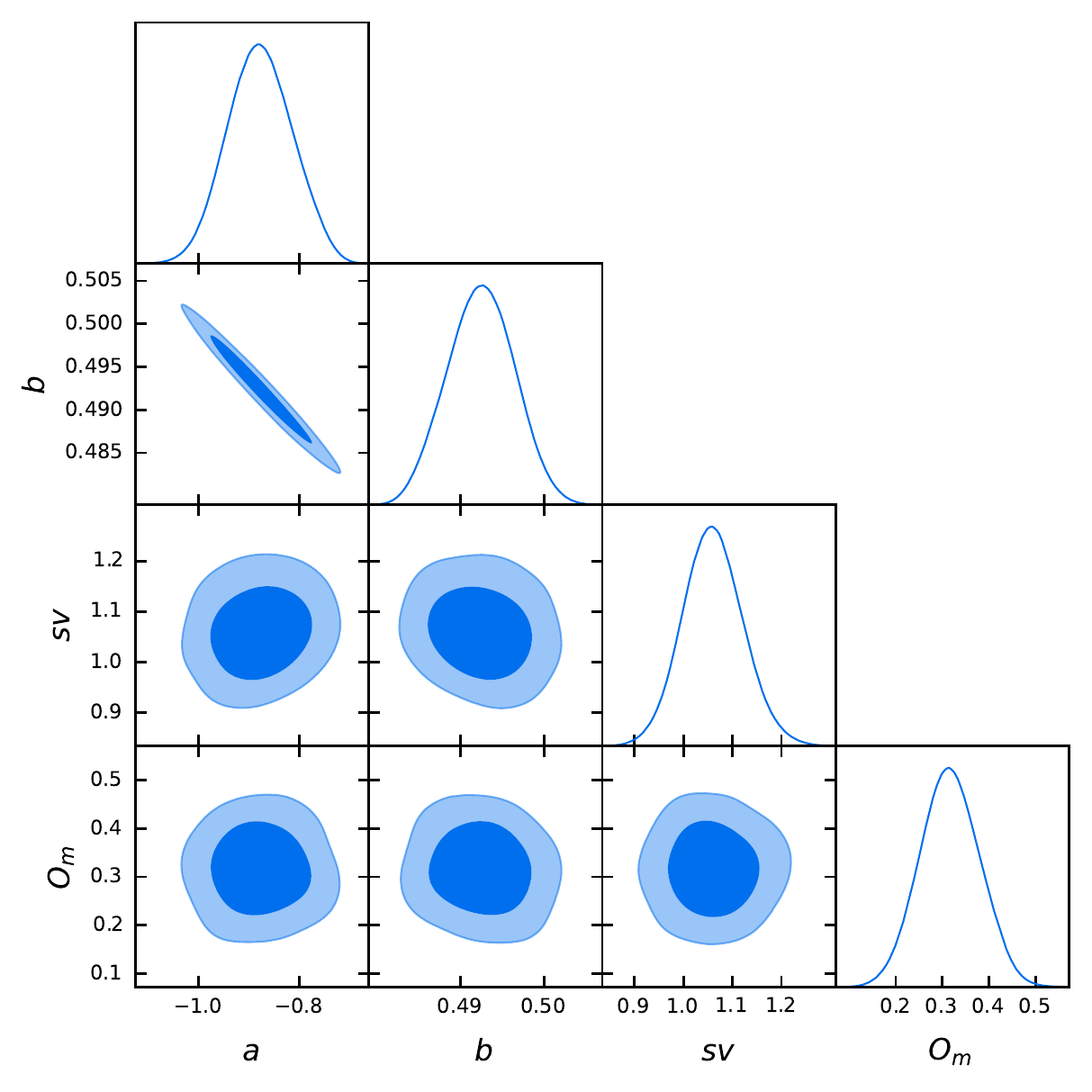}}
\subfloat[Varying only $H_0$ without correction for evolution]{\label{fig12_b}
\includegraphics[width=0.50\hsize,height=0.50\textwidth,angle=0,clip]{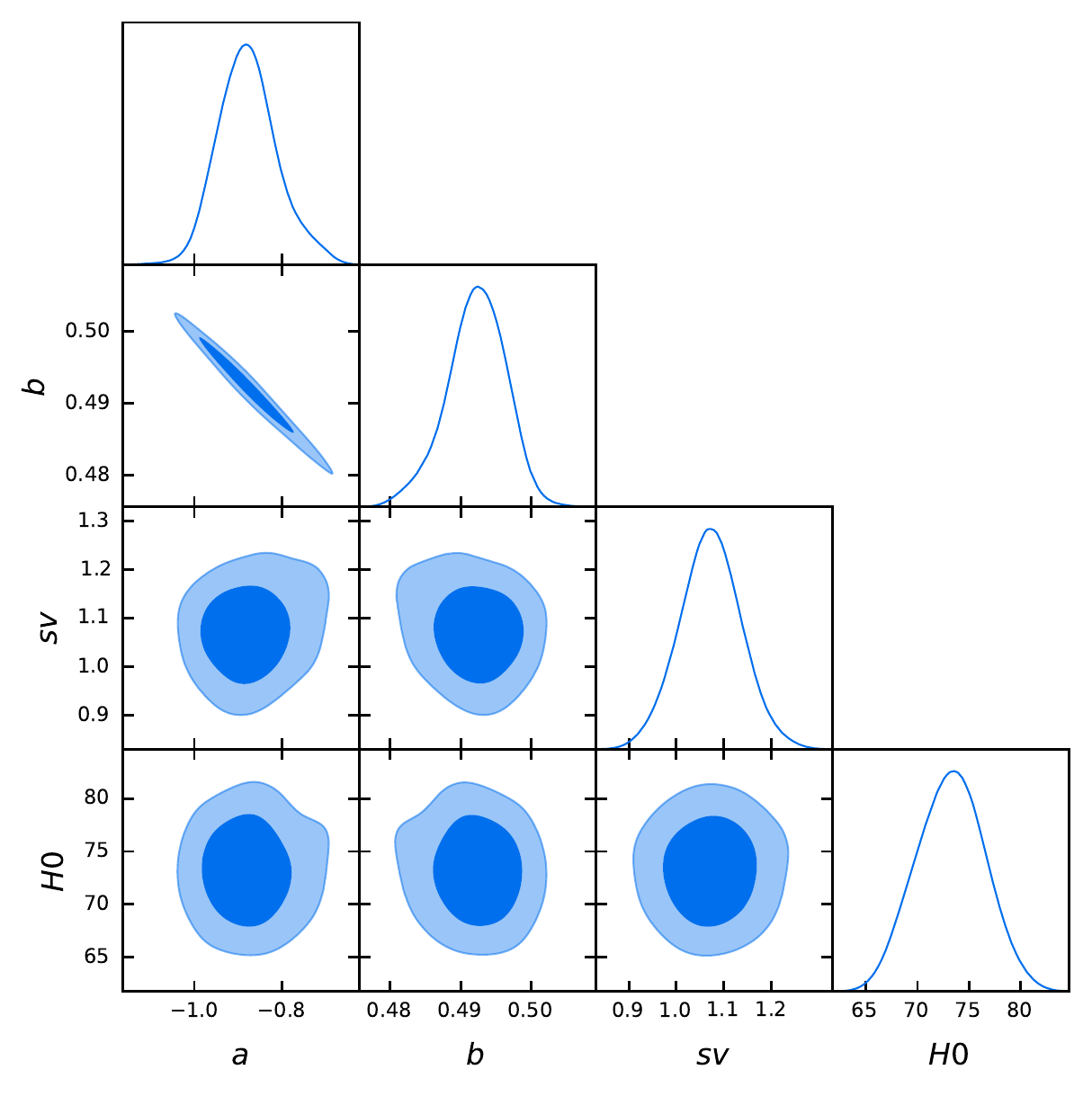}}\\
\subfloat[Varying both $\Omega_M$ and $H_0$ without correction for evolution]{\label{fig12_c}
\includegraphics[width=0.50\hsize,height=0.50\textwidth,angle=0,clip]{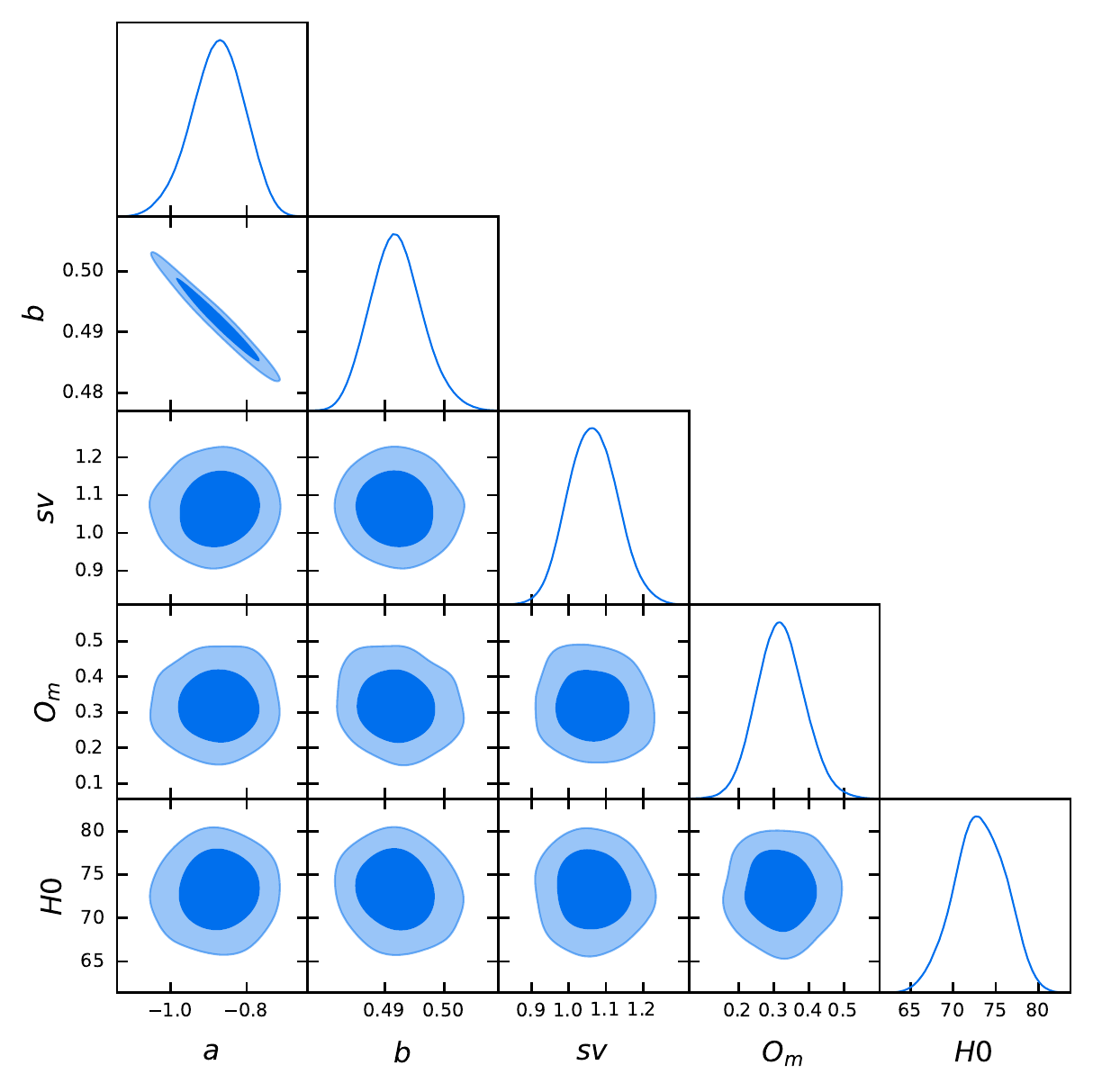}}
\subfloat[Varying only $w$ without correction for evolution]{\label{fig12_d}
\includegraphics[width=0.50\hsize,height=0.50\textwidth,angle=0,clip]{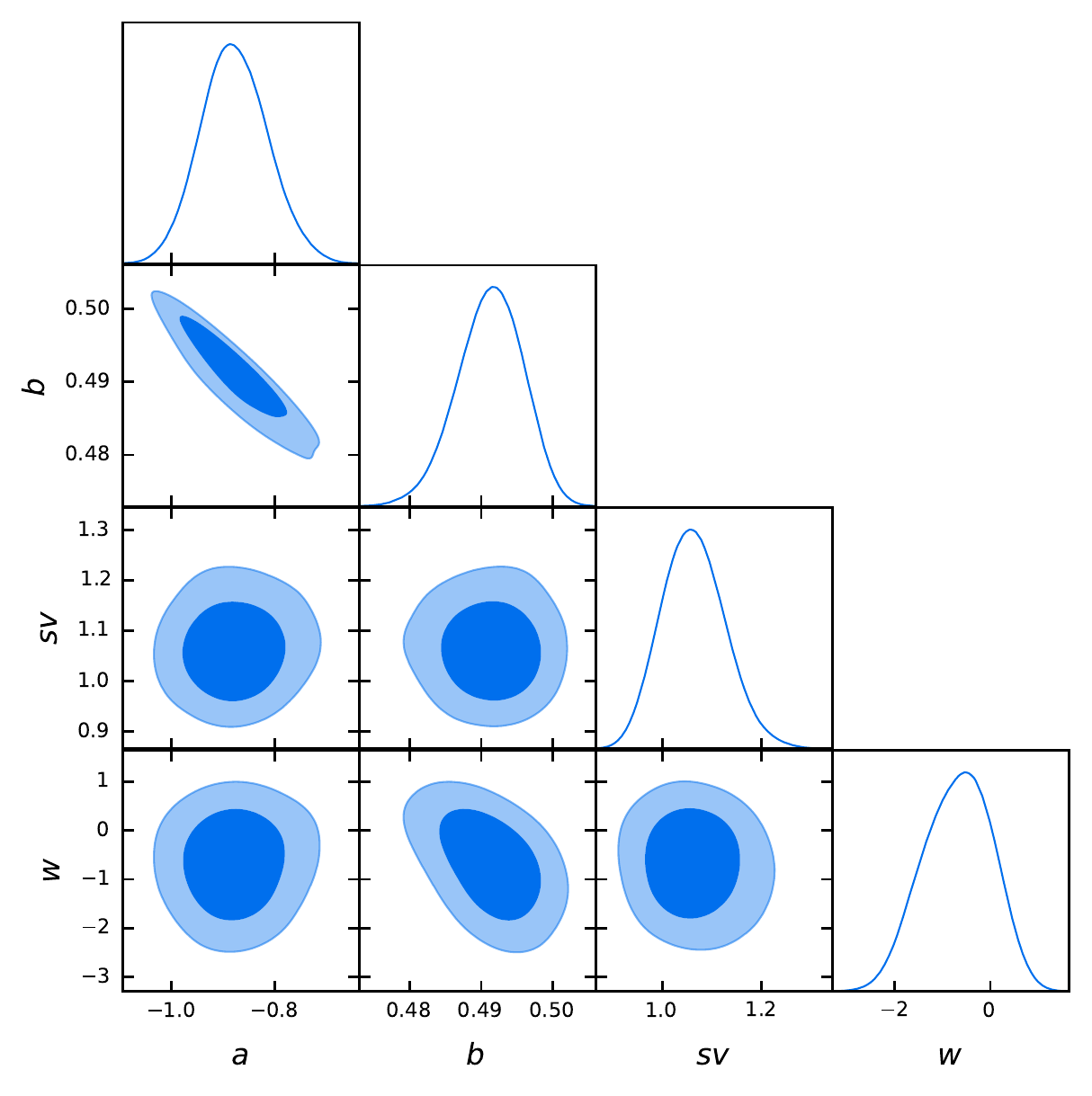}}
\caption{Cosmological results for the GRBs alone (with no calibration) with $\mu_{GRB}$, see Equation \ref{equmu}, without evolution and the assumptions of 3 $\sigma$ Gaussian priors on the cosmological parameters investigated following \citet{Scolnic}. Panels a), b), c) and d) show the contours from the case (iv) for the case of $\Omega_M$, $H_0$, $\Omega_M$ and $H_0$ together, and $w$, respectively. }
\label{fig10}
\end{figure}

\begin{figure} 
\centering
\subfloat[Varying only $\Omega_M$ with fixed evolution]{\label{fig13_a}
\includegraphics[width=0.50\hsize,height=0.50\textwidth,angle=0,clip]{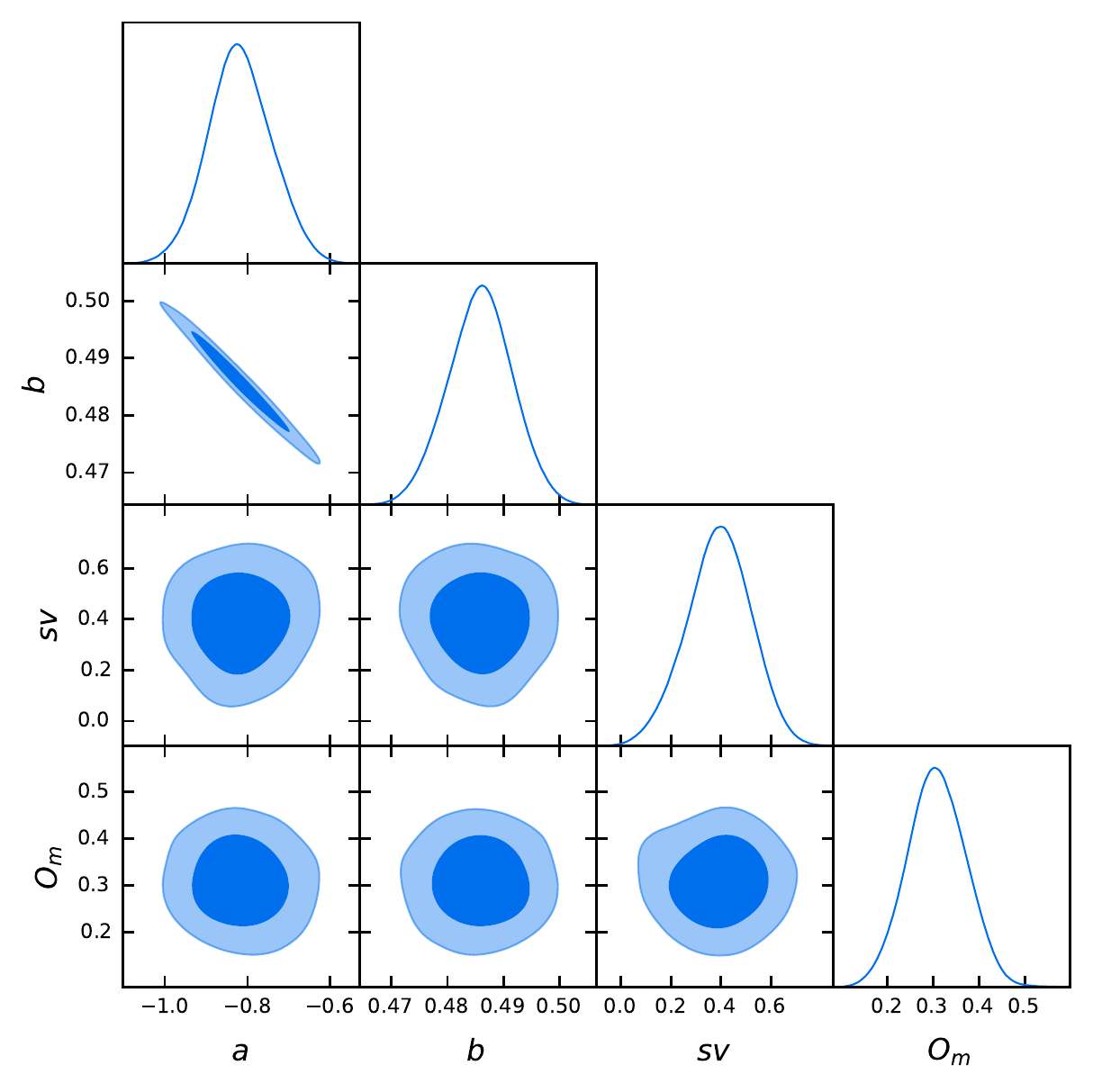}}
\subfloat[Varying only $H_0$ with fixed evolution]{\label{fig13_b}
\includegraphics[width=0.50\hsize,height=0.50\textwidth,angle=0,clip]{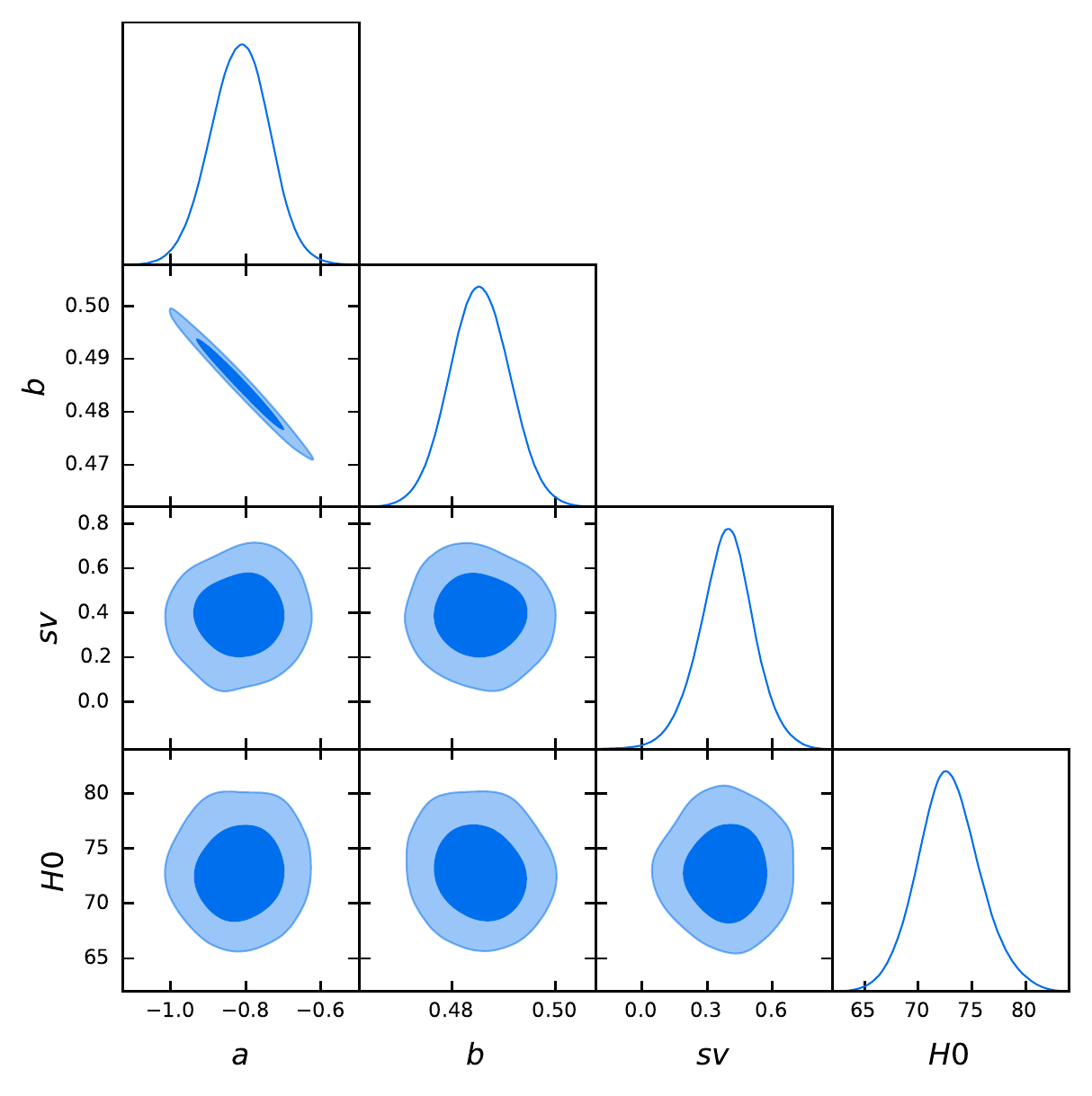}}\\
\subfloat[Varying both $\Omega_M$ and $H_0$ with fixed evolution]{\label{fig13_c}
\includegraphics[width=0.50\hsize,height=0.50\textwidth,angle=0,clip]{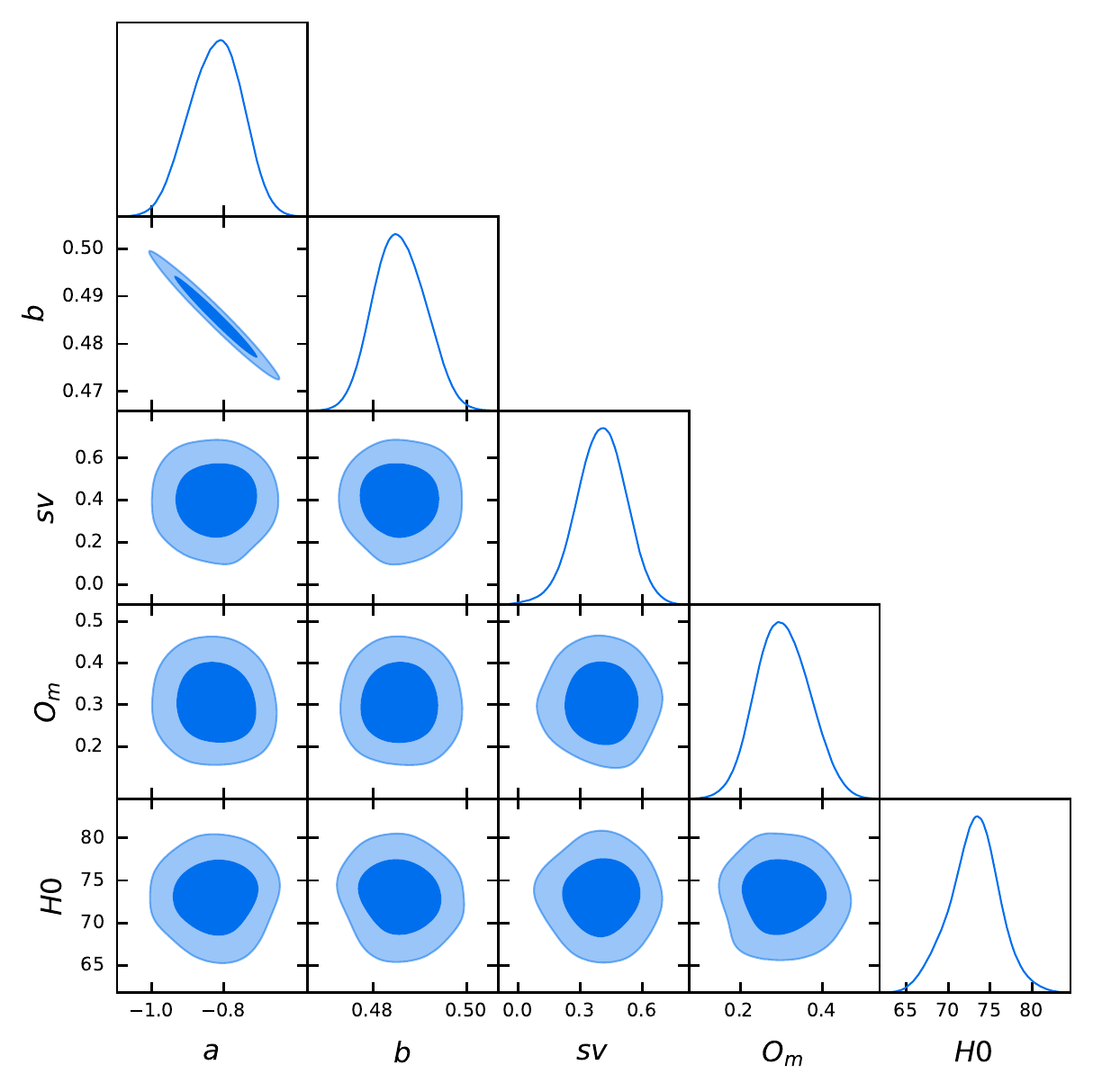}}
\subfloat[Varying only $w$ with fixed evolution]{\label{fig13_d}
\includegraphics[width=0.50\hsize,height=0.50\textwidth,angle=0,clip]{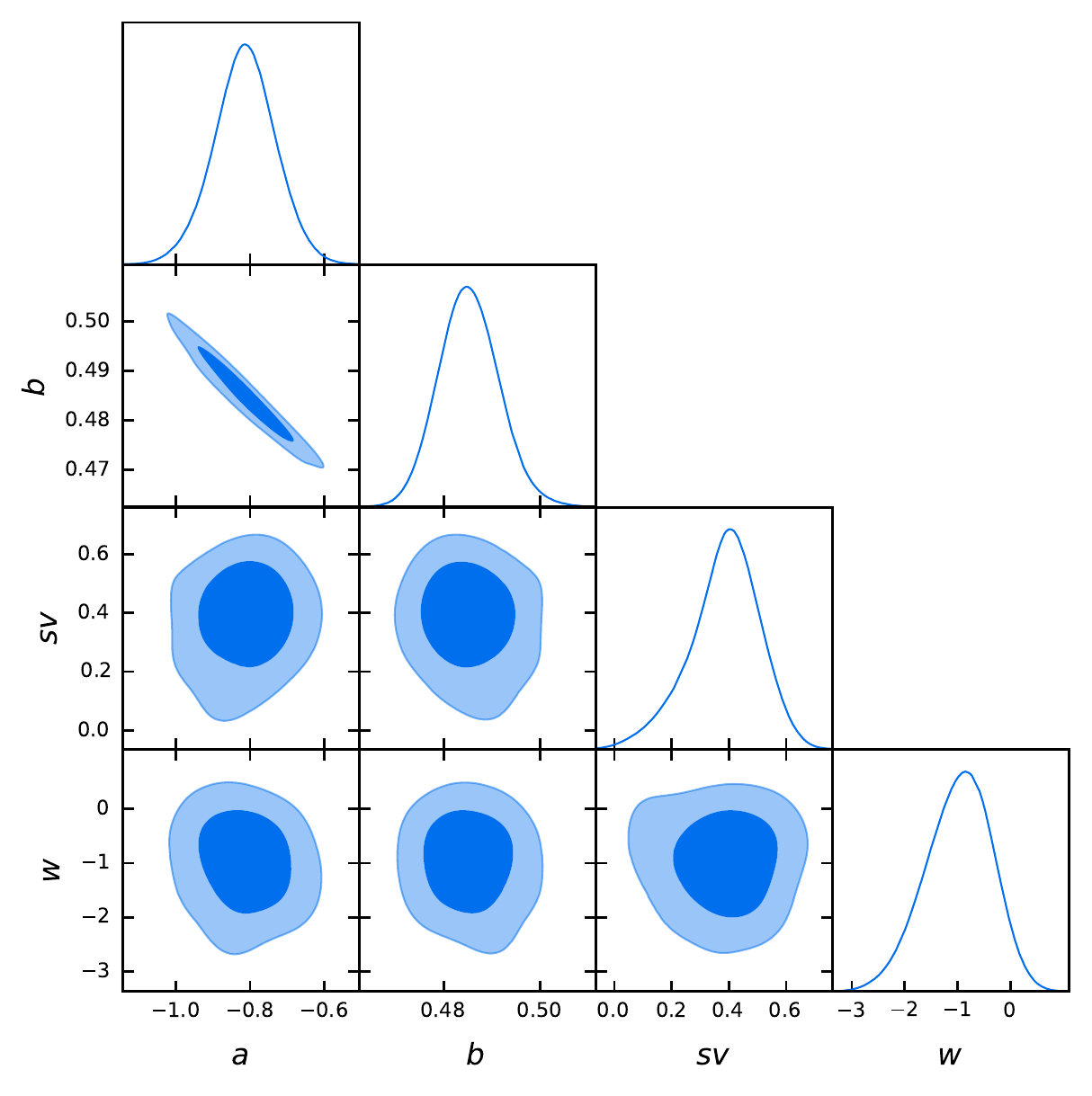}}
\caption{Cosmological results for the GRBs alone (with no calibration) with $\mu_{GRB}$, see Equation \ref{equmu}, using fixed evolution and the assumptions of 3 $\sigma$ Gaussian priors on the cosmological parameters investigated following \citet{Scolnic}. Panels a), b), c) and d) show the contours from case (v) for the case of $\Omega_M$, $H_0$, $\Omega_M$ and $H_0$ together, and $w$, respectively. }
\label{fig11}
\end{figure}

\begin{figure} 
\centering
\subfloat[Varying only $\Omega_M$ with evolutionary function]{\label{fig14_a}
\includegraphics[width=0.50\hsize,height=0.50\textwidth,angle=0,clip]{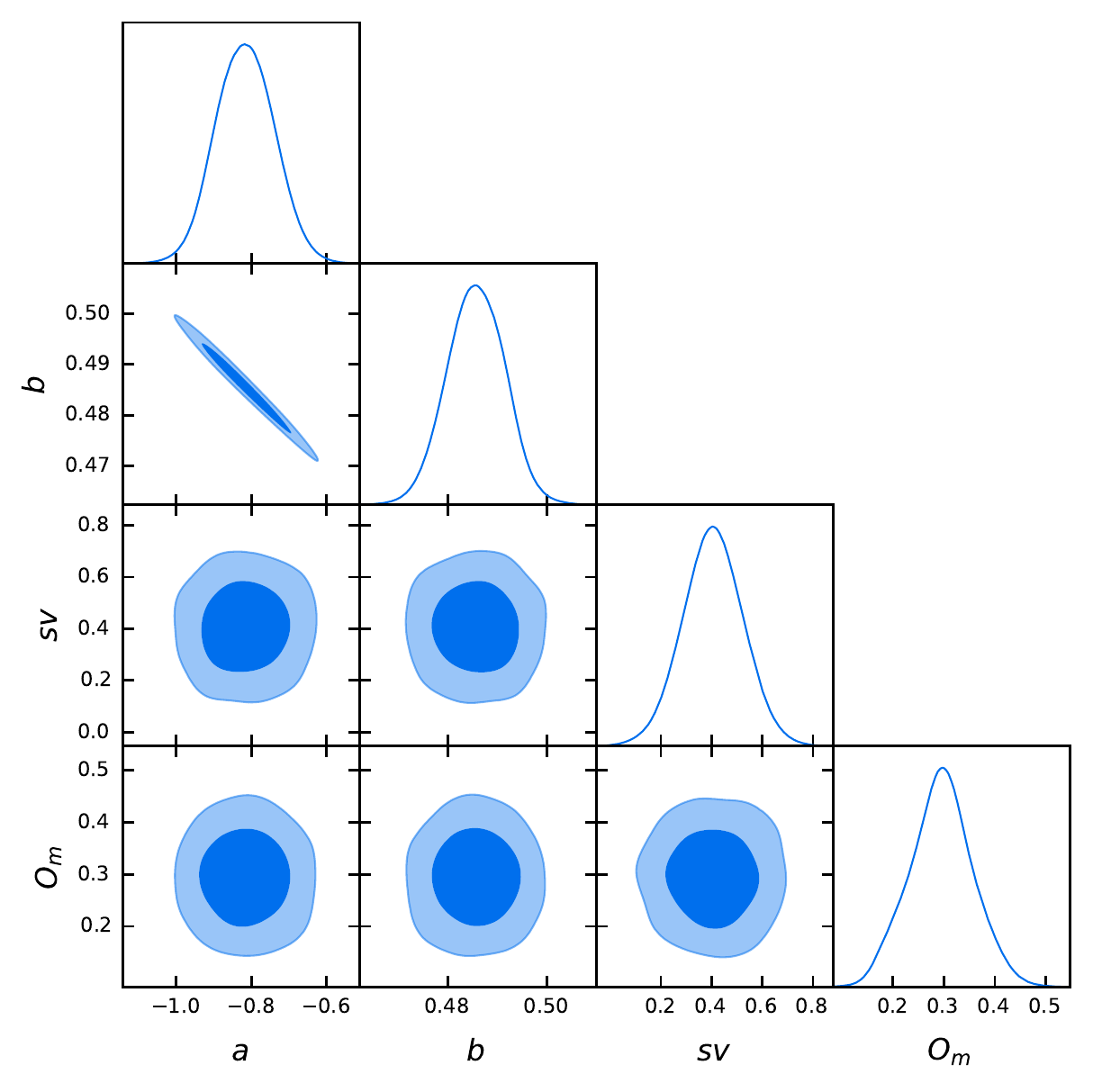}}
\subfloat[Varying both $\Omega_M$ and $H_0$ with evolutionary function]{
\includegraphics[width=0.50\hsize,height=0.50\textwidth,angle=0,clip]{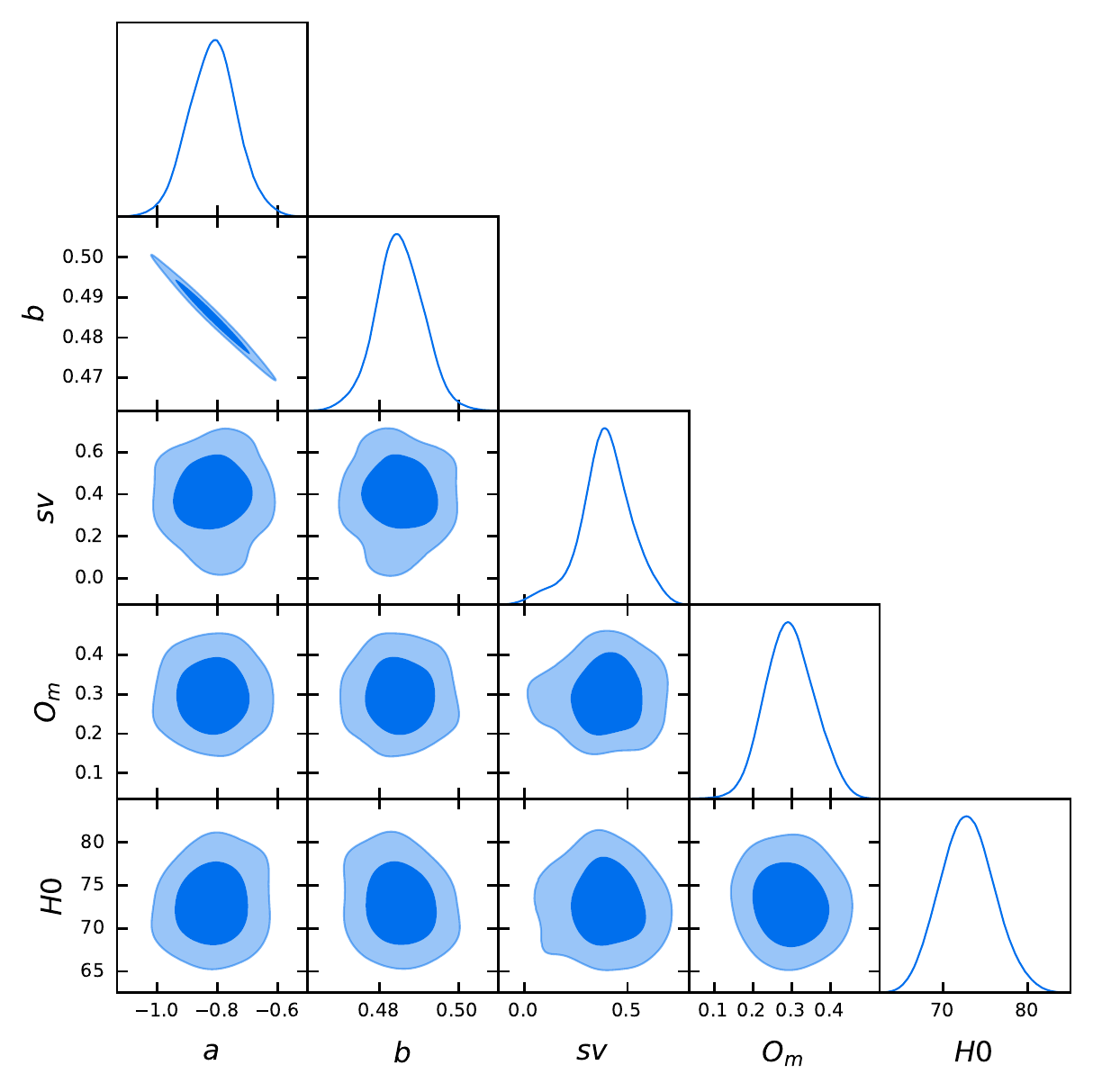}}
\caption{Cosmological results for the GRBs alone (with no calibration) with $\mu_{GRB}$ using evolutionary functions and the assumptions of 3 $\sigma$ Gaussian priors on the cosmological parameters investigated following \citet{Scolnic}. Panels a) and b) show the contours from the case (vi) for the case of $\Omega_M$ and the case of $\Omega_M$ and $H_0$ together, respectively. }
\label{fig12}
\end{figure}

\begin{enumerate}
 \item \hspace{1mm} We assume a likelihood with Equations \ref{Lpeak} and \ref{isotropic} without evolution, see Fig. \ref{fig7}.
 \item \hspace{1mm} We assume a likelihood with Equation \ref{planeev} with the evolution considering fixed values of $k_{L_{peak}}$, $k_{L_{a}}$ and $k_{T_{X}}$, see Fig. \ref{fig8}.
 \item \hspace{1mm} We assume a likelihood with Equation \ref{planeev} and considering a function for the evolutionary parameters $k_{L_{peak}}$ and $k_{L_{a}}$, since they vary together with the cosmological parameters, and we fix the $k_{T_{X}}$, since it does not depend on the cosmological models, see Fig. \ref{fig9}.
 \item \hspace{1mm} The likelihood from Equation \ref{equmu}, $\mu_{GRB}$, without evolution, see Fig. \ref{fig10}.
 \item \hspace{1mm} The likelihood from Equation \ref{equmu}, $\mu_{GRB}$, for the fixed evolutionary parameters, see Fig. \ref{fig11}.
 \item \hspace{1mm} The likelihood from Equation \ref{equmu}, $\mu_{GRB}$, for the evolutionary parameters $k_{L_{a}}$ and $k_{L_{peak}}$ as functions of $\Omega_M$, see Fig. \ref{fig12}.
\end{enumerate}

The Gaussian priors are justified by the fact that the underlying physics of the fundamental plane is not expected to vary within any given cosmology since it relies on a fundamental physical process, the magnetar emission, which gives rise to the plateau and, in turn, naturally drives the anti-correlation between $L_X$ and $T_a$ \citep{Rowlinson,Rea2015,Stratta}, and the prompt kinetic energy is positively correlated with the kinetic power in the afterglow \citep{dainotti2011,dainotti2015b}, as it is predicted within the standard fireball model assuming microphysical variations \citep{van Eerten2014a, van Eerten2014b}.

\begin{figure} 
\centering
\subfloat[Plane parameters without evolution]{\label{fig15_a}
\includegraphics[width=0.440\hsize,height=0.4365\textwidth,angle=0,clip]{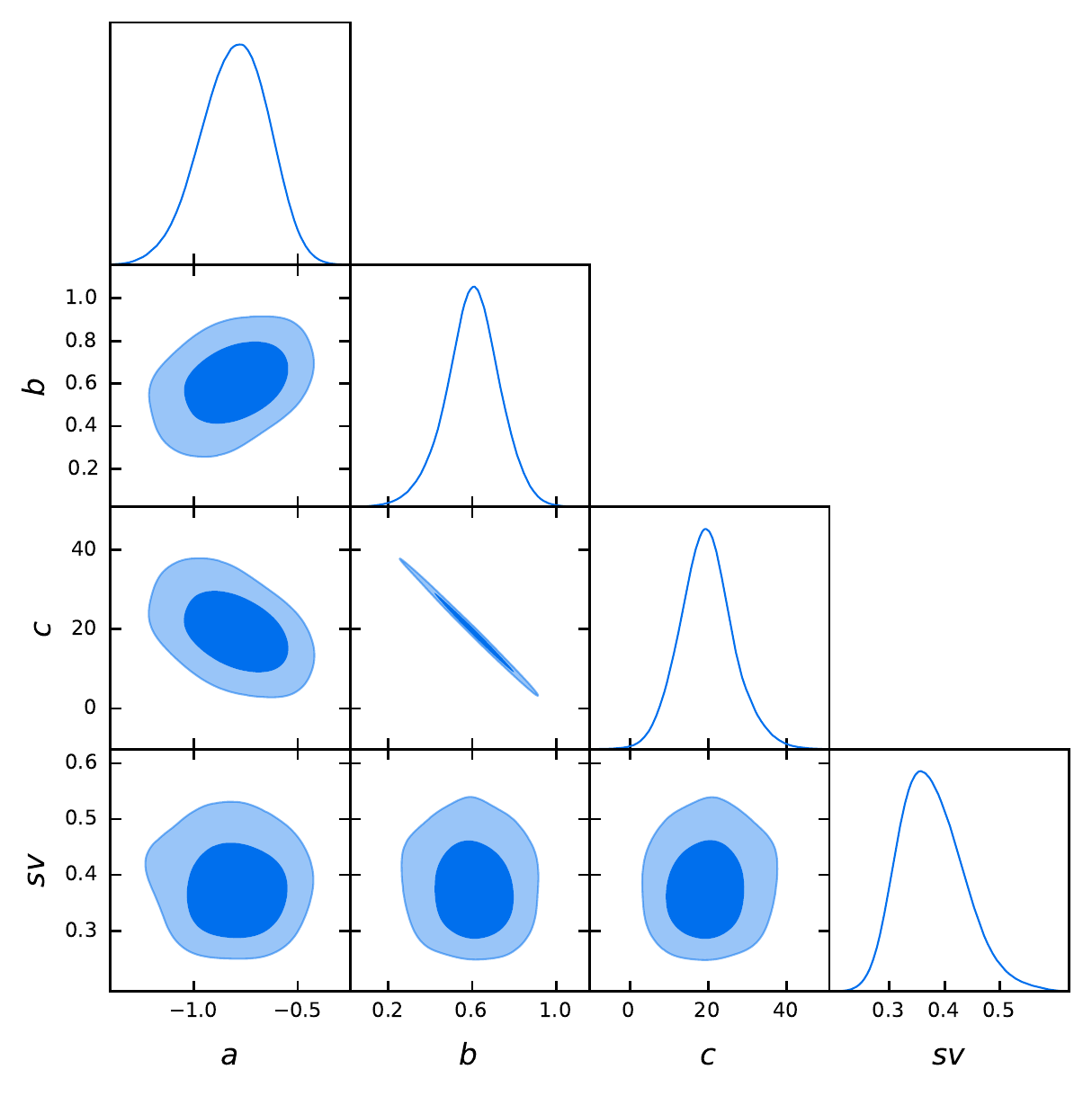}}
\subfloat[Plane parameters with fixed evolution]{\label{fig15_b}
\includegraphics[width=0.440\hsize,height=0.4365\textwidth,angle=0,clip]{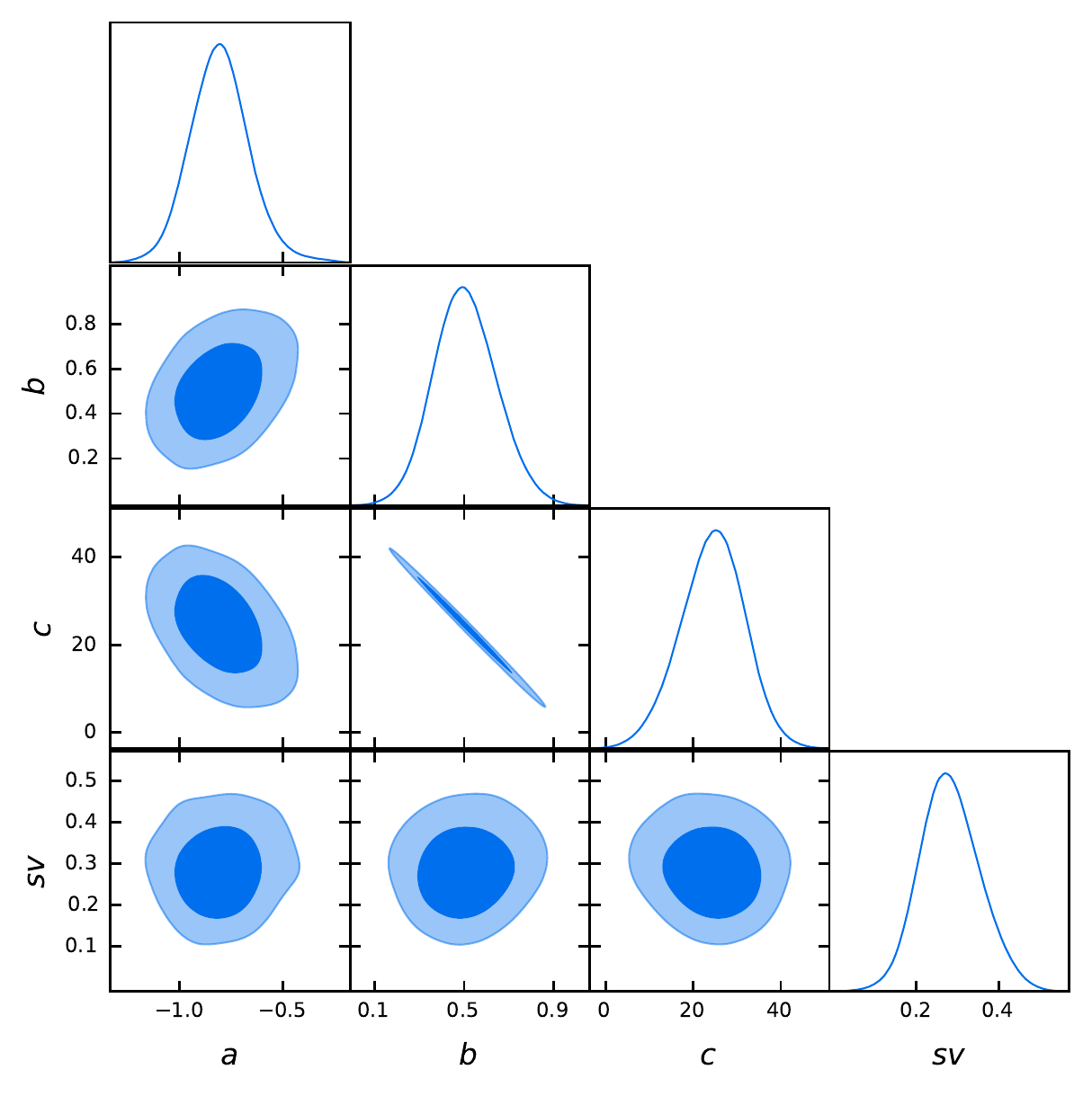}}
\caption{The Fundamental plane relation parameters with the nearest 25 GRBs used to calibrate them on SNe Ia using $\mu_{GRB}$, Equation \ref{equmu}. Panels a) and b) show the contours of the plane fitting parameters without evolution and with fixed evolution respectively.}
\label{fig13}
\end{figure}
\begin{figure}
\centering

\subfloat[Using $\mu_{GRB}$ on full sample varying only\\ $\Omega_M$ without correction for evolution]{\label{fig16_a}
\includegraphics[width=0.40\hsize,height=0.36\textwidth,angle=0,clip]{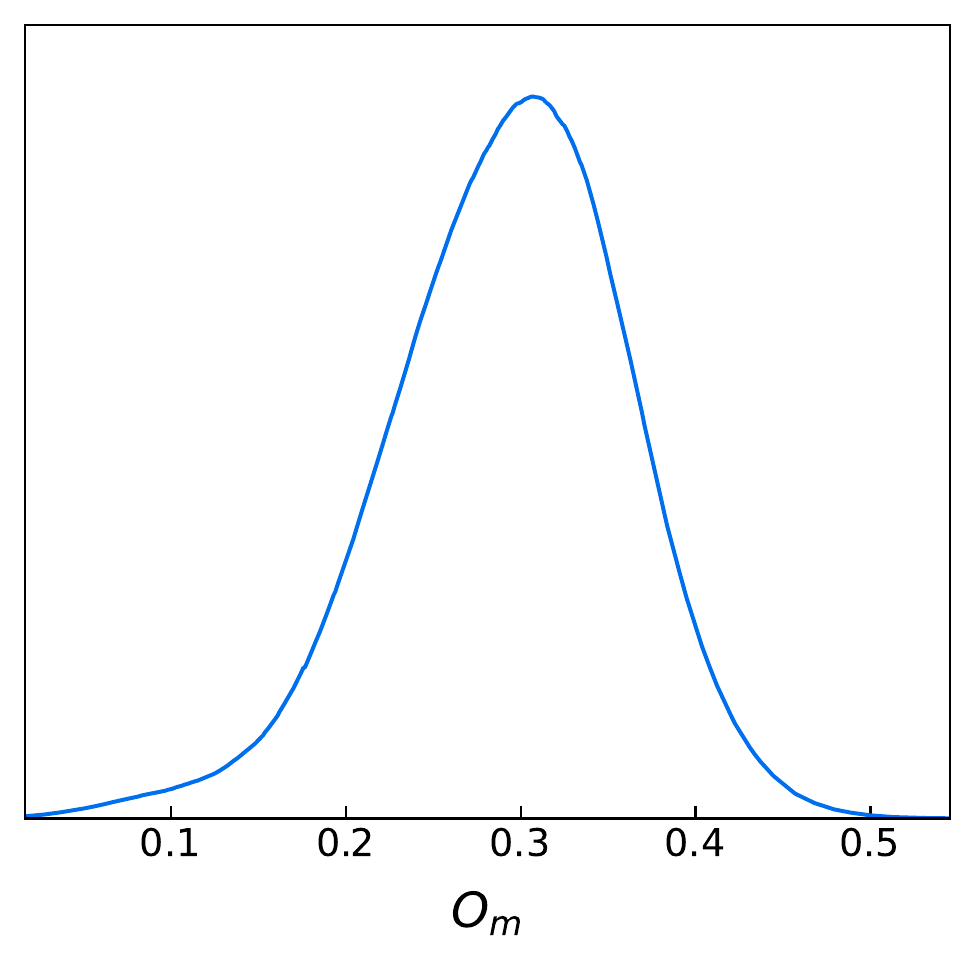}}\hspace{5mm}
\subfloat[Using $\mu_{GRB}$ on full sample varying only\\ $H_0$ without correction for evolution]{\label{fig16_b}
\includegraphics[width=0.40\hsize,height=0.36\textwidth,angle=0,clip]{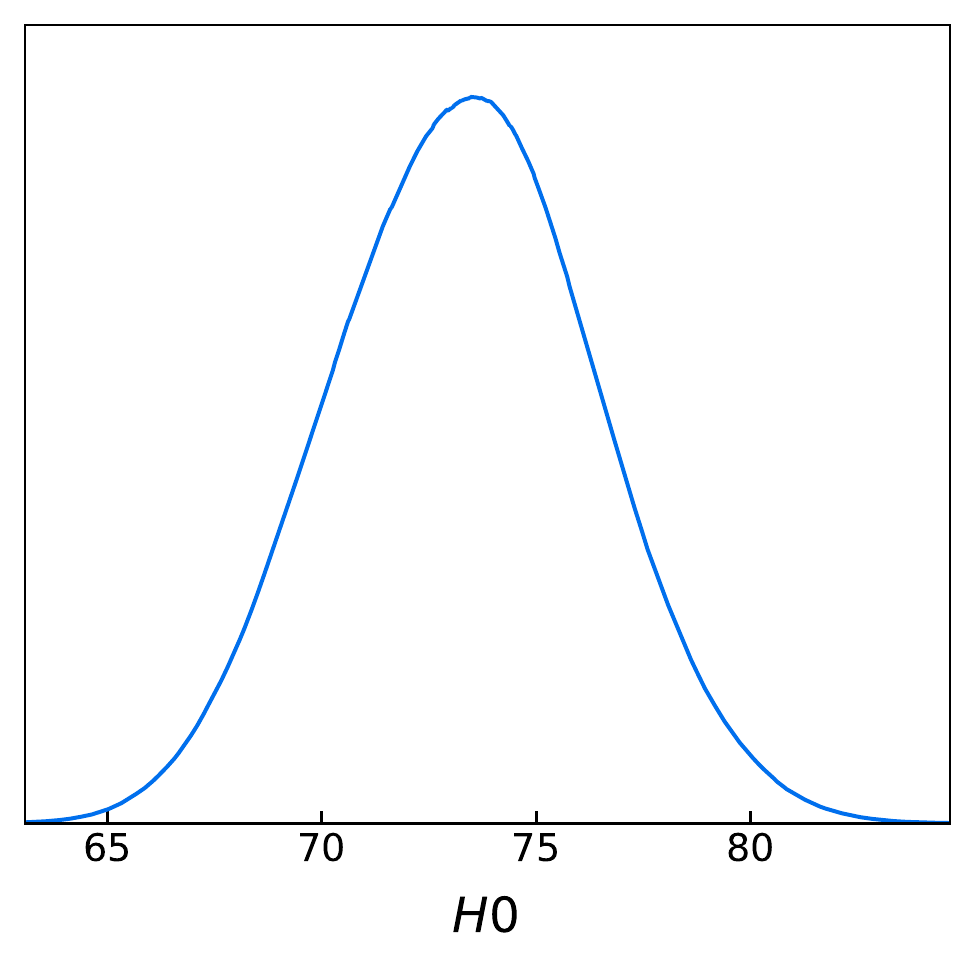}}\\
\subfloat[Using $\mu_{GRB}$ on full sample varying both $\Omega_M$\\ and $H_0$ without correction for evolution]{\label{fig16_c}
\includegraphics[width=0.40\hsize,height=0.36\textwidth,angle=0,clip]{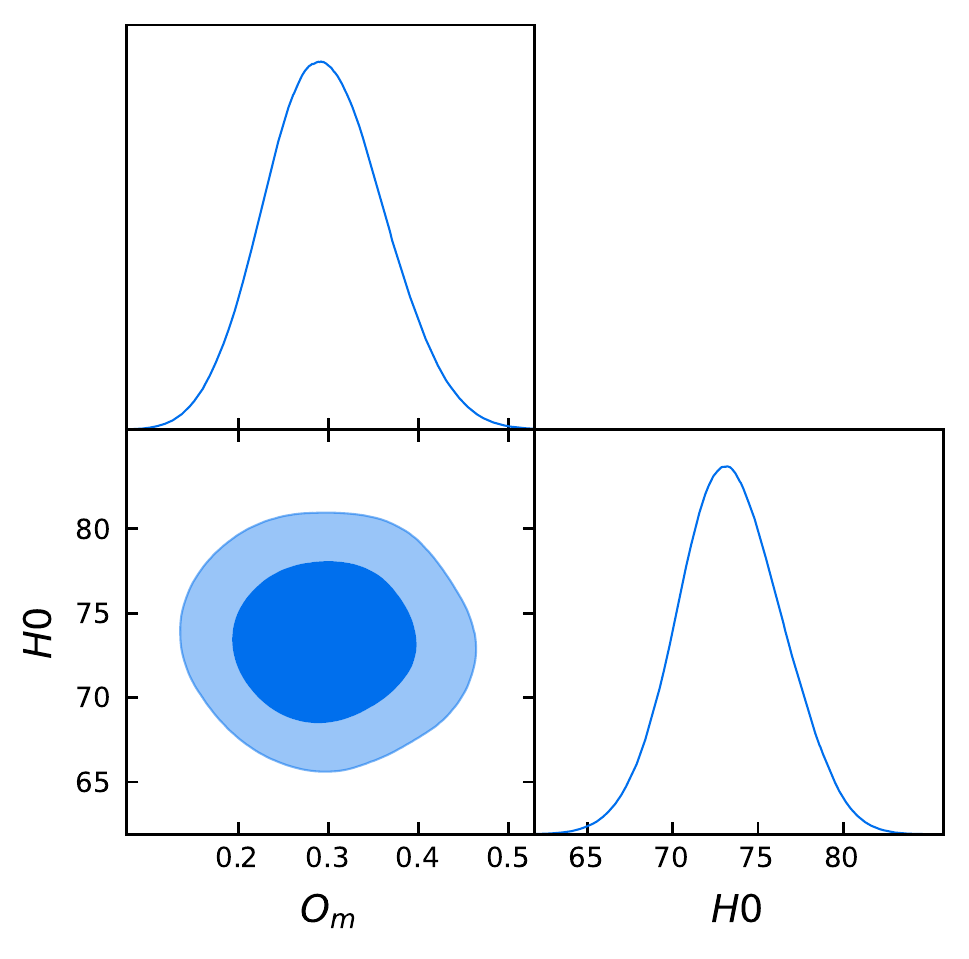}}\hspace{5mm} 
\subfloat[Using $\mu_{GRB}$ on full sample varying only\\ $w$ without correction for evolution]{\label{fig16_d}
\includegraphics[width=0.40\hsize,height=0.36\textwidth,angle=0,clip]{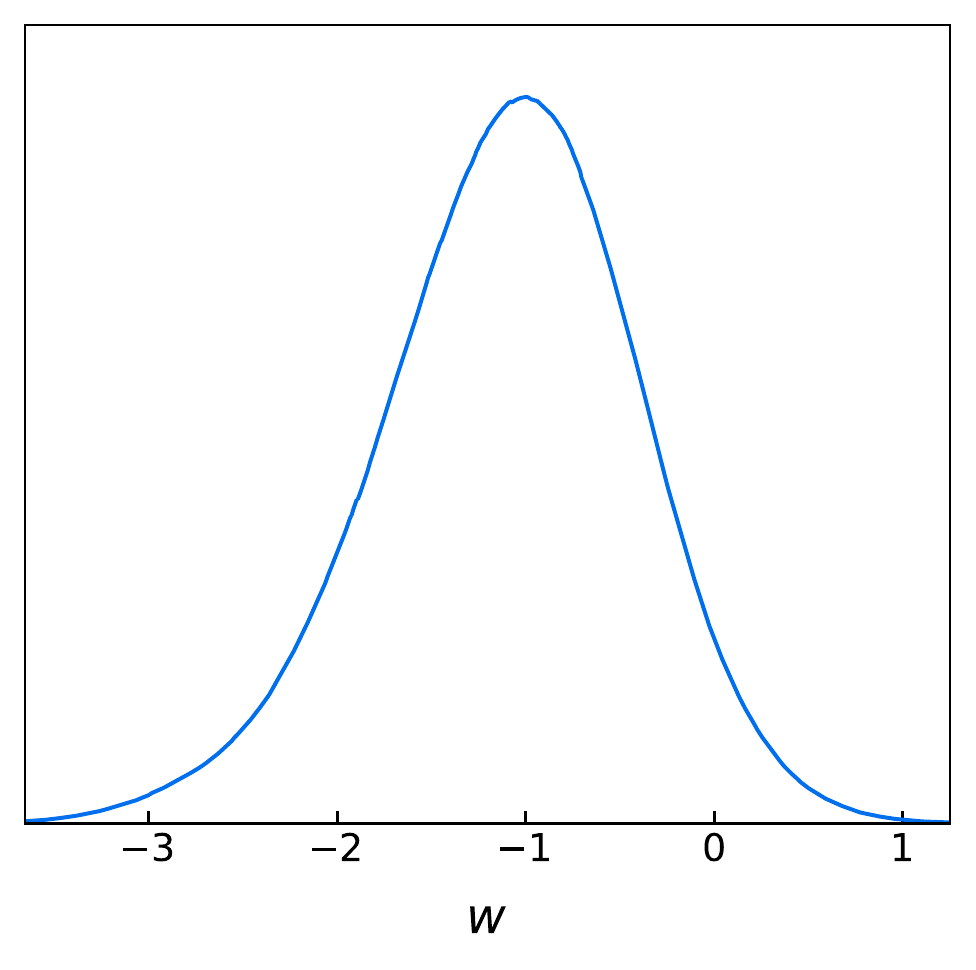}}
\caption{Cosmological results for the GRBs alone (with calibration on SNe Ia) with $\mu_{GRB}$ using no correction for the evolution and the assumptions of 3 $\sigma$ Gaussian priors on the cosmological parameters investigated following \citet{Scolnic}. Panels a), b), c) and d) show the contours from the case (vi) for the case of $\Omega_M$, $H_0$, $\Omega_M$ and $H_0$ together, and $w$, respectively.}
\label{fig14}
\end{figure}

\subsection{GRBs alone calibrated with SNe Ia with Gaussian priors}\label{GRB alone with calibration}

When it comes to observational cosmology, one can set up a standard candle by calibrating it with other well-known cosmological probes. This method is widely used in the literature with different approaches (it is, indeed, the main procedure to build the so-called cosmological ladder on which the most updated cosmological late-type results are based upon). To investigate the case of GRBs calibrated with SNe Ia we performed the following procedure:

\begin{enumerate}
 \item \hspace{1mm}First we fit a set of $a$, $b$, $c$, and $\sigma_{int}$ parameters related to the GRBs fundamental plane with the part of our GRBs sample whose redshift overlaps with the redshift range of SNe Ia (up to $z=2.25$), which corresponds to 25 GRBs. We fix the $H_{0}$ and $\Omega_{M}$ parameters considering the values obtained using SNe Ia alone (for simplicity, we use $H_{0}=70$ $km \hspace{1ex} s^{-1} Mpc^{-1}$ and $\Omega_{M}=0.3$), but using Gaussian priors with 3 $\sigma$ based on the values of \citet{Scolnic}.
 \item \hspace{1mm} We have then performed the same steps i)-vi) as in the previous Sec. \ref{GRBs alone without calibration} with the only difference that the parameters of the plane are fixed to the sample of the 25 GRBs with $z<2.25$ overlapping with the SNe Ia.
 
\end{enumerate}
\begin{table}
\centering
\scalebox{0.885}{
\begin{tabular}{p{35mm}|l|c|c|c|c|c|c|c}
 \toprule[1.2pt]
 \toprule[1.2pt]
\textbf{No calibration with SNe Ia with uniform priors, Equation \ref{equmu}} & {\textbf{parameters varied}} & \textbf{Model} & $<\boldsymbol{\Omega_{M}}>$ & $<\boldsymbol{H_{0}}>$ & $<\boldsymbol{w}>$ & $\boldsymbol{\Delta^{GRB_{U}}_{GRB_{G}}}\%$ & $z-score_{SN}$ & $z-score_{SN+BAO}$ \\ 
\midrule
without evolution &$\Omega_{M}$ & $\Lambda$CDM & $0.58 \pm 0.27$ & \bf{70} & \bf{-1} & 328.57 & 1.040 & 1.022 \\\hline
without evolution & $H_0$ & $\Lambda$CDM & \bf{0.30} & $77.51\pm 14.14$ & \bf{-1} & 327.58 & 0.533 & 0.534 \\\hline
without evolution & $w$ & $w$CDM & \bf{0.30} & \bf{70} & $-0.91 \pm 0.58$ & -19.11 & 0.155 & 0.184 \\\hline
\midrule
with fixed evolution & $\Omega_{M}$ & $\Lambda$CDM & $0.56 \pm 0.27$ & \bf{70} & \bf{-1} & 328.57 & 0.966 & 0.948 \\\hline
with fixed evolution & $H_0$ & $\Lambda$CDM & \bf{0.30} & $77.48 \pm 14.15$ & \bf{-1} & 384.42 & 0.531 & 0.531 \\\hline
with fixed evolution & $w$ & $w$CDM & \bf{0.30} & \bf{70} & $-0.93 \pm 0.58$ & -6.45 & 0.121 & 0.150 \\\hline
\midrule
with $k=k(\Omega_{M})$ & $\Omega_{M}$ & $\Lambda$CDM & $0.56 \pm 0.27$ & \bf{70} & \bf{-1} & 335.48 & 0.966 & 0.948 \\\hline
\toprule[1.2pt]
 \toprule[1.2pt]

\textbf{No Calibration with uniform priors, Equation \ref{isotropic} and \ref{planeev}} & {\textbf{parameters varied}} & \textbf{Model} & $<\boldsymbol{\Omega_{M}}>$ & $<\boldsymbol{H_{0}}>$ & $<\boldsymbol{w}>$ & $\boldsymbol{\Delta^{GRB_{U}}_{GRB_{G}}}\%$ & $z-score_{SN}$ & $z-score_{SN+BAO}$ \\ 
\midrule
without evolution & $\Omega_{M}$ & $\Lambda$CDM & $0.53 \pm 0.28$ & \bf{70} & \bf{-1} & 359.01 & 0.825 & 0.807 \\\hline
without evolution & $H_0$ & $\Lambda$CDM & \bf{0.30} & $75.00\pm 14.17$ & \bf{-1} & 355.19 & 0.355 & 0.356 \\\hline

without evolution & $w$ & $w$CDM & \bf{0.30} & \bf{70} & $-0.98 \pm 0.58$ & -44.65 & 0.034 & 0.064 \\\hline
\midrule
with fixed evolution & $\Omega_{M}$ & $\Lambda$CDM & $0.53 \pm 0.27$ & \bf{70} & \bf{-1} & 315.38 & 0.855 & 0.837 \\\hline
with fixed evolution & $H_0$ & $\Lambda$CDM & \bf{0.30} & $74.99 \pm 14.27$ & \bf{-1} & 356.49 & 0.352 & 0.352 \\\hline

with fixed evolution & $w$ & $w$CDM & \bf{0.30} & \bf{70} & $-0.98 \pm 0.58$ & -12.39 & 0.034 & 0.064 \\\hline
\midrule
with $k=k(\Omega_{M})$ & $\Omega_{M}$ & $\Lambda$CDM & $0.55 \pm 0.27$ & \bf{70} & \bf{-1} & 328.57 & 0.929 & 0.911 \\\hline
\end{tabular}}
\caption{Averaged cosmological parameters of 100 runs with no calibration using GRBs alone assuming uniform priors (indicated with the subscript U) and with $\mu_{GRB}$ (first part of the Table) and with Fundamental plane equation, see Equation \ref{isotropic} (2nd part) without evolution, and with the evolution correction as a function of $\Omega_{M}$. In the header we use the notation: "$<>$" to distinguish results obtained in this analysis from the ones for which Gaussian priors have been considered. The third column before the last corresponds to the percentage change in errors computed comparing the current results obtained with the GRBs alone with Gaussian priors (indicated with the subscript G) without calibration taking as a reference point GRB values from Table \ref{Table1}. The last column represents the z-score from the SNe Ia taking the SNe Ia as a reference point.}
\label{Table3}
\end{table}

\begin{figure}
\centering
\subfloat[Using $\mu_{GRB}$ on full sample varying only $\Omega_M$ with\\ correction for evolution with fixed parameters]{\label{fig17_a}
\includegraphics[width=0.40\hsize,height=0.40\textwidth,angle=0,clip]{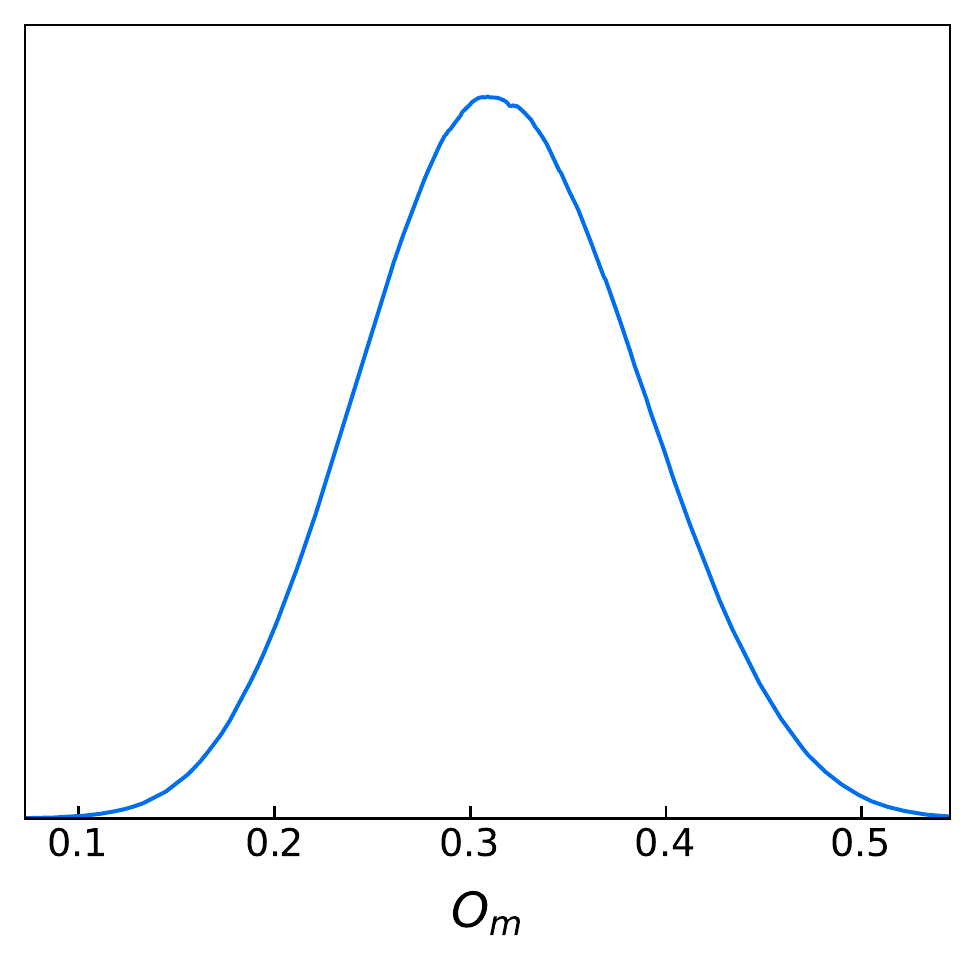}}\hspace{5mm}
\subfloat[Using $\mu_{GRB}$ on full sample varying only $H_0$ with\\ correction for evolution with fixed parameters]{\label{fig17_b}
\includegraphics[width=0.40\hsize,height=0.40\textwidth,angle=0,clip]{figures/FP_DM_withoutEvol/DMfull_H0.pdf}}\\
\subfloat[Using $\mu_{GRB}$ on full sample varying both $\Omega_M$ and $H_0$\\ with correction for evolution with fixed parameters]{\label{fig17_c}
\includegraphics[width=0.40\hsize,height=0.40\textwidth,angle=0,clip]{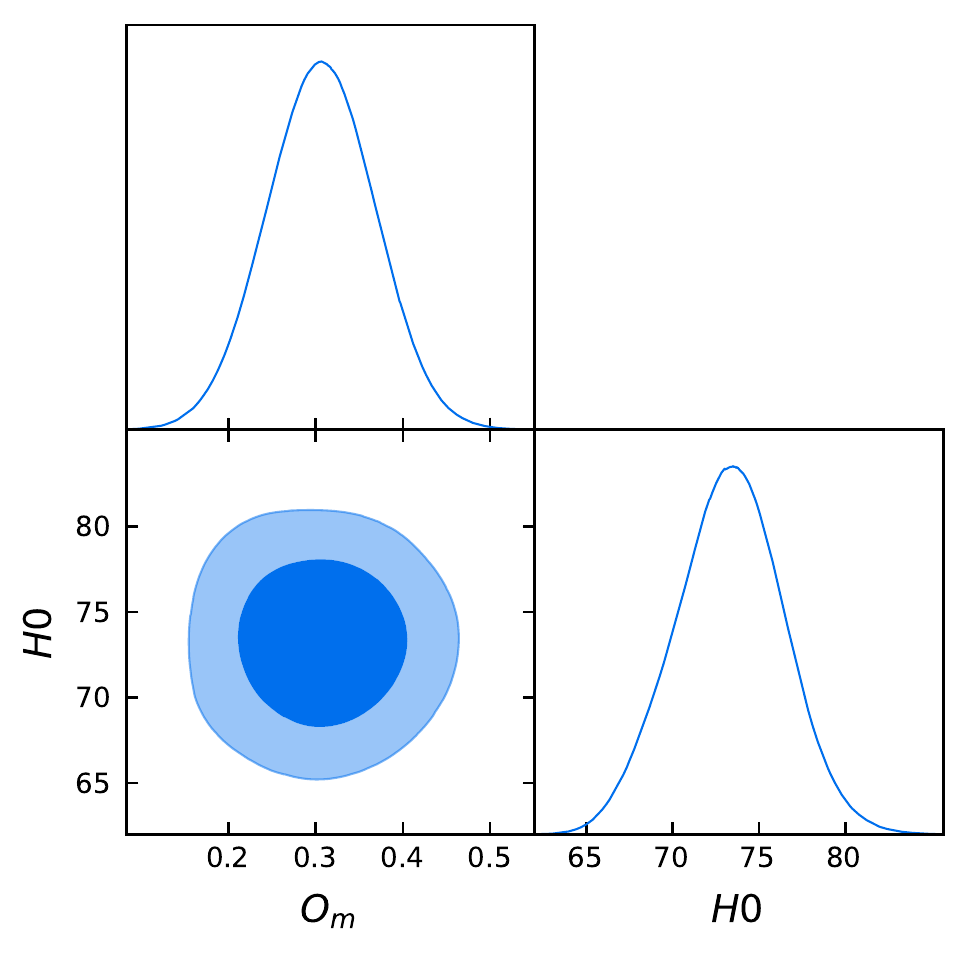}}\hspace{5mm} 
\subfloat[Using $\mu_{GRB}$ on full sample varying only $w$ with\\ correction for evolution with fixed parameters]{\label{fig17_d}
\includegraphics[width=0.40\hsize,height=0.40\textwidth,angle=0,clip]{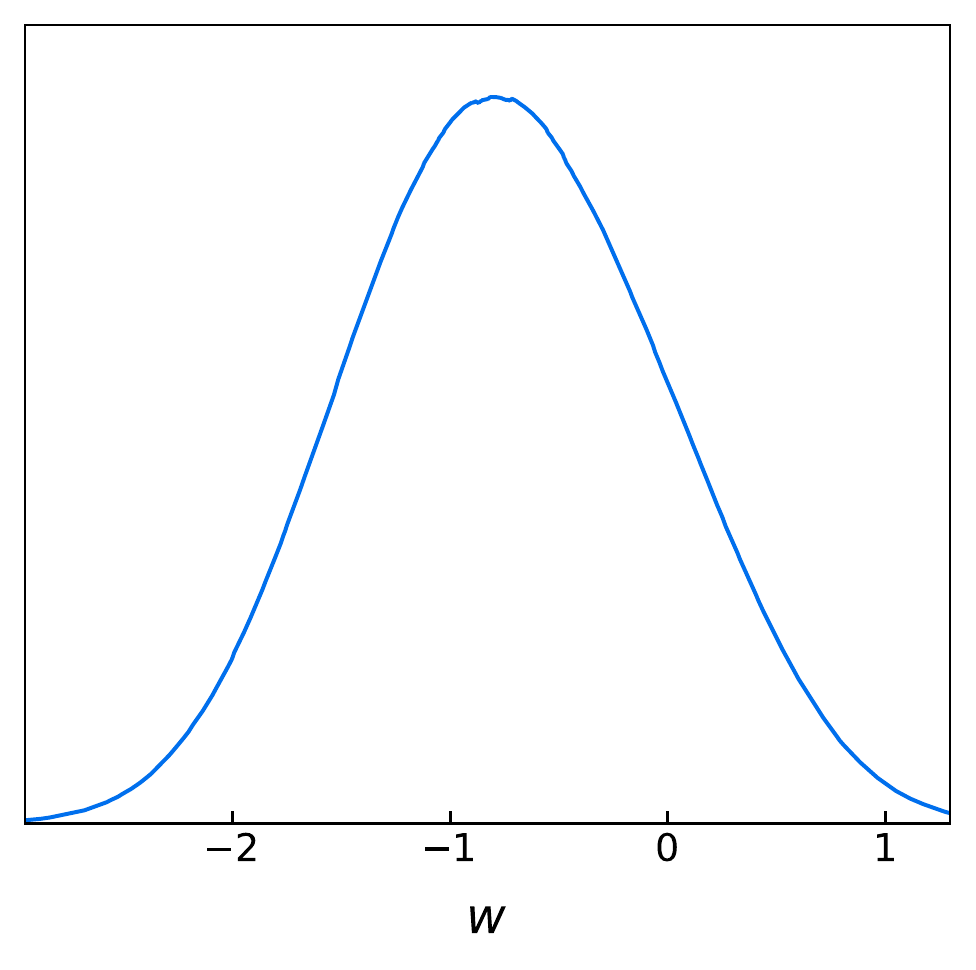}}
\caption{Cosmological results for the GRBs alone (with calibration on SNe Ia) with $\mu_{GRB}$ using correction for evolution and the assumptions of 3 $\sigma$ Gaussian priors on the cosmological parameters investigated following \citet{Scolnic}. Panels a), b), c) and d) show the contours from the case (vi) for the case of $\Omega_M$, $H_0$, $\Omega_M$ and $H_0$ together, and $w$, respectively.}
\label{fig15}
\end{figure}

\begin{figure} 
\centering
\subfloat[Using $\mu_{GRB}$ on full sample varying only\\ $\Omega_M$ with evolutionary function correction]{\label{fig18_a}
\includegraphics[width=0.44\hsize,height=0.44\textwidth,angle=0,clip]{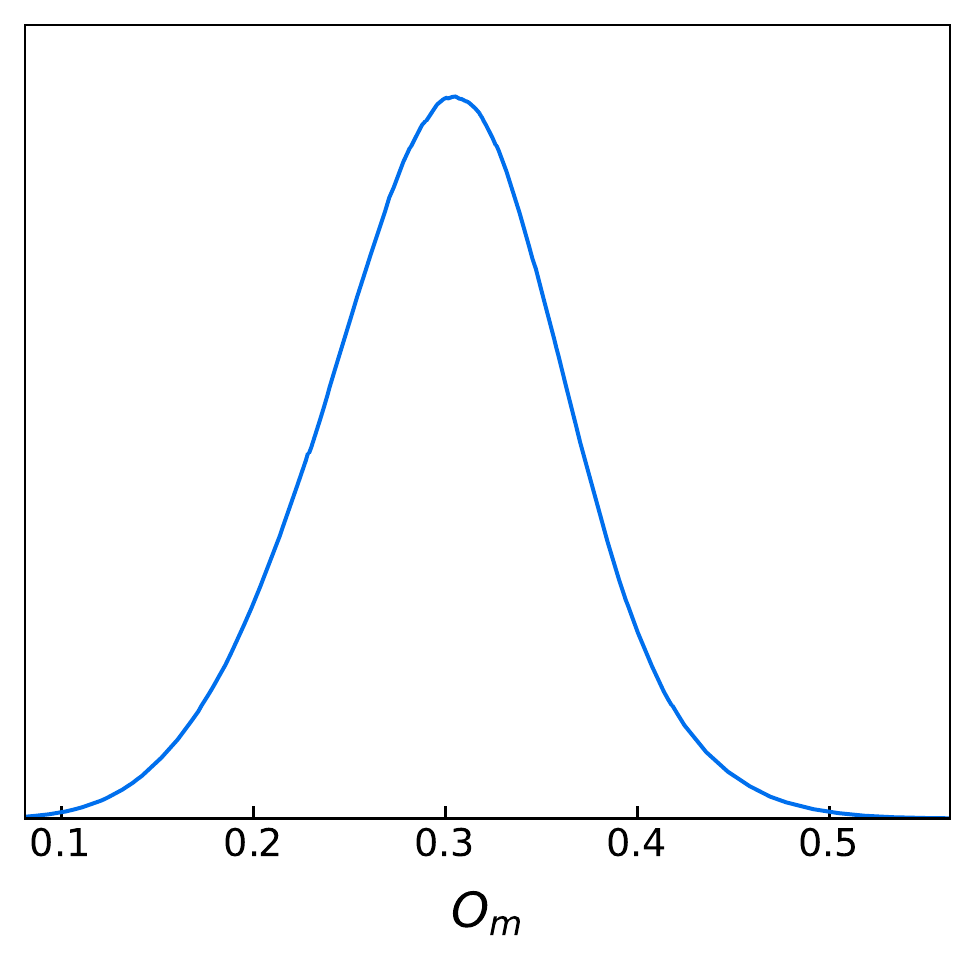}}\hspace{5mm}
\subfloat[Using $\mu_{GRB}$ on full sample varying both $\Omega_M$\\ and $H_0$ with evolutionary function correction]{\label{fig18_c}
\includegraphics[width=0.44\hsize,height=0.44\textwidth,angle=0,clip]{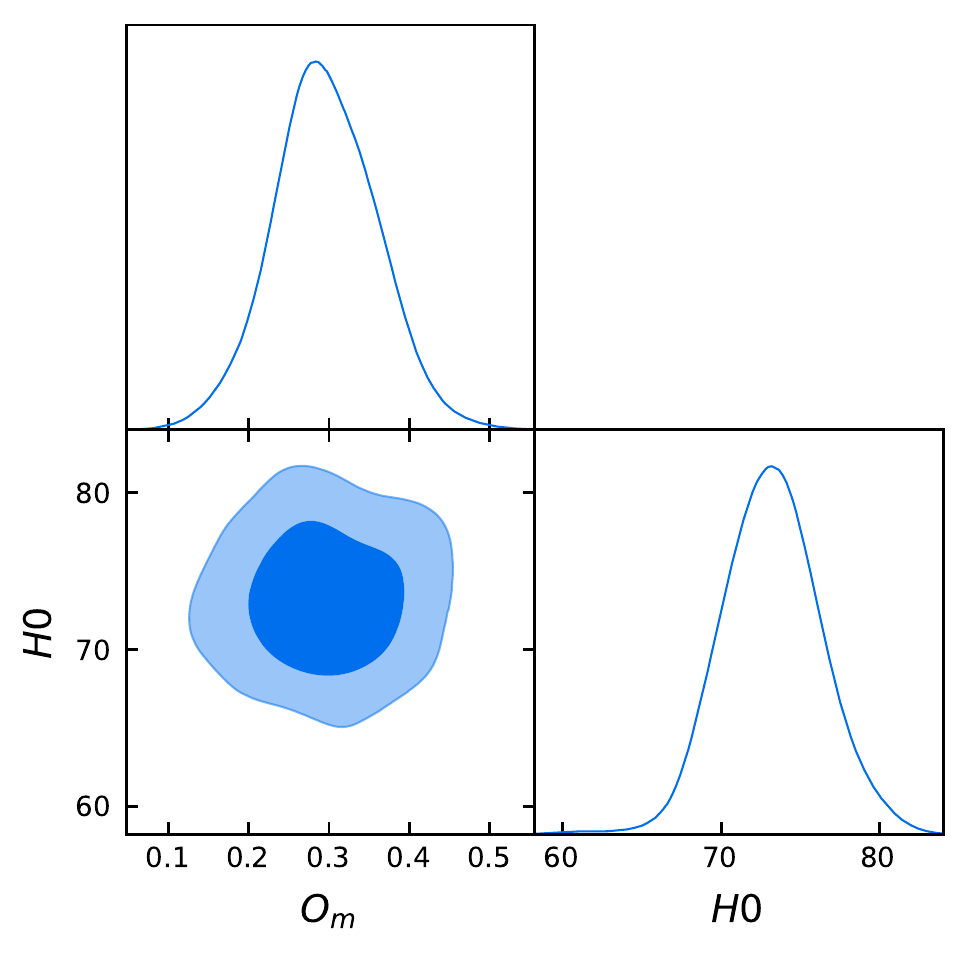}}\hspace{5mm} 
\caption{Cosmological results for the GRBs alone (with calibration on SNe Ia) with $\mu_{GRB}$ correcting with the evolutionary functions and the assumptions of 3 $\sigma$ Gaussian priors on the cosmological parameters investigated following \citet{Scolnic}. Panels a) and b) show the contours from the case (vi) for the case of $\Omega_M$ and the case of $\Omega_M$ and $H_0$ together, respectively.}
\label{fig16}
\end{figure}

\begin{figure} 
\centering
\subfloat[Plane parameters without evolution]{\label{fig19_a}
\includegraphics[width=0.44\hsize,height=0.4326\textwidth,angle=0,clip]{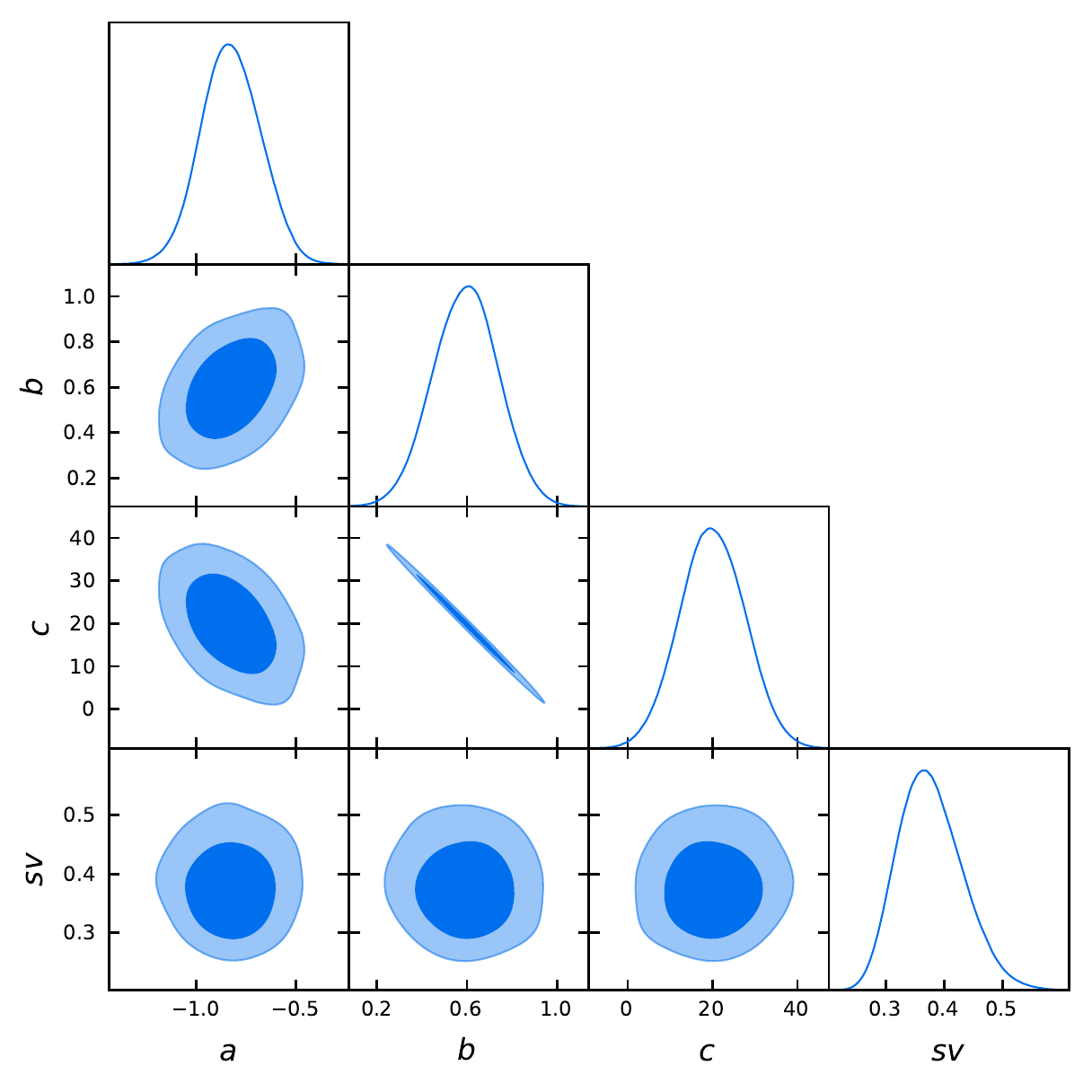}}
\subfloat[Plane parameters with fixed evolution]{\label{fig19_b}
\includegraphics[width=0.44\hsize,height=0.4326\textwidth,angle=0,clip]{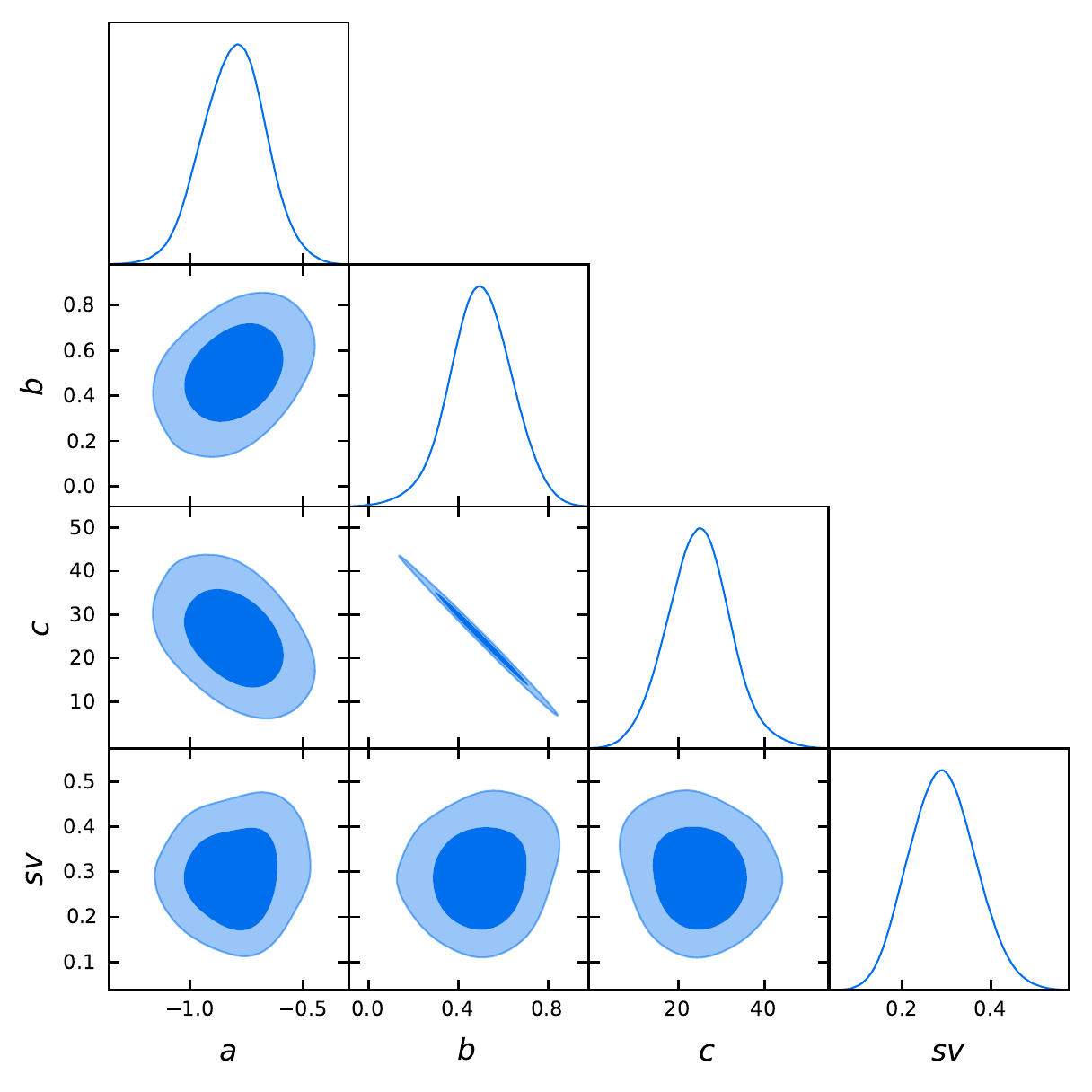}}
\caption{The Fundamental plane relation parameters obtained with the nearest 25 GRBs used to calibrate them on SNe Ia using the Equation \ref{isotropic}. Panels a) and b) show the contours of the plane fitting parameters without evolution and with fixed evolution respectively.}
\label{fig17}
\end{figure}
\begin{figure}
\centering
\subfloat[Using fundamental plane on full sample varying\\ only $\Omega_M$ without correction for evolution]{\label{fig20_a}
\includegraphics[width=0.40\hsize,height=0.35\textwidth,angle=0,clip]{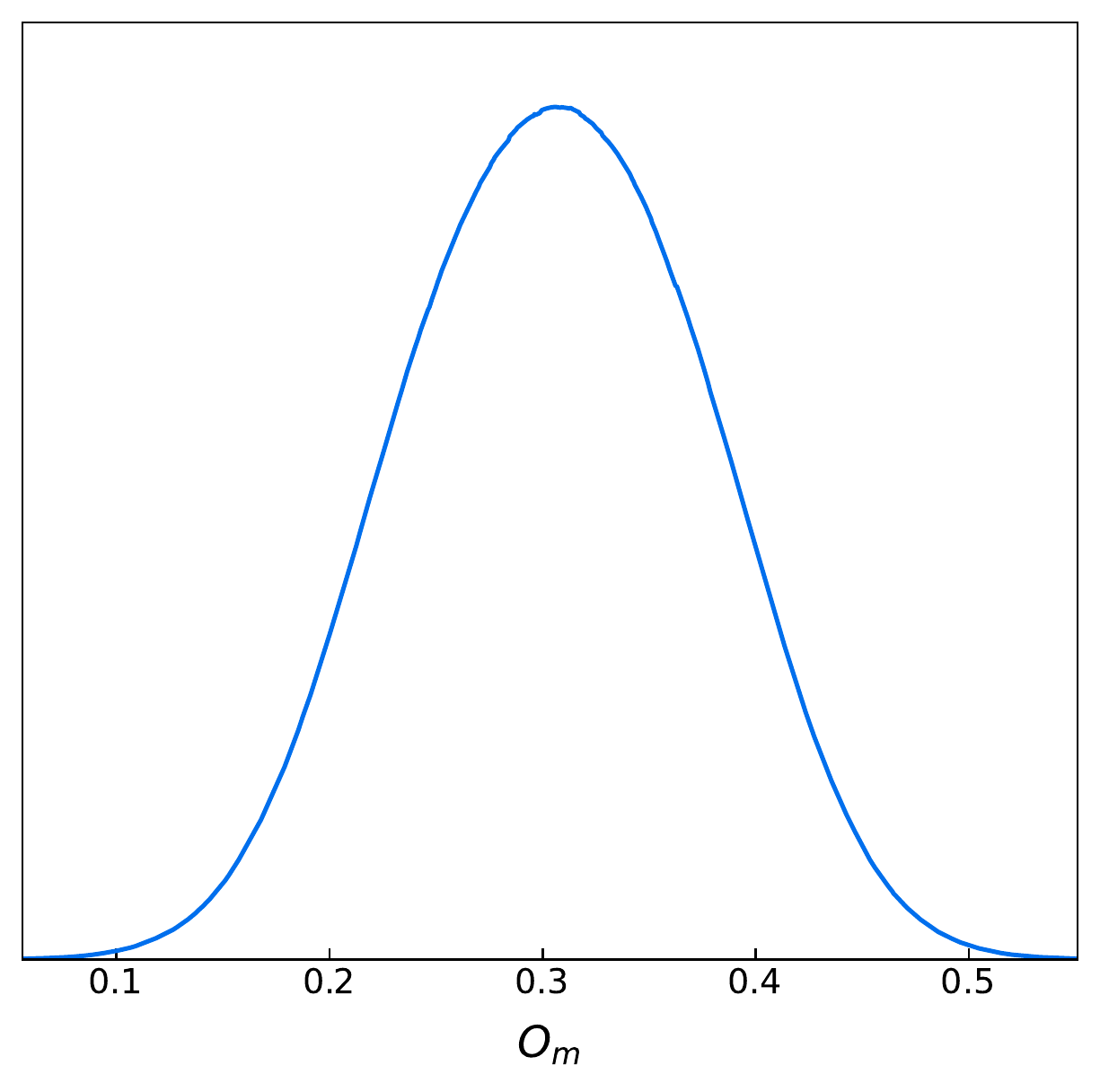}}\hspace{5mm}
\subfloat[Using fundamental plane on full sample varying\\ only $H_0$ without correction for evolution.]{\label{fig20_b}
\includegraphics[width=0.40\hsize,height=0.35\textwidth,angle=0,clip]{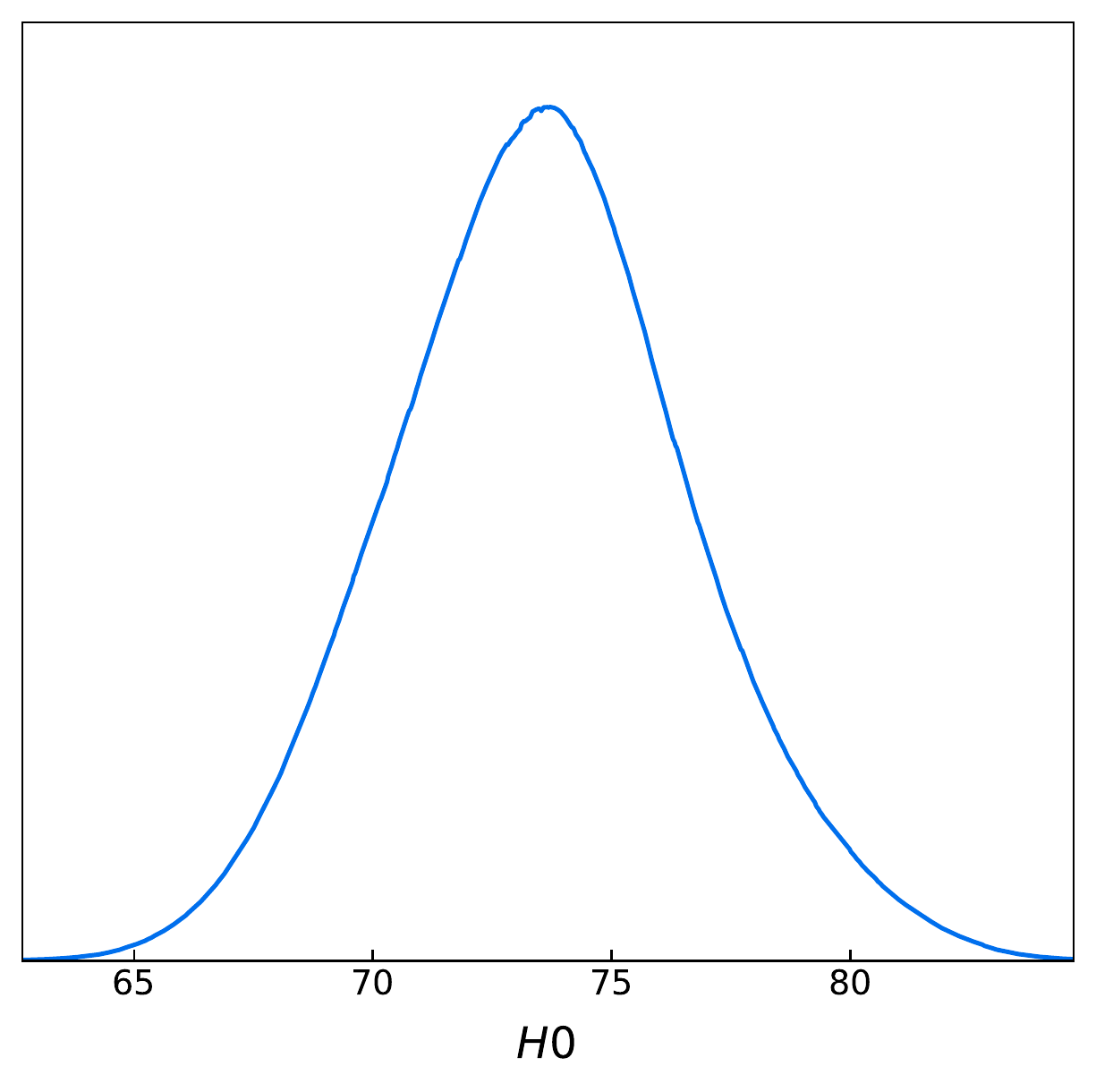}}\\
\subfloat[Using fundamental plane on full sample varying both\\ $\Omega_M$ and $H_0$ without correcting for the evolution.]{\label{fig20_c}
\includegraphics[width=0.40\hsize,height=0.35\textwidth,angle=0,clip]{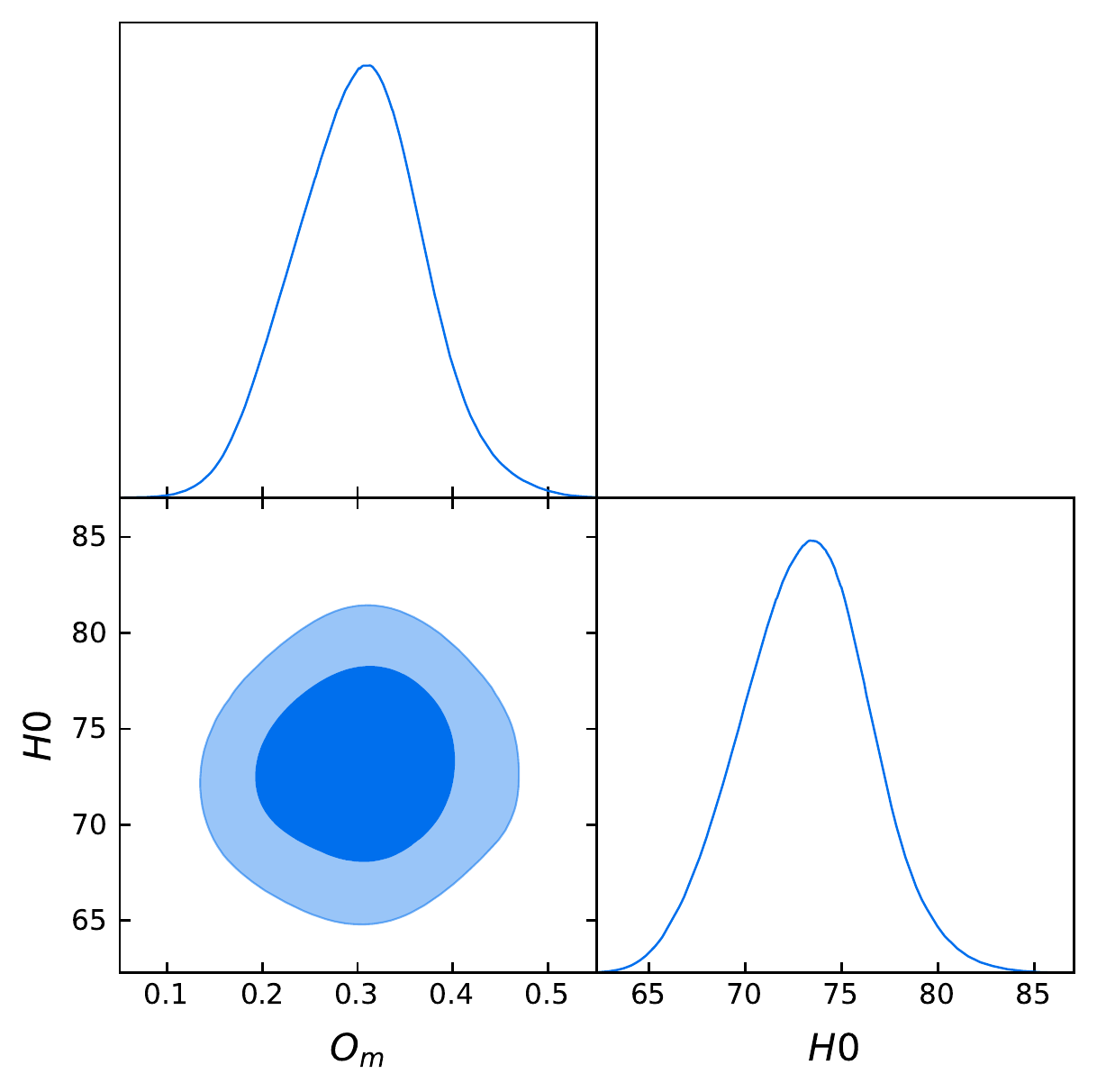}} \hspace{5mm}
\subfloat[Using fundamental plane on full sample varying\\ only $w$ without correction for evolution.]{\label{fig20_d}
\includegraphics[width=0.40\hsize,height=0.35\textwidth,angle=0,clip]{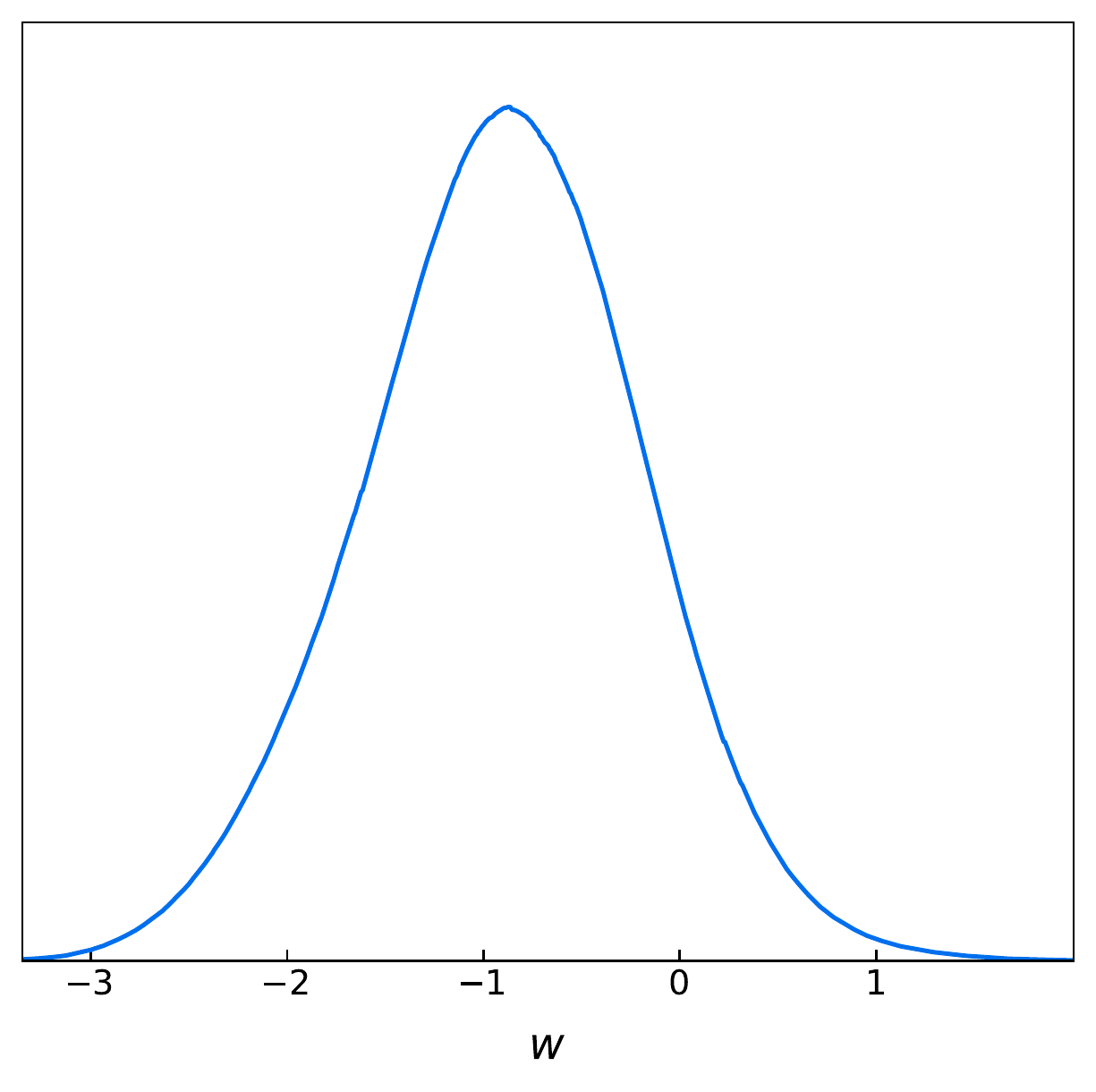}}
\caption{Cosmological results for the GRBs alone (with calibration) using the fundamental plane, Equation \ref{isotropic} with no evolution and the assumptions of 3 $\sigma$ Gaussian priors on the investigated cosmological parameters following \citet{Scolnic}. Panels a), b), c) and d) show the contours from the case (vi) for the case of $\Omega_M$, $H_0$, $\Omega_M$ and $H_0$ together, and $w$, respectively.}
\label{fig18}
\end{figure}

\begin{figure} 
\centering
\subfloat[Using fundamental plane on full sample varying only\\ $\Omega_M$ correcting with the fixed parameters of the evolution]{\label{fig21_a}
\includegraphics[width=0.40\hsize,height=0.40\textwidth,angle=0,clip]{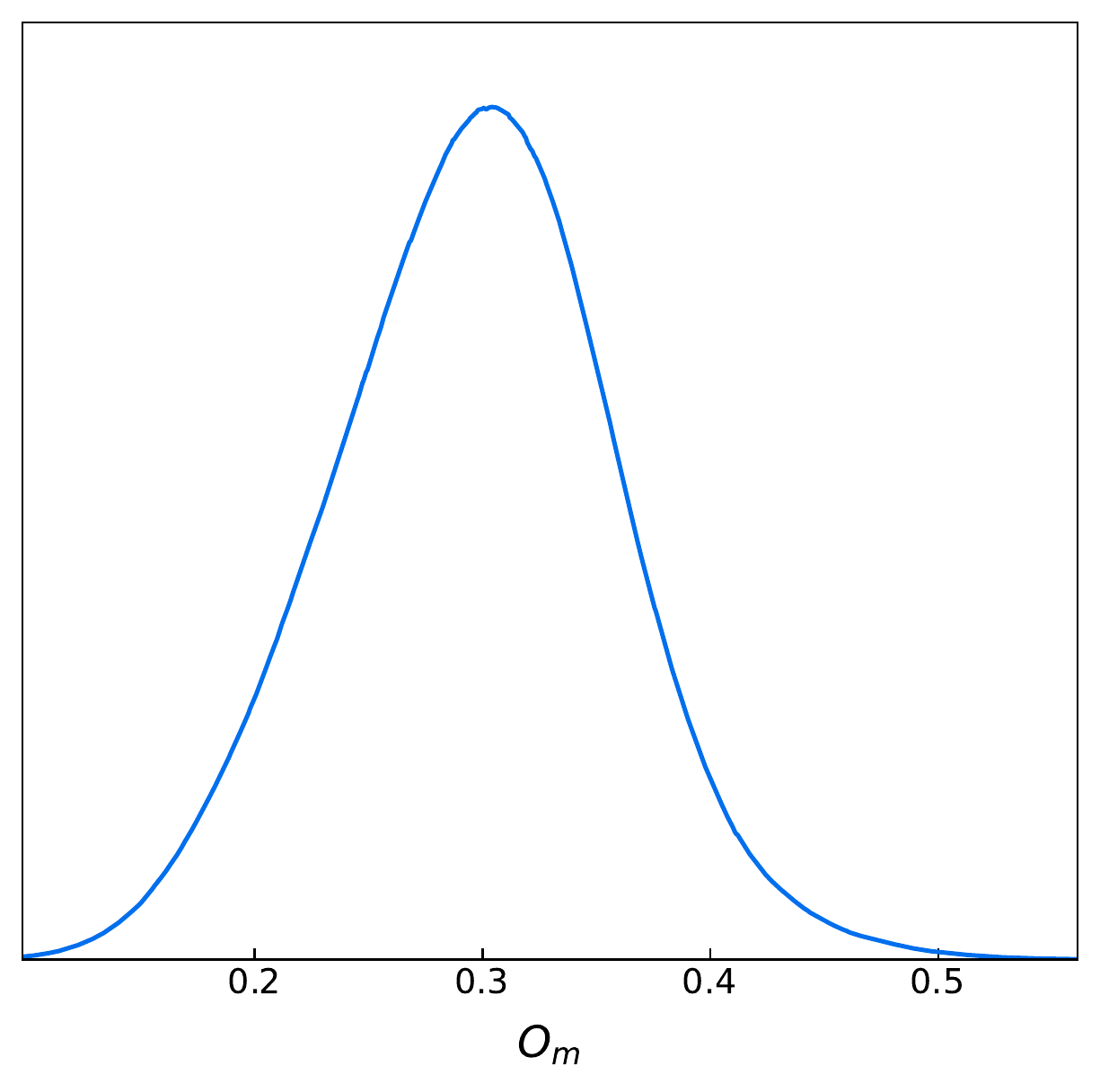}}\hspace{5mm}
\subfloat[Using fundamental plane on full sample varying only\\ $H_0$ correcting with the fixed parameters of the evolution]{\label{fig21_b}
\includegraphics[width=0.40\hsize,height=0.40\textwidth,angle=0,clip]{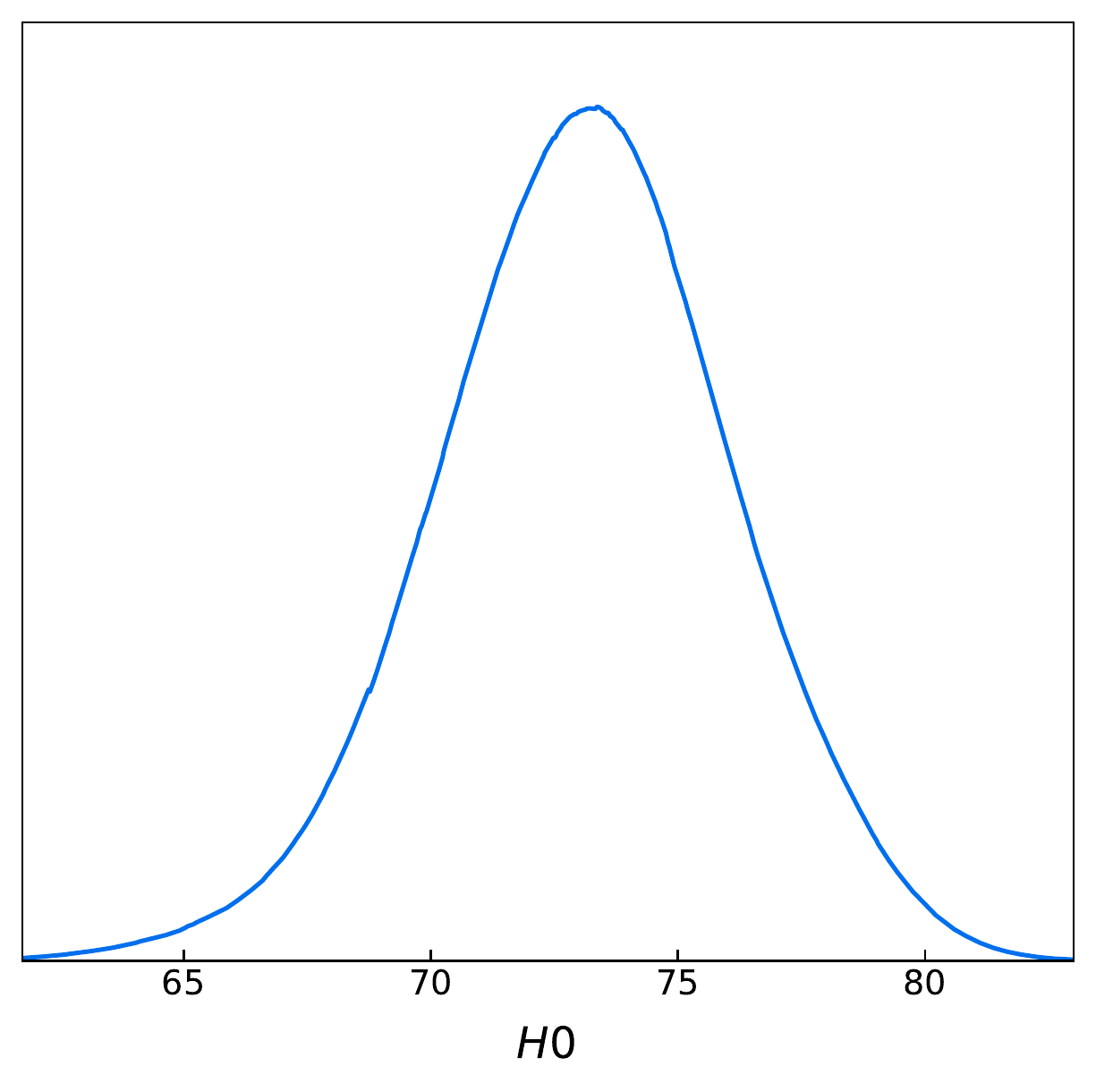}}\\
\subfloat[Using fundamental plane on full sample varying both $\Omega_M$\\ and $H_0$ correcting with the fixed parameters of the evolution]{\label{fig21_c}
\includegraphics[width=0.40\hsize,height=0.40\textwidth,angle=0,clip]{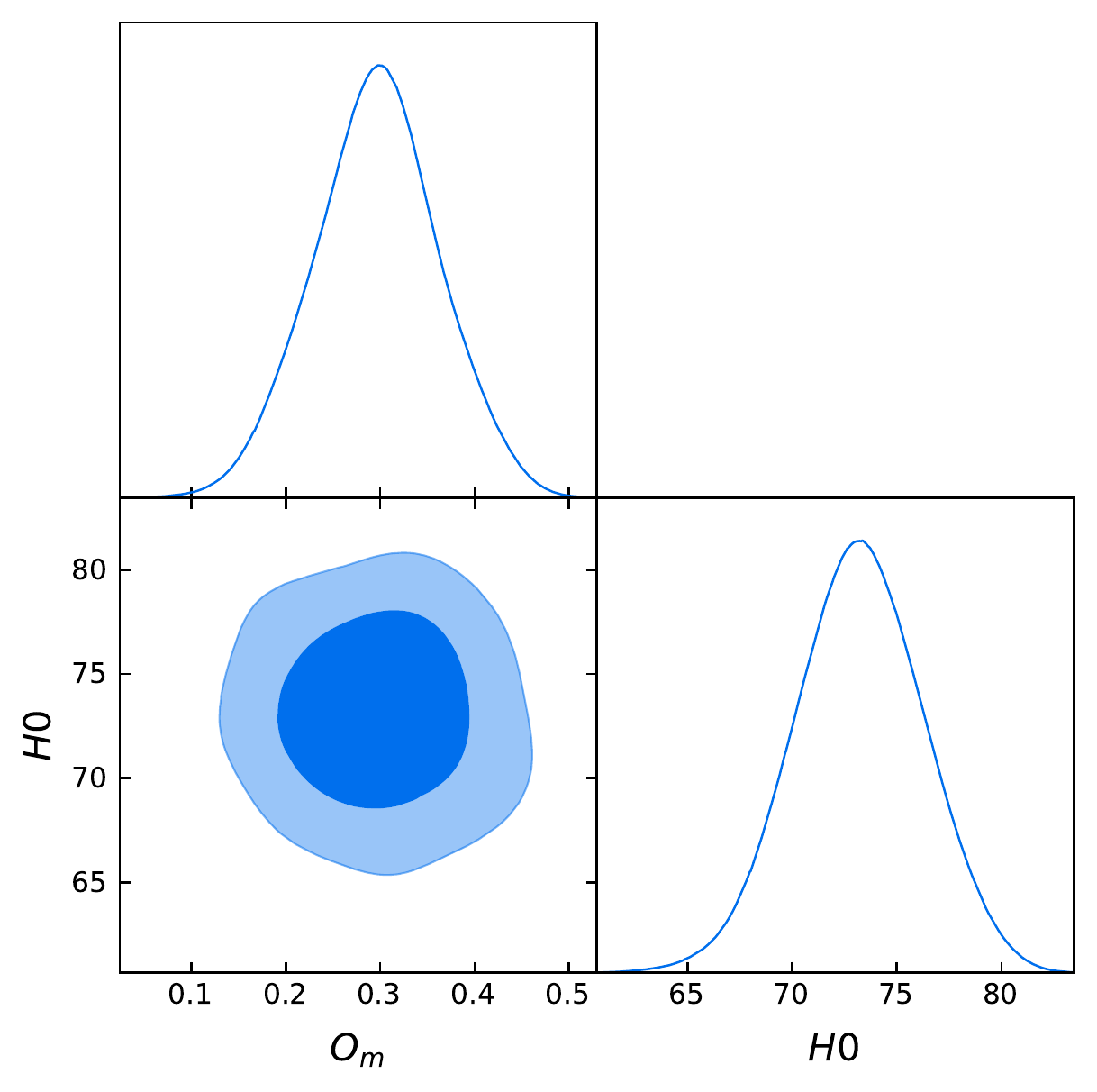}} \hspace{5mm}
\subfloat[Using fundamental plane on full sample varying only\\ $w$ correcting with the fixed parameters of the evolution]{\label{fig21_d}
\includegraphics[width=0.40\hsize,height=0.40\textwidth,angle=0,clip]{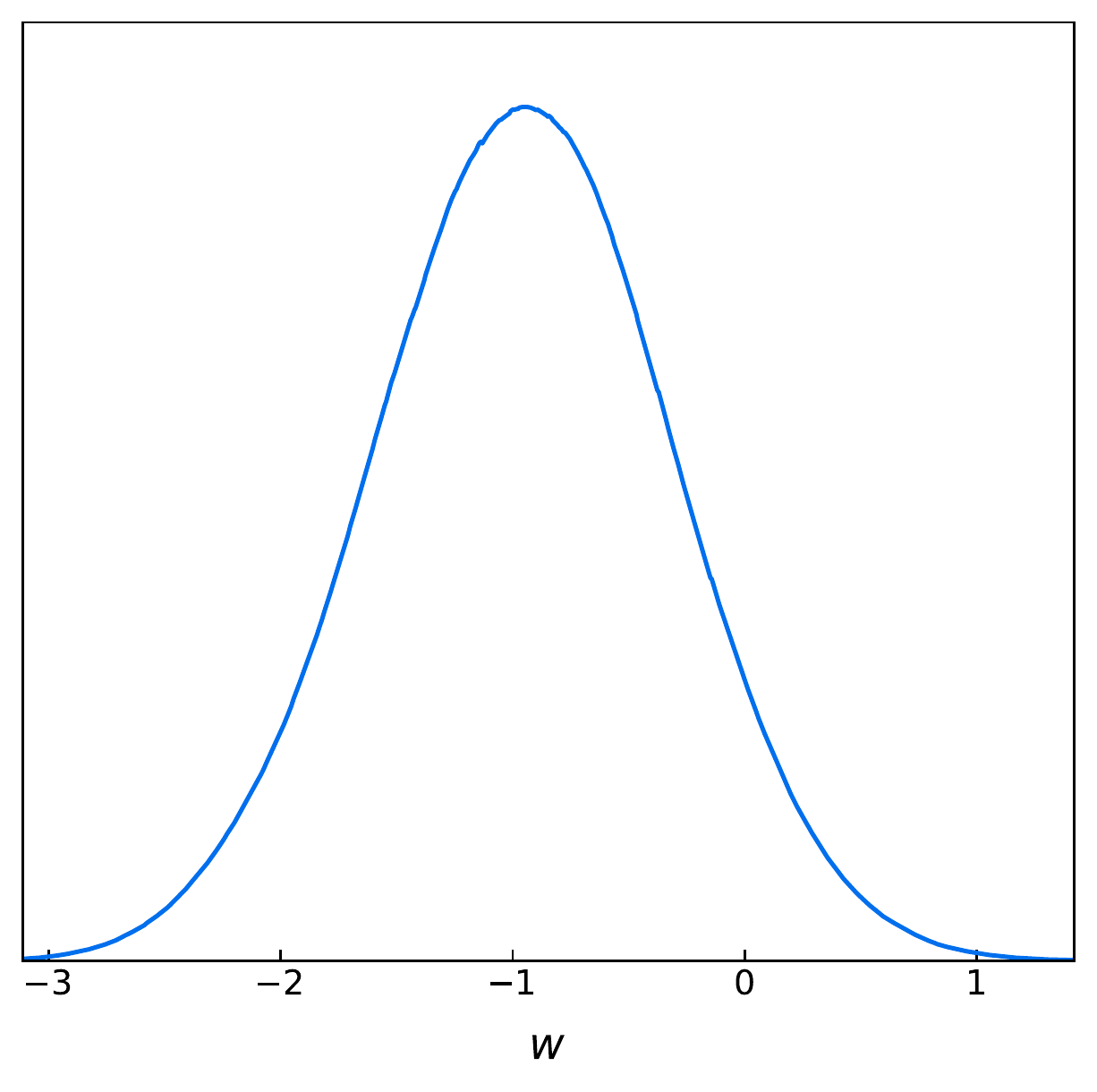}}
\caption{Cosmological results for the GRBs alone (with calibration on SNe Ia) using Fundamental plane, Equation \ref{isotropic} with fixed evolution and the assumptions of 3 $\sigma$ Gaussian priors for the cosmological parameters investigated following \citet{Scolnic}. Panels a), b), c) and d) show the contours from the case (vi) for the case of $\Omega_M$, $H_0$, $\Omega_M$ and $H_0$ together, and $w$, respectively.}
\label{fig19}
\end{figure}
\begin{figure} 
\centering

\subfloat[Using fundamental plane on full sample varying only $\Omega_M$]{\label{fig22_a}
\includegraphics[width=0.44\hsize,height=0.44\textwidth,angle=0,clip]{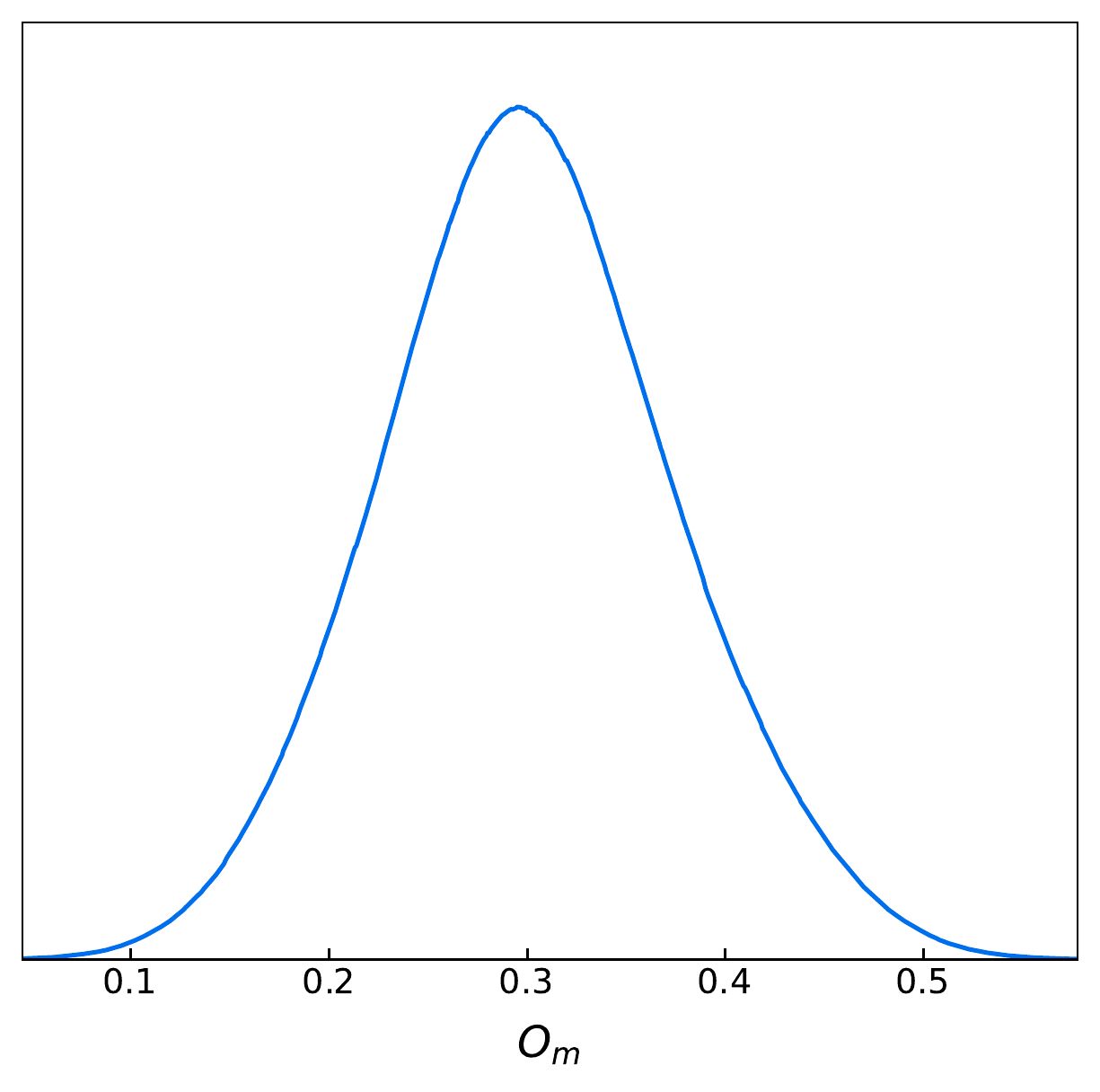}}\hspace{5mm}
\subfloat[Using fundamental plane on full sample varying both $\Omega_M$ and $H_0$]{\label{fig22_c}
\includegraphics[width=0.44\hsize,height=0.44\textwidth,angle=0,clip]{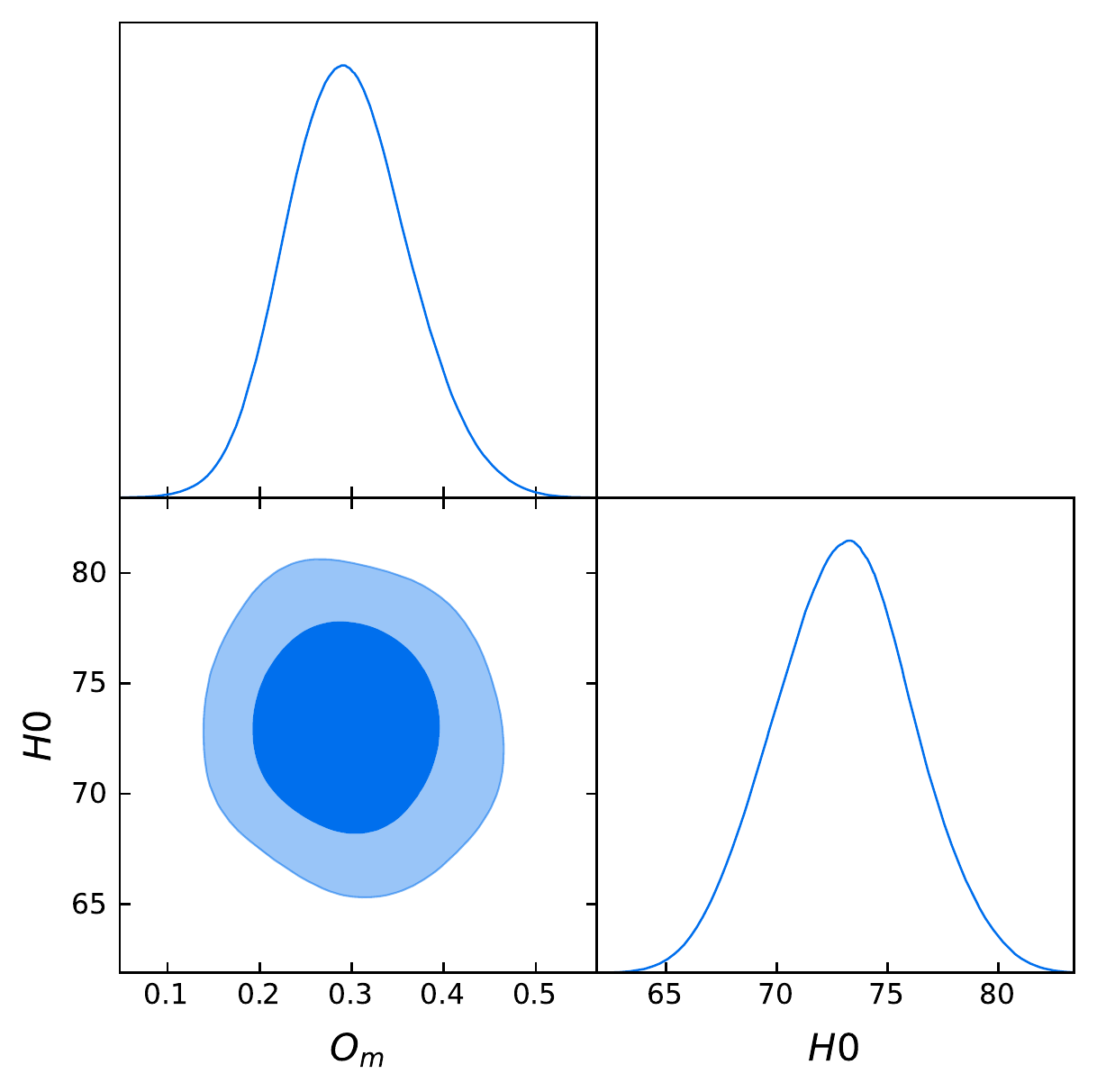}} \hspace{5mm}
\caption{Cosmological results for the GRBs alone (with calibration on SNe Ia) using Fundamental plane, Equation \ref{isotropic} with evolutionary functions and the assumptions of 3 $\sigma$ Gaussian priors on the cosmological parameters investigated following \citet{Scolnic}. Panels a) and b) show the contours from the case (vi) for the case of $\Omega_M$ and the case of $\Omega_M$ and $H_0$ varied together respectively.}
\label{fig20}
\end{figure}

The results for this analysis are presented without the corrections for evolution in Figs. \ref{fig14} and \ref{fig18}; with the correction for the evolutionary effects and selection biases in Figs. \ref{fig15} and \ref{fig19}; and considering these corrections, but with the evolutionary parameter computed as a function of $\Omega_{M}$, $k=k(\Omega_{M})$, in Figs. \ref{fig16} and \ref{fig20}.

We find that all the results of GRBs alone without calibration lie within 1 $\sigma$ with respect to
the cosmological results obtained with calibration on SNe Ia. The percentage change in the uncertainty values of the cosmological parameters is shown in the seventh column of Table \ref{Table2}. We also compare cosmological parameters obtained by GRBs alone with calibration using Gaussian priors and the results obtained with SNe Ia alone. We have found out that the results, both using Equation regarding $\mu_{GRB}$, \ref{equmu}, as well as the Fundamental plane Equations \ref{isotropic} and \ref{planeev} fall within 1 $\sigma$ as shown by the $z$-score in Table \ref{Table2}. The only exception to this result is the case of GRBs calibrated with SNe Ia using the fundamental plane for the case without evolution and with fixed evolution varying $H_0$. Indeed, we note that the z-score for this case is slightly larger than 1 ($z=1.140$). All the z-score with respect to the GRB results are shown in the last two columns of Table \ref{Table2}.

\begin{figure} 
\centering

\subfloat[Varying only $\Omega_M$ without evolution]{
\includegraphics[width=0.32\hsize,height=0.27\textwidth,angle=0,clip]{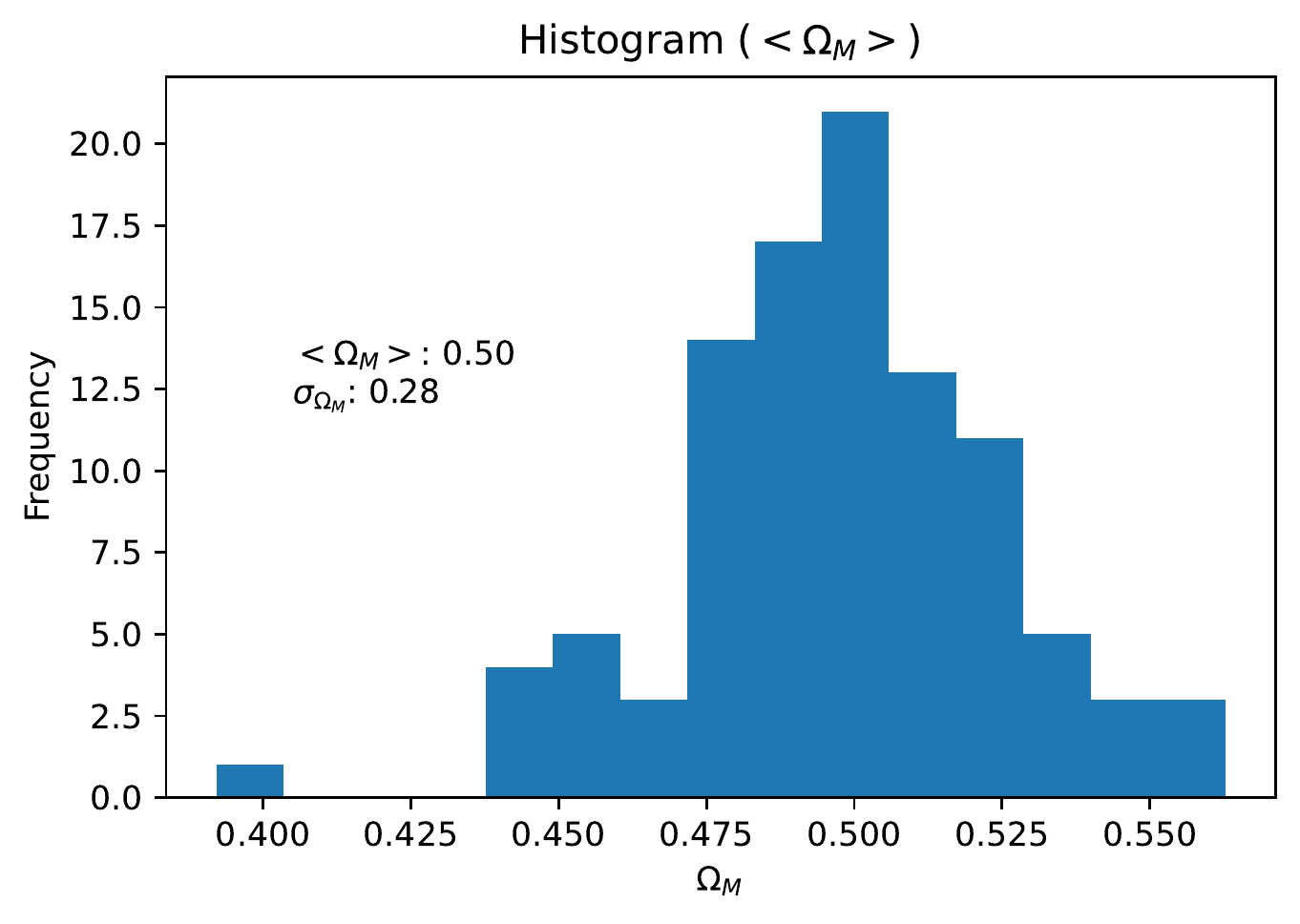}}
\subfloat[Varying only $H_0$ without evolution]{
\includegraphics[width=0.32\hsize,height=0.27\textwidth,angle=0,clip]{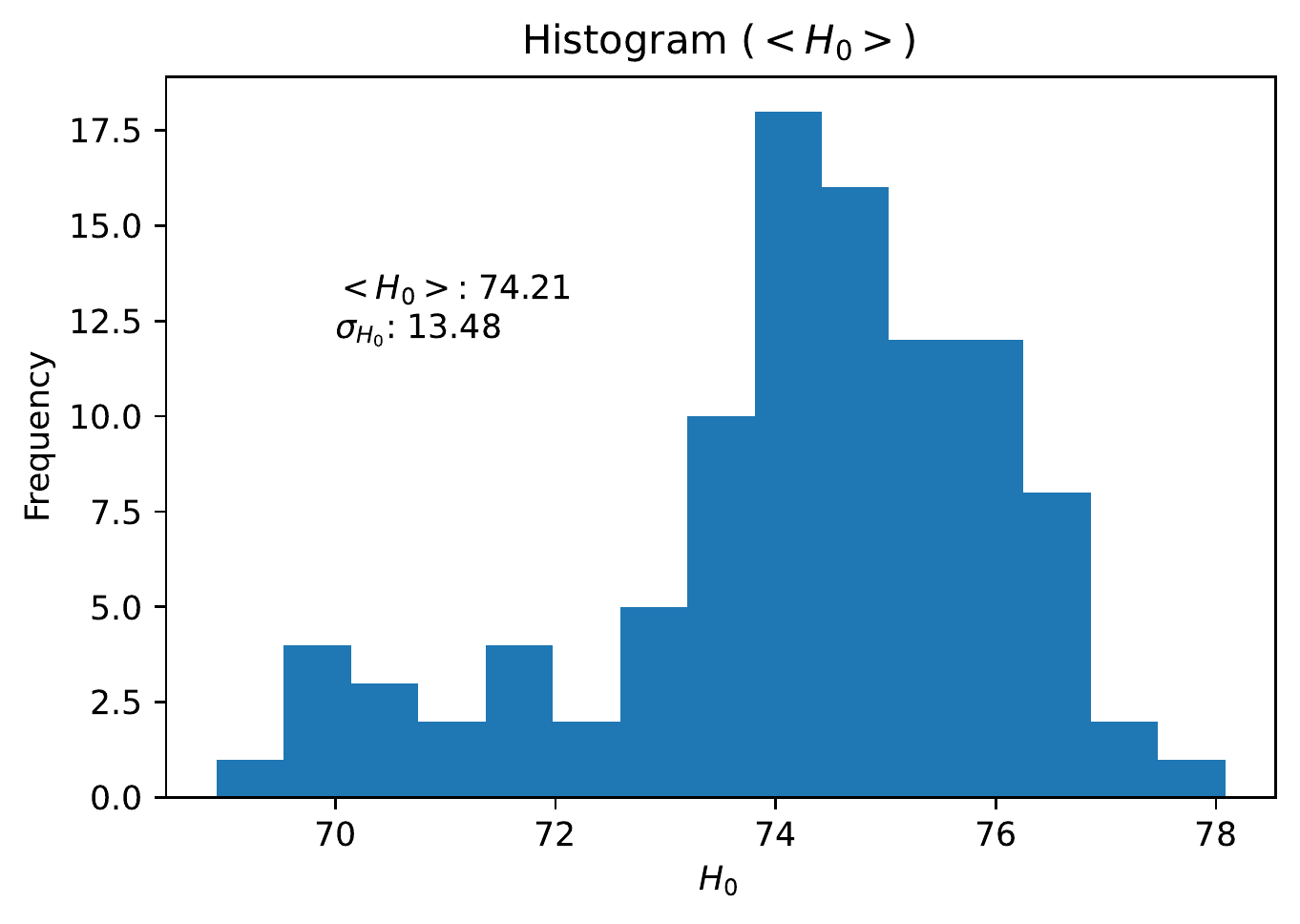}}
\subfloat[Varying only $w$ without evolution]{
\includegraphics[width=0.32\hsize,height=0.27\textwidth,angle=0,clip]{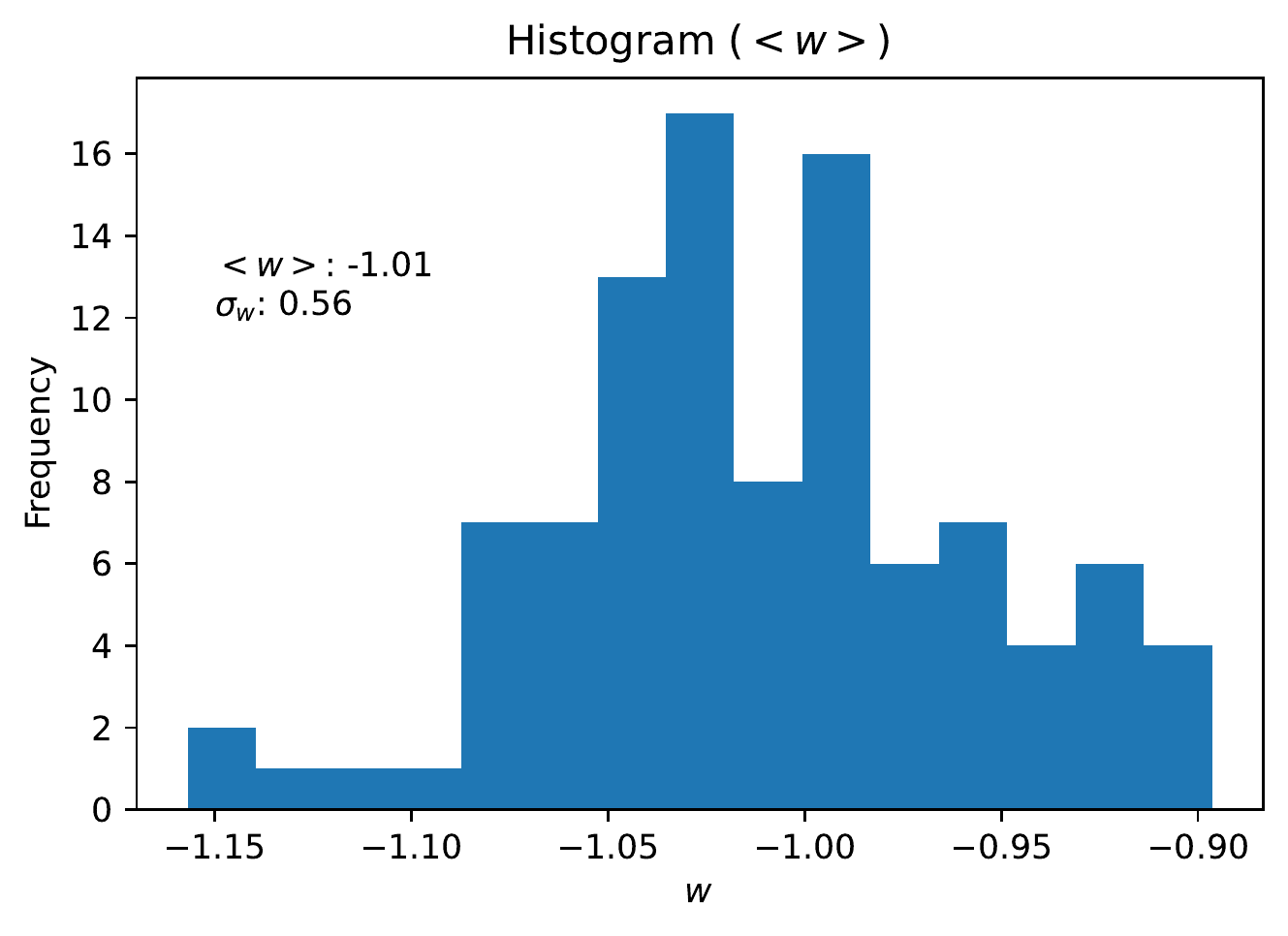}}\\

\subfloat[Varying only $\Omega_M$ with fixed evolution]{
\includegraphics[width=0.32\hsize,height=0.27\textwidth,angle=0,clip]{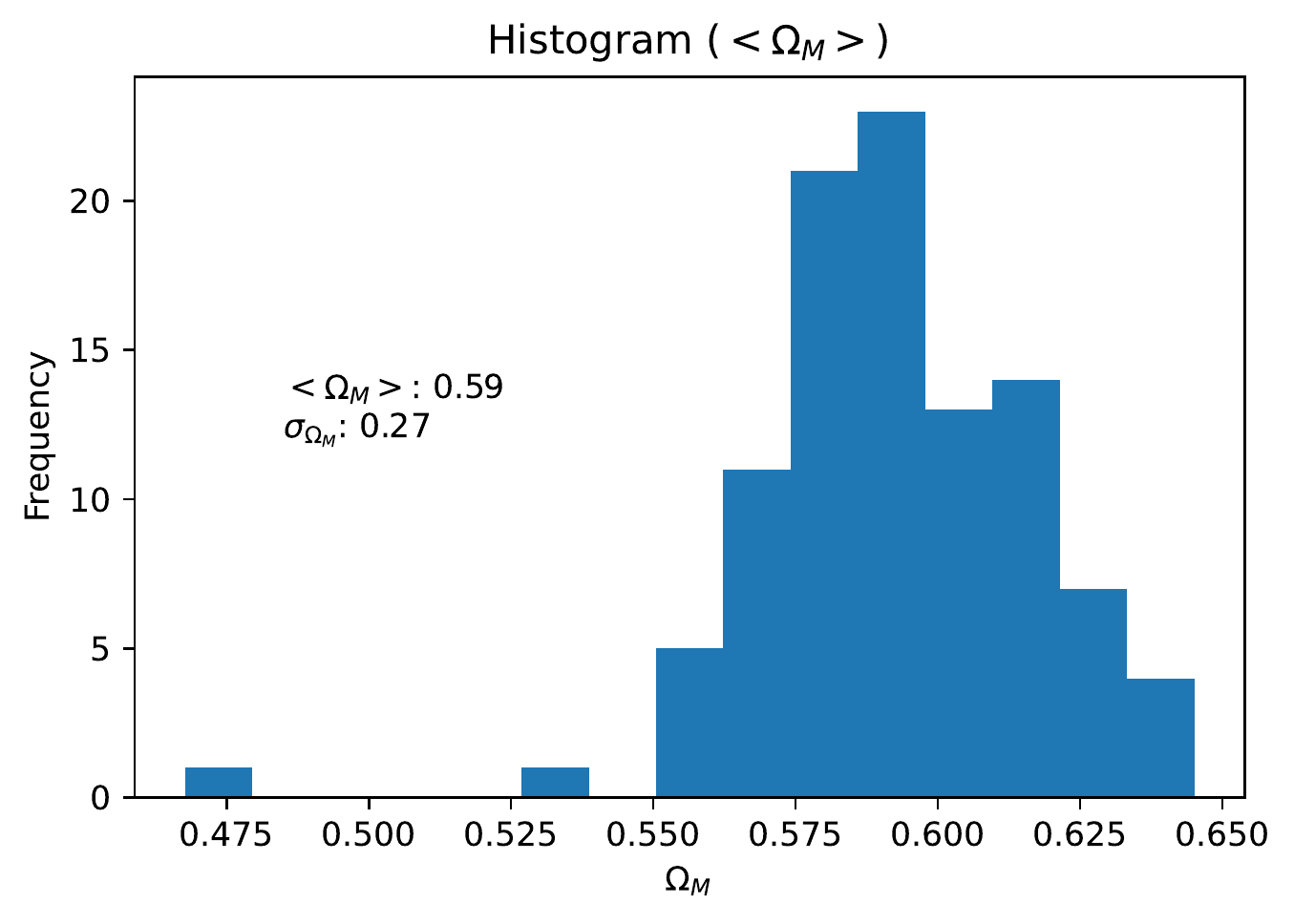}}
\subfloat[Varying only $H_0$ with fixed evolution]{
\includegraphics[width=0.32\hsize,height=0.27\textwidth,angle=0,clip]{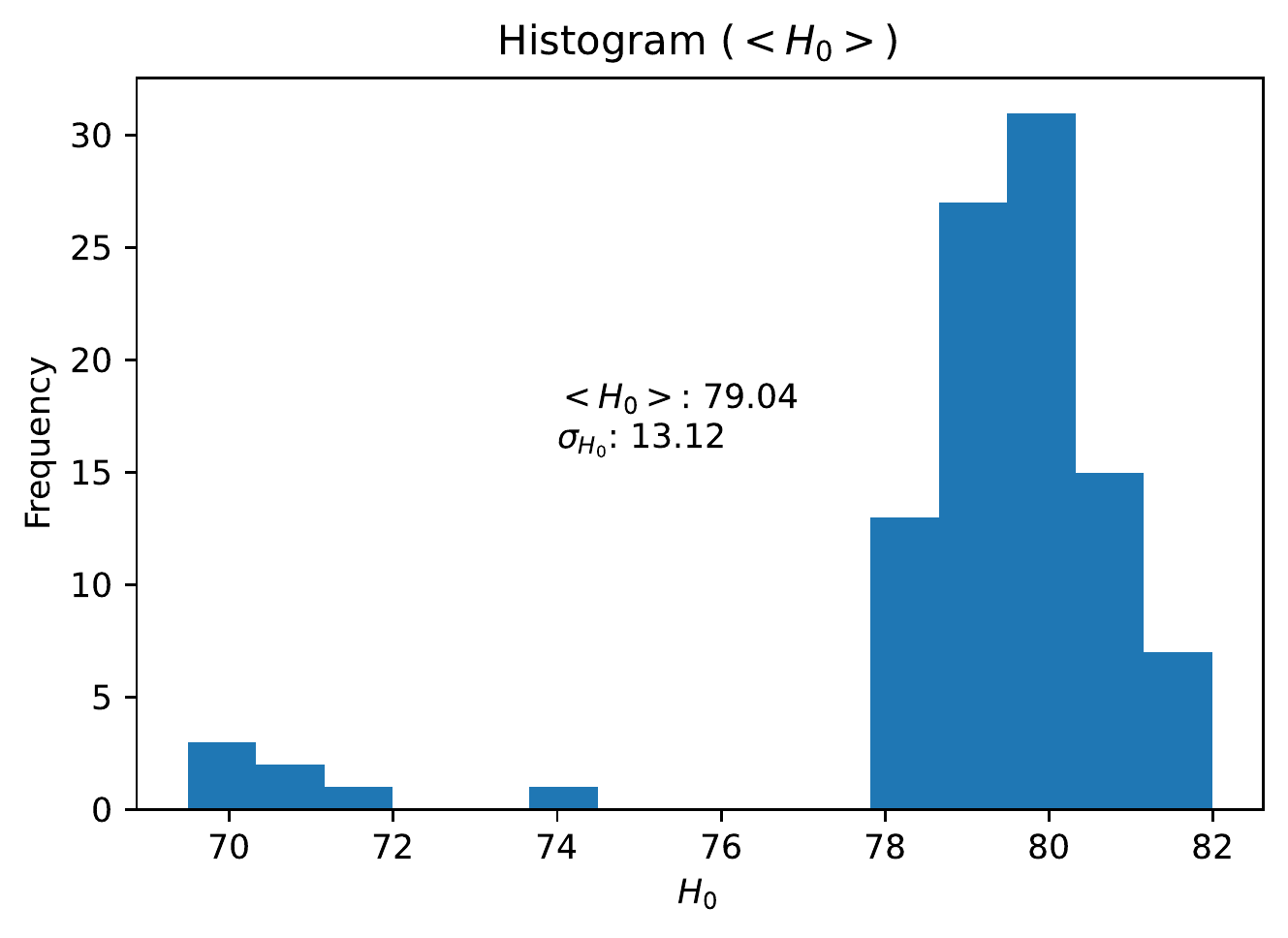}}
\subfloat[Varying only $w$ with fixed evolution]{
\includegraphics[width=0.32\hsize,height=0.27\textwidth,angle=0,clip]{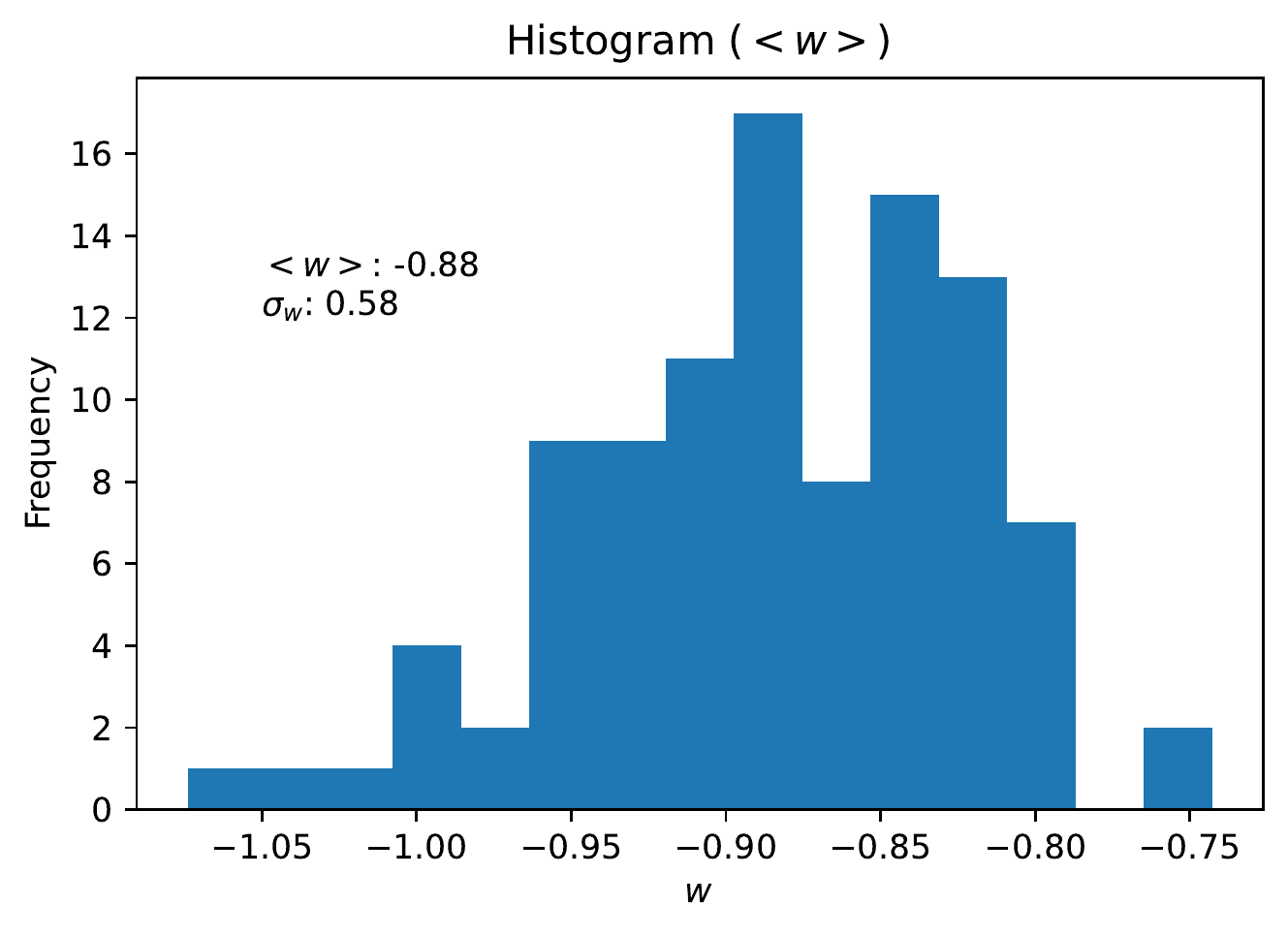}}\\

\subfloat[Varying only $\Omega_M$ with evolutionary function]{
\includegraphics[width=0.32\hsize,height=0.27\textwidth,angle=0,clip]{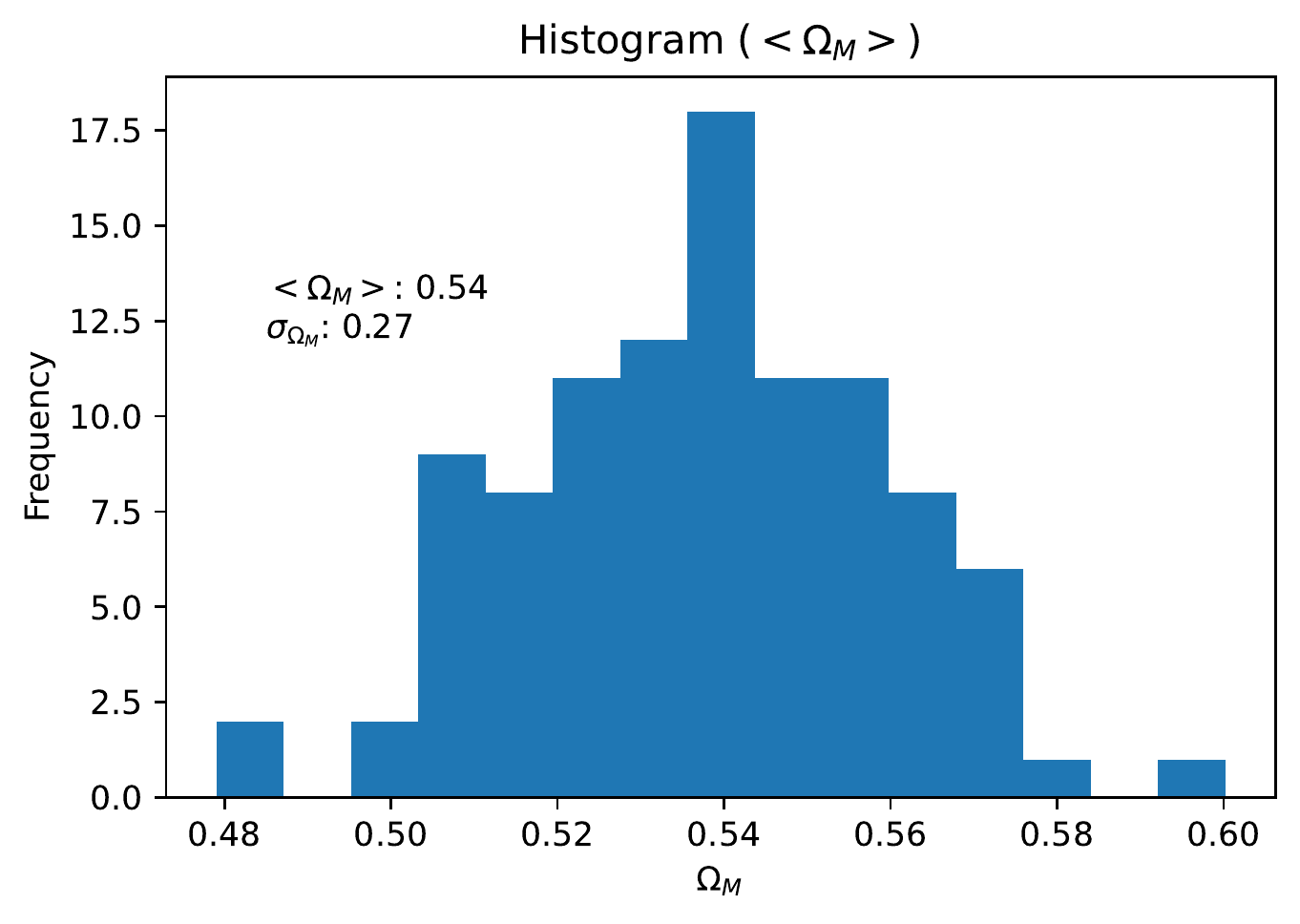}}
\caption{The distributions of the cosmological parameters for the GRBs with calibration and with $\mu_{GRB}$ using the assumptions of uniform priors of 100 runs of MCMC. Panels a), b), c) show the contours from case with no evolution (upper panel), evolution with fixed parameters (central panel) and with the evolutionary functions (lower panel) for the case of $\Omega_M$, $H_0$, and $w$, respectively.}
\label{fig21}
\end{figure}

\begin{figure} 
\centering

\subfloat[Varying only $\Omega_M$ without evolution]{
\includegraphics[width=0.32\hsize,height=0.27\textwidth,angle=0,clip]{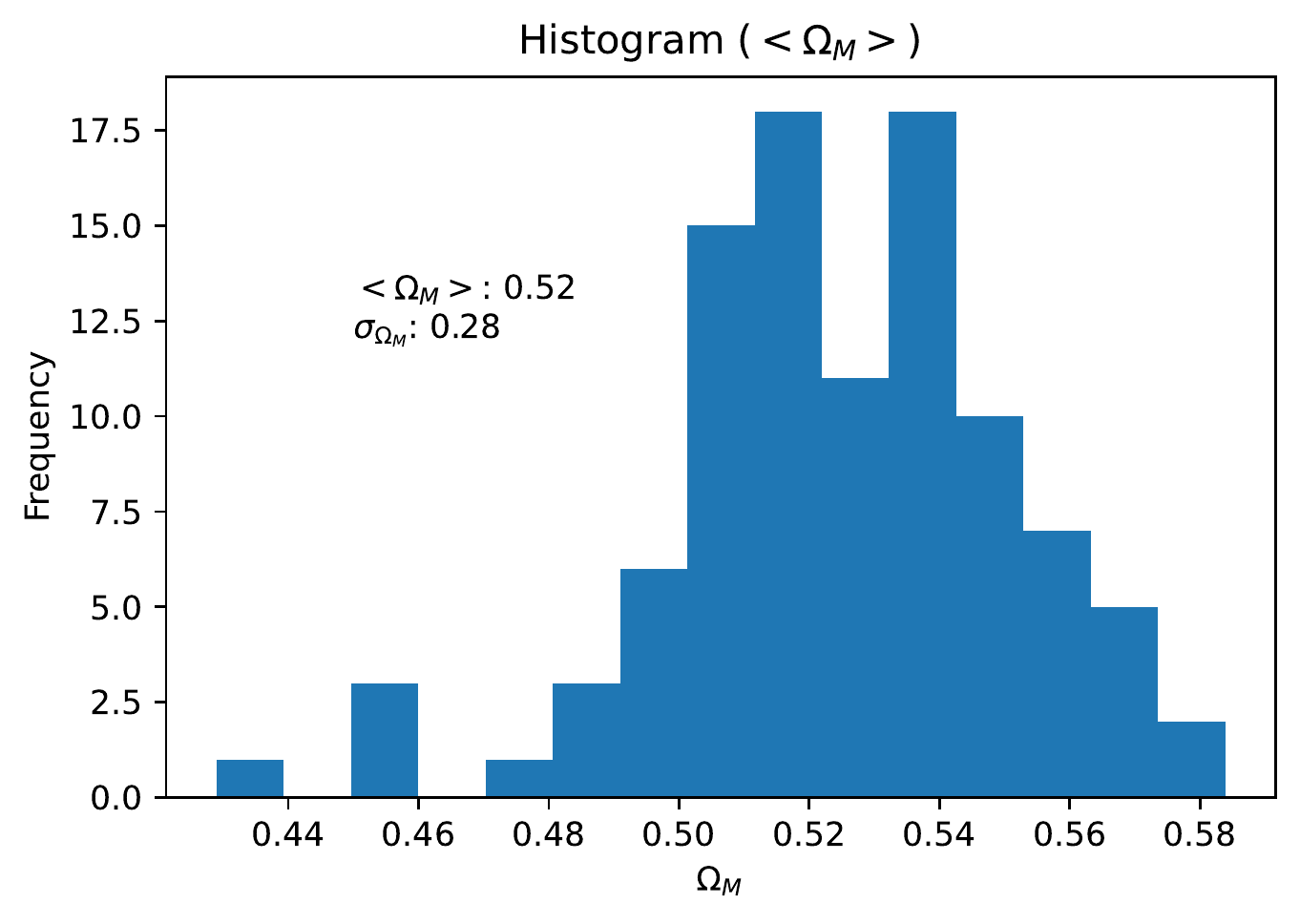}}
\subfloat[Varying only $H_0$ without evolution]{
\includegraphics[width=0.32\hsize,height=0.27\textwidth,angle=0,clip]{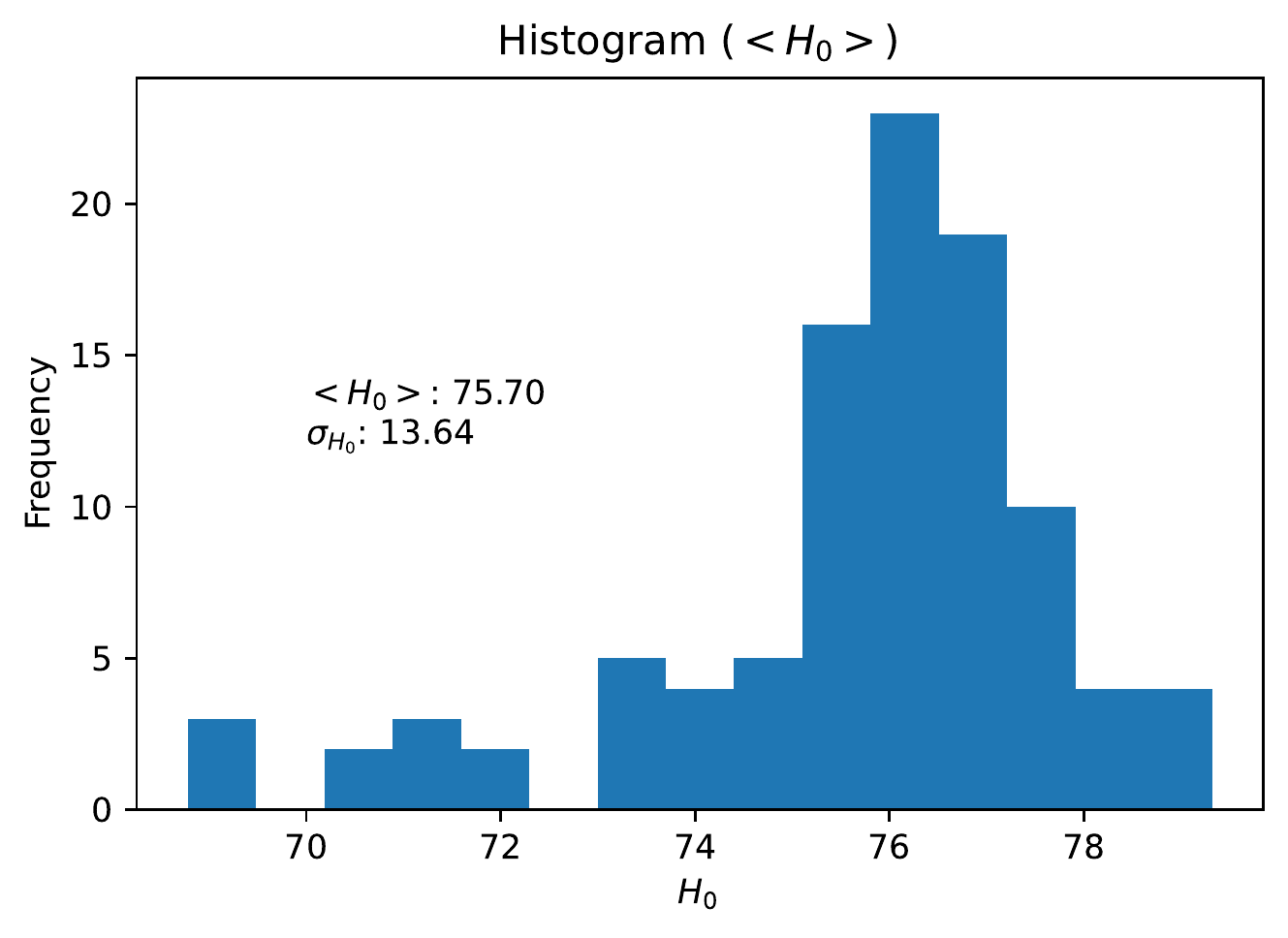}}
\subfloat[Varying only $w$ without evolution]{
\includegraphics[width=0.32\hsize,height=0.27\textwidth,angle=0,clip]{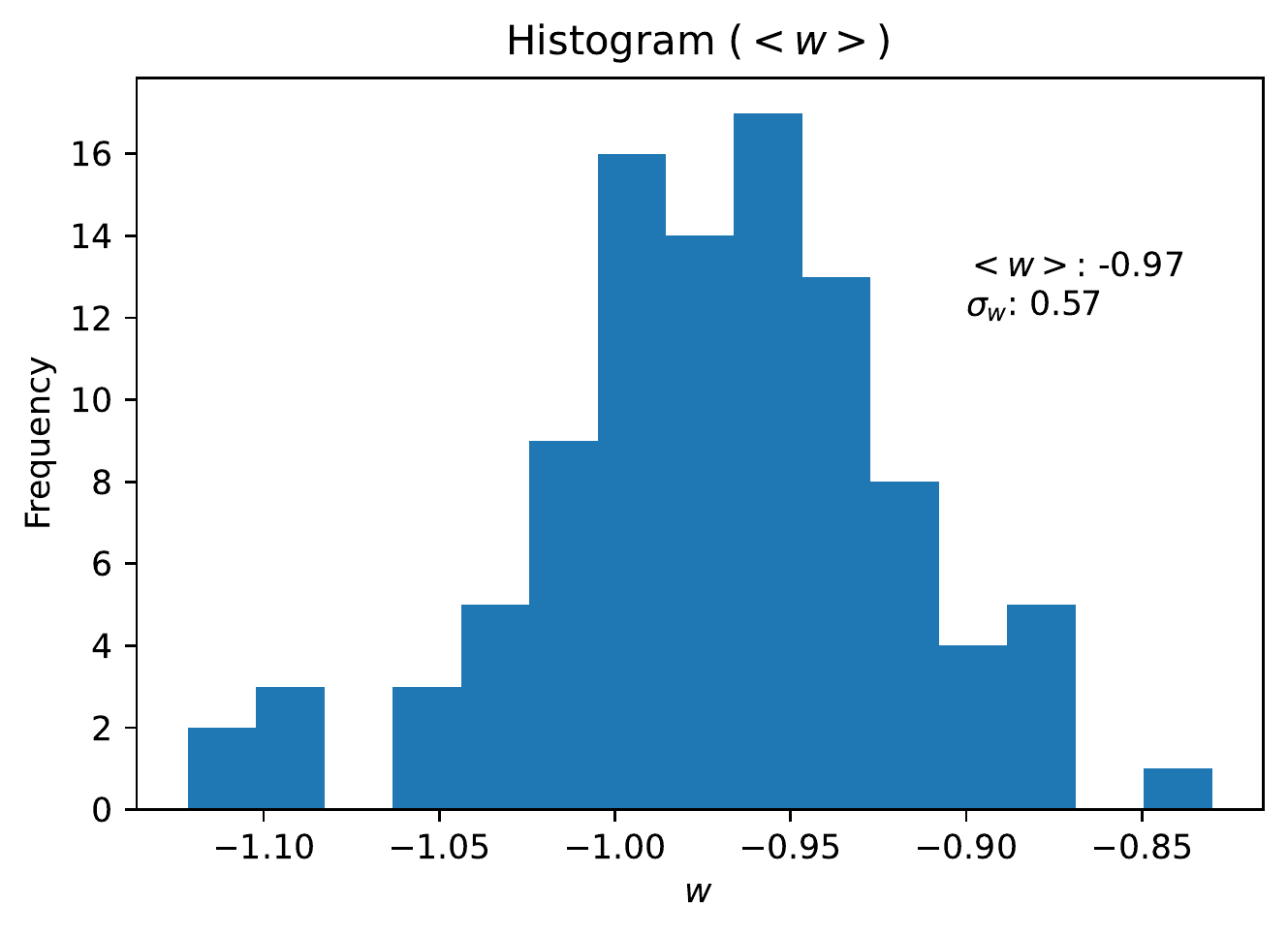}}\\

\subfloat[Varying only $\Omega_M$ with fixed evolution]{
\includegraphics[width=0.32\hsize,height=0.27\textwidth,angle=0,clip]{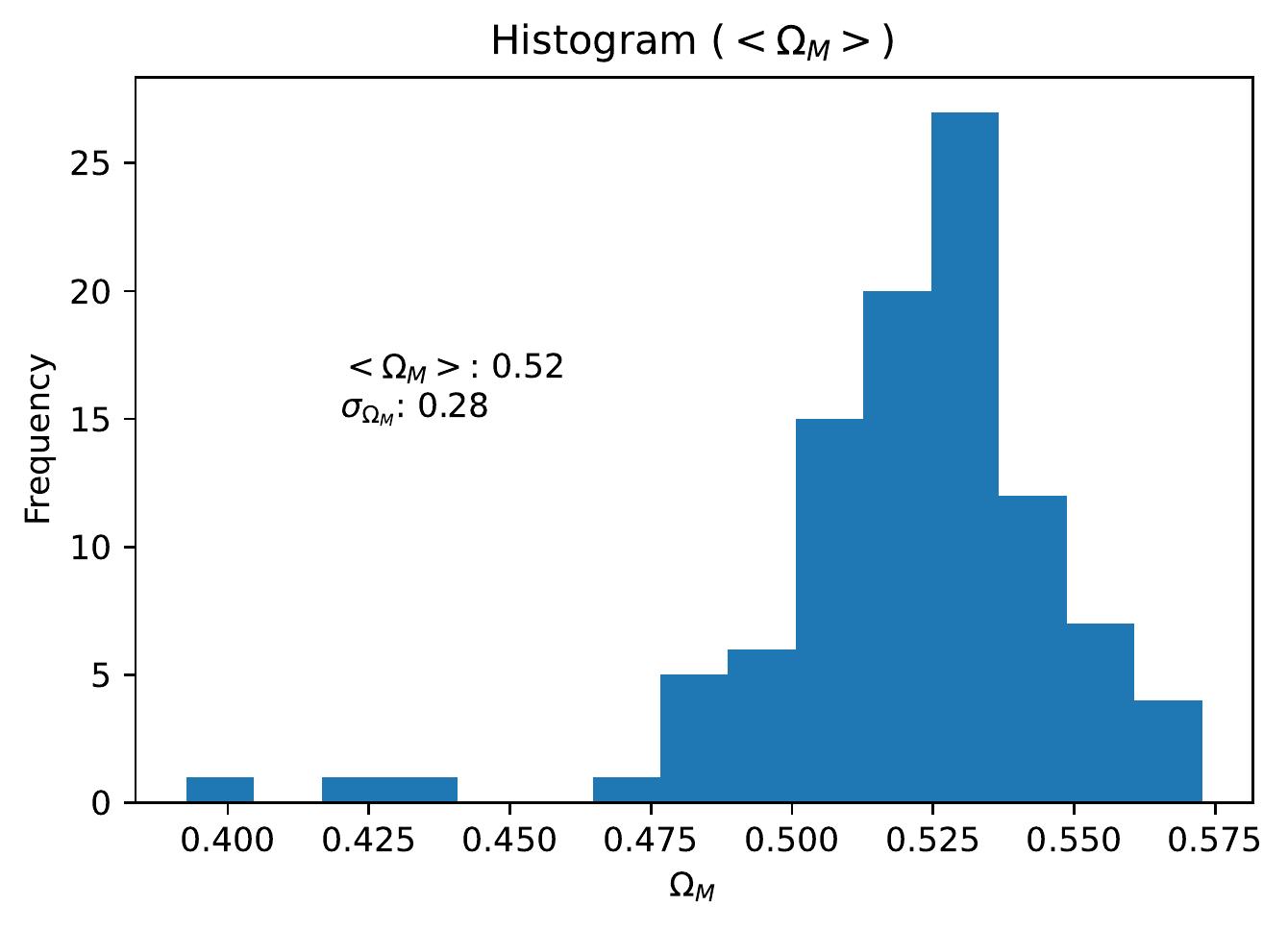}}
\subfloat[Varying only $H_0$ with fixed evolution]{
\includegraphics[width=0.32\hsize,height=0.27\textwidth,angle=0,clip]{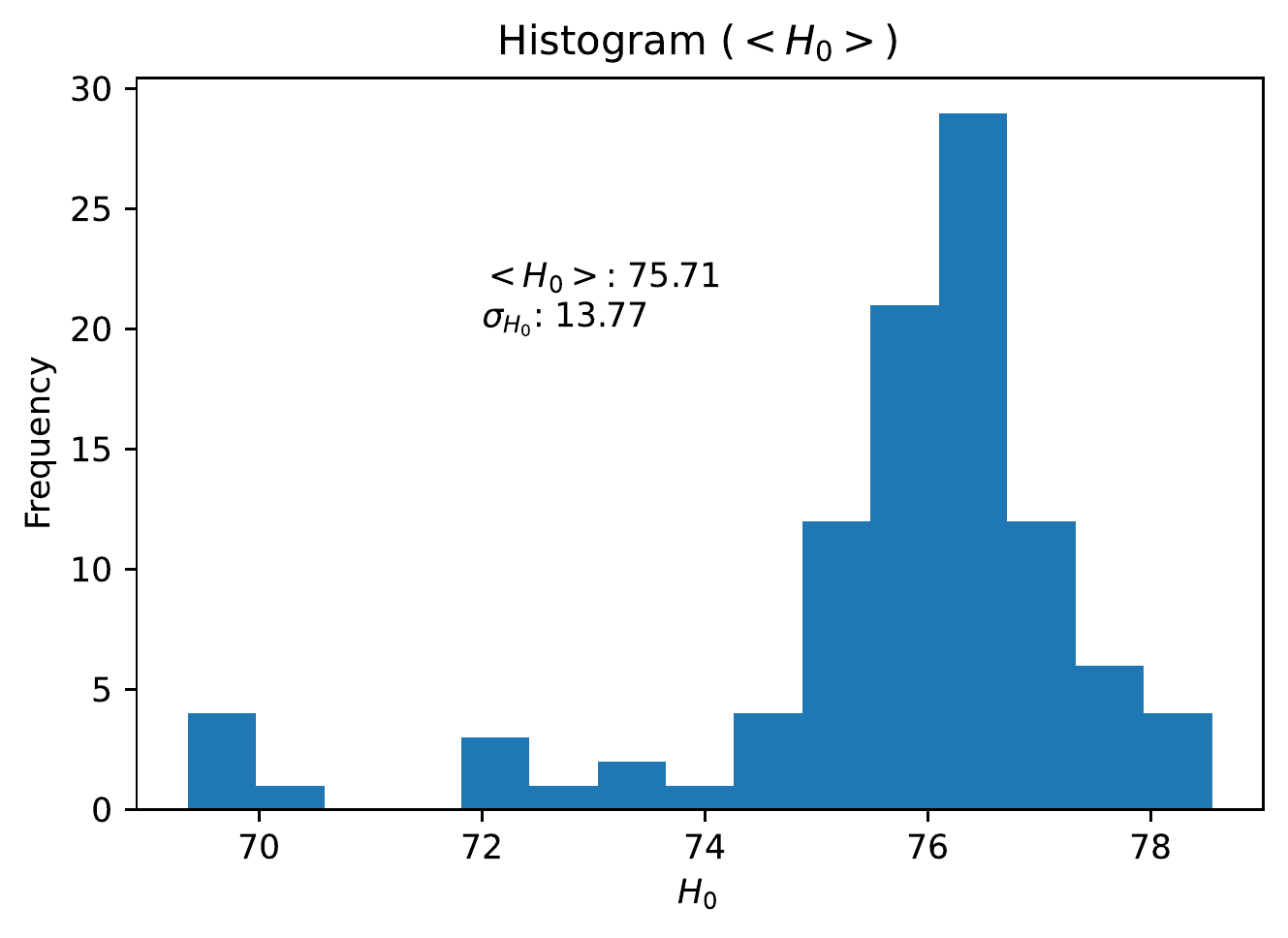}}
\subfloat[Varying only $w$ with fixed evolution]{
\includegraphics[width=0.32\hsize,height=0.27\textwidth,angle=0,clip]{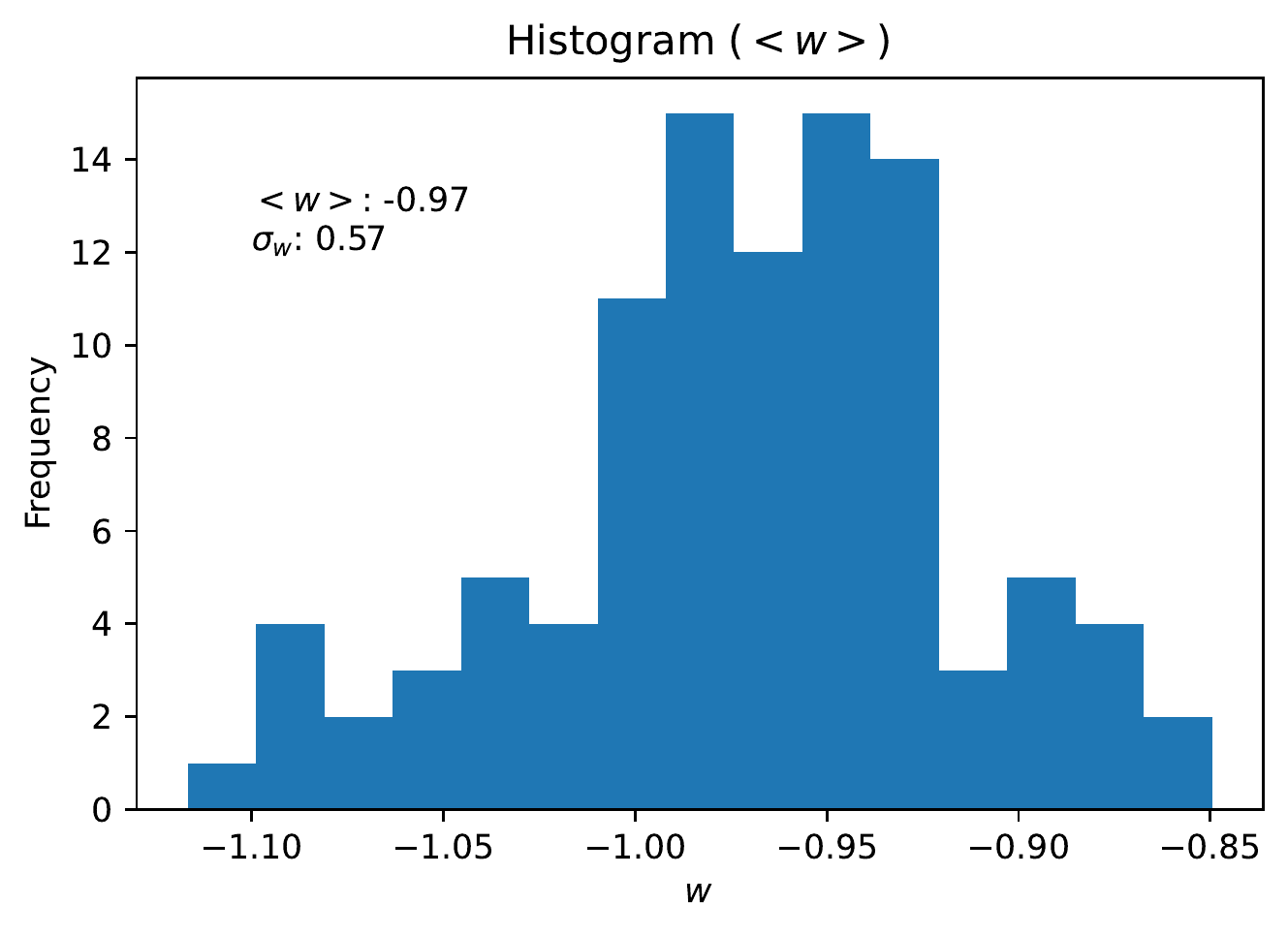}}\\

\subfloat[Varying only $\Omega_M$ with evolutionary function]{
\includegraphics[width=0.32\hsize,height=0.27\textwidth,angle=0,clip]{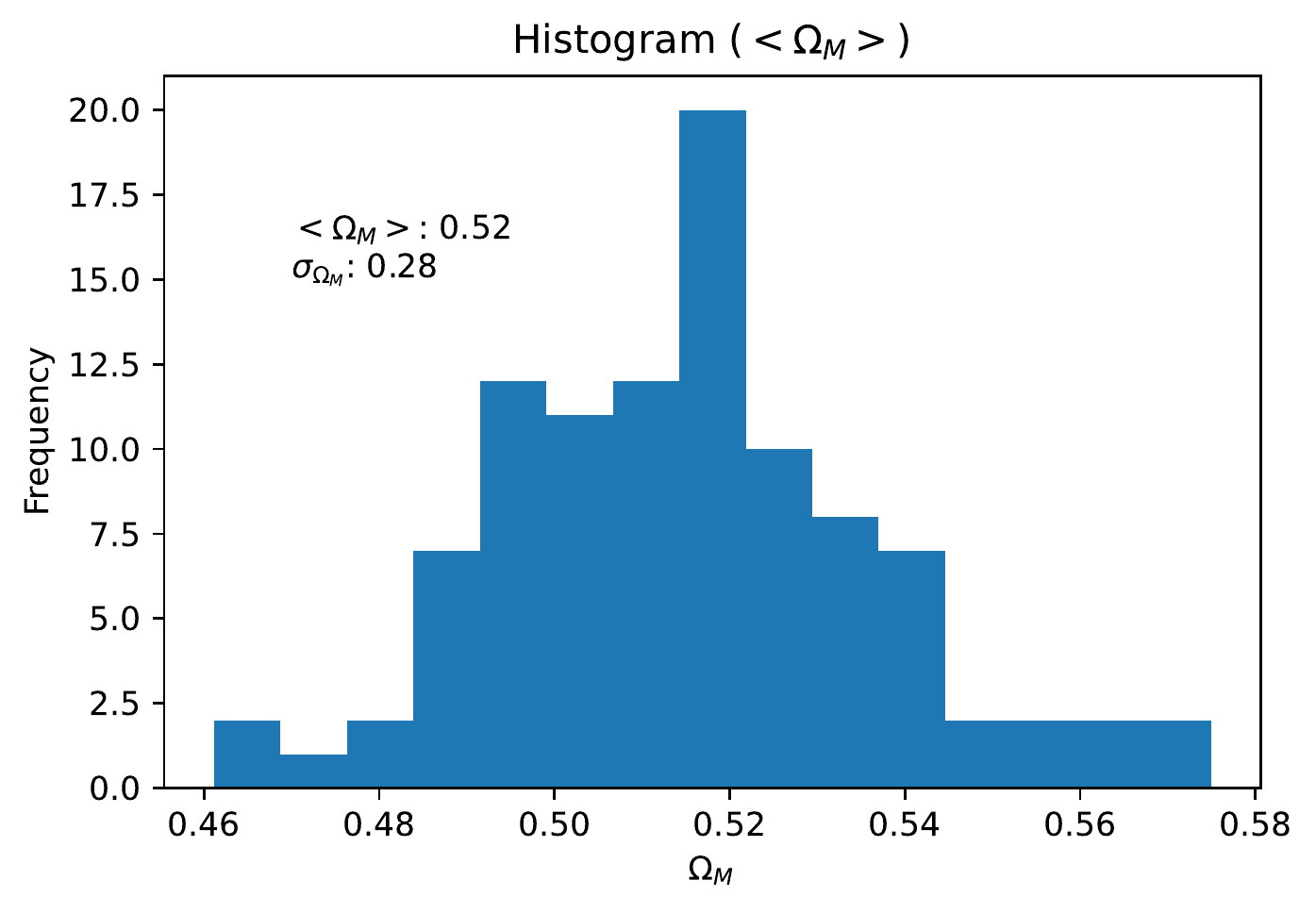}}

\caption{The distributions of the cosmological parameters for the GRBs with calibration and with $L_X$, Equations \ref{isotropic} and \ref{planeev} using the assumptions of uniform priors and over 100 runs of MCMC. Panels a), b), c) show the contours from case with no evolution (upper panel), evolution with fixed parameters (central panel) and with the evolutionary functions (lower panel) for the case of $\Omega_M$, $H_0$, and $w$, respectively.}
\label{fig22}
\end{figure}

\begin{figure} 
\centering


\subfloat[Varying only $\Omega_M$ without evolution]{
\includegraphics[width=0.32\hsize,height=0.27\textwidth,angle=0,clip]{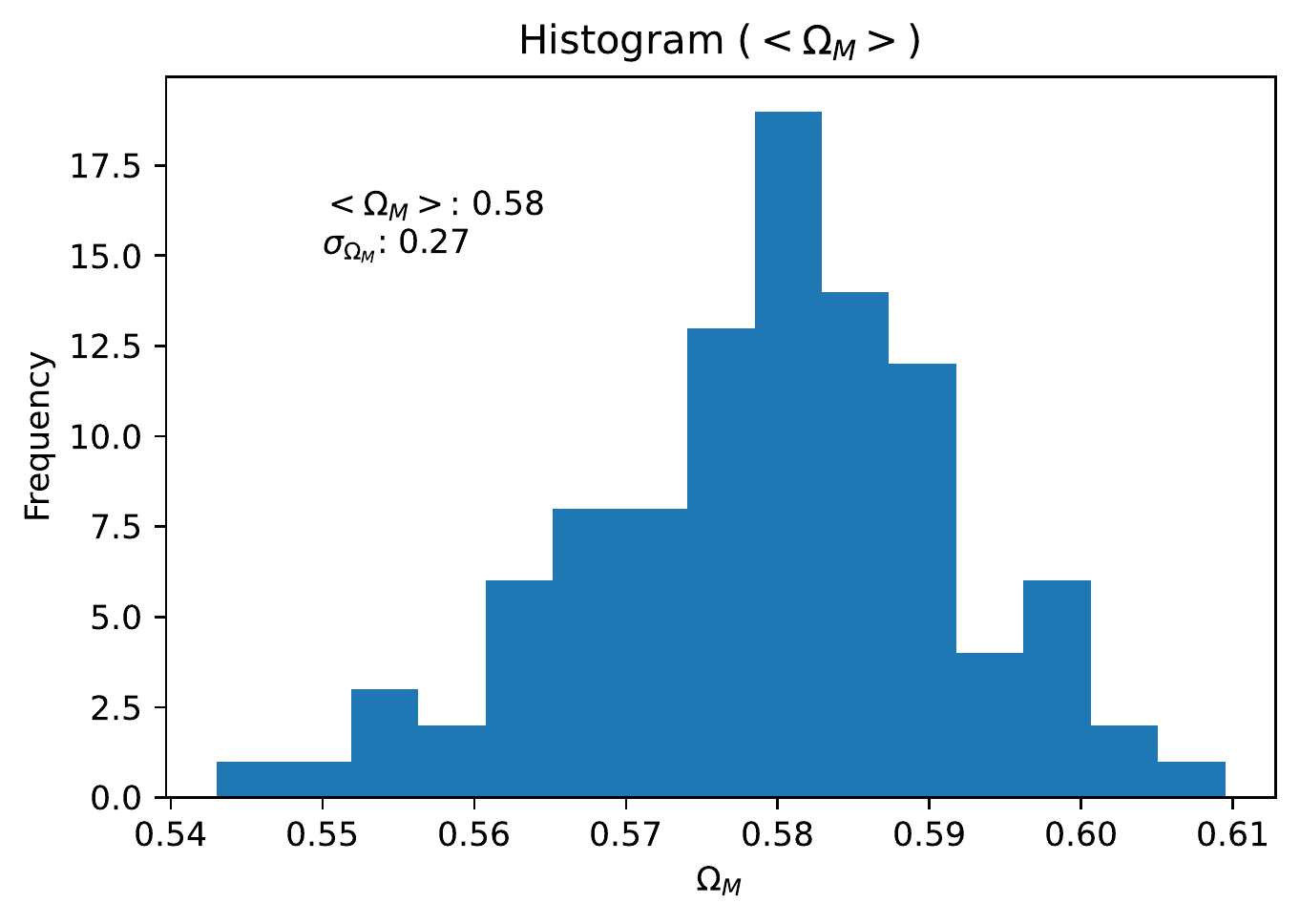}}
\subfloat[Varying only $H_0$ without evolution]{
\includegraphics[width=0.32\hsize,height=0.27\textwidth,angle=0,clip]{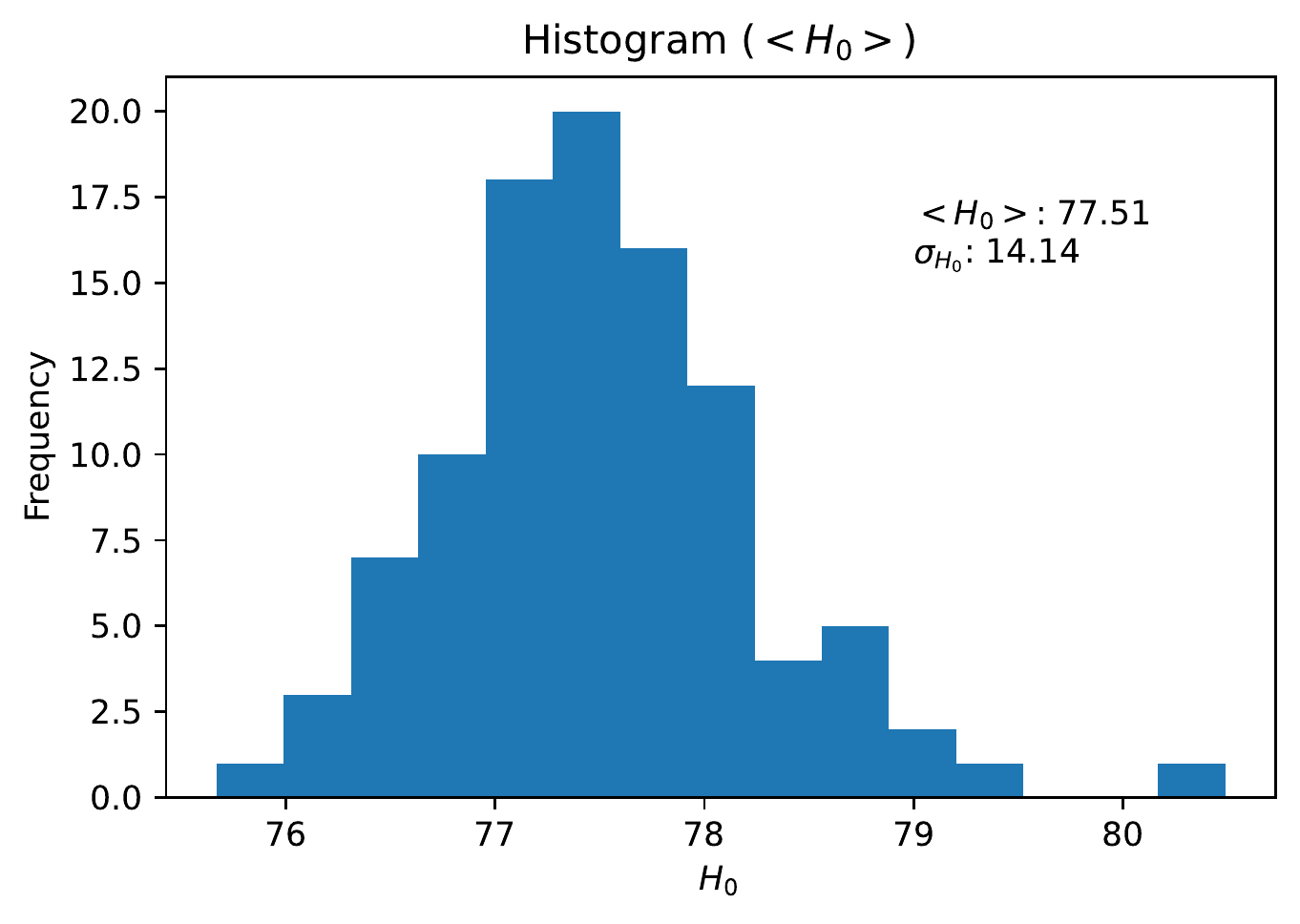}}
\subfloat[Varying only $w$ without evolution]{
\includegraphics[width=0.32\hsize,height=0.27\textwidth,angle=0,clip]{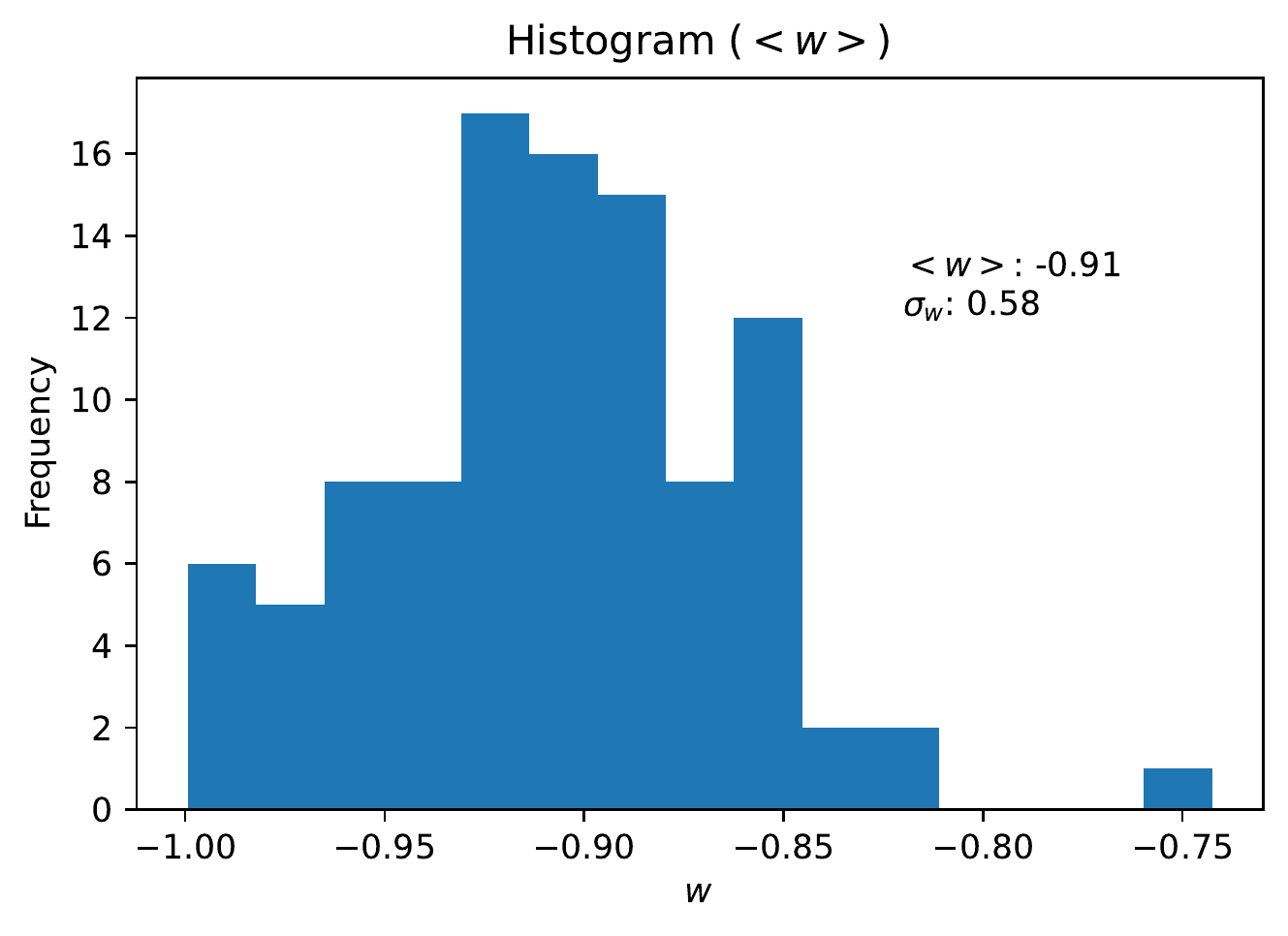}}\\

\subfloat[Varying only $\Omega_M$ with fixed evolution]{
\includegraphics[width=0.32\hsize,height=0.27\textwidth,angle=0,clip]{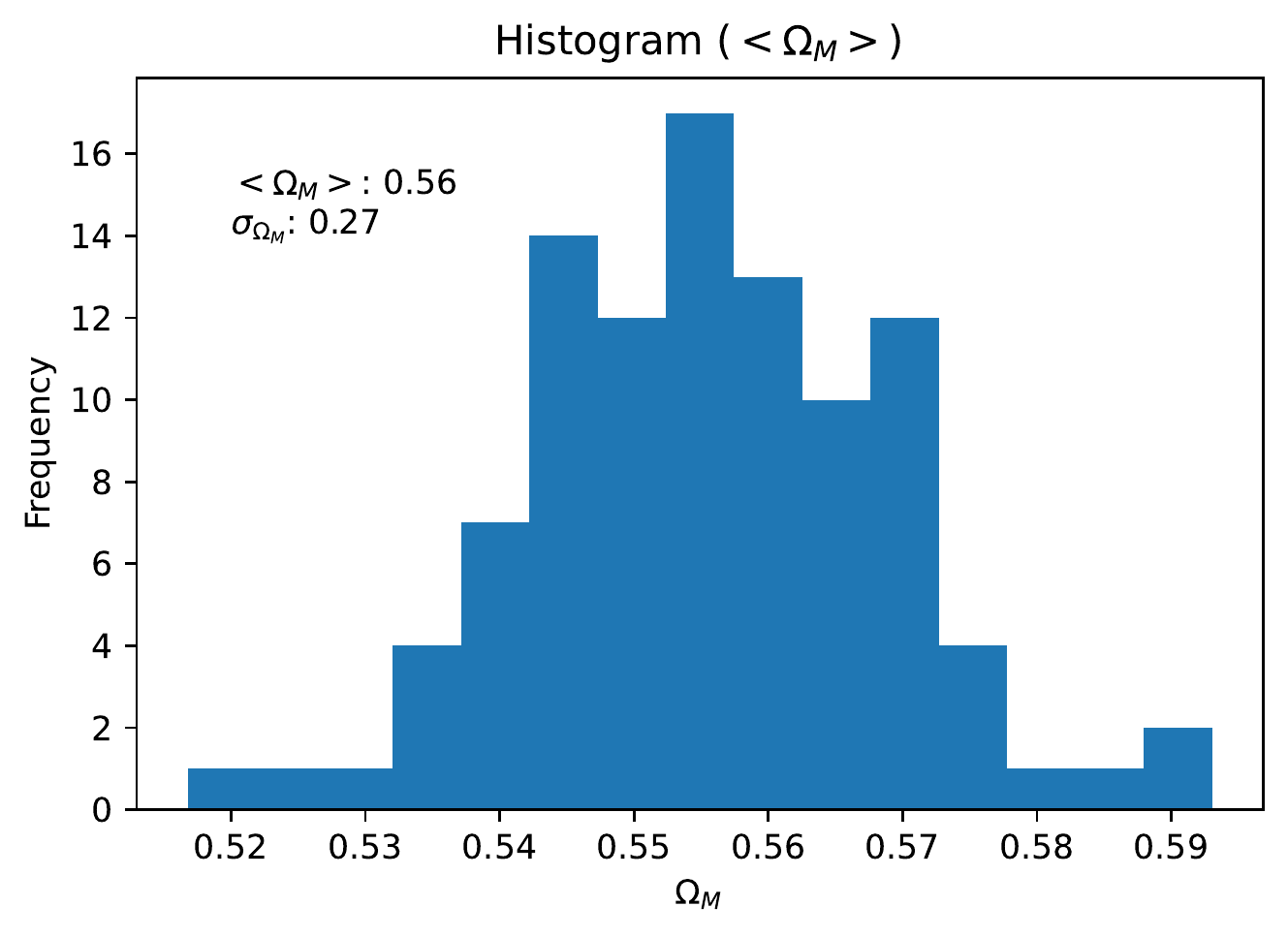}}
\subfloat[Varying only $H_0$ with fixed evolution]{
\includegraphics[width=0.32\hsize,height=0.27\textwidth,angle=0,clip]{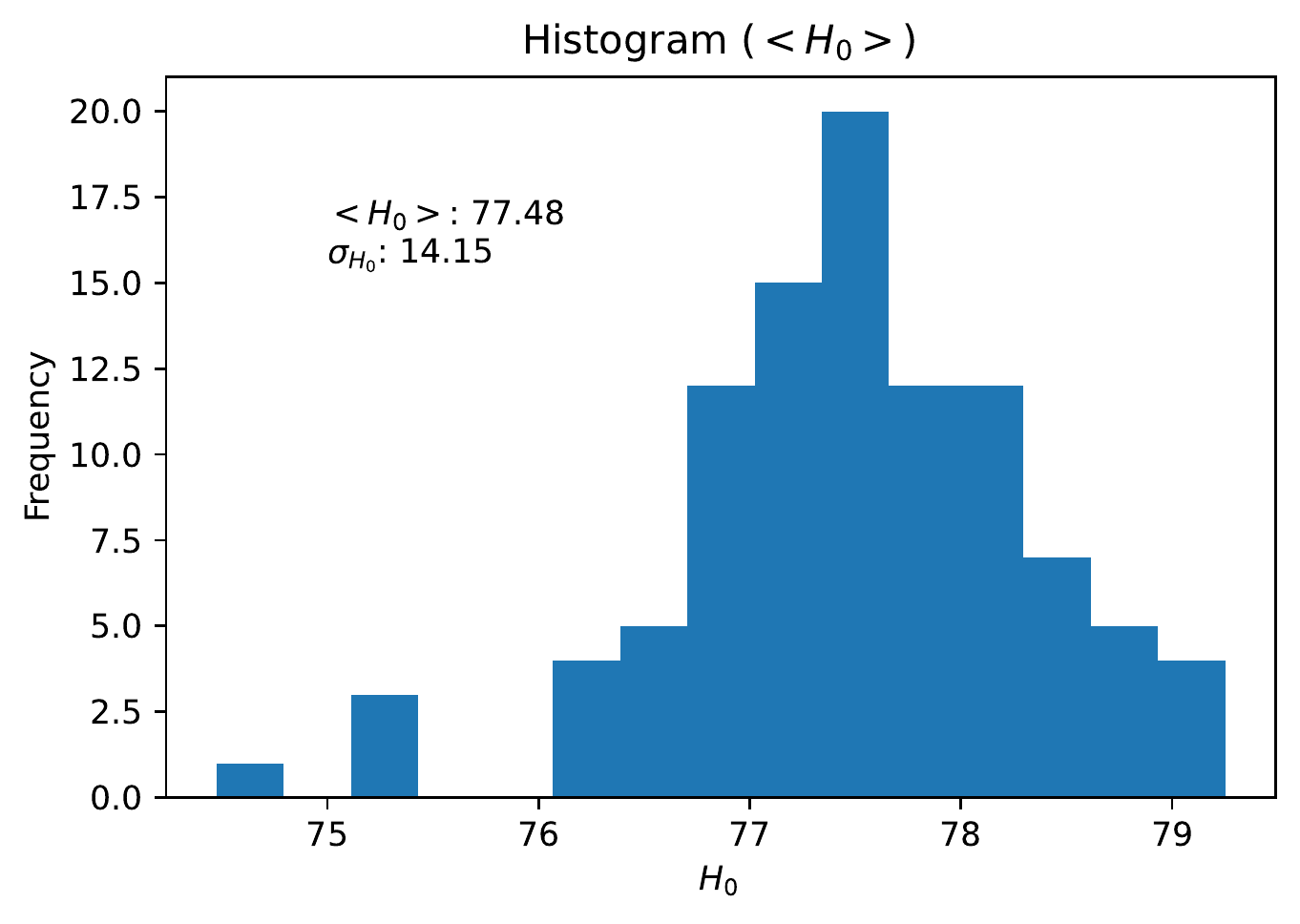}}
\subfloat[Varying only $w$ with fixed evolution]{
\includegraphics[width=0.32\hsize,height=0.27\textwidth,angle=0,clip]{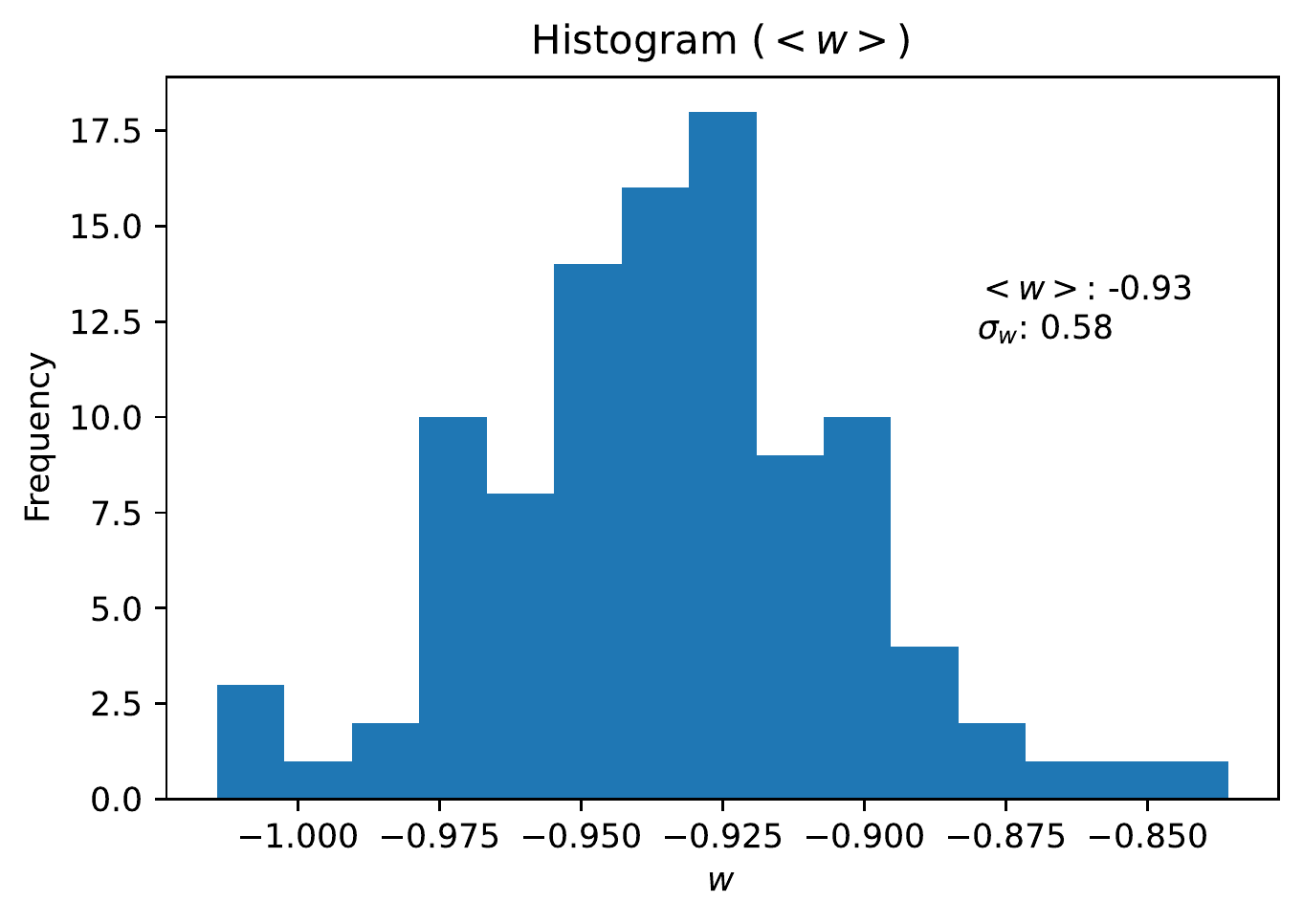}}\\

\subfloat[Varying only $\Omega_M$ with evolutionary function]{
\includegraphics[width=0.32\hsize,height=0.27\textwidth,angle=0,clip]{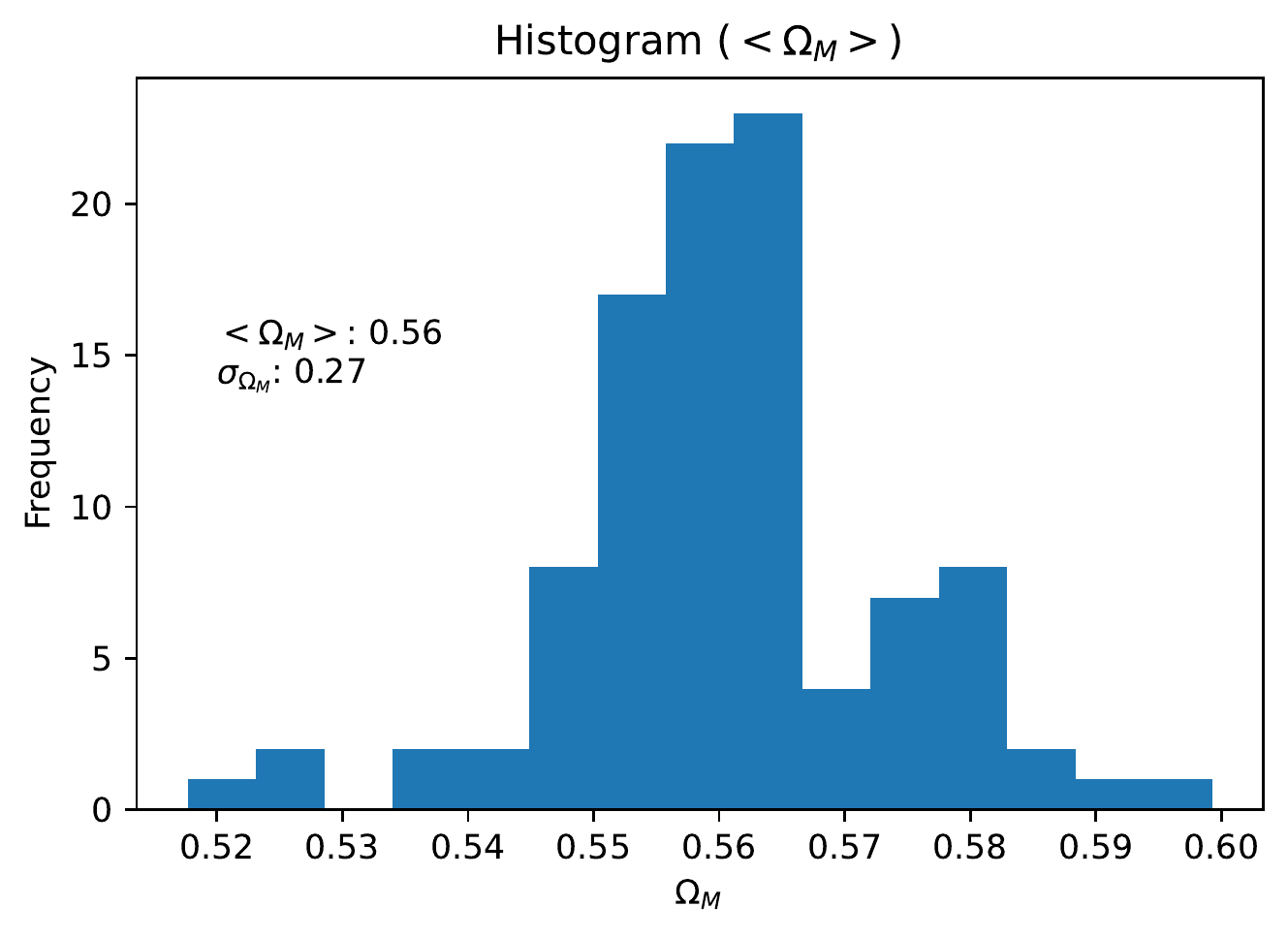}}
\caption{The distributions of the cosmological parameters for the GRBs without calibration and with $\mu_{GRB}$ using the assumptions of uniform priors of 100 runs of MCMC. Panels a), b), c) show the contours from case with no evolution (upper panels), evolution with fixed parameters (central panels) and with the evolutionary functions (lower panel) for the case of $\Omega_M$.}
\label{fig23}
\end{figure}

\begin{figure} 
\centering

\subfloat[Varying only $\Omega_M$ without evolution]{
\includegraphics[width=0.32\hsize,height=0.27\textwidth,angle=0,clip]{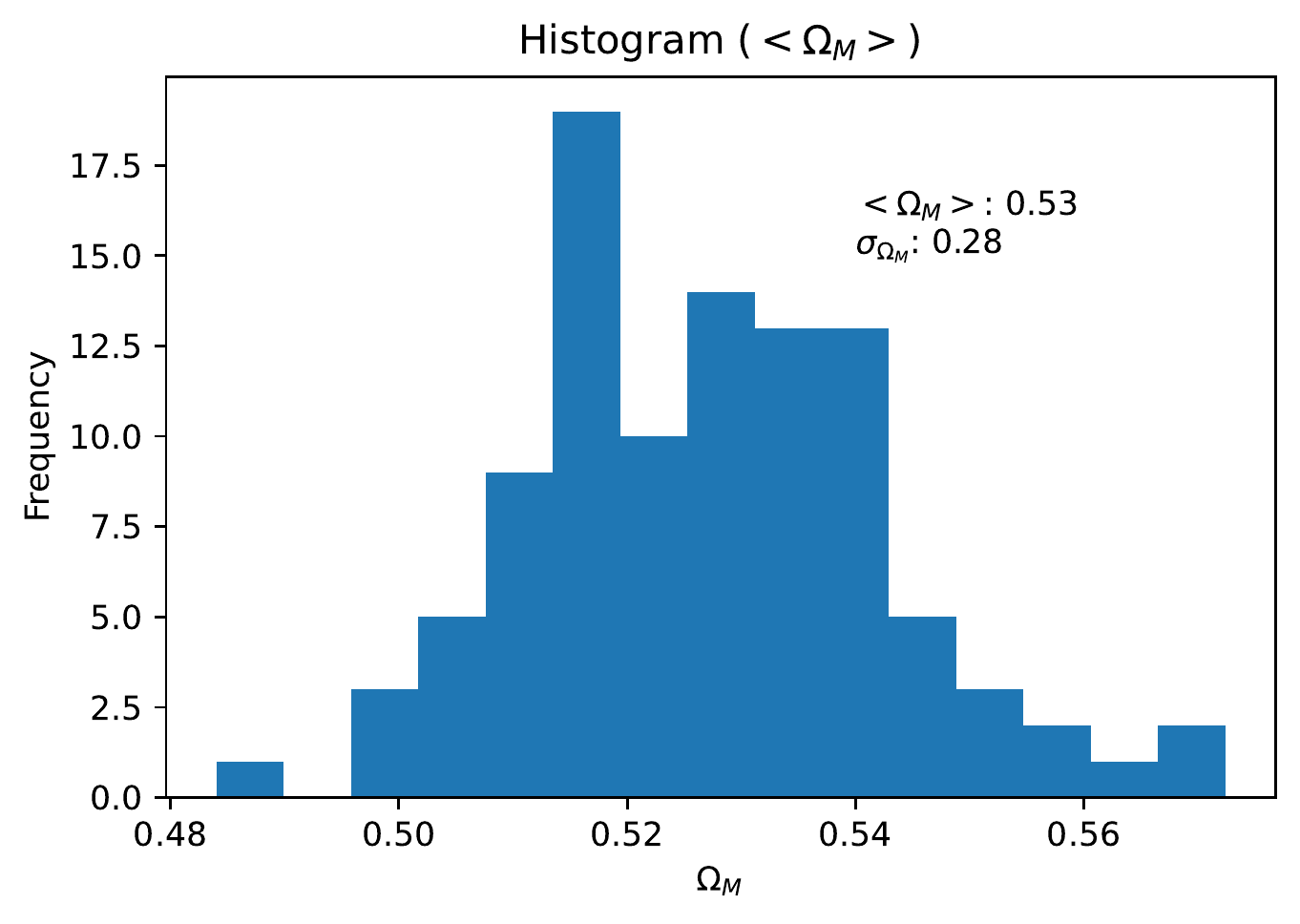}}
\subfloat[Varying only $H_0$ without evolution]{
\includegraphics[width=0.32\hsize,height=0.27\textwidth,angle=0,clip]{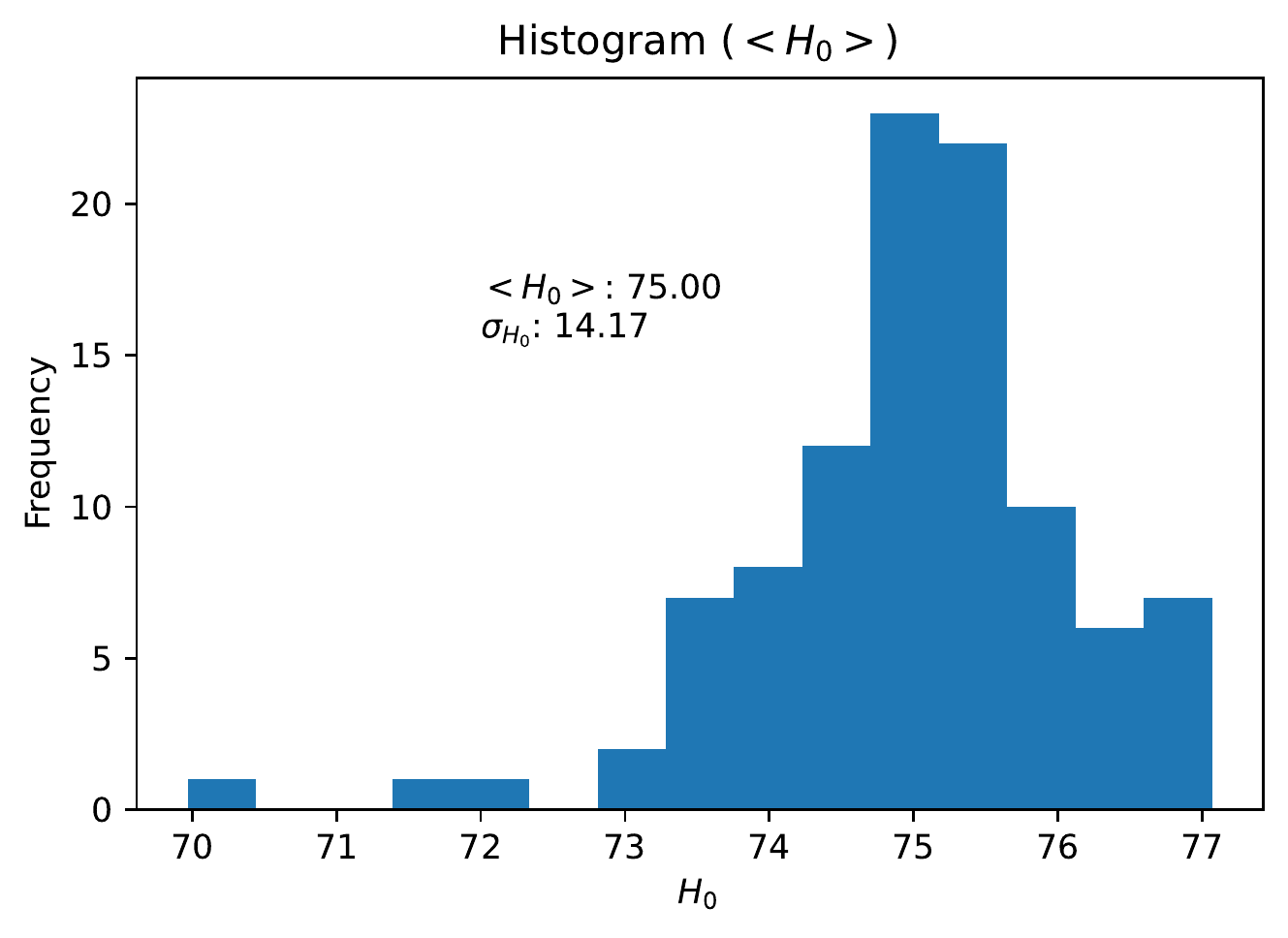}}
\subfloat[Varying only $w$ without evolution]{
\includegraphics[width=0.32\hsize,height=0.27\textwidth,angle=0,clip]{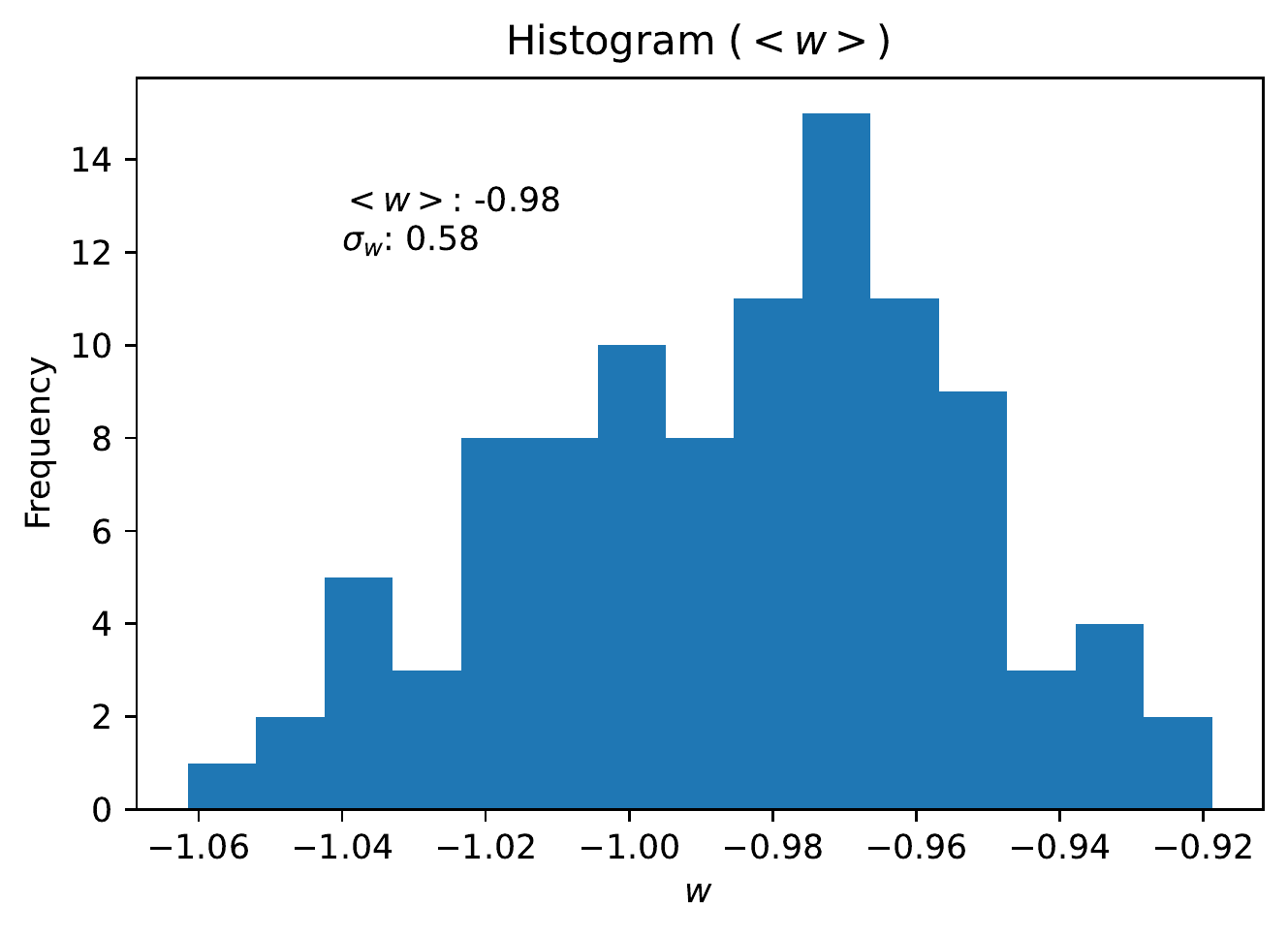}}\\

\subfloat[Varying only $\Omega_M$ with fixed evolution]{
\includegraphics[width=0.32\hsize,height=0.27\textwidth,angle=0,clip]{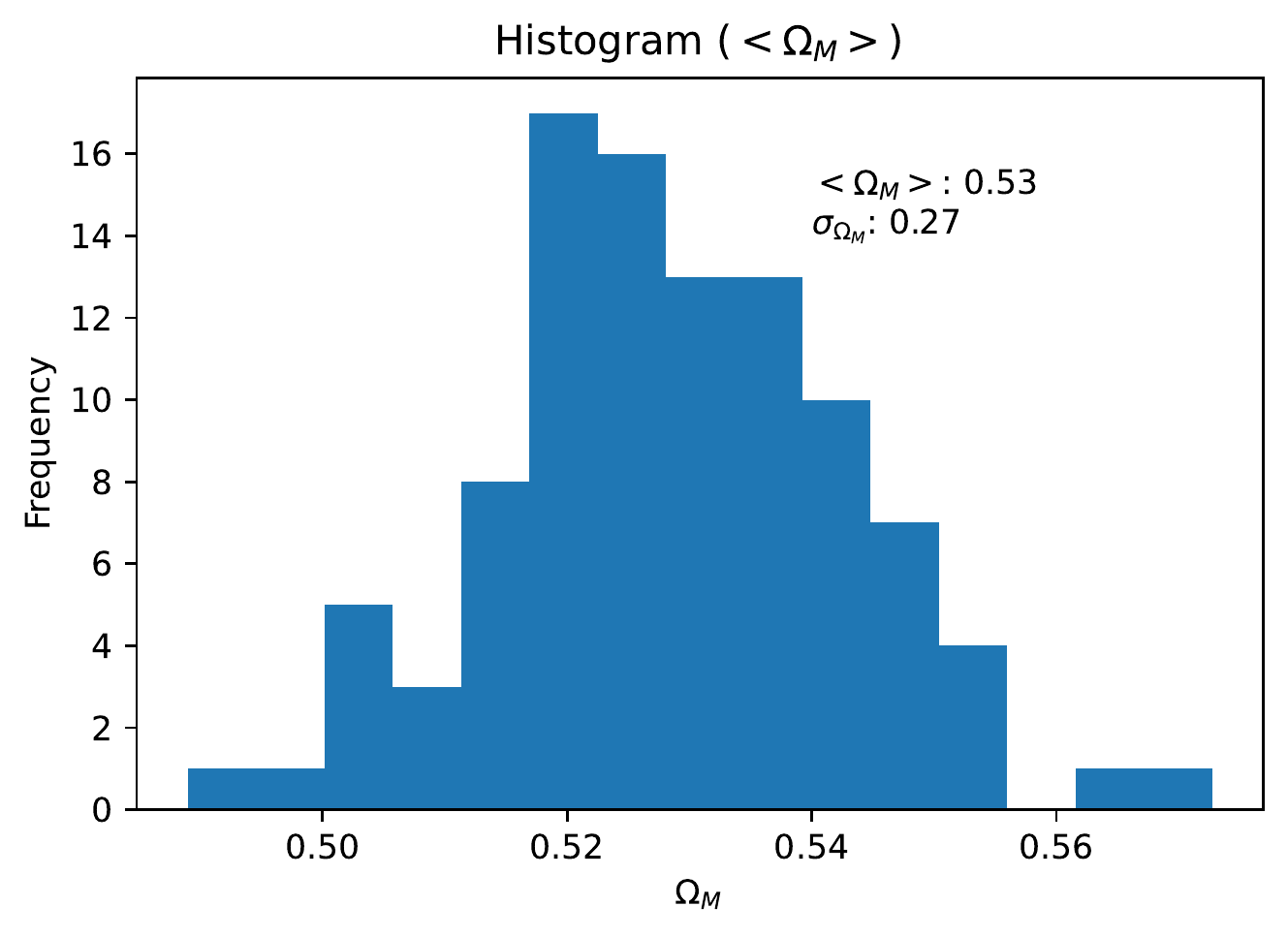}}
\subfloat[Varying only $H_0$ with fixed evolution]{
\includegraphics[width=0.32\hsize,height=0.27\textwidth,angle=0,clip]{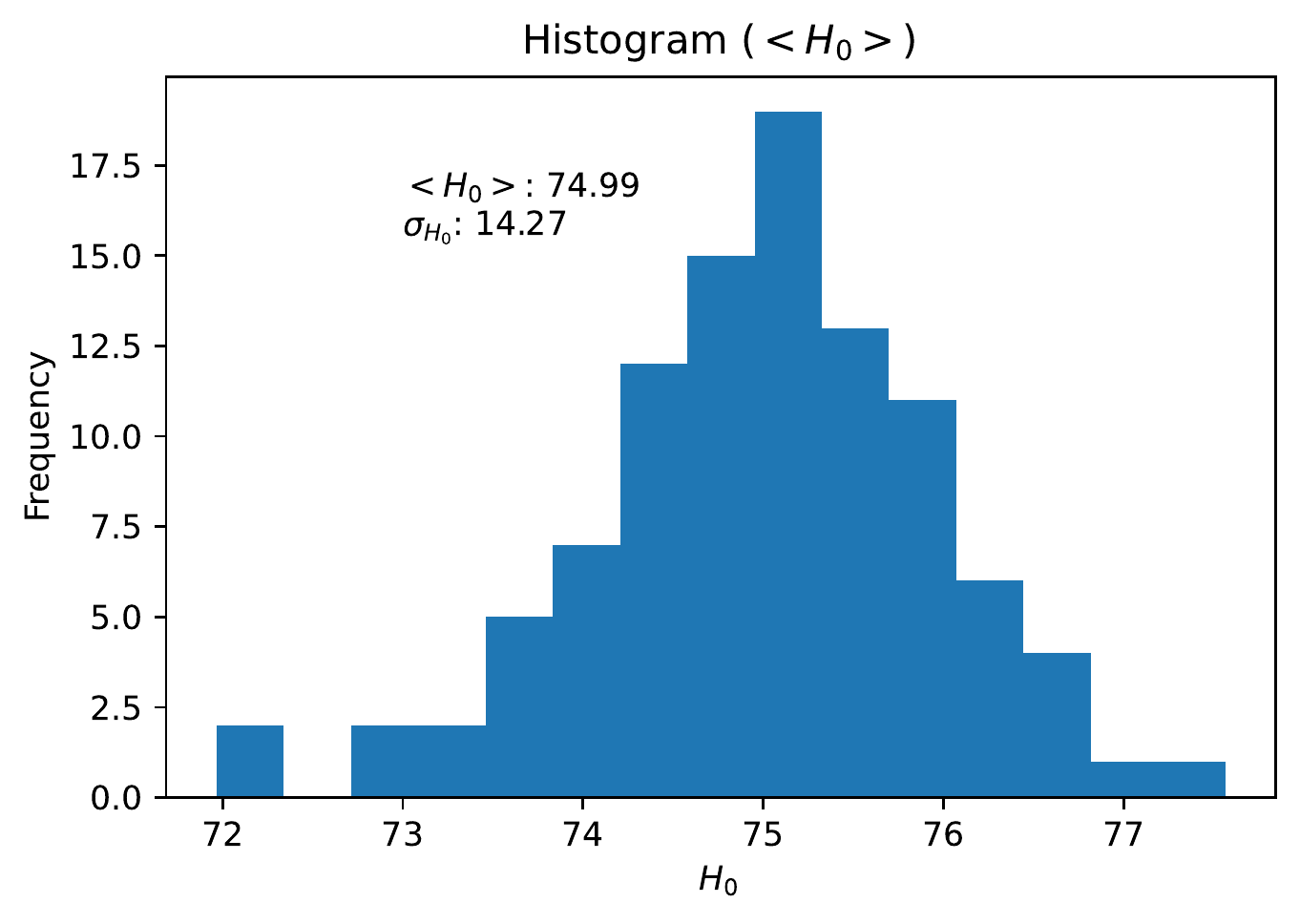}}
\subfloat[Varying only $w$ with fixed evolution]{
\includegraphics[width=0.32\hsize,height=0.27\textwidth,angle=0,clip]{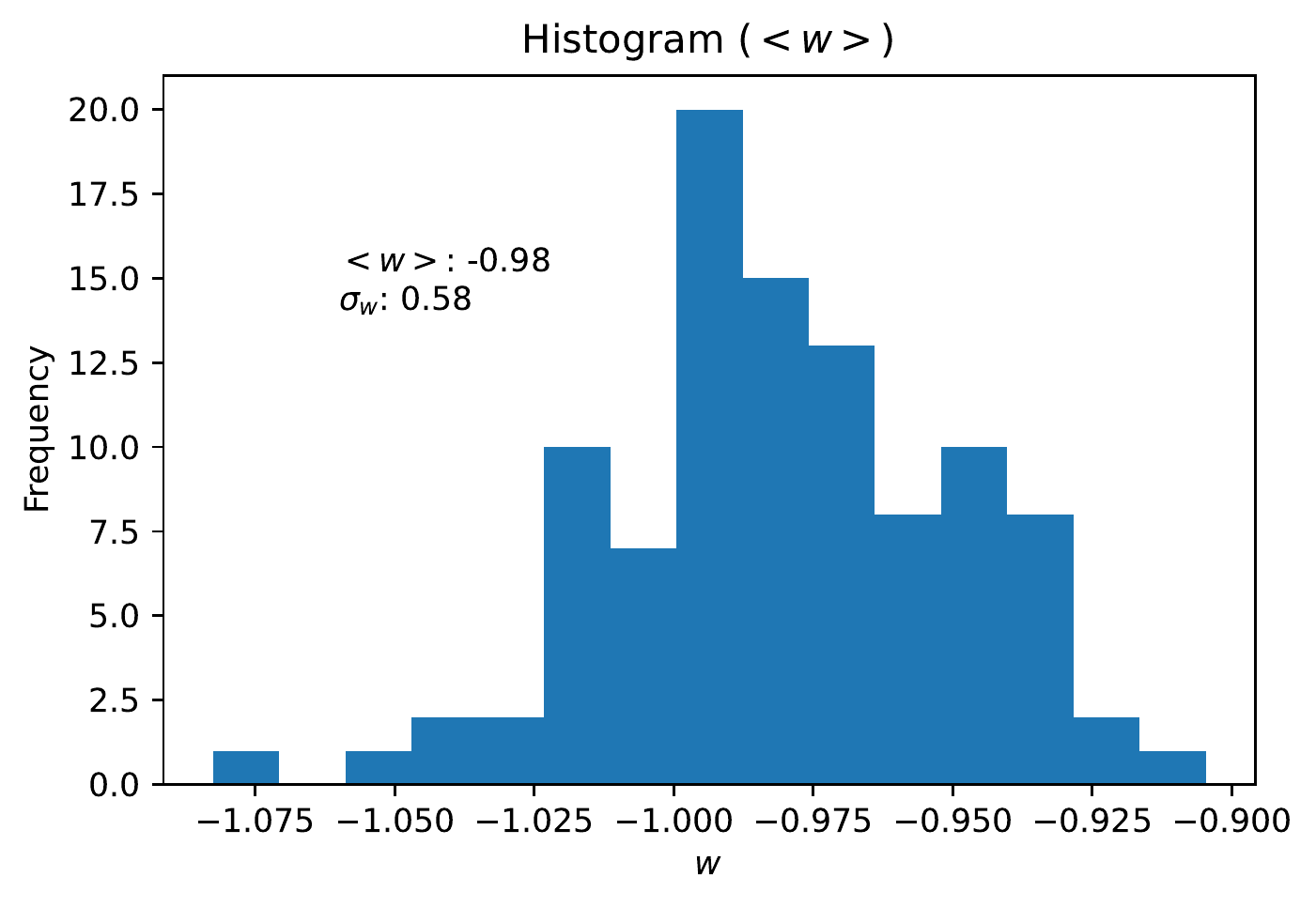}}\\

\subfloat[Varying only $\Omega_M$ with evolutionary function]{
\includegraphics[width=0.32\hsize,height=0.27\textwidth,angle=0,clip]{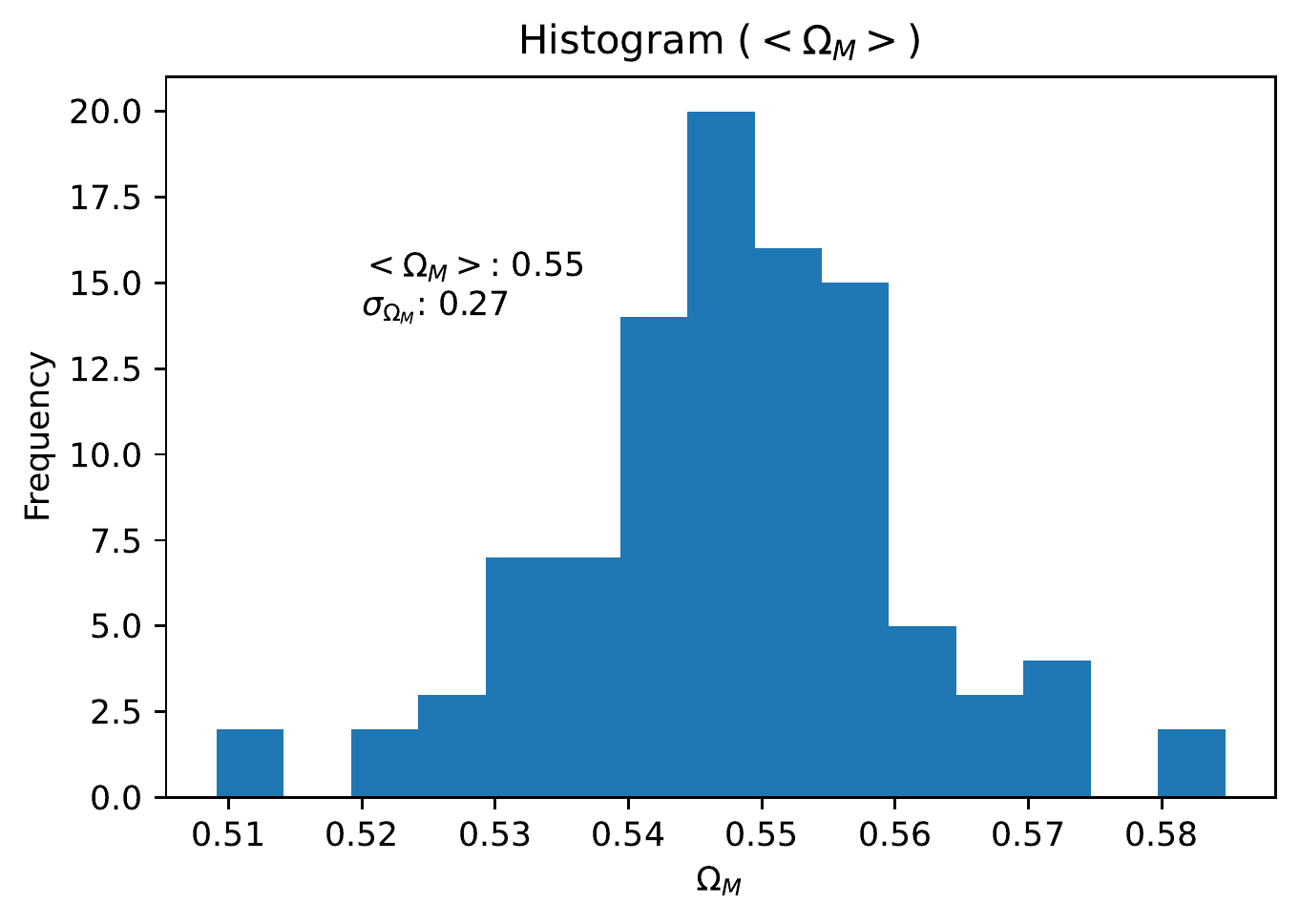}}
\caption{The distributions of the cosmological parameters for the GRBs without calibration and with $L_X$, Equations \ref{isotropic} and \ref{planeev} using the assumptions of uniform priors and over 100 runs of MCMC. Panels a), b), c) show the contours from case with no evolution (upper panels), evolution with fixed parameters (central panels) and with the evolutionary functions (lower panel) for the case of $\Omega_M$.}
\label{fig24}
\end{figure}

\subsection{GRBs alone with and without calibration with the SNe Ia with uniform priors}\label{uniformprior}


\begin{table}
\centering
\scalebox{0.88}{
\begin{tabular}{p{35mm}|l|c|c|c|c|c|c|c}
 \toprule[1.2pt]
 \toprule[1.2pt]
\textbf{Calibration with SNe Ia with uniform priors, Equation \ref{equmu}} & {\textbf{parameters varied}} & \textbf{Model} & $<\boldsymbol{\Omega_{M}}>$ & $<\boldsymbol{H_{0}}>$ & $<\boldsymbol{w}>$ & $\boldsymbol{\Delta^{GRB_{U}}_{GRB_{G}}}\%$ & $z-score_{SN}$ & $z-score_{SN+BAO}$ \\
\midrule
without evolution &$\Omega_{M}$ & $\Lambda$CDM & $0.50 \pm 0.28$ & \bf{70} & \bf{-1} & 326.47 & 0.718 & 0.700 \\\hline
without evolution & $H_0$ & $\Lambda$CDM & \bf{0.30} & $74.11\pm 13.48$ & \bf{-1} & 348.29 & 0.307 & 0.308 \\\hline
without evolution & $w$ & $w$CDM & \bf{0.30} & \bf{70} & $-1.01 \pm 0.56$ & -16.42 & 0.018 & 0.012 \\\hline
\midrule
with fixed evolution & $\Omega_{M}$ & $\Lambda$CDM & $0.59 \pm 0.27$ & \bf{70} & \bf{-1} & 297.06 & 1.077 & 1.059 \\\hline
with fixed evolution & $H_0$ & $\Lambda$CDM & \bf{0.30} & $79.04 \pm 13.12$ & \bf{-1} & 306.57 & 0.691 & 0.692 \\\hline
with fixed evolution & $w$ & $w$CDM & \bf{0.30} & \bf{70} & $-0.88 \pm 0.58$ & -16.43 & 0.207 & 0.236 \\\hline
\midrule
with $k=k(\Omega_{M})$ & $\Omega_{M}$ & $\Lambda$CDM & $0.54 \pm 0.27$ & \bf{70} & \bf{-1} & 328.57 & 0.892 & 0.874 \\\hline
\toprule[1.2pt]
 \toprule[1.2pt]

\textbf{Calibration with SNe Ia with uniform priors, Equations \ref{isotropic} and \ref{planeev}} & {\textbf{parameters varied}} & \textbf{Model} & $<\boldsymbol{\Omega_{M}}>$ & $<\boldsymbol{H_{0}}>$ & $<\boldsymbol{w}>$ & $\boldsymbol{\Delta^{GRB_{U}}_{GRB_{G}}}\%$ & $z-score_{SN}$ & $z-score_{SN+BAO}$ \\ 
\midrule
without evolution & $\Omega_{M}$ & $\Lambda$CDM & $0.52 \pm 0.28$ & \bf{70} & \bf{-1} & 305.80 & 0.789 & 0.771 \\\hline
without evolution & $H_0$ & $\Lambda$CDM & \bf{0.30} & $75.70\pm 13.64$ & \bf{-1} & 337.32 & 0.420 & 0.421 \\\hline
without evolution & $w$ & $w$CDM & \bf{0.30} & \bf{70} & $-0.97 \pm 0.57$ & -18.22 & 0.053 & 0.082 \\\hline
\midrule
with fixed evolution & $\Omega_{M}$ & $\Lambda$CDM & $0.52 \pm 0.28$ & \bf{70} & \bf{-1} & 366.67 & 0.789 & 0.771 \\\hline
with fixed evolution & $H_0$ & $\Lambda$CDM & \bf{0.30} & $75.71 \pm 13.77$ & \bf{-1} & 338.11 & 0.417 & 0.418 \\\hline
with fixed evolution & $w$ & $w$CDM & \bf{0.30} & \bf{70} & $-0.97 \pm 0.57$ & -9.67 & 0.053 & 0.082 \\\hline
\midrule
with $k=k(\Omega_{M})$ & $\Omega_{M}$ & $\Lambda$CDM & $0.52 \pm 0.28$ & \bf{70} & \bf{-1} & 283.56 & 0.789 & 0.771 \\\hline
\end{tabular}}
\caption{Averaged Cosmological parameters over 100 runs of the MCMC derived from the calibration on the SNe Ia using GRBs alone assuming uniform priors (indicated with the subscript U) and with $\mu_{GRB}$ (first part of the Table) and with Fundamental plane equation, see Equation \ref{isotropic} (2nd part) without evolution, and with the evolution correction, see Equation \ref{planeev} and as a function of $\Omega_{M}$. Columns' content is analogous to Table \ref{Table2}. The third column before the last corresponds to the percentage change in errors computed comparing the current results obtained with the GRBs alone with Gaussian priors (indicated with the subscript G) with no calibration taking as reference point values from Table \ref{Table2}. The last two columns represent the z-score from the SNe Ia taking the SNe Ia as a reference point.}
\label{Table4}
\end{table}


The aim of the previous sections is to explore the possibility of using GRBs as standalone standard candles up to redshift 5. Indeed, the aim is not to explore parameter spaces which may lead to exotic scenarios, but rather to consider the reliability of GRBs as cosmological probes. This is the reason why Gaussian priors up to 3 $\sigma$ around the best-fit values of the SNe Ia computations have been investigated.
It is clear that given the sample size and the large scatter, the applicability of GRBs as cosmological probes nowadays is not definitive, but one of our goals is to show that the complementarity of using GRBs in combination with SNe Ia is beneficial for exploring the cosmological setting at high-z. This can allow access to the universe in principle up to $z=9.4$, which is the redshift of the furthest GRB ever detected, and, in our case regarding the platinum sample, up to $z=5$. If we consider Gaussian priors, we can recover similar results without the need of exploring the full parameter space.
However, to explore how strong the impact of the Gaussian priors on our results is, and for comparing also with the results of GRBs and SNe Ia+BAO together, in which uniform priors have been used, we consider the steps i)-vi) with Gaussian priors as well. We here vary $\Omega_M$, $w$, and $H_0$ alone to appreciate the differences of the parameters computing them one by one, with the only difference that the priors are uniform and are the following : $0.0 \leq \Omega_M \leq 1$, $50 \leq H_0 \leq 100$ and $-2 \leq w \leq 0$. 
We have repeated each MCMC run 100 times, and plotted the distribution of $\Omega_M$, $w$, and $H_0$ both by calibrating the results on SNe Ia for $\mu_{GRB}$, see Fig. \ref{fig21} and Fig. \ref{fig22} for Equations \ref{isotropic} and \ref{planeev}, respectively. The same MCMC run 100 times is adopted without calibration on SNe Ia for $\mu_{GRB}$, see Fig. \ref{fig23} and Fig. \ref{fig24} for Equations \ref{isotropic} and \ref{planeev}, respectively. \\

We consider the scenarios of no evolutionary correction (upper panels of Figs. \ref{fig21}, \ref{fig22}, \ref{fig23}, \ref{fig24}), with evolution correction using fixed parameters (middle panel of Figs. \ref{fig21}, \ref{fig22}, \ref{fig23}, \ref{fig24}), and with evolutionary functions (bottom panels of Figs. \ref{fig21}, \ref{fig22}, \ref{fig23}, \ref{fig24}). Both without and with calibration on SNe Ia using $\mu{GRB}$, the $\Omega_M$ and $H_0$ uncertainty values tend to be higher by $328.57 \%$, $327.58 \%$ (without evolution and calibration); $328.57 \%$, $384.42 \%$ (with fixed evolution and without calibration); $326.47 \%$, $348.29 \%$ (without evolution and with calibration); $297.06 \%$, $306.57 \%$ (with fixed evolution and calibration), respectively, than the results obtained using Gaussian prior of 3 $\sigma$. Also, without and with calibration on SNe Ia using $\mu{GRB}$, uncertainties on the value of $\Omega_M$ using the $k(\Omega_M)$ evolutionary function are higher by $335.48 \% $ and $328.57 \%$, respectively, than the results obtained using Gaussian priors of 3 $\sigma$.

Now, considering Fundamental plane Equations \ref{isotropic} and \ref{planeev}, the uncertainties on the values for both $\Omega_M$ and $H_0$ obtained by GRBs alone are higher by $359.01 \%$, $355.19 \%$ (without evolution and calibration); $315.38 \%$, $356.49 \%$ (with fixed evolution and without calibration); $305.80 \%$, $337.32 \%$ (without evolution and with calibration); $366.67 \%$, $338.11 \% $ (with fixed evolution and calibration), respectively, than the results obtained using Gaussian prior of 3 $\sigma$. Also, without and with calibration on SNe Ia using Equations \ref{isotropic} and \ref{planeev}, uncertainties on the value of $\Omega_M$ using the $k(\Omega_M)$ evolutionary function are higher by $328.57 \% $ and $283.56 \%$, respectively, than the results obtained using Gaussian priors of 3 $\sigma$.

Surprisingly, the uncertainty values of $w$ using uniform priors are always smaller for both without and with calibration on SNe Ia and both with $\mu_{GRB}$ and Equations \ref{isotropic} and \ref{planeev} by $19.11 \%$ (without evolution and calibration), $6.45 \%$ (with fixed evolution and without calibration), $16.42 \%$ (without evolution and with calibration), $16.43 \%$ (with fixed evolution and calibration), compared to the values obtained with Gaussian priors. Considering the Fundamental plane Equations \ref{isotropic} and \ref{planeev}, the uncertainties on the value of $w$ using uniform priors for the case of no evolution and fixed evolution for no calibration are smaller than the values of $w$ with Gaussian priors by $44.65 \%$ and $12.39 \%$, respectively, and in the case of calibration on SNe Ia they are smaller in the cases of without evolution ($18.22 \%$) and with evolution ($9.67 \%$).

We note that the values of $\Omega_M$ and $H_0$ obtained considering the Gaussian priors are generally larger than the respective computations taking into account uniform priors.
However, as we have pointed out this is not the case for $w$, thus it is not clear yet why the uniform priors would enlarge the uncertainties only on $\Omega_{M}$ and $H_{0}$. On the other hand, the values of $w$ have narrow ranges and the difference between Gaussian and uniform priors may be less appreciable. 
This trend is slightly mitigated when we correct for the evolution and with the evolutionary functions, but the large uncertainties prevent to see a net difference. 
In the future, when we have more data with smaller uncertainties, the trend noted using the evolutionary functions can be important to reduce such a bias against higher values of $\Omega_M$.
On the other hand, we have seen a trend of increasing values for $\Omega_M$ when we consider cosmological computations with other high redshift probes \citep{Colgain2022}. 
This trend, however, does not appear when we simulate many GRBs based on the features of the 10 closest GRBs to the fundamental plane \citep{Dainotti2022c}.
Results are shown in Sec. \ref{standalone}. All averaged cosmological results over 100 MCMC runs using uniform priors lie within 1 $\sigma$ with respect to the cosmological results obtained using Gaussian priors, both considering calibration and no-calibration on the SNe Ia, see Table \ref{Table3} and \ref{Table4}.


\subsection{GRBs in combination with SNe Ia and BAO with and without the correction for selection biases and redshift evolution}\label{SN+BAO+GRB}

\begin{figure*} 
\centering
\subfloat[Without evolution]{\label{fig2_a}
\includegraphics[width=0.32\hsize,height=0.32\textwidth,angle=0,clip]{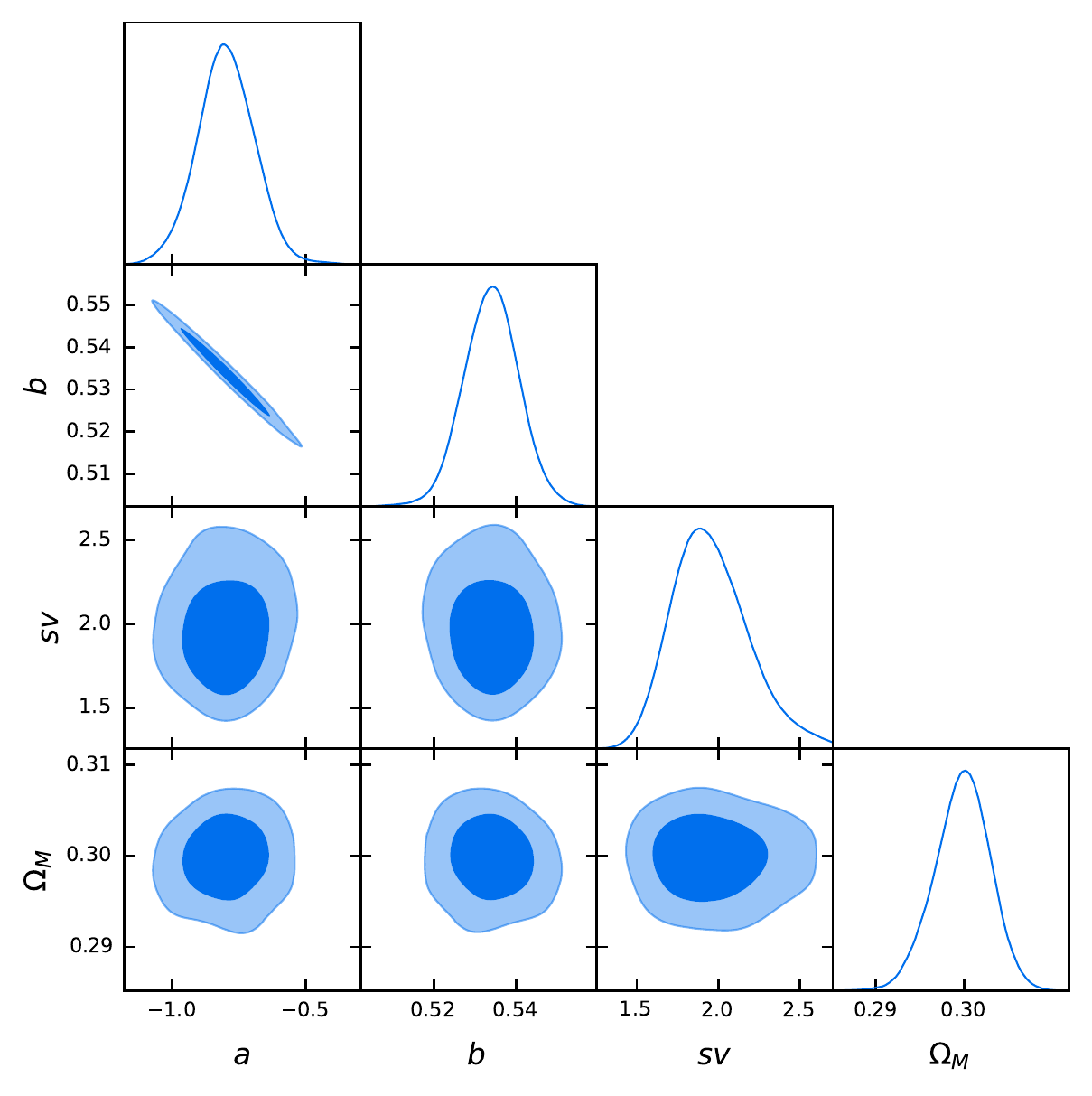}}
\subfloat[Without evolution]{\label{fig2_b}
\includegraphics[width=0.32\hsize,height=0.32\textwidth,angle=0,clip]{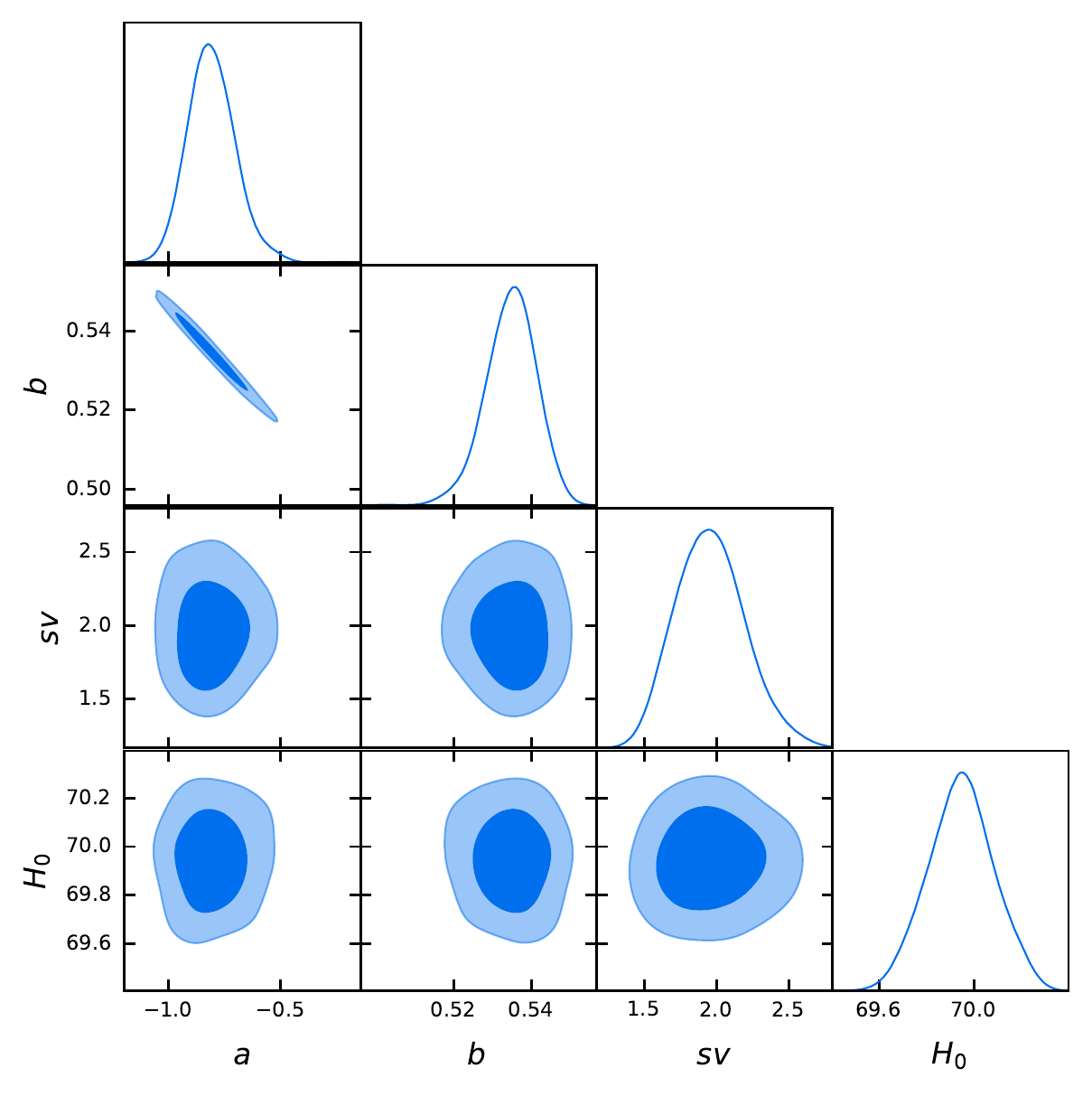}}
\subfloat[Without evolution]{\label{fig2_c}
\includegraphics[width=0.32\hsize,height=0.32\textwidth,angle=0,clip]{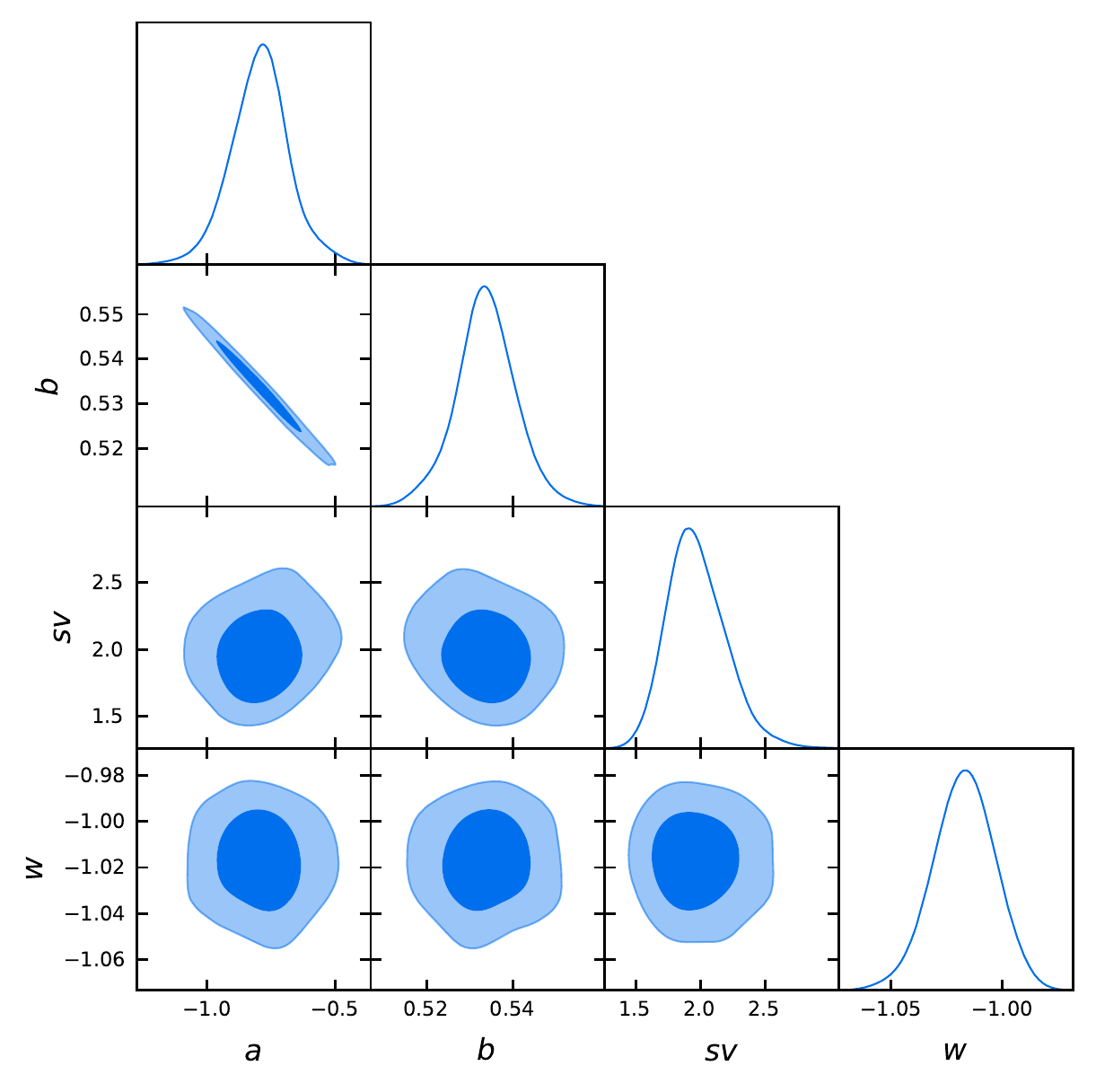}}\\\hspace{0cm}
\subfloat[With evolution]{\label{fig2_d}
\includegraphics[width=0.32\hsize,height=0.32\textwidth,angle=0,clip]{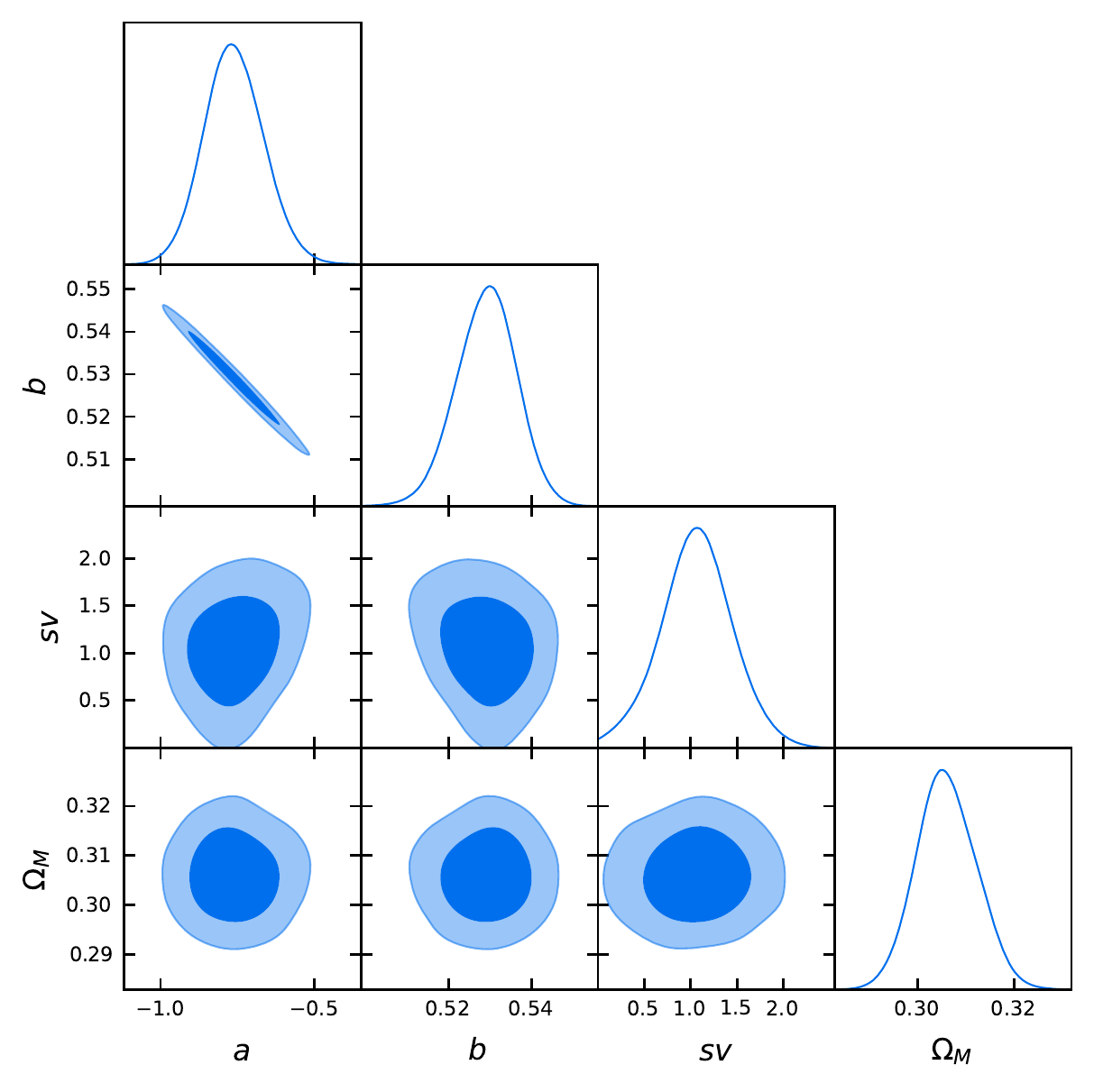}}
\subfloat[With evolution]{\label{fig2_e}
\includegraphics[width=0.32\hsize,height=0.32\textwidth,angle=0,clip]{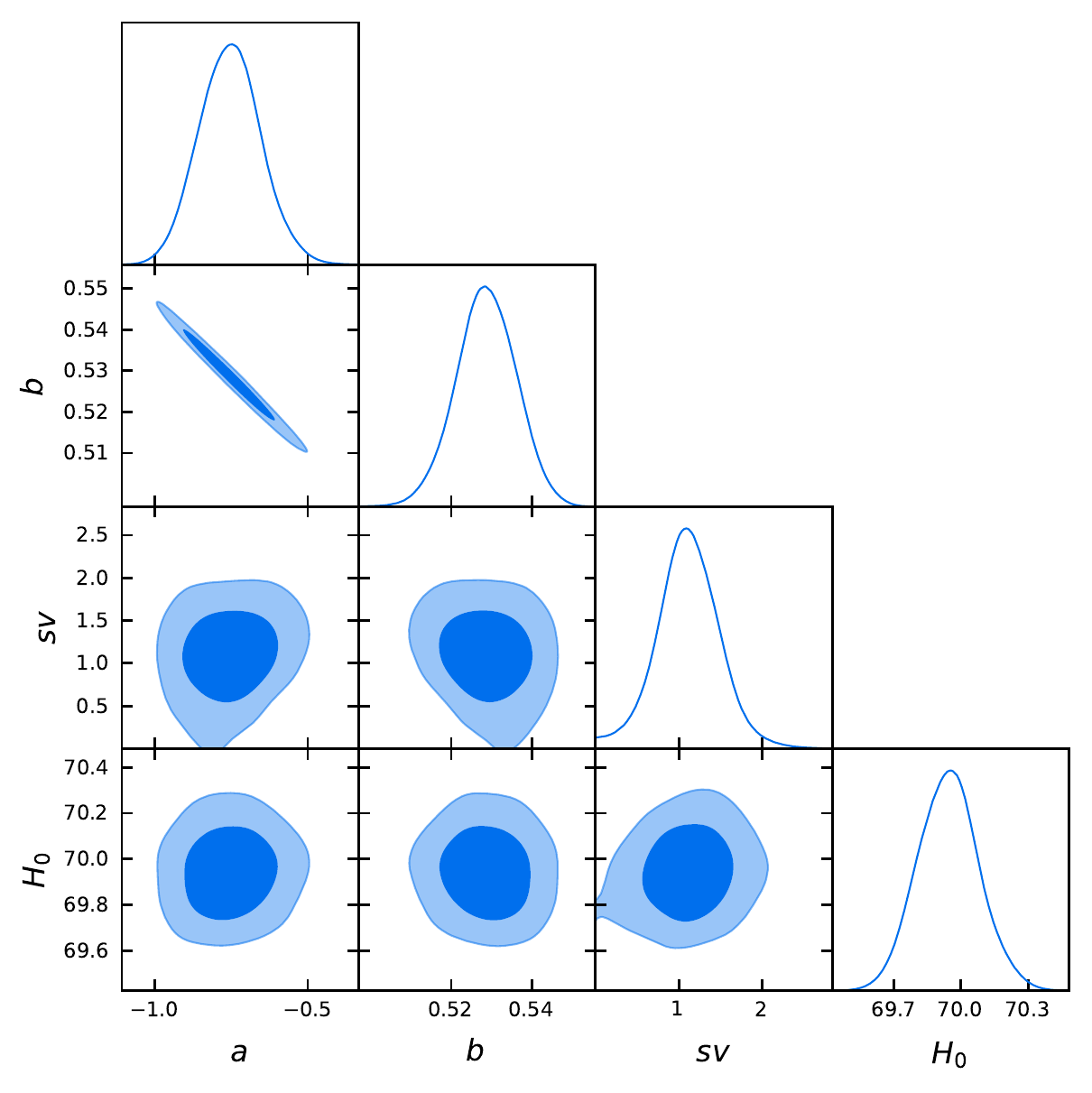}}
\subfloat[With evolution]{\label{fig2_f}
\includegraphics[width=0.32\hsize,height=0.32\textwidth,angle=0,clip]{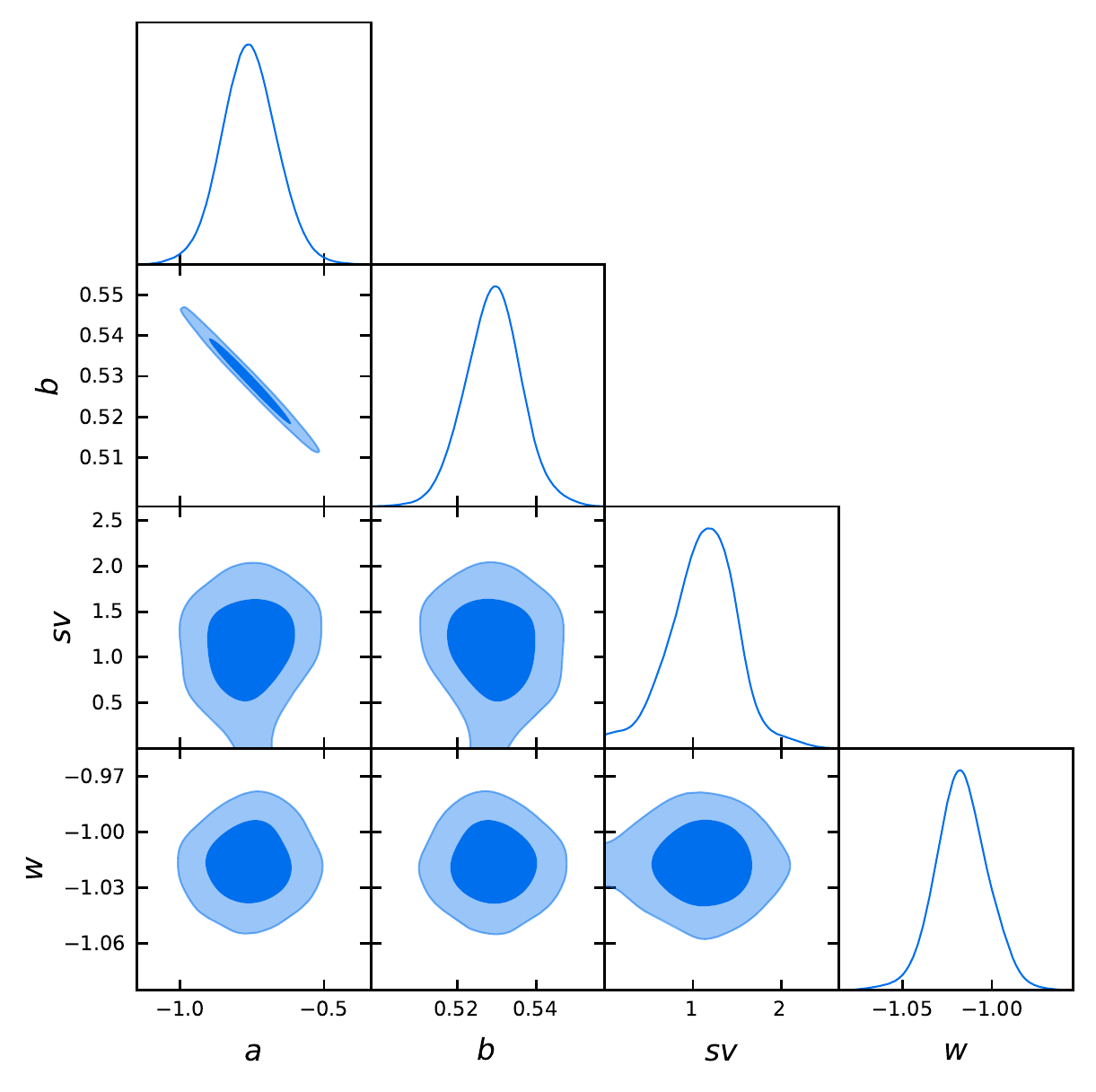}}\\\hspace{0cm}
\subfloat[Without evolution]{\label{fig2_g}
\includegraphics[width=0.49\hsize,height=0.49\textwidth,angle=0,clip]{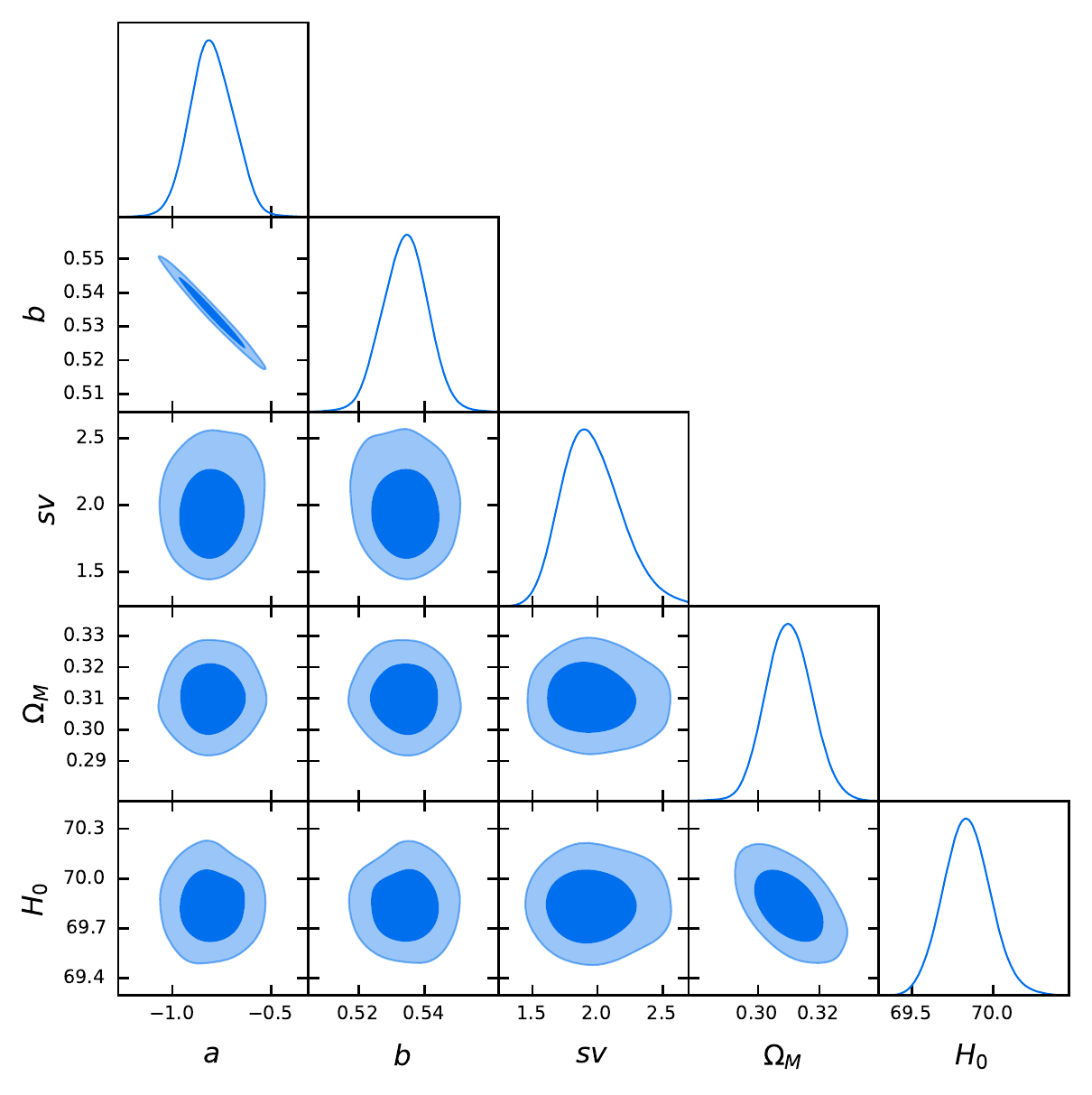}}
\subfloat[With evolution]{\label{fig2_h}
\includegraphics[width=0.49\hsize,height=0.49\textwidth,angle=0,clip]{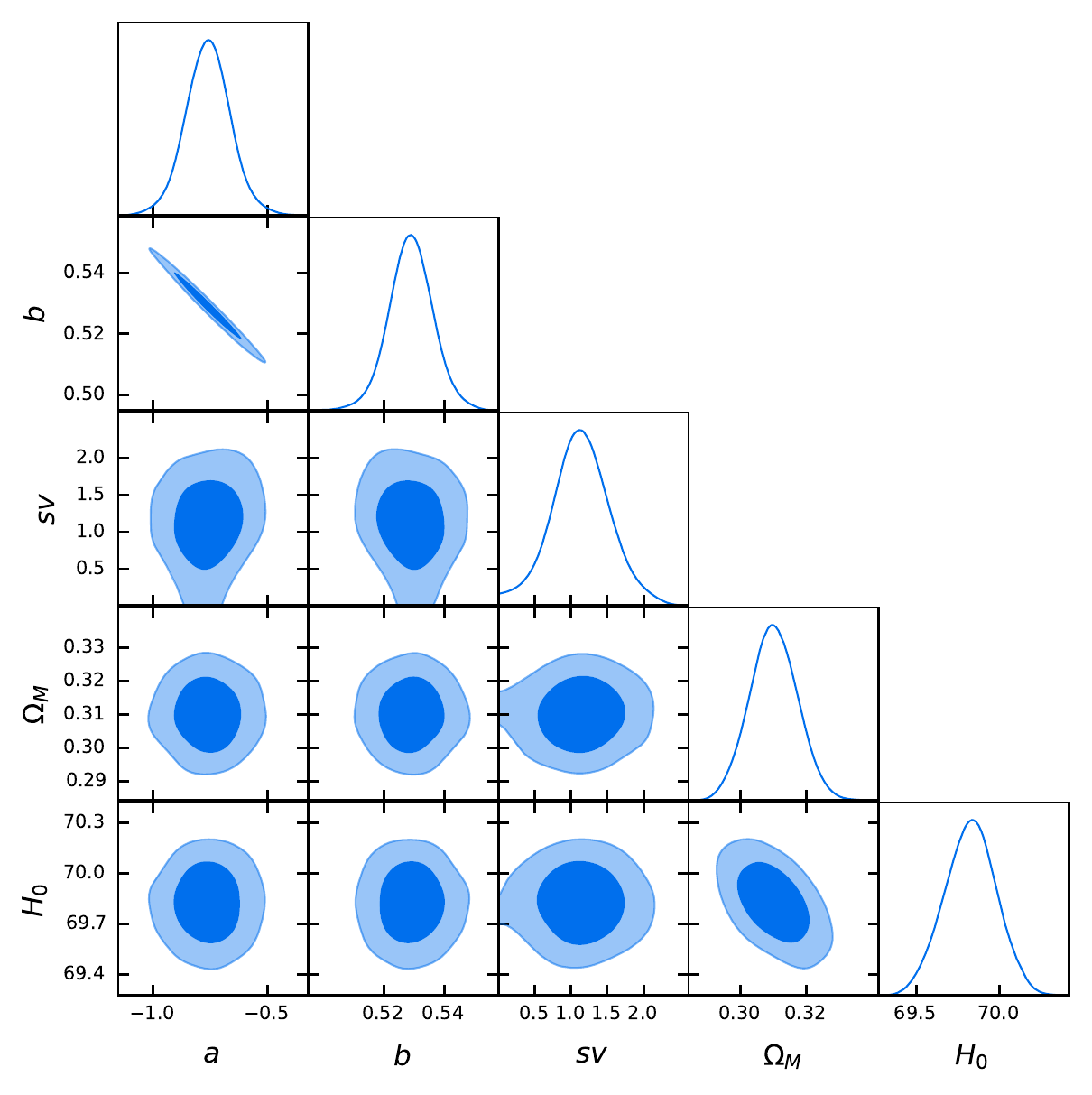}}

\caption{Cosmological results for the GRBs and SNe Ia data with the BAO constraints with uniform priors.}
\label{fig25}
\end{figure*}

\begin{figure*} 
\centering
\subfloat[SNe Ia]{\label{fig4_a}
\includegraphics[width=0.32\hsize,height=0.32\textwidth,angle=0,clip]{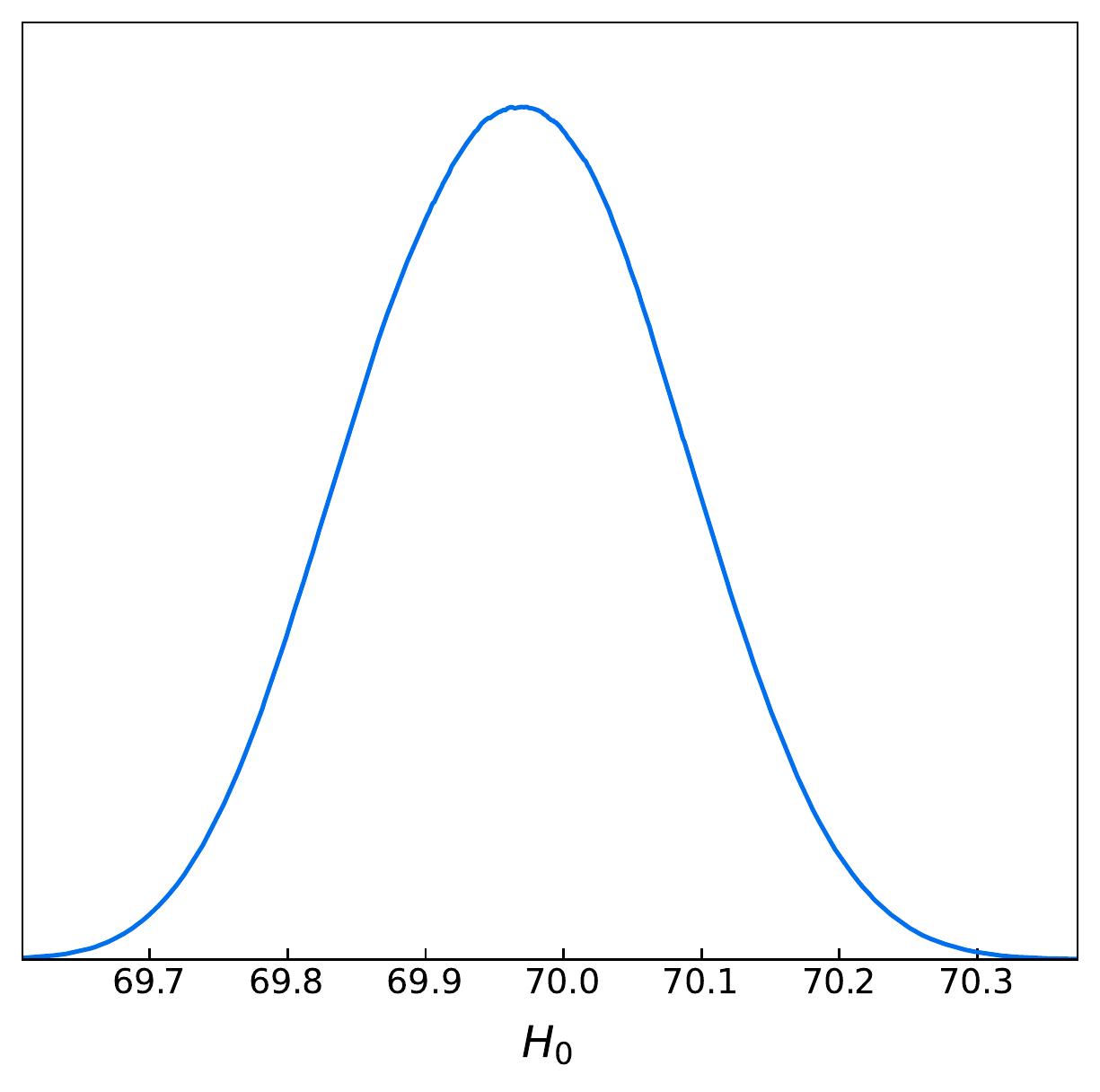}}
\subfloat[SNe Ia]{\label{fig4_b}
\includegraphics[width=0.32\hsize,height=0.32\textwidth,angle=0,clip]{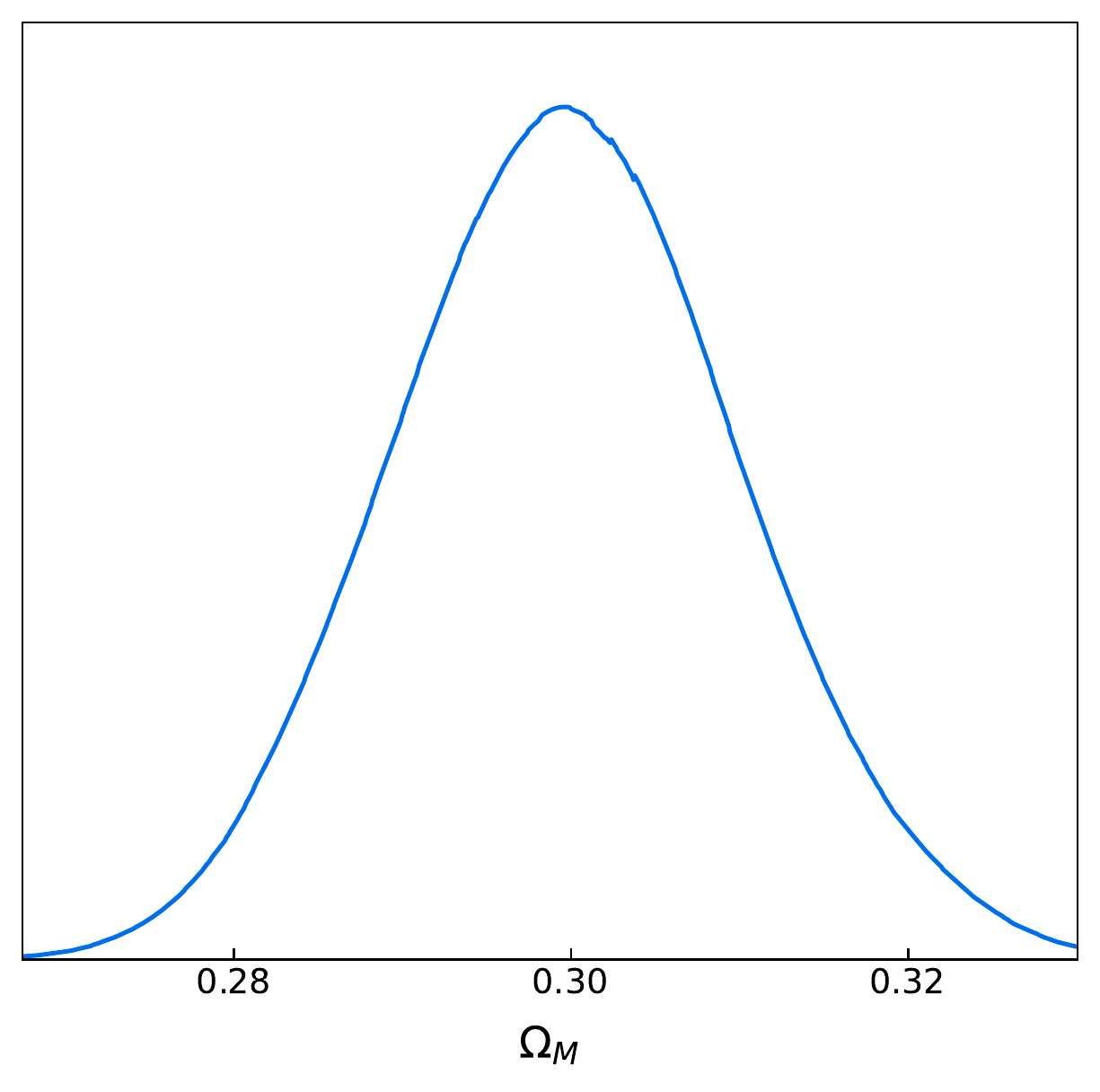}}
\subfloat[SNe Ia]{\label{fig4_c}
\includegraphics[width=0.32\hsize,height=0.32\textwidth,angle=0,clip]{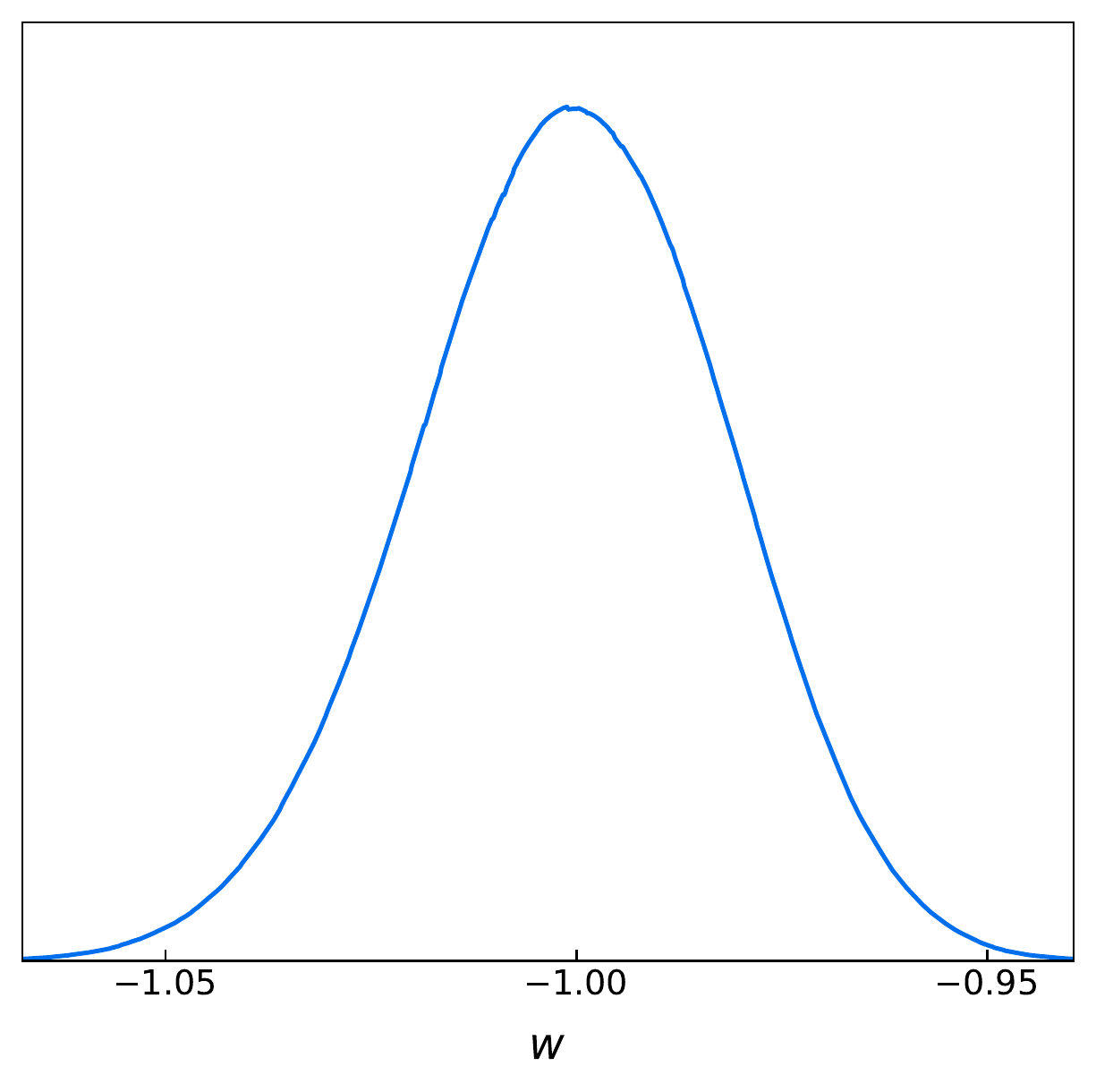}}\\\hspace{0cm}
\subfloat[SNe Ia + BAO]{\label{fig4_d}
\includegraphics[width=0.32\hsize,height=0.32\textwidth,angle=0,clip]{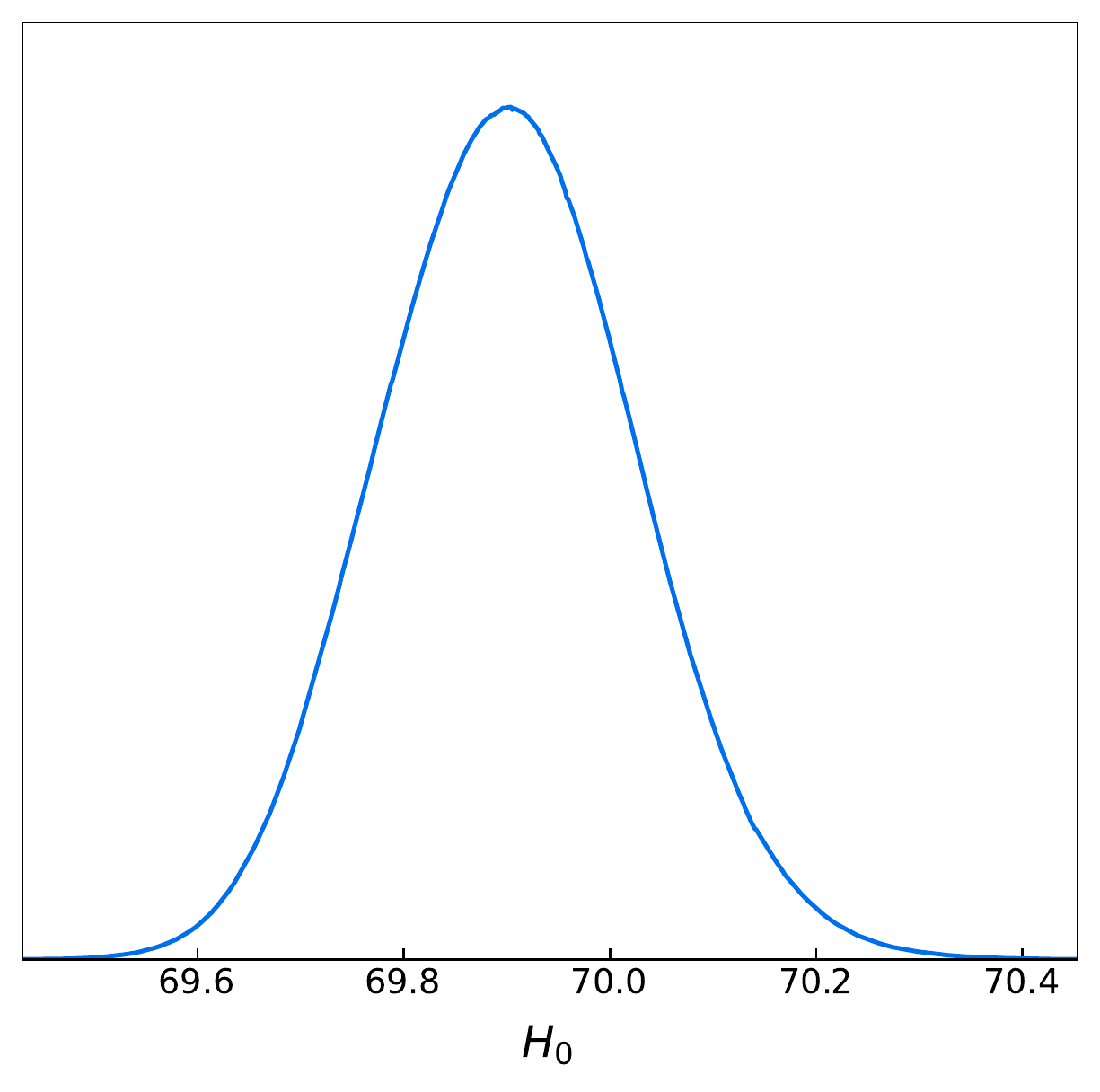}}
\subfloat[SNe Ia + BAO]{\label{fig4_e}
\includegraphics[width=0.32\hsize,height=0.32\textwidth,angle=0,clip]{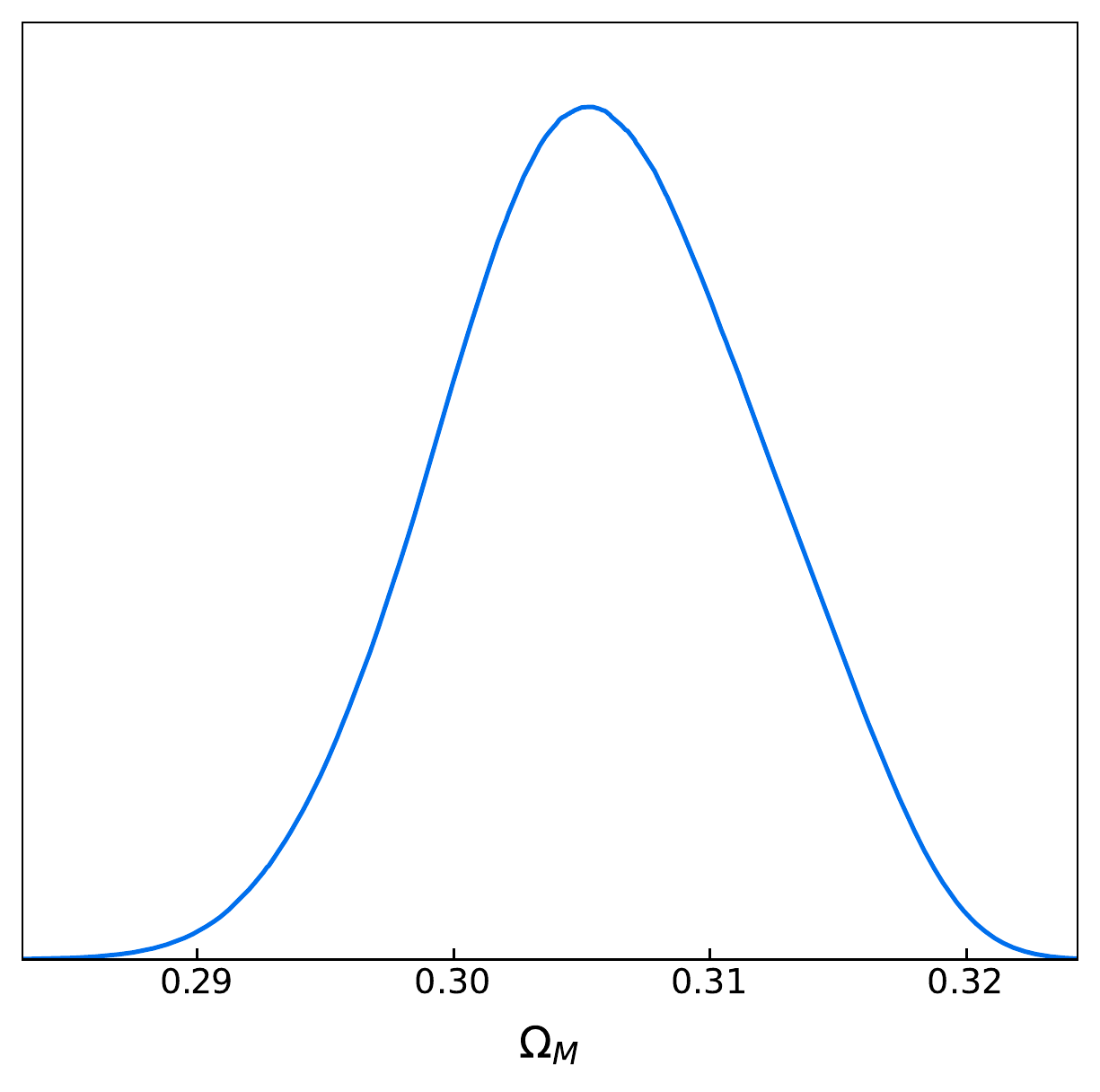}}
\subfloat[SNe Ia + BAO]{\label{fig4_f}
\includegraphics[width=0.32\hsize,height=0.32\textwidth,angle=0,clip]{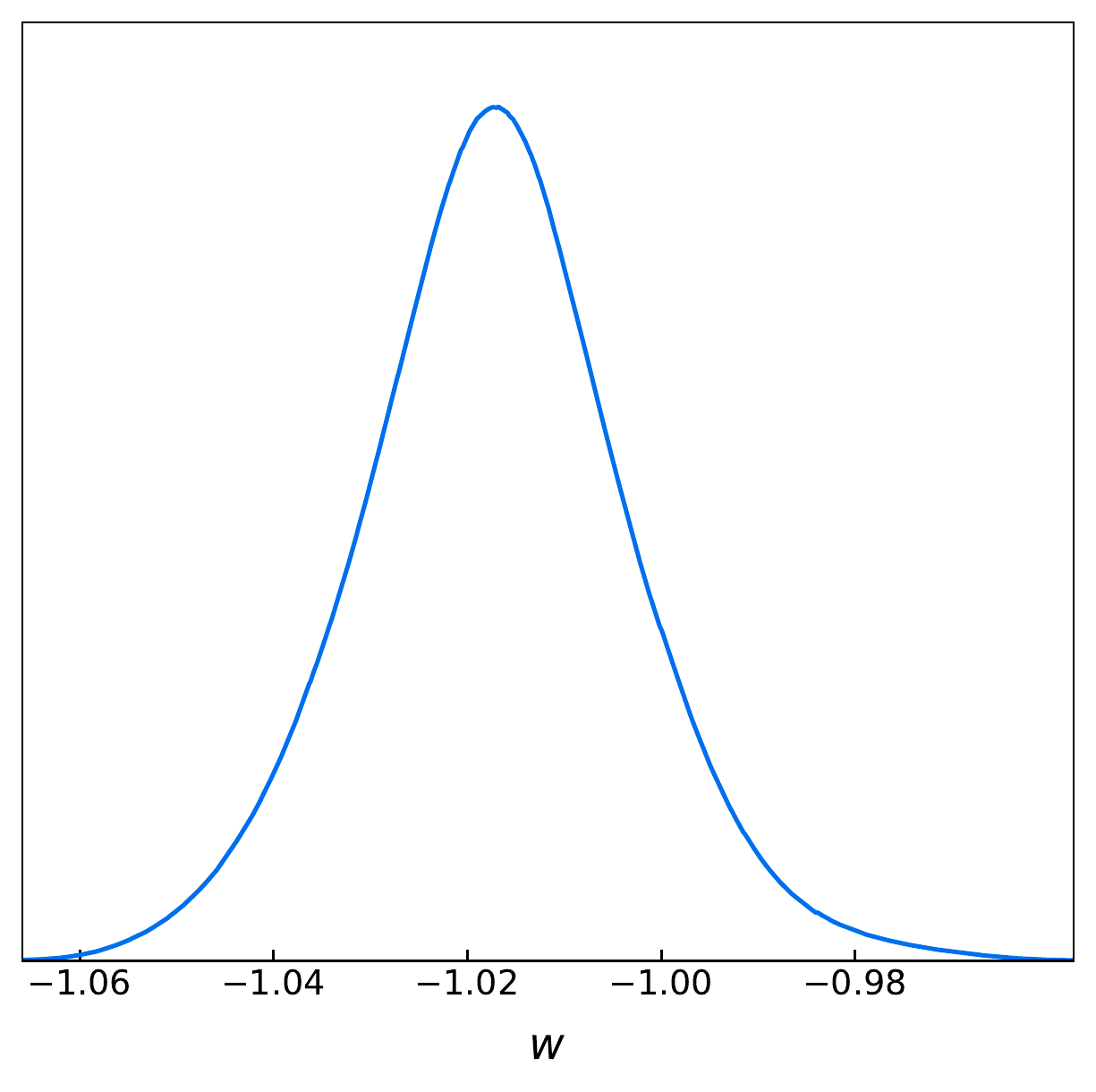}}\\\hspace{0cm}
\subfloat[SNe Ia]{\label{fig4_g}
\includegraphics[width=0.49\hsize,height=0.49\textwidth,angle=0,clip]{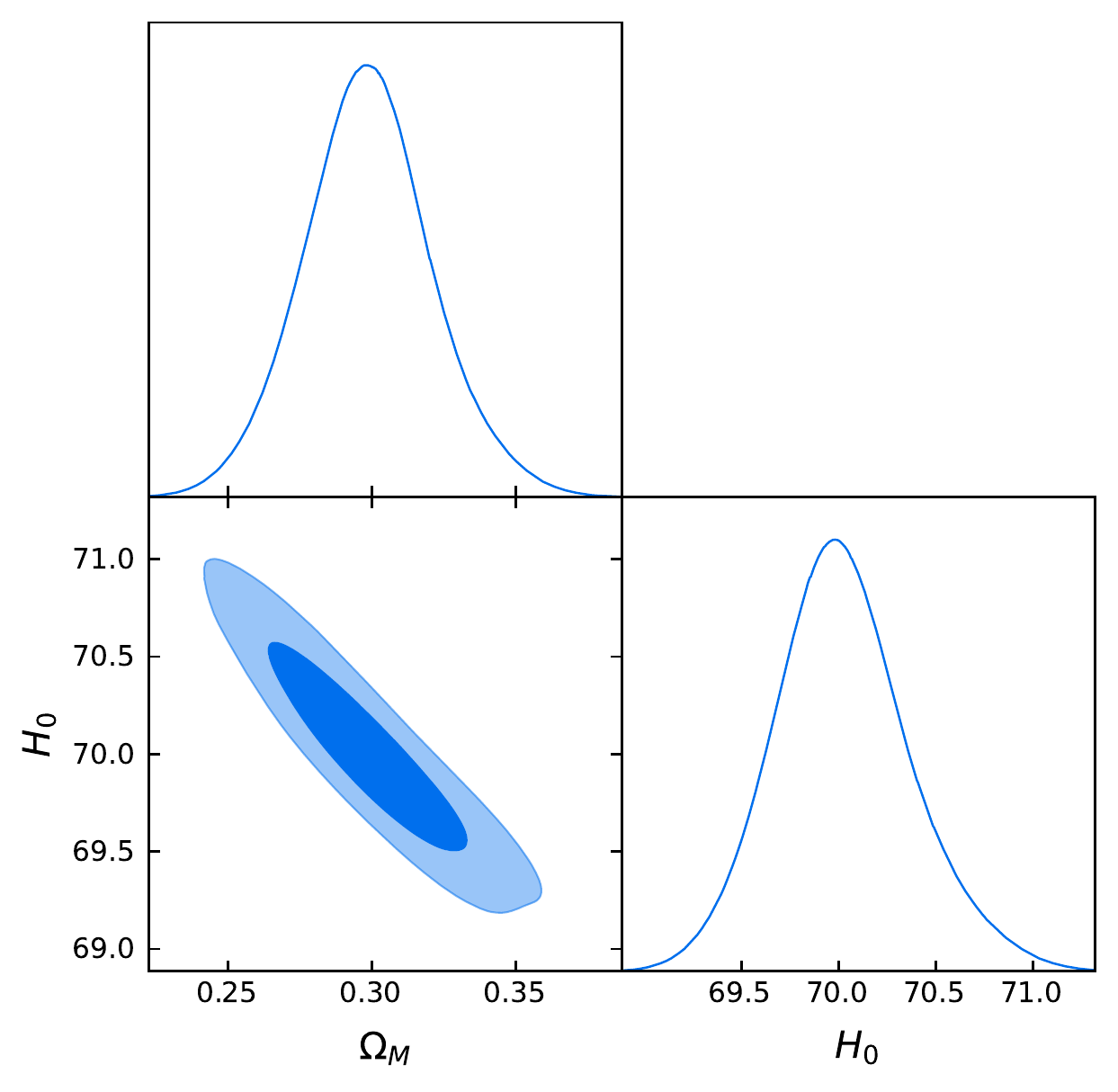}}
\subfloat[SNe Ia + BAO]{\label{fig4_h}
\includegraphics[width=0.49\hsize,height=0.49\textwidth,angle=0,clip]{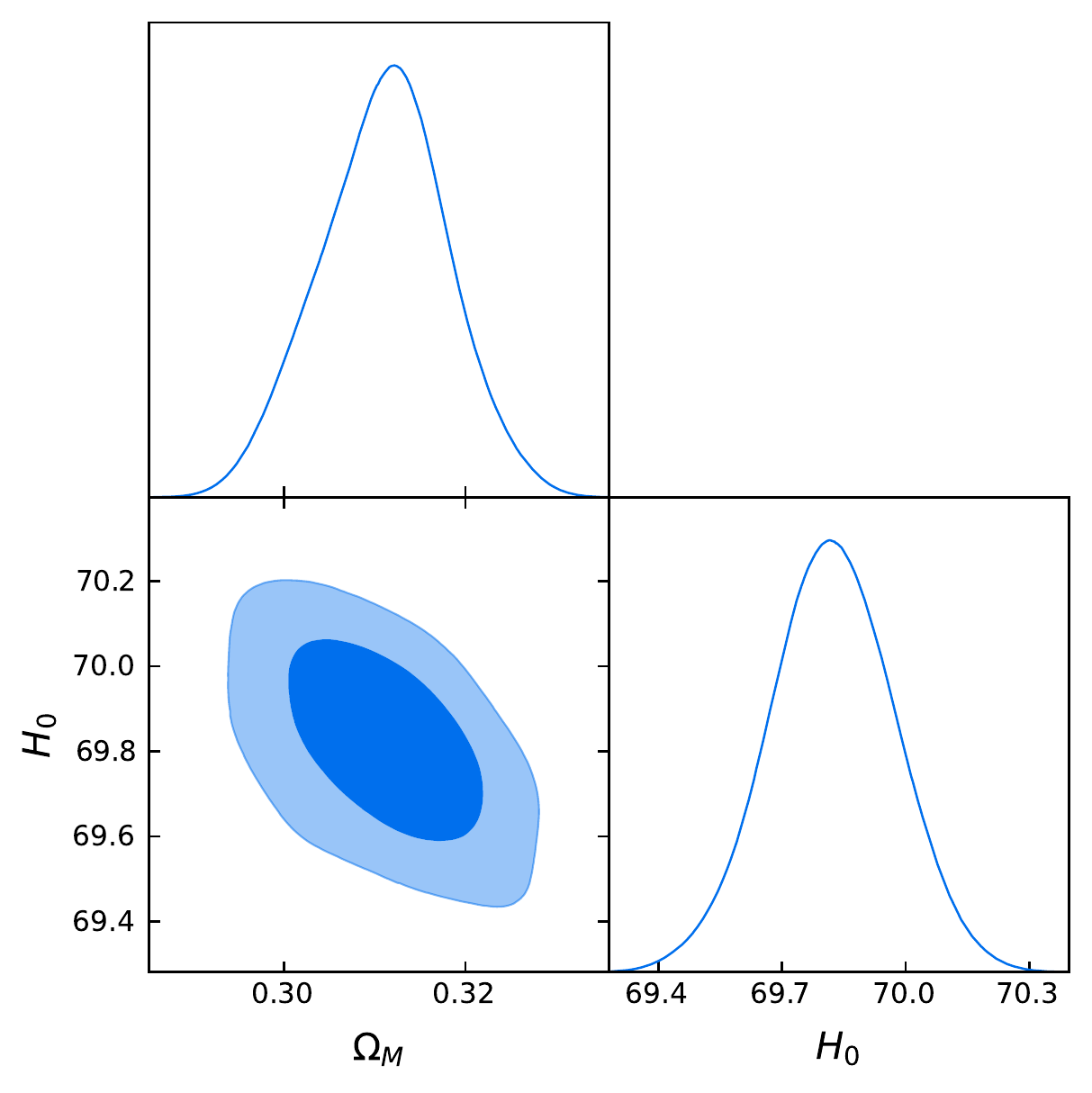}}
\caption{Cosmological results considering only SNe Ia and SNe Ia+BAO using uniform priors. In a) we fix $H_0$, in b) we fix $\Omega_M$ in c) we fix $\Omega_M$ and $H_0$. The central panel is the same as the upper panel but adding BAO.}
\label{fig26}
\end{figure*}

Results for the cosmological computations using GRBs+BAO+SNe Ia are shown in Fig. \ref{fig25} and for SNe Ia alone and SNe +BAO are shown in \ref{fig26}. All results are summarized in Table \ref{Table5}. We show the cosmological parameters obtained using: 1) SNe Ia alone 2) SNe Ia+BAO 3) SNe Ia+BAO+GRBs without correction for evolutionary effects; 4) SNe Ia+BAO+GRBs with these corrections.
The contour plots at 68\% (dark blue) and 95\% (light blue) are computed for each case. In all figures, the following fiducial values (at which the parameters are fixed) have been taken into account: $\Omega_M=0.30$, $H_0=70$ $km$ $s^{-1} Mpc^{-1}$ and $w=-1$ (bold-faced in Table \ref{Table5}).

In Fig. \ref{fig25} we present the cosmological results of SNe Ia+GRBs+BAO both with (panels d, e, f, h) and without evolution (panels a, b, c, and g) for GRBs. 
When we consider SNe Ia vs SNe Ia+BAO+GRBs with evolution, we observe smaller computed uncertainties for the cosmological values (see Table \ref{Table5}) once more probes are taken into account. More specifically, we see a decrease of 14.3\% for $\Omega_M$, and 16.7\% for $w$. When $\Omega_{M}$ and $H_0$ are varied contemporaneously, we obtain a decrease in the scatter of 68.2\%, and 52.9\%, respectively. When we compare SNe Ia+BAO vs SNe Ia+BAO+GRBs with evolution, we reproduce the precision on $\Omega_{M}$ with no reduction of the uncertainties. However, we note an increase of the scatter on $H_{0}$ of 14.3\% when both $\Omega_{M}$ and $H_{0}$ are varied contemporaneously. Furthermore, we see an increase of the scatter when ${H_0}$ and $w$ are varied alone of 7.7 \% and 7.1 \%, respectively. For completeness, all the percentage variations with respect to the SNe Ia and SNe Ia+BAO results are shown in the last two columns of Table \ref{Table5}. We stress that to check the numerical errors on the computation of the MCMC chain, we ran the computation of $H_0$ 100 times and we found that the uncertainty on the scatter on $H_0$ is $0.004$, which is two orders of magnitude smaller than the error in the results. Similarly, we find the scatter is two orders of magnitude smaller for $w$ ($0.0006$), while it is one order smaller for $\Omega_M$ ($0.0002$). 

\begin{figure} 
\centering
\subfloat[SNe Ia]{
\includegraphics[width=0.48\hsize,height=0.48\textwidth,angle=0,clip]{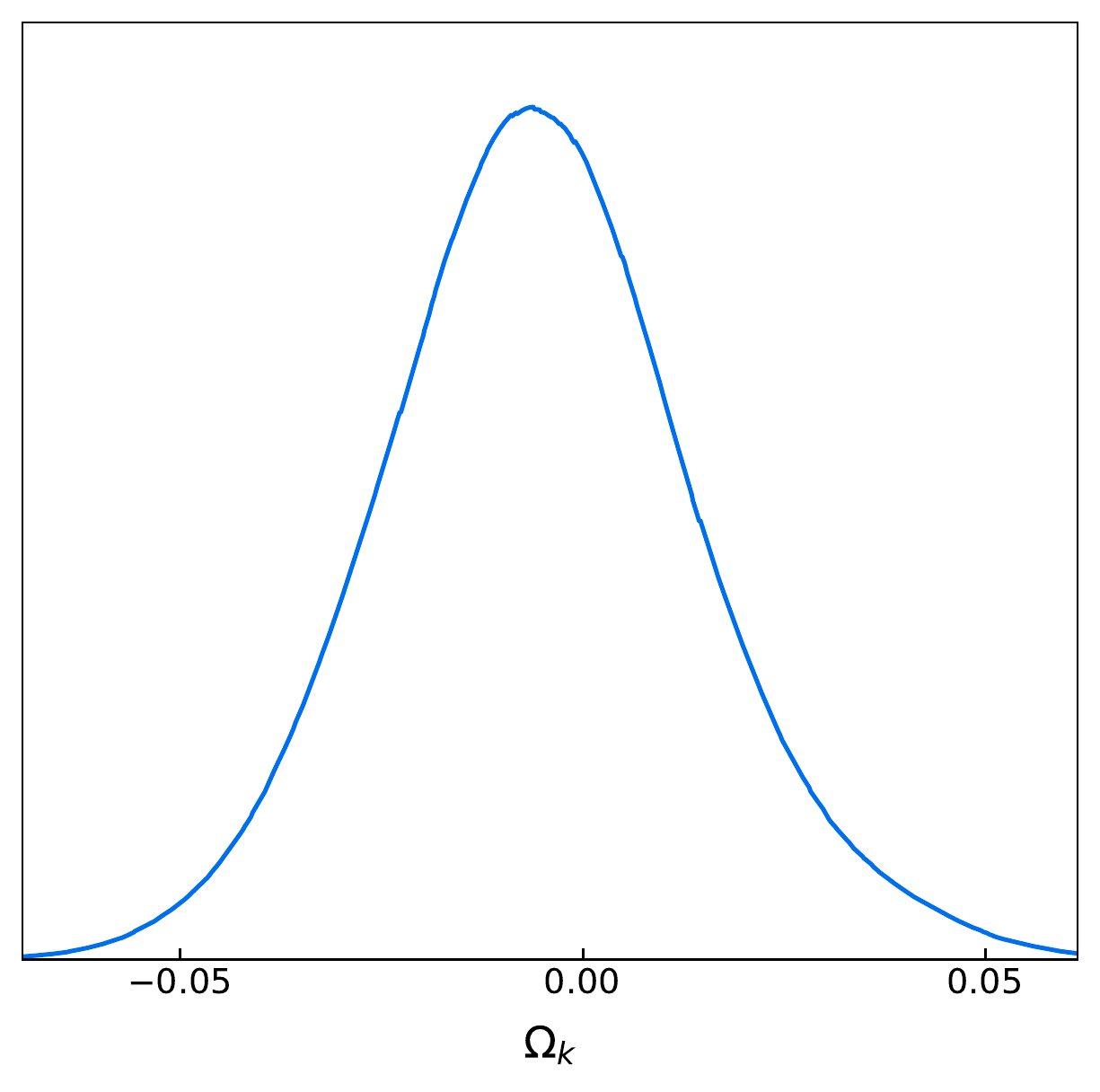}}
\subfloat[SNe Ia + BAO]{\label{figOk_b}
\includegraphics[width=0.48\hsize,height=0.48\textwidth,angle=0,clip]{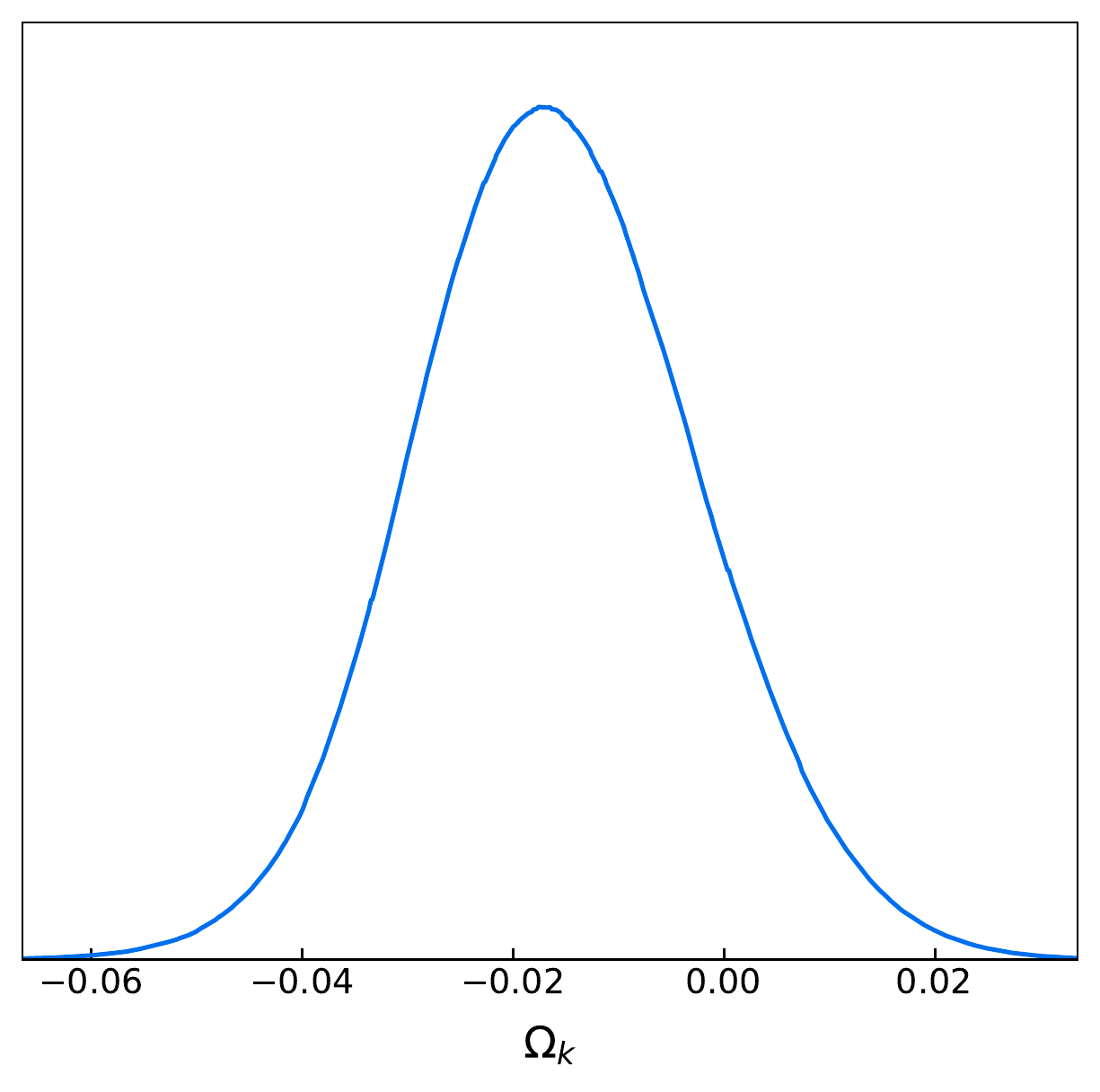}}\\\hspace{0cm}
\subfloat[SNe Ia + BAO + GRB without evolution]{\label{figOk_c}
\includegraphics[width=0.48\hsize,height=0.48\textwidth,angle=0,clip]{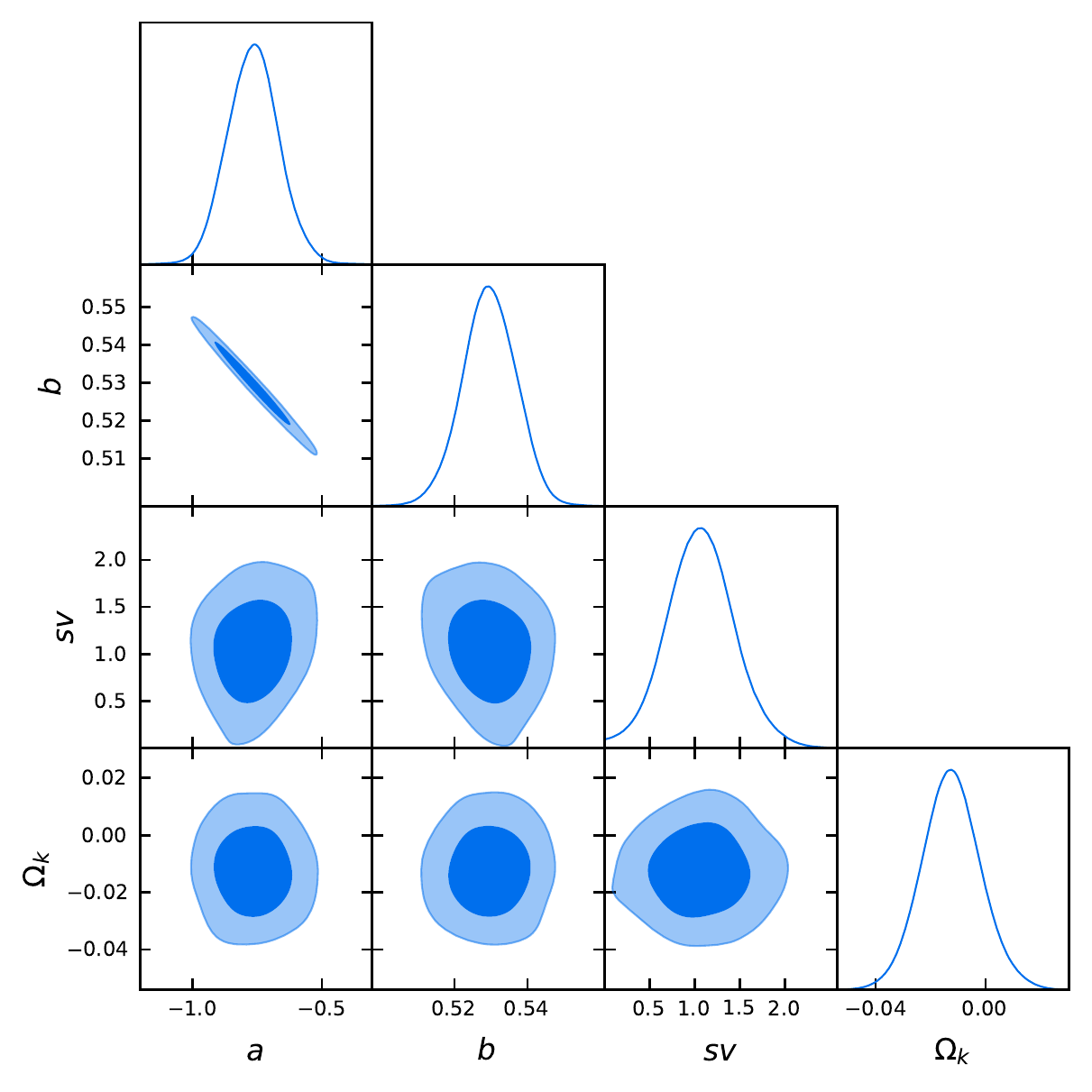}}
\subfloat[SNe Ia + BAO + GRB with evolution]{\label{figOk_d}
\includegraphics[width=0.48\hsize,height=0.48\textwidth,angle=0,clip]{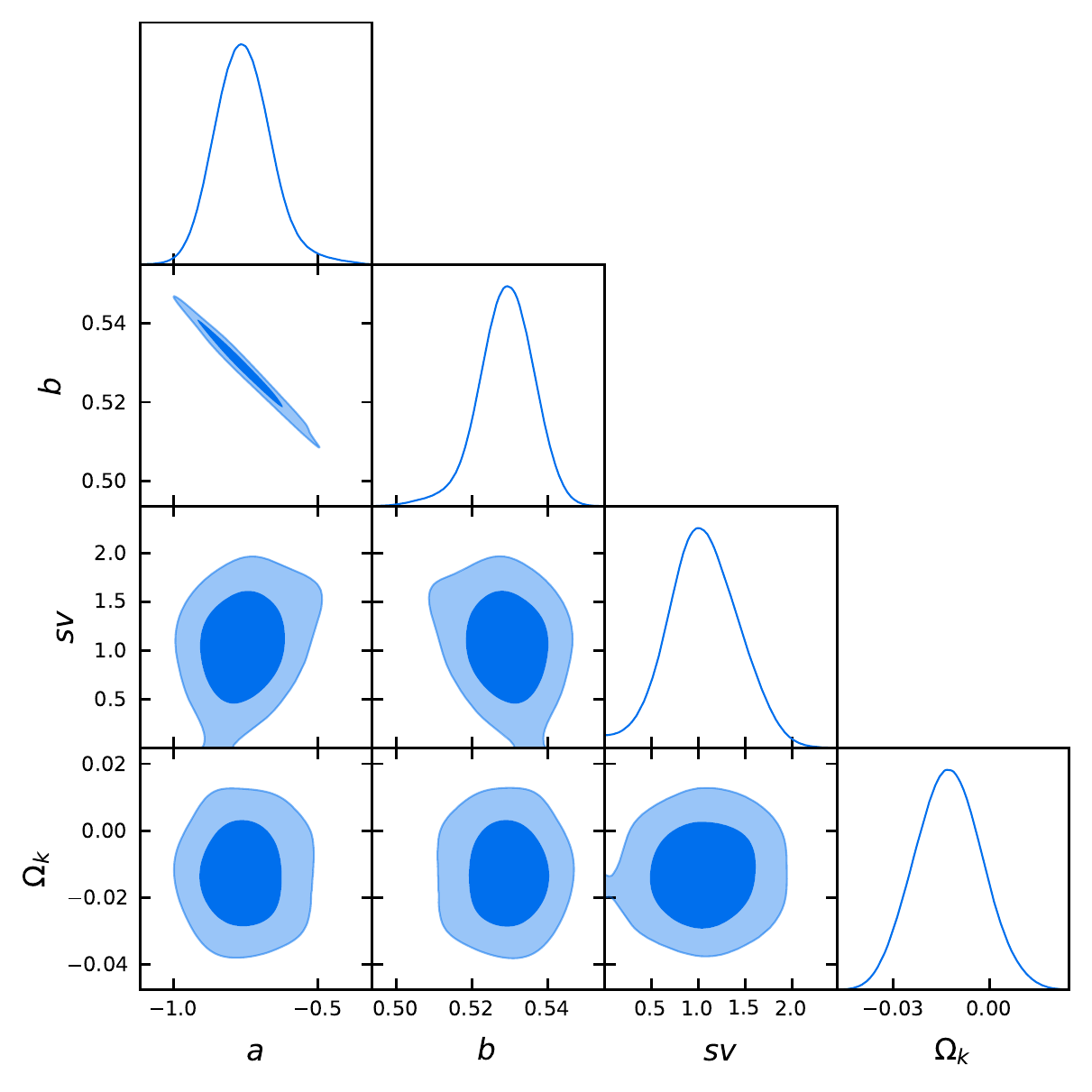}}

\caption{The upper left and right panels show the values of $\Omega_k$ fixing $H_0$ and $\Omega_M$ for the SNe Ia and SNe Ia +BAO, respectively. The lower right and left panel shows again the values of $\Omega_k$, but for the SNe Ia+BAO +GRBs with no evolution and SNe Ia+BAO +GRBs with evolution with fixed parameters, respectively. The priors on all probes are uniform.}

\label{fig27}
\end{figure}
 When we treat $k_{L_{peak}}$ and $k_{L_{a}}$ as a function of $\Omega_{M}$, the results remain unchanged. Both the best fit value and the uncertainties on parameters do not change within the computation accuracy of the MCMC algorithm in both cases, varying $\Omega_{M}$ alone and together with $H_{0}$.

In Fig. \ref{fig26} we show the results obtained using SNe Ia (panels a, b, c and g) and SNe Ia+BAO (panels d, e, f and h), to quantify how much the uncertainties on the cosmological parameters are changed when we add BAO. In the a) panel, we use SNe Ia only. We vary $H_0$ fixing $\Omega_M$ and $w$; in the b) panel we vary $\Omega_M$ and fix $H_0$ and $w$; in the c) panel we vary $H_0$ and $\Omega_M$ with $w$ fixed; in the d) panel we vary $w$ for the $w$CDM model and fix $H_0$ and $\Omega_M$.
In the bottom panels, the figures show the same quantities, considering BAO+SNe Ia.
The constraints derived when GRBs are added to the SNe Ia +BAO samples lead to a reduction or a confirmation of the scatter when using SNe Ia only, with the additional advantage that GRBs span up to $z=5$ in our sample.

\begin{table*}
\addtolength{\tabcolsep}{-2pt}
\centering
\begin{tabular}{c|l|c|c|c|c|l}
\toprule[1.2pt]
\toprule[1.2pt]
{\textbf{SNe Ia sample}} & \textbf{Model} & $\boldsymbol{w}$ & $\boldsymbol{\Omega_{M}}$ & $\boldsymbol{H_{0}}$ & $\boldsymbol{-}$ & $\boldsymbol{\Delta_{SNe+BAO}^{SNe}\%}$ \\ 
\midrule
 varying $\Omega_{M}$ & $\Lambda$CDM & {-1} & $0.299 \pm 0.007$ & \bf{70} & - & 16.7 \\\hline
 varying $H_0$ & $\Lambda$CDM & {-1} & \bf{0.30} & $69.97\pm 0.13$ & - & 0\%\\\hline
 varying $\Omega_{M}$ and $H_0$ & $\Lambda$CDM & {-1} & $0.298 \pm 0.022 $ & $70.02\pm 0.34$ & - & 214.3 \%, 142.3 \% \\\hline
 varying $w$ & $w$CDM & $-1.000\pm 0.018$ & \bf{0.30} & \bf{70} & - & 28.6 \% \\
\hline
\hline
 \textbf{SNe Ia + BAO sample} & \textbf{Model} & $\boldsymbol{w}$ & $\boldsymbol{\Omega_{M}}$ & $\boldsymbol{H_{0}}$ & $\boldsymbol{\Delta_{SNe}^{SNe+BAO}\%}$ & $\boldsymbol{-}$ \\
\midrule
varying $\Omega_{M}$ & $\Lambda$CDM & {-1} & $0.304 \pm 0.006$ & \bf{70} & -14.3 \% & - \\\hline
varying $H_0$ & $\Lambda$CDM & {-1} & \bf{0.30} & $69.96 \pm 0.13$ & 0 \% & - \\\hline
varying $\Omega_{M}$ and $H_0$ & $\Lambda$CDM & {-1} & $0.311 \pm 0.007 $ & $69.82 \pm 0.14$ & -68.2 \%, -58.8 \% & - \\\hline
varying $w$ & $w$CDM & $-1.017\pm 0.014$ & \bf{0.30} & \bf{70} & -22.2 \% & - \\
\hline
\hline
 \textbf{SNe Ia + BAO + GRB sample NO EV} & \textbf{Model} & $\boldsymbol{w}$ & $\boldsymbol{\Omega_{M}}$ & $\boldsymbol{H_{0}}$ & $\boldsymbol{\Delta_{SNE}^{SNe+BAO+GRBNOEV}\%}$ &$\boldsymbol{\Delta_{SNE+BAO}^{SNe+BAO+GRBNOEV}\%}$ \\
\midrule
varying $\Omega_{M}$ & $\Lambda$CDM &{-1} & $0.306 \pm 0.006$ & {70} & -14.3 \% & 0 \% \\\hline
varying $H_0$ & $\Lambda$CDM & {-1} & \bf{0.30} & $69.94 \pm 0.13$ & 0 \% & 0 \% \\\hline
varying $\Omega_{M}$ and $H_0$ & $\Lambda$CDM & {-1} & $0.310 \pm 0.007$ & $69.84 \pm 0.15$ & -68.2\%, -55.9 \% & 0\%, 7.1 \% \\\hline
varying $w$ & $w$CDM & $-1.017 \pm 0.014$ & \bf{0.30} & \bf{70} & -22.2 \% & 0\%\\
\hline
\hline
\textbf{SNe Ia + BAO + GRB sample EV} & \textbf{Model} & $\boldsymbol{w}$ & $\boldsymbol{\Omega_{M}}$ & $\boldsymbol{H_{0}}$ & $\boldsymbol{\Delta_{SNE}^{SNe+BAO+GRBEV}\%}$ &$\boldsymbol{\Delta_{SNE+BAO}^{SNe+BAO+GRBEV}\%}$ \\
\midrule
varying $\Omega_{M}$ & $\Lambda$CDM &{-1} & $0.306 \pm 0.006$ & \bf{70} & -14.3 \% & 0 \% \\\hline
varying $H_0$ & $\Lambda$CDM & {-1} & \bf{0.30} & $69.94 \pm 0.14$ & 7.7 \% & 7.7\%\\\hline
varying $\Omega_{M}$ and $H_0$ & $\Lambda$CDM & {-1} & $0.310 \pm 0.007$ & $69.83 \pm 0.16$ & -68.2 \%, -52.9 \% & 0\%,14.3 \% \\\hline
varying $w$ & $w$CDM & $-1.017 \pm 0.015$ & \bf{0.30} & \bf{70} & -16.7 \% & 7.1\% \\
\bottomrule
\bottomrule
\end{tabular}

\caption{Results of the fitting of the cosmological parameters using the SNe Ia (the upper part of the Table), SNe Ia+BAO (the second part), and SNe Ia+BAO+GRBs using GRBs without calibration on SNe Ia and with uniform priors without (the third part) and with (the bottom) the correction for the evolution indicated with EV, using together the platinum sample, the Pantheon sample for SNe Ia and the BAO measurements of \citet{SharovBAO}. The values in bold are fixed at fiducial values. We also show two columns with the percentage increase/decrease between the uncertainties of SNe Ia only (the ${\Delta_{SNE}^{considered\, sample} \%}$) versus the other results, and the BAO+SNe Ia (${\Delta_{SNe+BAO}^{considered\, sample} \%}$) versus the other results. The formula is: $\frac{\Delta_{\text{comparing}}-\Delta_{\text{reference}}}{\Delta_{\text{reference}}}$, where $\Delta_{\text{reference}}$ is the uncertainty obtained with SNe Ia (6th column) and SNe Ia + BAO (7th column) samples. With the negative sign we indicate a percentage decrease on the uncertainty compared to the reference sample (indicated with a lower subscript), while with the positive one a percentage increase.}
\label{Table5}
\end{table*}

\subsection{Comparing our results to the other cosmological computations with GRBs in the literature}
Many scientists have tried to address the problem of using GRBs as cosmological tools for almost two decades from now and we here discuss only a few papers which are more closely related to the fundamental plane relation or when a similar study about the comparison with and without the calibration on SNe Ia has been performed.
As an example, cosmological computations have been performed in \citet{Moresco22} involving GRBs both with and without calibration against SNe Ia considering uniform priors based on the Amati correlation between the peak energy in the $\nu F_{\nu}$ of the prompt emission spectrum and the isotropic prompt emission \citep{Amati2008}. 
For their sample composed of 70 GRBs without calibration, they have found $\Omega_{M}=0.27_{-0.18}^{+0.38}$.
To compare their results to ours, we symmetrize their results and obtain: $\Omega_{M}=0.27 \pm 0.28$, which is similar to the variance obtained by us for an analogous case in Table \ref{Table3}. In particular, for the case without correction for evolution and with the likelihood based on the Equation \ref{isotropic} without calibration and with uniform priors, we reproduce the same precision (line 10 in Table \ref{Table3} and the Figures \ref{fig24}, \ref{fig22}), while for all the other cases we have slightly smaller variance (0.27). 
For the sample of 208 GRBs, whose size is four times bigger than the one of our sample, they found $\Omega_{M}=0.26_{-0.12}^{+0.23}$ and $\Omega_{M}=0.30_{-0.06}^{+0.06}$ for the cases without and with calibration, respectively.
Both results hold a higher precision than our computations. 
We remind that our results without evolution and without and with calibration yield $\Omega_{M}=(0.53\pm 0.28)$ and $\Omega_{M}=(0.52\pm 0.28)$.
In \cite{Liu22b}, cosmological computations have been performed with 220 GRBs calibrated with SNe Ia considering uniform priors based on the Amati correlation \citep{Amati2008} and improved the Amati correlation \citep{Liu22a} using the copula function, a multivariate cumulative distribution function. They have found $\Omega_M > 0.651$ with the Amati correlation and $\Omega_M = 0.308 \pm 0.192$ with the improved Amati correlation. The result with the improved Amati correlation is consistent with the ones obtained by us for GRBs alone calibrated with SNe Ia using uniform priors and the distance modulus Equation, \ref{equmu}, see Table \ref{Table4}. Though in this paper, the variance of the measurement is smaller than ours (without evolution: $0.28$, with evolution fixed and $k=k(\Omega_{M}): 0.27$) it is probably due to the fact that their GRB sample size is more than four times the size of our sample.
\cite{Wang07} performed cosmological computations with 69 GRBs without calibration on SNe Ia using the distance modulus Equation, \ref{equmu}, and $\chi^2$ minimization technique to obtain $\Omega_M = 0.34 \pm 0.10$. Their results are consistent with the ones we obtained for GRBs alone without calibration using uniform priors and the distance modulus Equation, \ref{equmu}, see Table \ref{Table3}. The variance in their measurements is smaller than the one in our cosmological calculations as in the comparisons above (without evolution: $0.28$, with evolution fixed and $k=k(\Omega_{M}): 0.27$).

\cite{Cao22a} performed a set of computations involving GRBs using the Dainotti Fundamental plane correlation with 3 different sets consisting of 60 events altogether (one set of 5 GRBs only, one of 24, and the other composed of 31 GRBs) and using also the Amati relation for 118 GRBs. The considered likelihood corresponds to the one derived from equation \ref{isotropic}. In the analysis uniform priors were applied. \cite{Cao22a} obtained closed contours for only two cases: 115 GRBs (3 events are removed due to the overlap with the other sample) and for 5 GRBs. They obtained using the Fundamental plane alone: $\Omega_{M} = 0.630^{+0.352}_{-0.135}$ and $\Omega_{M} = 0.520^{+0.379}_{-0.253}$ for the two samples, respectively. We symmetrize those results in order to compare them with ours, obtaining: $\Omega_{M} = 0.630\pm 0.244$ and $\Omega_{M} = 0.520\pm 0.316$. In the first case the results are slightly more precise than the one obtained by us in Table \ref{Table3}, but the considered sample is more than 2 times larger, while the others are slightly less precise, but for a very small sample size of GRBs. In the newer paper \cite{Cao22b} used the Platinum sample for the Dainotti Fundamental plane correlation and 118 events using Amati correlation, out of which 17 overlap with the Platinum sample, thus using 101 events in addition to the Platinum sample and obtained: $\Omega_M > 0.411$ using the Platinum sample alone, $\Omega_M > 0.256$ using the 118 GRBs sample alone, and $\Omega_M = 0.614 \pm 0.255$ using both samples together ($101 + 50$).
The variance of those measurements ($0.255$) is slightly smaller than the ones of our results ($0.28 - 0.27$) when both the Amati and the Dainotti relations are combined. The difference in the results when the only Platinum sample is used is due to the fact that we run the analysis 100 times and we average our results, instead in the mentioned paper the results are computed only one time.

\subsection{The flatness of the universe}\label{flatnessUniverse}

\begin{table}
 \centering
 \begin{tabular}{|l|c|c|}
 \toprule[1.2pt]
 \toprule[1.2pt]
 \textbf{Sample} & $\boldsymbol{\Omega_{k}}$ & $\boldsymbol{z}$\textbf{-score}\\
 \midrule
 SNe Ia sample & $-0.003 \pm 0.018$ & $0.17$\\\hline
 SNe Ia + BAO sample & $-0.016 \pm 0.012$ & $1.33$ \\\hline
 SNe Ia + BAO + GRB sample NO EV & $-0.017 \pm 0.012$ & $1.42$ \\\hline
 SNe Ia + BAO + GRB sample EV & $-0.013 \pm 0.011$ & 1.18 \\
 \bottomrule
 \bottomrule
 \end{tabular}
 \caption{Results of the fitting of the $\Omega_{k}$ parameter for different samples and probes. All the obtained values are compatible within 1.5 $\sigma$ with $\Omega_{k}=0$ corresponding to the flat universe. To compare the results with the flat cosmology we use the z-score, defined as: $ z = \frac{|\Omega_{k, flat}-\Omega_{k}|}{\Delta_{\Omega_{k}}}\, =\, \frac{|\Omega_{k}|}{\Delta_{\Omega_{k}}}$.}\label{Table6}
\end{table}
Due to the recent results in which the flatness of the universe is questioned by \citet{Melchiorri} and several other authors we also consider scenarios accounting for its curvature.
To consider not-flat universe models we use the appropriate formula for the distance luminosity which reads as follows:
\begin{linenomath*}
\begin{equation}
 d_{L} = (1+z) \times d_{M},
\end{equation}
\end{linenomath*}
where $d_{M}$ is the transverse comoving distance given by the formula:
\begin{linenomath*}
\begin{equation}
d_{M} = 
\left\{\begin{matrix}
\frac{c}{H_{0}\,\sqrt{\Omega_{k}}} \sinh\left(\sqrt{\Omega_{k}} \times \int_{0}^{z} \frac{dz'}{\sqrt{\Omega_{M}(1+z')^{3}+\Omega_{k}(1+z')^{2}+\Omega_{\Lambda}}} \right) & \; \Omega_{k}>0 \\ \\

\frac{c}{H_{0}} \times \int_{0}^{z} \frac{dz'}{\sqrt{\Omega_{M}(1+z')^{3}+\Omega_{k}(1+z')^{2}+\Omega_{\Lambda}}} & \; \Omega_{k}=0 \\ \\

\frac{c}{H_{0}\,\sqrt{|\Omega_{k}|}} \sin\left(\sqrt{|\Omega_{k}|} \times \int_{0}^{z} \frac{dz'}{\sqrt{\Omega_{M}(1+z')^{3}+\Omega_{k}(1+z')^{2}+\Omega_{\Lambda}}} \right) & \; \Omega_{k}<0.
\end{matrix}\right.
\end{equation}
\end{linenomath*}

\begin{table}
\centering
\begin{tabular}{c|l|c|c|c|c}
 \toprule[1.2pt]
 \toprule[1.2pt]

\textbf{Likelihood based on $\mu_{GRB}$, Equation \ref{equmu}} & {\textbf{parameters varied}} & \textbf{Model} & $\boldsymbol{\Delta_{SN}^{GRB}}\%$& $\boldsymbol{\Delta_{SN+BAO}^{GRB}}\%$\\ 
\midrule
without evolution & $\Omega_{M}$ & $\Lambda$CDM & 800.00 & 950.00 \\\hline
without evolution & $H_0$ & $\Lambda$CDM & 2443.85 & 2443.85 \\\hline
without evolution & $\Omega_{M}$ and $H_0$ & $\Lambda$CDM & 209.09, 790.00 & 871.43, 2061.43 \\\hline
without evolution & $w$ & $w$CDM & 3883.33 & 5021.43 \\\hline
\midrule
with fixed evolution & $\Omega_{M}$ & $\Lambda$CDM & 800.00 & 950.00 \\\hline
with fixed evolution & $H_0$ & $\Lambda$CDM & 2146.92 & 2146.92\\\hline
with fixed evolution & $\Omega_{M}$ and $H_0$ & $\Lambda$CDM & 190.91, 784.71 & 814.29, 2048.57 \\\hline
with fixed evolution & $w$ & $w$CDM & 3344.44 & 4328.57 \\\hline
\midrule
with $k=k(\Omega_{M})$ & $\Omega_{M}$ & $\Lambda$CDM & 785.71 & 933.33\\\hline
with $k=k(\Omega_{M})$ & $\Omega_{M}$ and $H_0$ & $\Lambda$CDM & 195.46, 823.24 & 828.57, 2142.14 \\\hline
 \toprule[1.2pt]
 \toprule[1.2pt]

\textbf{Likelihood based on $L_X$, Equations \ref{isotropic} and \ref{planeev}} & {\textbf{parameters varied}} & \textbf{Model} & $\boldsymbol{\Delta_{SN}^{GRB}}\%$& $\boldsymbol{\Delta_{SN+BAO}^{GRB}}\%$ \\ 
\midrule
without evolution & $\Omega_{M}$ & $\Lambda$CDM & 771.43 & 916.67 \\\hline
without evolution & $H_0$ & $\Lambda$CDM & 2294.62 & 2294.62\\\hline
without evolution & $\Omega_{M}$ and $H_0$ & $\Lambda$CDM & 190.91, 825.00 & 814.29, 2146.43 \\\hline
without evolution & $w$ & $w$CDM & 5722.22 & 7385.71 \\\hline
\midrule
with fixed evolution & $\Omega_{M}$ & $\Lambda$CDM & 828.57 & 983.33 \\\hline
with fixed evolution & $H_0$ & $\Lambda$CDM & 2304.62 & 2304.62\\\hline
with fixed evolution & $\Omega_{M}$ and $H_0$ & $\Lambda$CDM & 195.46, 796.77 & 828.57, 2077.86 \\\hline
with fixed evolution & $w$ & $w$CDM & 3577.78 & 4628.57 \\\hline
\midrule
with $k=k(\Omega_{M})$ & $\Omega_{M}$ & $\Lambda$CDM & 800.00 & 950.00 \\\hline
with $k=k(\Omega_{M})$ & $\Omega_{M}$ and $H_0$ & $\Lambda$CDM & 190.91, 812.06 & 9042.86, 2115.00 \\\hline

\end{tabular}
\caption{The table of comparison between the cosmological parameters derived from SNe Ia and from SNe Ia + BAO with GRBs alone using Gaussian priors without calibration on SNe (Table \ref{Table1}). The table is divided into two parts. The first part compares SNe Ia and SNe Ia + BAO with GRBs alone with no calibration (in particular, the percentage decrease of the errors on the measurements) when the computation of the cosmological parameters with GRBs has been performed using Equations \ref{isotropic} and \ref{planeev}. The second part is related to the comparison with SNe Ia and SNe Ia +BAO with GRBs alone, but using $\mu_{GRB}$, Equation \ref{equmu}. The uncertainty percentage change is computed with formulas: $ \Delta_{SN}^{GRB}\% = \frac{\Delta_{GRB}-\Delta_{SN}}{\Delta_{SN}} $ and $ \Delta_{SN+BAO}^{GRB}\% = \frac{\Delta_{GRB}-\Delta_{SN+BAO}}{\Delta_{SN+BAO}}$. Those quantities measure how much SNe Ia alone and SNe Ia+BAOs have smaller scatter than the GRBs alone. With the negative sign we indicate a percentage decrease on the uncertainty compared to the reference sample (indicated with a lower subscript), while with the positive one a percentage increase. References for comparisons are SNe Ia and SNe Ia + BAO.}
\label{Table7}

\end{table}

\begin{table}
\centering
\begin{tabular}{c|l|c|c|c|c}
 \toprule[1.2pt]
 \toprule[1.2pt]

\textbf{Calibration with SNe Ia, Equation \ref{equmu}} & \textbf{parameters varied} & \textbf{Model} & $\boldsymbol{\Delta_{SN}^{GRB}}\%$& $\boldsymbol{\Delta_{SN+BAO}^{GRB}}\%$\\ 
\midrule
without evolution &$\Omega_{M}$ & $\Lambda$CDM & 871.43& 1033.33\\\hline
without evolution & $H_0$ & $\Lambda$CDM & 2213.08 & 2213.07\\\hline
without evolution & $\Omega_{M}$ and $H_0$ & $\Lambda$CDM & 190.91, 784.11 & 814.29, 2047.14\\\hline
without evolution & $w$ & $w$CDM & 3638.89 & 4707.14 \\\hline
\midrule
with fixed evolution & $\Omega_{M}$ & $\Lambda$CDM & 871.43 & 1033.33 \\\hline
with fixed evolution & $H_0$ & $\Lambda$CDM & 2382.31 & 2382.31\\\hline
with fixed evolution & $\Omega_{M}$ and $H_0$ & $\Lambda$CDM & 172.72, 806.47 & 757.14, 2101.43 \\\hline
with fixed evolution & $w$ & $w$CDM & 3755.56
&4857.14 \\\hline
\midrule
with $k=k(\Omega_{M})$ & $\Omega_{M}$ & $\Lambda$CDM & 800.00 & 950.00 \\\hline
with $k=k(\Omega_{M})$ & $\Omega_{M}$ and $H_0$ & $\Lambda$CDM & 190.91, 821.76 & 814.29, 2138.57\\\hline
 \toprule[1.2pt]
 \toprule[1.2pt]
\textbf{Calibration with SNe Ia, Equation \ref{isotropic} and \ref{planeev}} & {\textbf{parameters varied}} & \textbf{Model} & $\boldsymbol{\Delta_{SN}^{GRB}}\%$& $\boldsymbol{\Delta_{SN+BAO}^{GRB}}\%$\\ 
\midrule
without evolution & $\Omega_{M}$ & $\Lambda$CDM & 885.71 & 1050.00\\\hline
without evolution & $H_0$ & $\Lambda$CDM & 2299.23 & 2299.23\\\hline
without evolution & $\Omega_{M}$ and $H_0$ & $\Lambda$CDM & 195.45, 856.17 & 828.57, 2222.14\\\hline
without evolution & $w$ & $w$CDM & 3772.22 & 4878.57\\\hline
\midrule
with fixed evolution & $\Omega_{M}$ & $\Lambda$CDM & 757.14 & 900\\\hline
with fixed evolution & $H_0$ & $\Lambda$CDM & 2317.69 & 2317.69\\\hline
with fixed evolution & $\Omega_{M}$ and $H_0$ & $\Lambda$CDM & 200.00, 800.59 & 842.86, 2087.14\\\hline
with fixed evolution & $w$ & $w$CDM & 3405.56 & 4407.14\\\hline
\midrule
with $k=k(\Omega_{M})$ & $\Omega_{M}$ & $\Lambda$CDM & 942.86 & 1116.67\\\hline
with $k=k(\Omega_{M})$ & $\Omega_{M}$ and $H_0$ & $\Lambda$CDM & 190.91, 803.82 & 814.29, 2095.00 \\\hline
\end{tabular}
\caption{The Table of comparison between the cosmological parameters derived from SNe Ia and from SNe Ia + BAO with GRBs alone using Gaussian priors with calibration on SNe Ia. The first part compares SNe Ia and SNe Ia +BAO with GRBs alone calibrated with SNe Ia (in particular, the percentage decrease of the uncertainties on the measurements) when the computation of the cosmological parameters with GRBs has been performed using Equations \ref{isotropic} and \ref{planeev}. The second part is related to the comparison with SNe Ia and SNe Ia +BAO with GRBs alone, but using $\mu_{GRB}$, Equation \ref{equmu}. The percentage decrease is computed with formulas: $ \Delta_{SN}^{GRB}\% = \frac{\Delta_{GRB}-\Delta_{SN}}{\Delta_{SN}} $ and $ \Delta_{SN+BAO}^{GRB}\% = \frac{\Delta_{GRB}-\Delta_{SN+BAO}}{\Delta_{SN+BAO}}$. Those quantities measure how much GRBs alone have a larger scatter than SNe alone and SNe Ia + BAO. With the negative sign we indicate a percentage decrease on the uncertainty compared to the reference sample (indicated with a lower subscript), while with the positive one a percentage increase. References for comparisons are SNe alone and SNe Ia + BAO.}
\label{Table8}
\end{table}
The results shown in Table \ref{Table6} correspond to the computation of $\Omega_{k}$ parameter with the other parameters fixed: $\Omega_{M} = 0.30$, $H_{0} = 70\; km\,s^{-1}\,Mpc^{-1}$. All the obtained values are compatible within 1.5 $\sigma$ with $\Omega_{k}=0$ corresponding to the flat universe. To compare the results we use the z-score, defined as:
\begin{linenomath*}
\begin{equation}
 z = \frac{|\Omega_{k, flat}-\Omega_{k}|}{\Delta_{\Omega_{k}}}\, =\, \frac{|\Omega_{k}|}{\Delta_{\Omega_{k}}}.
\end{equation}
\end{linenomath*}

The z-score results are presented in the last column of Table \ref{Table6}. We note an interesting trend. When we add more probes to the SNe Ia sample, the value of $\Omega_{k}$ becomes less close to $\Omega_k=0$ (the flat universe). The addition of the correction for evolution is indeed leading to a more compatible value (1.18 for z-score) with the flat universe. We present the results of the following computation in Fig. \ref{fig27}.

\begin{figure}
 \centering
 \includegraphics[width=0.95\hsize,height=0.95\textwidth,angle=0,clip]{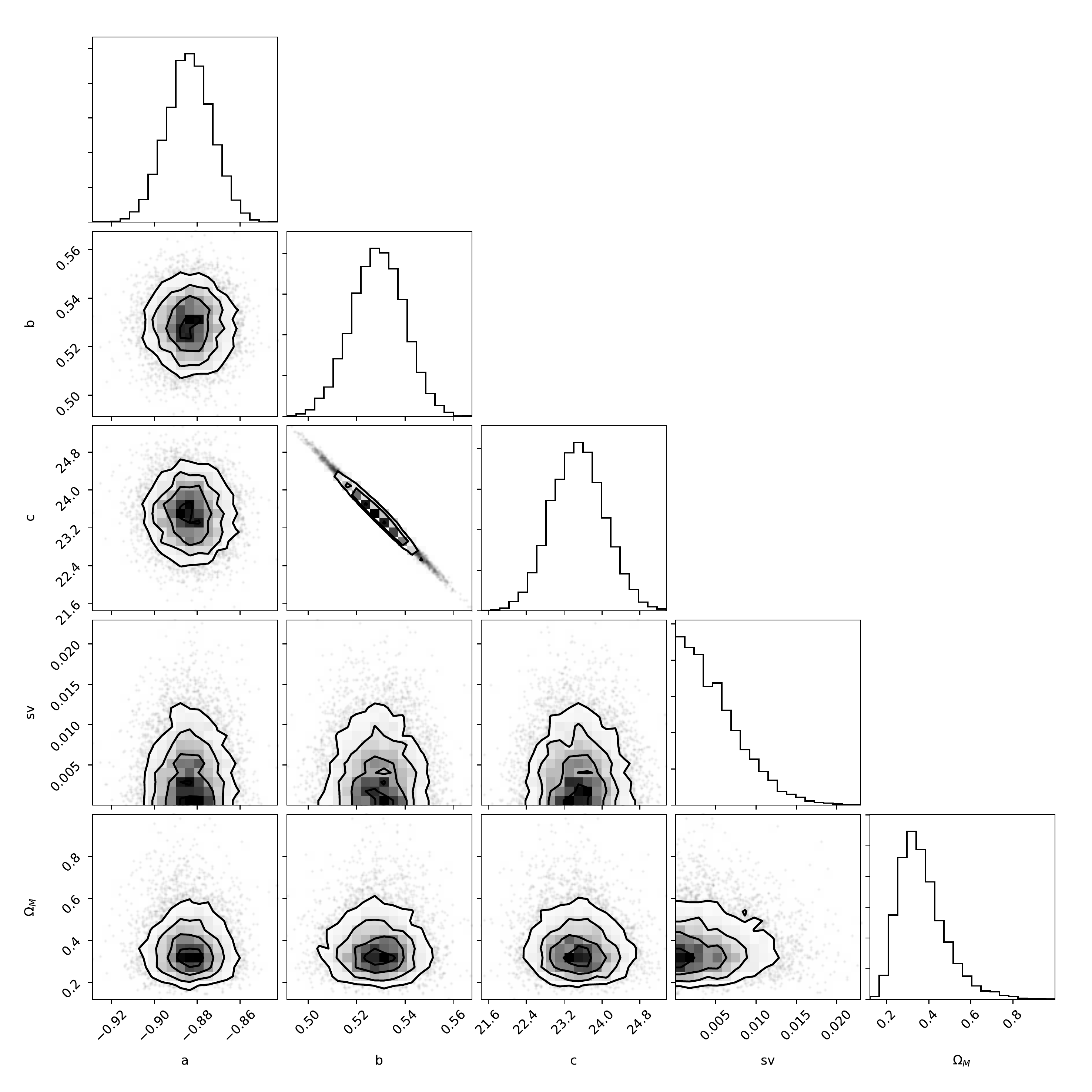}
 \caption{{ Posterior contours for $\Omega_M$ considering 800 simulated GRBs on the platinum fundamental plane.}}
 \label{GRBsimulation}
\end{figure}

\section{The comparison of GRBs alone and SNe Ia alone and SNe Ia+BAO} \label{comparison}
In this section, we evaluate the comparison between the cosmological parameters and their uncertainties of GRBs alone with the SNe Ia alone, and then SNe Ia + BAO. This analysis is done using GRBs for which we consider Gaussian priors of 3 $\sigma$ on the values of the cosmological parameters from SNe Ia taken from \cite{Scolnic}.
We divide our comparison between GRBs without and with calibration on SNe Ia, see Tables \ref{Table7} and \ref{Table8}, respectively.
Specifically, we compare the results in the i)-vi) cases as detailed in Sec. \ref{GRBs alone without calibration} without calibration, and in Sec. \ref{GRB alone with calibration} for i)-vi) cases with the calibration. In this way, it will be clear which analysis brings smaller uncertainties.
Looking specifically at the distance in terms of $\sigma$ to the SNe Ia and SNe Ia+BAO, we refer to Table \ref{Table1} for the comparison of GRBs without calibration with the Gaussian priors.

We first consider the comparison with SNe Ia alone. All the cosmological parameter results, except for the $H_0$ one when the likelihood of $\mu_{GRB}$ (see the upper part of Table \ref{Table1}) is considered in the cases of no evolution and fixed evolution, fall within 1 $\sigma$. For the cases with and without evolution the z-score is 1.024 and 1.025, respectively, when varying both $\Omega_M$ and $H_0$. 

When we look instead at the second part of Table \ref{Table1} we consider the likelihood for Equations \ref{isotropic} and \ref{planeev}, and the only case that fall outside 1 $\sigma$ is the one for which $H_0$ is computed without accounting for the evolution (z-score = 1.021). 

When comparing with SNe Ia + BAO and considering the likelihood of GRBs with $\mu_{GRB}$, all cases fall within 1 $\sigma$ with the exception again of $H_0$ without evolution (z-score = 1.09), with fixed evolution (z-score = 1.10) and with the $k = k(\Omega_M)$ evolutionary function (z-score = 1.02). When comparing with SNe Ia + BAO and considering the likelihood for Equations \ref{isotropic} and \ref{planeev}, all cases fall within 1 $\sigma$ with the exception again of $H_0$ without evolution when varying only $H_0$ (z-score = 1.024) and when varying both $\Omega_M$ and $H_0$ (z-score = 1.09), and with $k = k(\Omega_M)$ evolutionary function (z-score = 1.065).


In Table \ref{Table2}, we show results obtained for GRBs alone calibrated with SNe Ia with Gaussian priors compared with SNe Ia alone and SNe Ia + BAO.
We start the comparison with the SNe Ia only first. We see that the results which are not compatible within 1 $\sigma$ with the likelihood of $\mu_{GRB}$ (upper part of Table \ref{Table2}) are the following: $H_{0}$ varied alone without correction for evolution is compatible in $1.102$ $\sigma$, $H_{0}$ varied together with $\Omega_{M}$ without correction for evolution is compatible in $1.103$ $\sigma$, and $H_{0}$ varied together with $\Omega_{M}$ with fixed correction for evolution is compatible in $1.046$ $ \sigma$. 
Results that are not compatible in 1 $\sigma$ with SNe Ia when we consider again calibrated GRBs alone with Gaussian priors, but using instead the likelihood derived from the Fundamental plane equation (Equations \ref{planeev} and \ref{isotropic}) (see the lower part of Table \ref{Table2}), are the following: $H_{0}$ varied alone without correction for evolution is compatible in 1.137 $\sigma$, $H_{0}$ varied alone with fixed correction for evolution is compatible in 1.050 $\sigma$, and $H_{0}$ varied together with $\Omega_{M}$ with fixed correction for evolution is compatible in 1.033 $\sigma$.

When we compare SNe Ia + BAO results with the ones obtained with calibrated GRBs alone using the likelihood of $\mu_{GRB}$ (see the upper part of Table \ref{Table2}) we obtain that the following are not compatible in 1 $\sigma$: $H_{0}$ varied alone without correction for evolution is compatible in 1.105 $\sigma$, $H_{0}$ varied together with $\Omega_{M}$ without correction for evolution is compatible in 1.176 $ \sigma$, $H_{0}$ varied together with $\Omega_{M}$ with fixed correction for evolution is compatible in 1.116 $\sigma$, $H_{0}$ varied together with $\Omega_{M}$ with the correction for evolution as $k=k(\Omega_{M})$ is compatible in 1.064 $\sigma$.

When we compare SNe Ia + BAO results with the ones obtained with calibrated GRBs alone using the likelihood derived from the Fundamental plane equation (Equations \ref{planeev} and \ref{isotropic}) (see the lower part of Table \ref{Table2}) we obtain, that the following results are not compatible in 1 $\sigma$: $H_{0}$ varied alone without correction for evolution is compatible in 1.140 $\sigma$, $H_{0}$ varied together with $\Omega_{M}$ without correction for evolution is compatible in 1.004 $\sigma$, $H_{0}$ varied alone with fixed correction for evolution is compatible in 1.140 $\sigma$, $H_{0}$ varied together with $\Omega_{M}$ with fixed correction for evolution is compatible in 1.103 $\sigma$, $H_{0}$ varied together with $\Omega_{M}$ with the correction for evolution as $k=k(\Omega_{M})$ is compatible in 1.042 $\sigma$.

We now check the compatibility of the results obtained with GRBs alone with uniform priors in Tables \ref{Table3}, \ref{Table4}. In Table \ref{Table3} where GRBs have not been calibrated with SNe Ia and with the likelihood based on $\mu_{GRB}$, the only case that exceeds 1 $\sigma$ limit for both SNe Ia alone and SNe Ia + BAO is when we vary $\Omega_{M}$ only without correction for the evolution. The z-scores for SNe Ia alone and SNe Ia + BAO in this case are 1.040 and 1.022, respectively.

In Table \ref{Table4} the only case of GRBs alone with calibration with uniform priors and with the likelihood based on $\mu_{GRB}$ that exceeds the 1 $\sigma$ limit for both SNe Ia alone and SNe Ia+ BAO is the one where we compute $\Omega_{M}$ only with fixed correction for evolution. The z-scores for SNe Ia alone and SNe Ia + BAO in this case are 1.077 and 1.059, respectively. When we look instead at the lower part of Tables \ref{Table3} and \ref{Table4} considering the likelihood Equations \ref{isotropic} and \ref{planeev}, all cases for SNe Ia alone and SNe Ia + BAO fall within 1 $\sigma$.

All the percentage variations of SNe Ia alone and SNe Ia+BAO results with respect to the GRB results (SNe Ia and SNe Ia + BAO are taken as reference in the percentage variation computation, respectively) both without and with calibration on SNe Ia are shown in the last two columns of Tables \ref{Table7} and \ref{Table8}, respectively. The percentage variation of SNe Ia + BAO with GRBs alone ($\Delta_{SN+BAO}^{GRB}\%$) is in general larger than the one of SNe Ia alone with GRBs ($\Delta_{SN}^{GRB}\%$). More specifically, looking at Table \ref{Table7} where GRBs have not been calibrated, the minimum percentage increase (190.91 \%) is when comparing with SNe Ia results and in two cases 1) when we vary $\Omega_M$ and $H_0$ together with fixed evolution using distance modulus $\mu{GRB}$, equation \ref{equmu} and 2) when we vary $\Omega_M$ and $H_0$ together with evolutionary functions, $k = k(\Omega_M)$ using fundamental plane equations, \ref{isotropic} and \ref{planeev}. The maximum percentage increase (5021.43 \%) is when comparing with SNe Ia + BAO results and in the case when we vary $w$ in the case without evolution using distance moduli $\mu{GRB}$, equation \ref{equmu}.

Now, considering the comparison of calibrated GRBs with SNe Ia and SNe Ia + BAOs (see Table \ref{Table8}), the percentage variation of the uncertainties spans from $172.72\%$ when comparing with SNe Ia only in the case of fixed evolutionary parameters and varying $\Omega_M$ and $H_0$ together using distance moduli $\mu{GRB}$, equation \ref{equmu}, to almost $4878.57\%$ when comparing with SNe Ia + BAOs and varying $w$ in the case without evolution using fundamental plane equations, \ref{isotropic} and \ref{planeev}. In general, the minimum percentage difference in the uncertainty has been found for the case when both $\Omega_M$ and $H_0$ are varied together, both for the calibration and no-calibration cases, and when comparing with SNe Ia alone results. The maximum percentage difference, instead, is found for the case where $w$ is varied both for the calibration and no-calibration cases, and when comparing with SNe Ia + BAO results. We can conclude that SNe Ia alone and SNe Ia + BAO have tighter constraints on the cosmological parameters than GRBs alone, which in turn has larger uncertainties, as expected.

This conclusion does not undermine the possibility of using GRBs as standalone cosmological probes. Indeed, in the very first papers regarding SNe Ia cosmology, \citep{Riess, Perlmutter1999}, the uncertainties on the $\Omega_M$ parameter was $0.28 \pm 0.09$, which, compared to the current analysis, is $47.54\%$ larger than the same parameter computed using GRBs for the case without evolution and calibration. Further, when we compare this value to GRBs for the case in which evolutionary functions, $k = k(\Omega_M)$ are considered and with calibration on SNe Ia (the largest variance obtained in Tables \ref{Table1} and \ref{Table2} $\Omega_{M}=0.300\pm0.073$), our results have an uncertainty on $\Omega_M$ $23.29\%$ smaller. In addition, the current sample of GRBs is 100 times smaller than the current sample of SNe Ia. In the next section indeed we calculate the predictions on the uncertainties on the cosmological parameters if we simulate a sample of 800 GRBs.

\section{The future use of GRB cosmology as standalone probes}
\label{standalone}
 Regarding the use of GRBs as standalone standard candles, in \cite{Dainotti2022c}, we studied how many GRBs are required to obtain constraints on the cosmological parameters (in particular on ${\Omega_M}$) similar to the ones obtained in the literature using SNe Ia, thus giving a realistic forecast of the time required to reach these limits, taking into account present and future telescopes and missions. In order to do so, we simulated different numbers of GRBs starting from the physical features of the platinum sample, {\bf as well as an optical sample of GRBs, opening this studies also to the optical wavelength domain}. One of our simulations is shown in Fig. \ref{GRBsimulation}, where we note a closed contour for ${\Omega_M}$ for 800 GRBs. The number of 800 GRBs gives a satisfactory precision on ${\Omega_M}$. Indeed, we have also computed that the number of platinum GRBs which are needed to achieve the same precision reached by \cite{Conley} with 472 SNe Ia is 789 if no machine learning and lightcurve reconstructions are involved.
 
{\bf More specifically, in \citet{Dainotti2022c} we have studied, via simulations, the effects that the platinum and the optical samples will have on future studies when the sample size will be increased thanks to 1) new observations by SVOM, Theseus and other ground-based observations 2) the use of machine learning (Dainotti et al. in preparation), 3) the use of lightcurves reconstruction (Dainotti et al. in preparation). Going more-in depth in the analysis performed in the cited paper, the simulations were based on the best-fit fundamental planes obtained by considering different baseline samples. Indeed, not only the entire platinum and optical sets were considered but also subsets built by choosing the closest GRBs to the best-fit planes following two different methods. Then, the simulations were performed by creating GRBs observations lying on the best-fit planes, considering as starting distributions the ones provided by the observed GRBs used as a baseline. We have also run simulations in which we halved the errors on the quantities involved in the best-fit fundamental planes, simulating the increase in the precision which is expected to be achieved by future observations. The goal was to infer how many simulated GRBs are needed to reach pivotal thresholds on the precision on ${\Omega_M}$ achieved by the SNe Ia cosmological literature, namely \citet{Conley, Betoule2014}, and \citet{Scolnic} results, who found an error on $\Omega_M$ equal to $0.10$, $0.042$, and $0.022$, respectively, with the assumption that the ratio of the X-rays and optical plateau will be the same as currently. Indeed, this is a conservative estimate, since we expect to have more GRBs in the future at high redshift and more enhanced sensitivities in X-rays and optical so that more plateaus can indeed be observed. Once we have found the exact number of simulated GRBs necessary to achieve these limits, we have considered both present and future missions and observational campaigns (especially the future SVOM and Theseus missions), to give a realistic time period in which these precisions shall be reached by the standalone GRBs used as cosmological probes. This has been done both considering the current observed data, as well as taking into account machine learning and lightcurve reconstruction methods, which can be used for inferring the redshifts of GRBs observations and decreasing the errors on the observed quantities themselves, respectively. With the analysis of machine learning we will be able to double the optical plateaus, while with the lightcurve reconstruction analysis we will be able to obtain a lightcurve with almost half (47.5 $\%$) of the error carried by the lightcurves not reconstructed. When the machine learning, lightcurve reconstruction and errorbars divided by half will be applied for the optical sample, it is calculated in this analysis that the precision obtained by SNe Ia on ${\Omega_M}$ as the one in \cite{Conley} is reachable now, while the precision of \cite{Betoule2014} is reachable in 2026, and the precision of \cite{Scolnic} is reachable in 2042. All these results have been gathered in the tables 9 and 10 presented in \citet{Dainotti2022c}. This shows the increasing importance the GRBs in the next decades for cosmological applications. Thus, with these two tandem papers we have shown that GRBs have not only the credibility to be used reliably as standard candles, but they can be an important aid and complementary tool together with the SNe Ia to extend the Hubble diagram at high redshifts. For a more complete discussion, we refer to the paper itself \citep{Dainotti2022c}.}

\section{Summary, Discussion and Conclusions} \label{conclusions}
The fundamental plane relation carried a $\sigma_{int}=0.18 \pm 0.07$ when selection biases and redshift evolution is accounted for, which is the smallest intrinsic scatter in the current literature regarding GRB correlations involving the plateau emission.
In this paper, we first investigate the reliability of the 3D fundamental plane as an intrinsic correlation when we correct for selection biases and redshift evolution, then we also study its application as a cosmological tool.
To this end, we performed several tests to check the reproducibility of the parameters of the correlation, see Fig. \ref{fig1}, when we consider simulations of the evolutionary coefficients within a 1 $\sigma$ range.
We also tested the reliability of the intrinsic scatter on the fundamental plane by showing the distribution of this quantity obtained with the HyperFit online routine, which uses many fitting methods to derive the best-fit parameters of the fundamental plane both with and without the evolutionary corrections, see Fig. \ref{hist:paisim}. 
Before applying the fundamental plane relation corrected for the evolutionary effects as a cosmological tool, we investigate to which extent the parameters of the evolutionary functions, determined through statistical methods, depend on the cosmological ones, see Fig. \ref{fig:evoL}.
We find out that while the evolutionary parameters have no dependence on $H_0$, on the other end, they depend on $\Omega_M${\bf, $\Omega_{k}$ and $w$}.
This discovery opens the way to the application of the evolution to the fundamental plane both considering fixed parameters for the evolution, as well as the evolutionary function dependent on $\Omega_M$ {\bf and in the future also on $\Omega_{k}$ and $w$}.
Thus, the application of the fundamental plane corrected for selection biases and redshift evolution allows us to estimate cosmological parameters for GRBs alone for the first time considering such correction.

To use GRBs as standalone cosmological probes, we have adopted two methods: the Fundamental plane (Equations \ref{isotropic}, \ref{planeev}), and the distance moduli, $\mu_{GRB}$, variables (Equation \ref{equmu}), taking into account GRBs without calibration on SNe Ia, see \ref{GRBs alone without calibration}, as well as considering these calibrations on SNe Ia, see \ref{GRB alone with calibration}. We have obtained cosmological parameters by GRBs alone using Gaussian priors of 3 $\sigma$ based on the values of SNe Ia, see Tables \ref{Table1} and \ref{Table2}. We show the percentage change between the uncertainties of GRBs alone without and with calibration on SNe, see the seventh column of Table \ref{Table2}. We find that all the results lie within 1 $\sigma$ when comparing cases i-vi) both with and without calibration.

We, then, explored how strong the impact of the Gaussian priors is on our results and how much the parameter space of GRBs can be constrained. To do so, we obtain averaged cosmological parameter results over 100 runs performed by MCMC by GRBs alone using a uniform prior, see Sec. \ref{uniformprior}. Then, we compare the uncertainties on the results of GRBs alone using both Gaussian and the uniform priors, see the seventh column of Tables \ref{Table3} and \ref{Table4}. We find that all the results obtained by GRBs alone using uniform priors lie within 1 $\sigma$ with respect to the cosmological ones obtained by GRBs alone using Gaussian priors both with and without the calibration on SNe Ia. We used as Gaussian priors the parameters based on the results of SNe Ia.

The uncertainties on the cosmological parameter values using $\mu_{GRB}$ for both $\Omega_M$ and $H_0$ obtained by GRBs alone are higher by $328.57 \%$, $327.58 \%$ (without evolution and calibration, the upper part of Table \ref{Table3}); $328.57 \%$, $384.42 \%$ (with fixed evolution and without calibration, the upper part of Table \ref{Table3}); $326.47 \%$, $348.29 \%$ (without evolution and with calibration, the upper part of Table \ref{Table4}); $297.06 \%$, $306.57 \%$ (with fixed evolution and calibration, the upper part of Table \ref{Table4}), respectively, than the uncertainties on the results obtained using Gaussian priors of 3 $\sigma$. 

We then note that the uncertainties on the value of $\Omega_M$ using the $k(\Omega_M)$ evolutionary function, both without and with calibration on SNe Ia using $\mu{GRB}$, are higher by $335.48 \% $ and $328.57 \%$, respectively (Table \ref{Table3}) than the ones obtained using Gaussian priors of 3 $\sigma$. 

Surprisingly, the uncertainties on the values of $w$ using $\mu_{GRB}$ and uniform priors for the case of no evolution and fixed evolution and for no calibration are smaller than the values of $w$ with Gaussian priors by $19.11 \%$ and $6.45 \%$ (upper panel of Table \ref{Table3}), respectively, while for the case considering the calibration on SNe Ia they are smaller in both of without evolution ($16.42 \%$) and with evolution ($16.43 \%$) cases, the upper panel of Table \ref{Table4}. 
This difference may be due to the fact that the variation in $w$ is smaller than the variation of $H_0$ ($50<H_0<100$) and $\Omega_M$ ($0<\Omega_M<1$).

Now, considering the Fundamental plane Equations \ref{isotropic} and \ref{planeev}, the uncertainties on the values for both $\Omega_M$ and $H_0$ obtained by GRBs alone are higher by $359.01 \%$, $355.19 \%$ (without evolution and without calibration, the lower part of Table \ref{Table3}); $315.38 \%$, $356.49 \%$ (with fixed evolution and without calibration, the lower part of Table \ref{Table3}); $305.80 \%$, $337.32 \%$ (without evolution and with calibration, the lower part of Table \ref{Table4}); $366.67 \%$, $338.11 \% $ (with fixed evolution and with calibration, the lower part of Table \ref{Table4}), respectively, than the uncertainties on the correspondent results obtained using Gaussian priors of 3 $\sigma$ based on the results of SNe Ia. 

Also, when using $k(\Omega_M)$ evolutionary function without and with calibration on SNe Ia using Equations \ref{isotropic} and \ref{planeev}, uncertainties on the value of $\Omega_M$ are higher by $328.57 \% $ and $283.56 \%$, respectively, than the correspondent results obtained using Gaussian priors of 3 $\sigma$. 

Similarly to the results obtained with the $\mu_{GRB}$, when we compare the values of $w$ using the Fundamental plane Equations \ref{isotropic} and \ref{planeev} with and without evolution, respectively, in the case without calibration, the uncertainties on $w$ using uniform priors for the case without evolution and fixed evolution are smaller than the values of $w$ with Gaussian priors by $44.65 \%$ and $12.39 \%$, respectively (lower part of Table \ref{Table3}). In the case of calibration on SNe Ia, the values of $w$ are smaller in the cases of no evolution ($18.22 \%$) and with evolution ($9.67 \%$) (lower part of Table \ref{Table4}).
We have also computed the percentage change of the uncertainties between the results obtained by GRBs alone with Gaussian priors without calibration and with calibration on SNe Ia, see the last two columns of Tables \ref{Table7} and \ref{Table8}, respectively, with the results obtained by SNe Ia alone and SNe + BAOs. We have concluded from this analysis that SNe Ia alone and SNe Ia + BAO have tighter constraints on the cosmological parameters than GRBs alone.

To better understand what the advantage of using Gaussian priors is with respect to uniform priors in the comparison with other probes, we have then computed the z-scores of GRBs alone with respect to the SNe Ia alone and SNe Ia + BAO results, see the last two columns of Tables \ref{Table1}, \ref{Table2}, \ref{Table3}, \ref{Table4}. The last two columns of Tables \ref{Table1} and \ref{Table2}, present GRB results alone using Gaussian priors compared with the SNe Ia and SNe Ia + BAO with no calibration and with calibration, respectively, see Sec. \ref{comparison} for more details. The last two columns of Tables \ref{Table3} and \ref{Table4}, present the same comparison among the GRBs alone and SNe Ia and SNe Ia+BAO results, but using uniform priors.

More specifically, we first consider the comparison with SNe Ia using both the $\mu_{GRB}$ (upper panels of Tables \ref{Table1} and \ref{Table2}) and the fundamental plane equations \ref{isotropic} and \ref{planeev} (lower panels of Tables \ref{Table1} and \ref{Table2}) using Gaussian priors.
In relation to the cases of no calibration, we obtain that all cases have z-score $<1$ with the only exception of $H_0$, for which the maximum z-score $=1.025$ in the case of $\Omega_M$ and $H_0$ varied contemporaneously without considering evolution. 
In relation to the cases with calibration, we obtain that in all cases z-score $<1$ with the only exception of $H_{0}$ without evolution varied alone for which the maximum z-score is $1.137$. 

Now we consider the comparison between SNe Ia+ BAO and GRBs with both $\mu_{GRB}$ and the fundamental plane Equations \ref{isotropic} and \ref{planeev}, see Table \ref{Table1} and Table \ref{Table2} using Gaussian priors. 
In relation to the cases of no calibration, we obtain that all cases have z-score $<1$ with the only exception of $H_0$ for which the maximum z-score $=1.10$ in the case of $\Omega_M$ and $H_0$ varied contemporaneously with fixed evolution. In relation to the cases considering the calibration, we obtain that in all cases z-score$<1$, with the only exception of $H_0$ for which the maximum z-score is $1.176$ in the case of $\Omega_M$ and $H_0$ varied contemporaneously without considering evolution.

Now, we consider the comparison with SNe Ia using both the $\mu_{GRB}$ (upper panels of Tables \ref{Table3} and \ref{Table4}) and the fundamental plane equations \ref{isotropic} and \ref{planeev} (lower panels of Tables \ref{Table3} and \ref{Table4}) using uniform priors.
In relation to the cases of no calibration, we obtain that all cases have z-score $<1$ with the only exception of $\Omega_M$ for which the maximum z-score $=1.040$ in the case when $\Omega_M$ only is varied without considering evolution. 
In relation to the cases with calibration, we obtain that in all cases z-score $<1$ with the only exception of $\Omega_M$ with fixed evolution varied alone for which the maximum z-score is $1.077$. 

Now we consider the comparison between SNe Ia+ BAO and GRBs with both $\mu_{GRB}$ and the fundamental plane Equations \ref{isotropic} and \ref{planeev}, see Table \ref{Table3} and Table \ref{Table4} using uniform priors. 
In relation to the cases of no calibration, we obtain that all cases have z-score $<1$ with the only exception of $\Omega_M$ for which the maximum z-score$=1.022$ in the case when $\Omega_M$ only is varied without evolution. In relation to the cases considering the calibration, we obtain that in all cases z-score$<1$, with the only exception of $\Omega_M$ for which the maximum z-score is $1.059$ in the case when $\Omega_M$ only is varied contemporaneously with fixed evolution. 
The most notable comparison is achieved between SNe Ia, SNe Ia + BAO vs SNe Ia + BAO + GRBs using a uniform prior, for the cases of both without and with correction for evolution, see Table \ref{Table5}. For the case without correction for evolution, we reduced the scatter related to $\Omega_M$ by $14.3\%$, $\Omega_M$ and $H_0$ by $68.2\%$ and $55.9\%$ when varied together, and $w$ by $22.2 \%$, in comparison to the results obtained by SNe Ia alone. For the case of correction for evolution, we reduced the scatter related to $\Omega_M$ by $14.3\%$, $\Omega_M$ and $H_0$ by $68.2\%$ and $52.9\%$ when varied together, and $w$ by $16.7 \%$, in comparison to the results obtained by SNe Ia alone. All our results are consistent at the $68\%$ level with the $\Lambda$CDM model. The crucial points of our derivations are: 1) we have obtained cosmological parameters compatible with the $\Lambda$CDM model; 2) in all cases, except for $H_0$, we obtained a smaller intrinsic scatter by using GRBs + SNe Ia +BAO compared to SNe Ia alone, see subsection \ref{SN+BAO+GRB}.

From this analysis we have concluded that, even if uncertainties are smaller in the SNe Ia alone and SNe Ia+BAO cases, GRBs alone can still be used to verify if cosmological parameters are compatible with the ones inferred by the SNe Ia. Indeed, we may conclude that GRBs alone can be used as cosmological tools, which carry the great advantage of being observed up to $z=5$ in the Platinum sample case.

We have also computed the z-scores of GRBs alone using a uniform prior with respect to the SNe Ia alone and SNe Ia + BAO results, see last two columns of Tables \ref{Table3} and \ref{Table4}, to check the compatibility of the GRBs results using a uniform prior with the SNe Ia and SNe Ia + BAO ones, see Sec. \ref{comparison}. We obtained the maximum z-score of $1.077$ for the $\Omega_M$ parameter when it is varied with fixed evolution and when we compare SNe Ia results with the ones obtained with calibrated GRBs alone using the $\mu_{GRB}$ likelihood. Surprisingly, the average results of GRBs alone using uniform priors do not change even if we correct for evolution, both fixing the parameters as well as using a function for the evolution, see Tables \ref{Table3} and \ref{Table4}.
However, it is very likely that this result is due to the paucity of the sample size, given that when we simulate 800 GRBs starting from the ones closest to the fundamental plane \citep{Dainotti2022c} we do not recover these results.
On the other hand, the closest GRBs to the plane in the simulated data have been built with a given fiducial cosmology.
Thus, it is necessary to wait for additional data to come with future missions, such as SVOM and Theseus \citep{Wei2016,Amati2018} to cast light on this discrepancy.
Very interestingly, also other probes at high redshift, like the quasars, show a tendency to have a higher value of $\Omega_M$ \citep{Colgain2022}.
Although this discussion surely deserves much attention, it is far beyond the scope of the current paper. 

Although results related to GRB cosmology have been reached by previous studies for the prompt correlation \citep{Amati2008, Kodama, Wang, Demiansky, Luongo} and the Combo relation \citep{Izzo2015}, this is the first time cosmological constraints have been achieved considering a 3D correlation involving the plateau, and it is also the first time the evolutionary parameters have been included in the computation of cosmological ones, thus allowing a road-map for a new methodology to treat selection biases and redshift evolution contemporaneously inside the cosmological setting.

\section*{Funding} \label{Funding}
This work is supported by JSPS Grants-in-Aid for Scientific Research “KAKENHI” (A: Grant Number JP19H00693). N.F. acknowledges the support from UNAM-DGAPA-PAPIIT through grant IA102019.

\section*{Acknowledgments}

This work made use of data supplied by the UK Swift Science Data Centre at the University of Leicester. We are grateful to E. Rinaldi for helping in the discussion of the constants in the fundamental parameters and useful comments about the manuscript, to B. De Simone for the help in the analysis of the SNe Ia's likelihood, and to G. Srinivasaragavan, R. Wagner, L. Bowden, Z. Nguyen for the help in the fitting of the parameters of the LCs. Z. Kania and J. Fernandez for helping to run our cosmological computations. We are particularly grateful to Dr. Cuellar for managing the SULI program during the summer of 2020. S.N also acknowledges the support from the Pioneering Program of RIKEN for Evolution of Matter in the universe (r-EMU). We thank the RIKEN HOKUSAI BigWaterfall cluster system for the computational time. A. Lenart is greatful for the financial support from the NAOJ Division of Science.

\section{Data Availability Statement}
The data underlying this article will be shared upon reasonable request to the corresponding author.

\end{document}